%% file: main.tex
\documentclass{article}
\pdfoutput=1

\PassOptionsToPackage{nonumberlist}{glossaries}
\PassOptionsToPackage{automake,toc=false}{glossaries-extra}

\usepackage[oldstyle]{hep-paper}
\usepackage{subdepth}
\usepackage{needspace}

\bibliography{bibliography}
\graphics{figures}
\numberwithin{equation}{section}
\allowdisplaybreaks[2]

\usepackage{subfiles}
\usepackage{xr}
\newcommand{\subbib}{}
\newcommand{\subtoc}{}
\makeatletter\newcommand{\setsection}[1]{\ifx\@onlypreamble\@notprerr\else\setcounter{section}{#1-1}\def\subbib{\printbibliography}\def\subtoc{\tableofcontents\bigskip\@afterindentfalse\@afterheading}\fi}\makeatother

\usepackage{array}
\usepackage{makecell}

\include{plots}
\include{definitions}
\makeglossaries
\usepackage{glossary-list}
\usepackage{multicol}
\setglossarystyle{list}
\renewcommand{\glossarysection}[2][]{}

\preprint{CP3-21-13}

\title{Portal Effective Theories}

\subtitle{A framework for the model independent description\\of light hidden sector interactions}

\author[louvain]{Chiara Arina \footnote{\texttt{\href{mailto:chiara.arina@uclouvain.be}{chiara.arina@uclouvain.be}}}}
\author[louvain,basel]{Jan Hajer \footnote{\texttt{\href{mailto:jan.hajer@unibas.ch}{jan.hajer@unibas.ch}}}}
\author[louvain,bern]{Philipp Klose \footnote{\texttt{\href{mailto:pklose@itp.unibe.ch}{pklose@itp.unibe.ch}}}}

\affiliation[louvain]{Centre for Cosmology, Particle Physics and Phenomenology, Universit\'e catholique de Louvain, Louvain-la-Neuve B-1348, Belgium}
\affiliation[basel]{Department of Physics, Universität Basel, Klingelbergstraße 82, CH-4056 Basel, Switzerland}
\affiliation[bern]{Albert Einstein Center for Fundamental Physics, Institute for Theoretical Physics, Universität Bern, Sidlerstraße 5, CH-3012 Bern, Switzerland}

\begin{document}

\maketitle

\subfile{abstract}

\clearpage
\tableofcontents
\clearpage
\listoffigures
\listoftables
\clearpage
\section*{List of Acronyms}
\begin{multicols}{2}\setlist{noitemsep}\printglossary[type=main]\end{multicols}
\clearpage

\subfile{introduction}

\subfile{quantum-chromo-dynamics}

\subfile{portals}

\subfile{chiral-perturbation-theory}

\subfile{mesons}

\subfile{models}

\subfile{conclusion}

\appendix

\subfile{construction}

\subfile{electroweak}

\subfile{strong-scale-operators}

\subfile{expansion}

\printbibliography
\end{document}

%% file: plots.tex
\usepackage{pgfplots}
\usepackage{tikzscale}
\usetikzlibrary{external}

\usepackage{tikz-feynman}
\usepackage{environ}
\usepackage{xargs}
\newcommandx{\NewEnvironment}[5][2,3]{\expandafter\newcommandx\csname start#1\endcsname[#2][#3]{#4}\NewEnviron{#1}{\csname start#1\expandafter\endcsname\BODY #5}}
\NewEnvironment{feyn}[1][1=]{\begin{tikzpicture}\begin{feynman}[#1]}{\end{feynman}\end{tikzpicture}}
\usepackage{tikz-feynman}
\usetikzlibrary{positioning}
\usetikzlibrary{arrows}
\usetikzlibrary{fit}
\usetikzlibrary{shapes.misc}
\tikzset{cross/.style={cross out, draw, minimum size=2*(#1-\pgflinewidth), inner sep=2pt, outer sep=0pt,rotate=45}}
\tikzfeynmanset{with arrow/.style={decoration={markings,mark=at position 0.5 with {
\arrow[xshift=1pt]{>}
}},postaction=decorate}}
\newcommand{\width}{10em}
\newcommand{\height}{10ex}

%% file: definitions.tex
\acronym{QFT}{quantum field theory}[quantum field theories]
\acronym{QCD}{quantum chromodynamics}
\acronym{QED}{quantum electrodynamics}
\acronym{GD}{gluon dynamics}
\acronym{SM}{Standard Model}
\acronym{RGE}{renormalization group equation}
\acronym{VEV}{vacuum expectation value}
\acronym{CKM}{Cabibbo-\allowbreak Kobayashi-\allowbreak Maskawa}
\acronym{PMNS}{Pontecorvo–\allowbreak Maki–\allowbreak Nakagawa–\allowbreak Sakata}
\acronym[\chi PT]{cPT}{chiral perturbation theory}[chiral perturbation theories]
\acronym{NGB}{Nambu-\allowbreak Goldstone boson}
\acronym{PNGB}{pseudo Nambu-\allowbreak Goldstone boson}
\acronym{WZW}{Wess-\allowbreak Zumino-\allowbreak Witten}
\acronym{MC}{Maurer-\allowbreak Cartan}
\acronym{GM}{Gell-\allowbreak Mann}
\acronym{CS}{Chern-\allowbreak Simons}
\acronym{YM}{Yang-\allowbreak Mills}
\acronym{HNL}{heavy neutral lepton}
\acronym{ALP}{axion-like particle}
\acronym{LLP}{long-lived particle}
\acronym{EOM}{equation of motion}
\acronym{PI}{partial integration}
\acronym[1PI]{onePI}{one particle irreducible}
\acronym{BSM}{beyond the Standard Model}
\longacronym{WIMP}{weakly interacting massive particle}
\acronym{DM}{dark matter}
\acronym{UV}{ultraviolet}
\acronym{IR}{infrared}
\acronym{LEC}{low energy coefficient}
\acronym{LER}{low energy realisation}
\acronym{NP}{new physics}
\acronym{EM}{electromagnetic}
\acronym{EW}{electroweak}
\acronym{EWSB}{electroweak symmetry breaking}
\acronym[\overline{MS}]{MSb}{modified minimal subtraction}
\acronym{NDA}{naive dimensional analysis}
\acronym{DOF}{degree of freedom}[degrees of freedom]
\acronym{LO}{leading order}
\acronym{NLO}{next-to-leading order}
\acronym{NNLO}{next-to-next-to-leading order}
\acronym{LHC}{Large Hadron Collider}
\shortacronym{CMS}{compact muon solenoid}
\shortacronym{ATLAS}{a toroidal LHC apparatus}
\shortacronym{LHCb}{LHC beauty}
\shortacronym{NA}*{North Area experiment \textnumero~}
\shortacronym{KOTO}{K0 at Tokai}
\shortacronym{SHiP}{Search for Hidden Particles}
\shortacronym{MATHUSLA}{Massive Timing Hodoscope for Ultra-Stable neutraL pArticles}
\shortacronym{FASER}{ForwArd Search ExpeRiment}
\shortacronym[CODEX-b]{CODEXb}{compact detector for exotics at LHCb}
\shortacronym{CP}{charge-parity}
\acronym{PET}{\emph{portal effective theory}}[\emph{portal effective theories}]
\acronym{EFT}{effective field theory}[effective field theories]
\acronym{SMEFT}{Standard Model effective field theory}[Standard Model effective field theories]
\acronym{HEFT}{Higgs effective field theory}[Higgs effective field theories]
\acronym{LEFT}{light effective field theory}[light effective field theories]
\acronym{SCET}{soft-collinear effective theory}[soft-collinear effective theories]
\acronym{HQET}{heavy quark effective theory}[heavy quark effective theories]
\acronym{NRQCD}{non-relativistic quantum chromodynamics}
\acronym{NRQED}{non-relativistic quantum electrodynamics}
\acronym[2HDM]{THDM}{two Higgs-doublet model}
\acronym{SUSY}{supersymmetry}
\acronym{SUGRA}{supergravity}
\acronym{MSSM}{minimal supersymemtric Standard Model}
\acronym{NMSSM}{next-to-minimal supersymemtric Standard Model}
\longacronym{CC}{charged current}
\longacronym{NC}{neutral current}
\longacronym{LE}{low energy}[low energies]
\longacronym{HE}{high energy}[high energies]
\acronym{BR}{branching ratio}

\acronyms{Omit}

\acronym{SNF}{Schweizerischer Nationalfonds zur Förderung der wissenschaftlichen Forschung}
\acronym[exp]{ex}{experimental}
\acronym{lat}{lattice}
\acronym{UCLouvain}{Université catholique de Louvain}
\acronym{FSR}{Fonds Spéciaux de Recherche}

\crefname{type}{type}{types}
\crefalias{inlinelisti}{type}
\crefalias{enumi}{type}

\crefformat{footnote}{#2\footnotemark[#1]#3}

\eqcrefname{lag}{Lagrangian}
\eqcrefname{set}{stress-energy tensor}
\eqcrefname{cur}{current}
\eqcrefname{sym}{symmetry}
\eqcrefname{bilinear}{quark bilinear}

\newcommand{\makediff}[2]{\expandafter\NewDocumentCommand\csname#1\endcsname{ogd()}{\IfNoValueTF{##2}{\IfNoValueTF{##3}{#2\IfNoValueTF{##1}{}{^{##1}}}{\mathinner{#2\IfNoValueTF{##1}{}{^{##1}}\argopen(##3\argclose)}}}{\mathinner{#2\IfNoValueTF{##1}{}{^{##1}}##2} \IfNoValueTF{##3}{}{(##3)}}}}
\makediff{mD}{\mathcal D}
\DeclareMathOperator{\T}{T}
\DeclareMathOperator{\SU}{SU}
\mathdef{\U}{\operatorname{U}}
\mathdef{\P}{\operatorname{P}}

\newcommand{\overleft}[1]{\oset{\leftarrow}{#1}}
\newcommand{\overright}[1]{\oset{\rightarrow}{#1}}

\newcommand{\qn}[1]{\mathsf{#1}}
\newcommand{\fl}{{\qn{f}}}
\newcommand{\nfl}{n_\qn{f}}
\newcommand{\Nc}{n_c}

\newcommand{\param}{\mathord{\color{black!33}\circ}}%
\newcommand{\trf}[1]{\ev{#1}_{\!\fl}}
\newcommand{\trproj}[3]{\ev{#1}_{\!#2}^{\!#3}}
\newcommand{\trsd}[1]{\trproj{#1}{\strange}{\down}}
\newcommand{\trds}[1]{\trproj{#1}{\down}{\strange}}
\newcommand{\trc}[1]{\ev{#1}_{\!c}}
\newcommand{\trF}[1]{\left\langle\mkern-\medmuskip\middle\langle#1\middle\rangle\mkern-\medmuskip\right\rangle_{\!\fl}}

\newcommand{\particle}[1]{\text{#1}}
\newcommand{\strange}{\particle{s}}
\newcommand{\down}{\particle{d}}
\newcommand{\up}{\particle{u}}
\newcommand{\charm}{\particle{c}}
\newcommand{\topq}{\particle{t}}
\newcommand{\electron}{\particle{e}}
\newcommand{\muon}{\particle{\mu}}
\newcommand{\neutrino}{\particle{\nu}}
\newcommand{\tauon}{\particle{\tau}}
\newcommand{\photon}{\particle{A}}

\newcommand{\higgs}{\frac{h}{\sqrt 2}}
\newcommand{\etasinglet}{\frac{\eta_1}{\sqrt 3}}
\newcommand{\etaoctet}{\frac{\eta_8}{\sqrt 6}}
\newcommand{\etaprime}{\frac{\eta^\prime}{\sqrt 3}}
\newcommand{\etaparticle}{\frac{\eta}{\sqrt 6}}
\newcommand{\pion}{\frac{\pi_8}{\sqrt 2}}
\newcommand{\neutpion}{\frac{\pi^0}{\sqrt 2}}
\newcommand{\Klong}{\frac{K^0_L}{\sqrt2}}
\newcommand{\Kshort}{\frac{K^0_S}{\sqrt2}}

\newcommand{\light}{{\up\down}}
\newcommand{\octet}[1]{\mathbf{\mathsf{#1}}}
\newcommand{\nonet}[1]{\bm #1}
\newcommand{\flavour}[1]{{\bm{#1}}}
\newcommand{\tensor}[1]{\mathfrak{#1}}

\newcommand{\epsilonEW}{\epsilon_{\EW}}
\newcommand{\epsilonUV}{\epsilon_{\UV}}
\newcommand{\epsilonSM}{\epsilon_{\SM}}

\newcommand{\chromo}{\gamma}
\newcommand{\Chromo}{\Gamma}
\newcommand{\gluons}{\Upsilon}
\newcommand{\fermi}{h}
\newcommand{\Fermi}{H}
\newcommand{\scalar}{s}
\newcommand{\sym}{+}
\newcommand{\asym}{-}
\newcommand{\symasym}{{\pm}}

\newcommand{\strong}{s}
\newcommand{\betacoeff}{\beta_0}
\newcommand{\proj}{\lambda}
\newcommand{\charge}{q}
\newcommand{\rep}[1]{{#1}}
\newcommand{\qu}{\ensuremath{\text{\fontencoding{OT1}\fontfamily{cmfr}\selectfont\textit q}}}

\newcommand{\Left}{\widehat}
\newcommand{\octetm}{b}
\newcommand{\coupl}{\kappa}
\newcommand{\vacuum}[1]{#1^\prime}
\newcommand{\fp}{f_0}
\newcommand{\B}{b_0}
\newcommand{\mth}{m_0}
\newcommand{\m}{m}
\newcommand{\M}{M}
\renewcommand{\u}{g}
\renewcommand{\L}{u}
\newcommand{\VL}{\nonet V}
\newcommand{\VR}{\overline{\nonet V}}
\newcommand{\kfactor}{k}

\newcommand{\boson}[1]{\mathop{}#1\mathop{}}
\newcommand{\lhf}{\xi}
\newcommand{\lhfc}{\Xi}

\usepackage{tocloft}

\AtBeginEnvironment{tabular}{\tlstyle}

\let\oldfigure\figure
\def\figure{\oldfigure\small}
\let\oldtable\table
\def\table{\oldtable\small}

%% file: abstract.tex
\begin{abstract}
We present a framework for the construction of \PETs that couple effective field theories of the \SM to light hidden messenger fields.
Using this framework we construct electroweak and strong scale \PETs that couple the \SM to messengers carrying spin zero, one half, or one.
The electroweak scale \PETs encompass all portal operators up to dimension five, while the strong scale \PETs additionally contain all portal operators of dimension six and seven that contribute at leading order to quark-flavour violating transitions.
Using the strong scale \PETs, we define a set of portal currents that couple hidden sectors to $\QCD$, and construct portal \cPTs that relate these currents to the light pseudoscalar mesons.
We estimate the coefficients of the portal \cPT Lagrangian that are not fixed by \SM observations using non-perturbative matching techniques and give a complete list of the resulting one- and two-meson portal interactions.
From those, we compute transition amplitudes for three golden channels that are used in hidden sector searches at fixed target experiments:
\begin{inlinelist}
\item charged kaon decay into a charged pion and a spin zero messenger
\item charged kaon decay into a charged lepton and a spin one half messenger
\item neutral pion decay into a photon and a spin one messenger
\end{inlinelist}
Finally, we compare these amplitudes to specific expressions for models featuring
\begin{commalist}
\item light scalar particles
\item axion-like particles
\item heavy neutral leptons
\item dark photons.
\end{commalist}
\end{abstract}

\glsunset{PET}\glsunset{PET}
\glsunset{SM}\glsunset{SM}
\glsunset{cPT}\glsunset{cPT}

%% file: introduction.tex
\subtoc

\section{Introduction}

The search for physics \BSM is one of the most pursued research avenues in modern high-energy physics.
Models of \BSM physics can be constructed from the top down by postulating a novel set of first principles, as \eg in grand unified \cite{Pati:1974yy, Georgi:1974yf, Georgi:1974sy} or supersymmetric \cite{Golfand:1971iw, Gervais:1971ji, Neveu:1971rx, Ramond:1971gb} theories, or from the bottom up by augmenting the \SM with new particles and interactions that address specific hints for \BSM physics, such as \eg
\HNLs generating neutrino masses \cite{Schechter:1981cv, Schechter:1980gr, Yanagida:1980xy, Mohapatra:1979ia, GellMann:1980vs, Minkowski:1977sc, Shaposhnikov:2006nn, Asaka:2005pn},
axions addressing the strong \CP problem \cite{Wilczek:1977pj, Weinberg:1977ma, Peccei:1977ur, Peccei:1977hh, Alves:2017avw} or
little Higgs models addressing the hierarchy problem \cite{ArkaniHamed:2001nc, Chang:2003zn, Kaplan:2003uc, Chen:2017dwb}.
The new particles predicted in both approaches are constrained to be relatively heavy or rather weakly coupled in order to be consistent with bounds from past and current collider and intensity experiments, respectively.

\EFTs describe physics at a specific energy scale, with the impact of physics at other scales being contained within the free parameters of the theory \cite{Weinberg:1978kz, Leutwyler:1993iq}.
They can be used to describe the impact of \NP at energy scales well above the characteristic energy scale of the \EFT while remaining agnostic about the specific realisation of \NP in nature.
\EFTs are constructed by identifying the relevant fields and symmetries that determine the physics one intends to characterise.
The theory then contains all available operators constructed from these fields.
In particular, \EFTs typically contain an infinite tower of higher dimensional, non-renormaliseable operators that capture the impact of the heavy \DOFs.
At the \EW scale, there are two \EFTs that encompass the entire \SM and that are commonly used to include heavy \NP \cite{Brivio:2017vri}:
\SMEFT, which is composed of all the \SM fields including the Higgs doublet and restricted by the \SM gauge group \cite{Buchmuller:1985jz, Grzadkowski:2010es, Ellis:2018gqa, AguilarSaavedra:2018nen, Slade:2019bjo}, and
\HEFT, which lifts the restriction on the Higgs boson to be part of a doublet \cite{Barbieri:2007bh, Burgess:1999ha, Grinstein:2007iv, Feruglio:1992wf}.
\EFTs at lower energies, which encompass only a part of the \SM, account for the impact of the heavy \SM \DOFs via their higher dimensional operators.
Examples include
\LEFT, which describes the interactions of the \SM after integrating out its heavy particles \cite{Fermi:1934sk, Jenkins:2013zja, Jenkins:2013wua, Alonso:2013hga},
\cPT, which encompasses the interactions of light hadrons \cite{Weinberg:1966fm, Weinberg:1968de, Cronin:1967jq, Schwinger:1967tc, Wess:1967jq, Dashen:1969ez, Gasiorowicz:1969kn},
\HQET \cite{Isgur:1989vq, Isgur:1989ed, Shifman:1987rj, Grinstein:1990mj, Georgi:1990um, Falk:1990yz} and \NRQCD \cite{Caswell:1985ui, Bodwin:1994jh}, which capture the interactions of the hadrons containing heavy quarks, and
\SCET, which describes physics of highly energetic particles, appearing for instance in jets \cite{Bauer:2000ew, Bauer:2000yr, Bauer:2001ct, Bauer:2003pi, Beneke:2004in, Bosch:2004th, Bauer:2008jx}.

These \EFTs do not include the large class of \SM extensions that feature new feebly interacting particles, such as \ALPs, light scalar particles, dilatons, \HNLs, and novel gauge bosons, with masses at or below the energy scale of the \EFT.
In this paper, we address this gap by developing a framework for constructing \PETs, which couple \SM \DOFs to light hidden messenger particles.
To satisfy all existing experimental bounds, see \eg \cite{Alekhin:2015byh, Beacham:2019nyx, Agrawal:2021dbo}, the latter can couple only very weakly to the \SM fields.
Besides the high intensity data sets of \CMS \cite{Sirunyan:2018mgs, Sirunyan:2019sgo, Mukherjee:2019anz}, \ATLAS \cite{Aad:2020cje} and \LHCb \cite{Aaij:2020iew, Aaij:2017rft, Aaij:2018mea, Aaij:2019bvg, Aaij:2020ikh, Aaij:2020ovh, Borsato:2021aum}, and the high luminosity runs of the \LHC \cite{CidVidal:2018eel}, which are optimised for such searches, these particles could be produced in large quantities via meson decays in fixed target experiments such as \NA62 \cite{NA62:2017rwk, CortinaGil:2018fkc, CortinaGil:2020fcx,CortinaGil:2017mqf, NA62:2020mcv,CortinaGil:2019nuo,Drewes:2018gkc}, \KOTO \cite{Ahn:2018mvc}, SeaQuest \cite{Aidala:2017ofy}, or \SHiP \cite{Alekhin:2015byh}.
If the messenger particles are unstable and decay predominantly into \SM particles via the suppressed portal interactions, they are long-lived and can also be searched for in dedicated \LLP experiments \cite{Alimena:2019zri}, such as \MATHUSLA \cite{Curtin:2018mvb}, \FASER \cite{Ariga:2018uku} and \CODEXb \cite{Gligorov:2017nwh}.

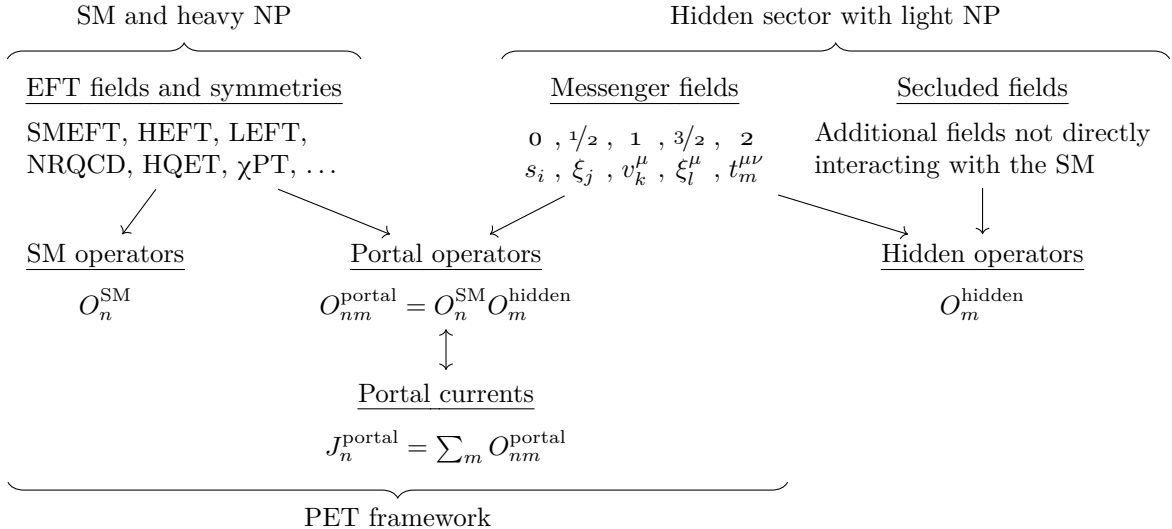
\begin{figure}
\tikzsetnextfilename{PET-framework}\input{PET-framework.pgf}
\caption[The $\PET$ framework]{
The \PET framework extends a given \EFT of the \SM by combining its operators with portal operators that couple the \SM \DOFs to messenger fields that are dynamic at the relevant energy scale.
The portal operators $O_{nm}^\text{portal}$ can be collected into a set of portal currents $J_n^\text{portal}$ that allow for a spurion analysis and for \eg model-independent bounds.
Here $n$ and $m$ symbolically label \SM and hidden sector operators, respectively, so that $\mathcal L_\text{portal} = O_n^{\SM} J_n^\text{portal}$.
The \PET framework is independent from additional secluded particles that do not interact directly with the \SM fields.
} \label{fig:PET framework}
\end{figure}

By extending the existing \EFTs of the \SM, the \PETs encompass all portal operators that conform with the symmetries of the relevant \EFT, and can be used to constrain the coupling of the \SM to light hidden sectors while remaining largely agnostic about the internal structure of the hidden sector.
The hidden sector can in general contain an arbitrary number of secluded fields that do not couple directly to the \SM but interact among themselves and with the messenger fields.
This setup, which is illustrated in \cref{fig:PET framework}, describes both heavy and light new particles, since heavy particles with masses well above the characteristic energy of the \EFT are captured by infinite towers of \SM, portal, and hidden operators.
Our comprehensive approach builds on previous works, in which \SM particles are coupled to specific hidden particles, see \eg \cite{Duch:2014xda, DeSimone:2016fbz, Brivio:2017ije, Dekens:2019ept, Contino:2020tix}, and is closely related to \EFTs describing non-relativistic \DM interactions \cite{Fitzpatrick:2012ix, Hoferichter:2015ipa, Hoferichter:2016nvd, Hoferichter:2018acd, Cirigliano:2012pq, DelNobile:2013sia, Bishara:2016hek, Bishara:2017pfq, Criado:2021trs}. 

To demonstrate the power of the \PET framework, we construct a number of \PETs and highlight the connections between them.
Extending \SMEFT, we first construct \EW scale \PETs that couple the \SM to a light messenger field of spin 0, \textfrac12, or 1 and encompass all available non-redundant portal operators up to dimension five.
To connect these portal \SMEFTs to \PETs that describe the interactions of hidden fields at the strong scale, where many high intensity experiments search for feebly interacting particles,
we subsequently construct portal \LEFTs, which additionally encompass quark-flavour violating portal operators up to dimension seven.
These additional operators capture \LO contributions to hidden sector induced, strangeness-violating kaon decays.
Since the perturbative description of \QCD breaks down at \LEs, it is not possible to compute transition amplitudes for meson decays using standard perturbative methods in \QCD, however, \cPT provides an appropriate framework.
In order to supply a complete toolkit for the computation of hidden sector induced meson transitions, we construct portal \cPTs, which couple the light pseudoscalar mesons to a messenger of spin 0, \textfrac12, or 1, and match them to the corresponding portal \LEFTs.
For this matching, we adapt to our framework a number of well-established non-perturbative techniques used to match \cPT to \QCD in the \SM, as in \eg \cite{Gasser:1984gg, Leutwyler:1989xj, Pich:1990mw, Pich:1995qp, Kaiser:2000gs, Pallante:2001he, Cirigliano:2003gt, Gerard:2005yk}.

Throughout this work, we encode the coupling to hidden sectors in terms of external currents, as depicted in \cref{fig:PET framework}.
We use these currents to derive the coupling of \cPT with the messenger particles via a spurion analysis, where we require that the \cPT path integral changes like the \QCD path integral under transformations of the external currents.
Besides simplifying the spurion analysis, the external current approach has two advantages:
First, it clarifies the discussion, as most of our work is independent of the specific content of the external currents.
Second, this formulation makes it easier to generalise our framework.
For instance, inclusive amplitudes do not encode any detailed information about the individual hidden sector particles.
Therefore, we expect that, when computing such amplitudes, it is possible to integrate out the hidden fields entirely.
In the resulting effective theory, the impact of hidden sectors would be encoded via an infinite tower of external current interactions, where the currents are space-time dependent functions of hidden sector parameters rather than being functionals of the hidden fields.
These currents can then serve as a source or drain of energy, angular momentum, or other conserved quantum numbers, which, after matching the effective theory to the full theory, should exactly mimic the impact of the hidden sector fields on inclusive scattering amplitudes.
\footnote{
This approach is inspired by a technique from non-equilibrium quantum field theory, where the impact of an external bath is captured by the von Neumann density matrix in the path integral, see \eg \cite[Section 3.2 in][]{Berges:2004yj}, and this density matrix can be recast as an infinite tower of external current interactions.
}
This means that the currents could be used to efficiently parameterise and therefore constrain the coupling to arbitrary hidden sectors in an extremely model independent way.

\subsection*{Organisation and novel contributions}

\begin{figure}
\tikzsetnextfilename{flowchart}\input{flowchart.pgf}
\caption[Procedure to derive the \PET Lagrangian coupling mesons to messengers]{
Overview of our procedure to derive the \PET Lagrangian that couples the light mesons to messengers of spin 0, \textfrac12, or 1.
In the final step, we apply the Feynman rules extracted from the portal Lagrangian to compute universal amplitudes for the three golden processes.
} \label{fig:flowchart}
\end{figure}
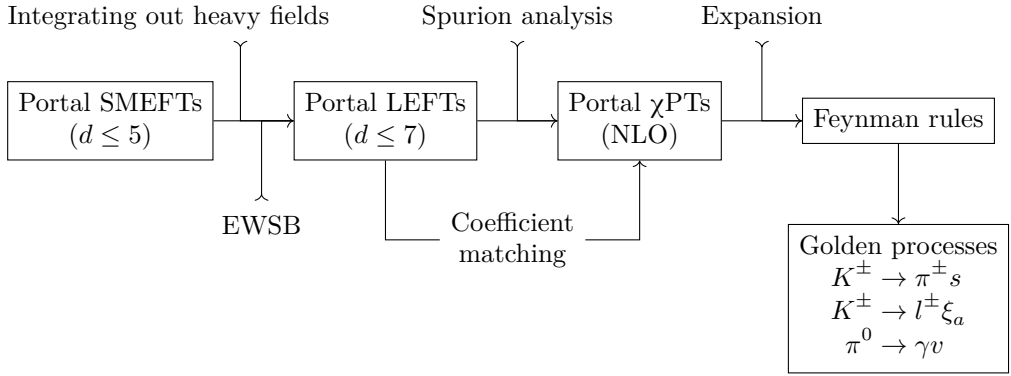

\Cref{fig:flowchart} visualises the structure of this paper, which is organised as follows.
In \cref{sec:quantum chromo dynamics}, we summarise aspects of \QCD at low energies that are pertinent to the discussion in the remainder of this work.
In particular, we focus on the axial anomaly, the large $\Nc$ expansion, and the impact of higher dimensional operators that result from integrating out the heavy \SM particles.
We use the readers familiarity with the topic to introduce a notation that lends itself to the transition from \QCD to \cPT.
In \cref{sec:portals}, we construct portal \SMEFTs and \LEFTs that couple the \SM to a single messenger field.
Furthermore, we construct the corresponding hidden currents and specify the interaction Lagrangian that couples the currents to the \SM fields.
In \cref{sec:chiral perturbation theory}, we use the external current approach to derive the coupling of \cPT to hidden sectors captured by the portal \LEFTs.
In \cref{sec:light mesons}, we list the \cPT portal interactions in terms of mesons and hidden fields, starting from the \cPT Lagrangian derived in \cref{sec:chiral perturbation theory}.
In \cref{sec:models}, we use the interactions derived in the previous section to compute smoking gun processes for meson decays into hidden fields, which are relevant for intensity experiments such as \NA62 and \KOTO.
We additionally connect our results to characteristic \BSM models, such as \ALPs, scalar portal models, \HNLs and dark photons.
\Cref{sec:conclusions} concludes the paper with a discussion of the results and an outlook to prospective future work.
Further details about the derivation of the main results of this paper are given in \cref{sec:redundant,sec:ewpetsapp,sec:gev_portals,sec:expansions}.

In the following list we summarise the main new results that we present throughout this paper.

\begin{description}[style=nextline]
\setlist[itemize]{topsep=0pt,before=\leavevmode\vspace{-3.5ex},leftmargin=0pt}

\item[\Cref{sec:quantum chromo dynamics}]\needspace{6ex}

\begin{itemize}
\item
We generalise the standard large $\Nc$ counting formula to also capture diagrams that contain higher-dimensional four-quark operators generated by virtual $W$-boson exchanges at the \EW scale.
\item
We construct an alternative basis for the four-quark operators that contains four independent octet operators and one 27-plet operator.
Compared to the standard basis, \cf \eqref{eq:four-quark operators}, which consists of six operators, this basis simplifies the matching between \cPT and \QCD.

\end{itemize}

\item[\Cref{sec:portals}]\needspace{6ex}

\begin{itemize}
\item
We develop the \PET framework and define the procedure for constructing general \PETs.
\item
We construct \EW scale \PETs that couple \SMEFT to a light messenger particle with spin 0, \textfrac12, or 1, which is neutral under the unbroken \SM gauge group $G_{\SM} = \SU(3)_c \times \SU(2)_L \times \U(1)_Y$.
These \PETs encompass all available portal operators up to dimension five, and are embedded into 21 portal currents.
We further derive the shape of the \EW portal Lagrangian after \EWSB in the unitary gauge, which is sufficient for computations at tree level.
\item
We construct strong scale \PETs that couple \LEFT to a light messenger particle with spin 0, \textfrac12, or 1 that is neutral with respect to the broken \SM gauge group $G_\text{sm} = \SU(3)_c \times \U(1)_{\EM}$.
These \PETs contain all available portal operators up to dimension five and additionally encompass all \LO quark-flavour violating portal operators up to dimension seven.
\item
We embed the portal \LEFTs into ten external portal currents $J \in \{S_\omega$, $\Theta$, $\nonet \M$, $\nonet L^\mu$, $\nonet R^\mu$, $\octet T^{\mu\nu}$, $\octet \Chromo$, $\tensor \Fermi_\scalar$, $\tensor \Fermi_l$, $\tensor \Fermi_r\}$
that parameterise the coupling of the messenger particles to \QCD.
\end{itemize}

\item[\Cref{sec:chiral perturbation theory}]\needspace{6ex}

\begin{itemize}
\item
We derive the coupling of \cPT to the scalar current $S_\omega$.
The \SM does not contain an external current that couples to \QCD like $S_\omega$, and hence this term is usually not included in \SM \cPT.
Our result generalises the \cPT Lagrangian in \cite{Leutwyler:1989xj}, where the authors derived the coupling of a light Higgs boson to \cPT,
which interacts with \QCD via an operator $h G_{\mu\nu} G^{\mu\nu}$ that is encompassed in $S_\omega$.
\item
Using the spurion technique, we derive the coupling of \cPT to the four external currents $\octet \Chromo$, $\tensor \Fermi_l$, $\tensor \Fermi_r$ and $\tensor \Fermi_\scalar$.
The coupling of \cPT to \emph{constant} currents $\octet \Chromo$ and $\tensor \Fermi_x$ is well-understood \cite{Cronin:1967jq, Kambor:1989tz, Ecker:1992de, Donoghue:1992dd, Buras:2018evv}.
Here, we generalise the description to account for \emph{spacetime dependent} external currents.
\item
The \EW sector of the portal \cPT Lagrangian contains 27 coefficients $\coupl$, 21 of which are not fixed completely by \SM observations.
We estimate the two coefficients $\coupl_\Chromo$ and $\coupl_\Chromo^{\M} + \coupl_\Chromo^{\M\prime}$, that measure the strength of the chromomagnetic current interactions,
the seven coefficients $\coupl_\omega^x$, which measure the strength of the scalar current interactions, and the 13 coefficients $\coupl_y^x$,
that measure the coupling of \cPT to the hidden currents $\octet \Fermi_x$ and $\tensor \Fermi_x$.
The authors of \cite{Leutwyler:1989xj} have estimated four out of the seven coefficients $\coupl_\omega^x$.
Here, we adapt their strategy to also estimate the remaining three coefficients.
Similarly, the coefficients $\coupl_y^x$ are known in the large $\Nc$ limit \cite{Pich:1990mw, Pich:1995qp, Pallante:2001he, Cirigliano:2003gt, Gerard:2005yk}.
Here, we adapt the strategies used in \cite{Leutwyler:1989xj, Pich:1990mw, Pich:1995qp, Gerard:2005yk} in order to obtain improved estimates for the $\kappa_y^x$ that incorporate corrections beyond the large $\Nc$ limit.
\end{itemize}

\item[\Cref{sec:light mesons}]\needspace{6ex}

\begin{itemize}
\item
We expand the \cPT Lagrangian in the meson matrix $\nonet \Phi$, and present a complete list of one- and two-meson interactions that couple \cPT to generic hidden sectors.
\end{itemize}

\item[\Cref{sec:models}]\needspace{6ex}

\begin{itemize}
\item
We compute the most general \LO transition amplitudes for three smoking-gun processes with hidden particles, relevant for searches at fixed target experiments such as \NA62 and \KOTO.
Specifically, we consider the following meson decays: $K^\pm \to \pi^\pm s_i$, $K^\pm \to \ell^\pm \lhf_a$, and $\pi^0 \to \gamma v$, where $s_i$, $\lhf$, and $v^\mu$ are a spin 0, spin \textfrac12, and spin 1 hidden field, respectively.
\end{itemize}

\end{description}

\subbib

\glsunset{SMEFT}
\glsunset{HEFT}
\glsunset{LEFT}
\glsunset{SCET}
\glsunset{HQET}
\glsunset{NRQCD}
\glsunset{NRQED}

%% file: PET-framework.pgf
\begin{tikzpicture}[
 node distance=8ex and 6em,
 align=left
]
\node (EFTs) [label={[name=EFTs label] \underline{$\EFT$ fields and symmetries}}] {$\SMEFT$, $\HEFT$, $\LEFT$,\\$\NRQCD$, $\HQET$, $\cPT$, \dots};
\node (messengers) [label={[name=messengers label] \underline{Messenger fields}}, right = of EFTs.north east, anchor = north west] {$\arraycolsep=1.5pt\begin{array}{*9c}
\text 0 & , & \textfrac12 & , & \text 1 & , & \textfrac32 & , & \text 2 \\
s_i & , & \xi_j & , & v^\mu_k & , & \xi^\mu_l & , & t^{\mu\nu}_m
\end{array}$};
\node (secluded) [right = of messengers.north, anchor = north west, label={[name=secluded label] \underline{Secluded fields}}] {Additional fields not directly\\interacting with the $\SM$};
\node (hidden) [fit=(messengers) (messengers label) (secluded) (secluded label)] {};

\node (temp1) [below = of secluded.south west, anchor = north west] {};
\node (temp2) at (temp1 -| messengers label.east) {};
\node (temp4) at (temp1 -| messengers label.west) {};
\node (temp3) at (temp1 -| EFTs.east) {};

\node (Hid) [below = of secluded.south, anchor = north, label={[name=Hid label] \underline{Hidden operators}}] {$O_m^\text{hidden}$};
\node (operators) at (Hid.north -| EFTs.west) [anchor=north west, align=center, label={[name=operators label] \underline{$\SM$ operators}}] {$O_n^{\SM}$\\\phantom{\underline{$\SM$ operators}}};
\node (portals) at ($(temp4.north)!.5!(temp3.north)$) [anchor=north, label={[name=portals label] \underline{Portal operators}}] {$O_{nm}^\text{portal} = O_n^\text{SM} O_m^\text{hidden}$};
\node (currents) [below = of portals, label={[name=currents label] \underline{Portal currents}\strut}] {$J_n^\text{portal} = \sum_m O_{nm}^\text{portal}$};

\node (framework) [fit=(operators) (operators label) (currents) (currents label) (portals) (portals label) (messengers) (messengers label)] {};
\node (SM) [fit=(EFTs) (EFTs label) (operators) (operators label)] {};

\draw [decorate,decoration={brace,amplitude=6pt}] (hidden.north west) -- (hidden.north east) node [midway, above, yshift=6pt, name=hidden] {Hidden sector with light $\NP$\strut};
\draw [decorate,decoration={brace,amplitude=6pt}] (SM.north west) -- (SM.north east) node [midway, above, yshift=6pt, name=hidden] {$\SM$ and heavy $\NP$\strut};
\draw [decorate,decoration={brace,amplitude=6pt,mirror}] (framework.south west) -- (framework.south east) node [midway,below, yshift=-6pt] {$\PET$ framework};

\draw [->] (EFTs) -- (operators label);
\draw [->] (messengers) -- (portals label);
\draw [<->] (portals) -- (currents label);
\draw [->] (EFTs) -- (portals label);
\draw [->] (messengers) -- (Hid label);
\draw [->] (secluded) -- (Hid label);

\end{tikzpicture}

%% file: flowchart.pgf
\begin{tikzpicture}[
 node distance=7ex and 3em,
 align=center,
 inner sep = 4pt
]
\node[draw] (EW) {Portal $\SMEFTs$\\($d\leq5$)};
\node[draw] (QCD) [right = of EW] {Portal $\LEFTs$\\($d\leq7$)};
\node (EWSB) [below = of $(EW.east)!0.6!(QCD.west)$] {$\EWSB$\strut};
\node (integrating) [above = of EW.west, anchor=south west] {\hspace{-\pgfkeysvalueof{/pgf/inner xsep}}Integrating out heavy fields};
\node[draw] (chiPT) [right = of QCD] {Portal $\cPTs$\\($\NLO$)};
\node (current) [above = of $(QCD.east)!0.5!(chiPT.west)$] {Spurion analysis};
\node (matching) [below = of $(QCD)!0.5!(chiPT)$] {Coefficient\\matching};
\node[draw] (feynman) [right = of chiPT] {Feynman rules};
\node (expansion) [above = of $(chiPT.east)!0.5!(feynman.west)$] {Expansion};
\node[draw] (example) [below = of feynman] {Golden processes\\$\begin{aligned}K^\pm &\to \pi^\pm s\\[-1ex] K^\pm &\to l^\pm \xi_a\\[-1ex] \pi^0 &\to \gamma v\end{aligned}$};
\draw [->] (EW) -- (QCD);
\draw [->] (QCD) -- (chiPT);
\draw [>-] (EWSB) |- (QCD);
\draw [>-] ($(integrating.south)!.4!(integrating.south east)$) |- (QCD);
\draw (QCD) |- (matching);
\draw [->] (matching) -| (chiPT);
\draw [>-] (current) |- (chiPT);
\draw [->] (chiPT) -- (feynman);
\draw [>-] (expansion) |- (feynman);
\draw [->] (feynman) -- (example);
\end{tikzpicture}

%% file: quantum-chromo-dynamics.tex
\subtoc

\section{Quantum chromodynamics} \label{sec:quantum chromo dynamics}

\QCD is a $\SU(\Nc)$ gauge theory, where $\Nc = 3$ is the number of colours.
It depends on \mbox{$\Nc^2 -1 = 8$} gluons $G_\mu$ as gauge fields and features $\nfl$ massive quark flavours~$\fl$.
Using the \QCD gauge coupling~$g_\strong$, we define the fine-structure constant and its inverse as
\begin{align} \label{eq:fine-structure constant}
\alpha_\strong &= \frac{g_\strong^2}{4\pi} \ , &
\omega &= \frac{2\pi}{\alpha_\strong} \ .
\end{align}
The inverse fine-structure constant $\omega$ is the natural parameter for describing the dependence of the gauge coupling on the renormalisation scale $\mu$.
In the \MSb scheme, it obeys the particularly simple \RGE \cite{tHooft:1973mfk,Weinberg:1951ss}
\begin{align} \label{eq:beta function}
\dv{\omega}{t} &= \beta_\strong \ , &
\beta_\strong &= \betacoeff + \order{\frac{1}{\omega}} \ , &
\betacoeff &= \frac{11}3 \Nc - \frac23 \nfl \ ,
\end{align}
where $t = \ln \nicefrac{\mu}{\Lambda}$ is the logarithm of the renormalisation scale, and $\betacoeff$ is the \LO coefficient of the $\beta$-function.
In this scheme, the heavier quark flavours have to be integrated out when they become inactive, so that $\nfl$ ranges from six above the top mass to three below the charm mass.
At \LEs, this prescription reveals an \IR divergence for the coupling strength at \cite{Aoki:2019cca,Bruno:2017gxd, Nakayama:2016atf, Bazavov:2014soa, Chakraborty:2014aca, McNeile:2010ji, Aoki:2009tf, Maltman:2008bx, Zafeiropoulos:2019flq}
\footnote{We label the errors of quantities calculated on the lattice with the subscript $\lat$.}
\begin{equation} \label{eq:Landau pole}
\Lambda_{\QCD}^{\MSb}(\omega) = \unit[(343 \pm 12_{\lat})]{MeV} \ ,
\end{equation}
which invalidates the perturbative expansion in the gauge coupling.
Working with $\omega$ simplifies the inclusion of flavour invariant external currents introduced in \cref{sec:portals}.
For the same reason, it is also convenient to normalise the gluon fields such that the covariant quark derivative $D^\mu = \partial^\mu - \i G^\mu$ is independent of $g_\strong$.
Then, the kinetic part of the \QCD Lagrangian is
\begin{align} \label[lag]{eq:Q kinetic Lagrangian}
\mathcal L_Q^\text{kin} &= \mathcal L_Q^\omega
+ \i q^\dagger \slashed D q
+ \i \overline q \slashed D \overline q^\dagger \ , &
\mathcal L_Q^\omega &= - \omega \gluons(x) \ , &
\gluons(x) &= \inv[2]{(4\pi)} \trc{G_{\mu\nu} G^{\mu\nu}} \ ,
\end{align}
where angle brackets $\trc{\param}$ indicate a trace in colour space, and the gauge singlet $\gluons(x)$ is normalised such that the gauge coupling does not explicitly appear in the anomalous contribution to the trace of the improved stress-energy tensor $\mathcal T$ introduced below.
Following \cite[appendix J of][]{Dreiner:2008tw}, we use two distinct left-handed Weyl fermions $q$ and $\overline q$ to describe each Dirac fermion $(q$, $\overline q^\dagger)$.
\footnote{\label{fn:bar}Note that the bar over the fermion does not denote a mathematical operation but is part of its definition.}
The kinetic Lagrangian is invariant under global flavour rotations
\begin{align} \label[sym]{eq:flavour symmetry}
q_a &\to \VL_a^b q_b \ , &
\overline q^{\dot a} &\to \overline q^{\dot b} \VR_{\dot b}^{\dot a} \ , &
(\VL, \VR) &\in G_{LR} = \U(\nfl)_L \times \U(\nfl)_R \ ,
\end{align}
where $\nfl = 3$ is the number of active quark flavours below the charm mass and boldface symbols indicate matrices in flavour space.
Lower \prefix{un}{dotted} indices denote objects that transform as members of the fundamental representations of $\U(\nfl)_L$ and $\U(\nfl)_R$, respectively, while upper indices denote objects that transform as members of the anti-fundamental representations.
\footnote{
The index-notation is inspired by the \prefix{un}{dotted} \emph{Greek} indices used in \SUSY to distinguish between left- and right-chiral \emph{spinor} indices.
In contrast to the \SUSY notation, the \emph{Latin} indices we use run over $\nfl$-tuples in the (\up, \down, \strange) \emph{flavour} space of \QCD.
We suppress flavour indices whenever the meaning is captured by the implicit boldface notation.
}

Various mechanisms, either spontaneously or explicitly, break the $G_{LR}$ symmetry of the kinetic Lagrangian.
First, the finite \VEVs of the light and strange quark condensates \cite{Aoki:2017paw, Cossu:2016eqs, Boyle:2015exm, Durr:2013goa, Borsanyi:2012zv, Bazavov:2010yq,Aoki:2019cca,Davies:2018hmw,McNeile:2012xh}
\begin{subequations} \label{eq:quark condensate}
\begin{align} \label{eq:chiral quark condensate}
\Sigma_\light = - \frac12 \vev{\overline \up\up + \overline \down\down}^{\MSb}_{\unit[2]{GeV}} + \text{h.c.} &= \unit[(272 \pm 5_{\lat})^3]{MeV}^3 \ , \\ \label{eq:strange quark condensate}
\Sigma_\strange = - \vev{\overline \strange\strange}^{\MSb}_{\unit[2]{GeV}} + \text{h.c.} &= \unit[(296 \pm 11_{\lat})^3]{MeV}^3 \ ,
\end{align}
\end{subequations}
spontaneously break $G_{LR}$ to the global vector symmetry $G_V \cong \U(\nfl)_V$ by causing the \QCD vacuum to change under the action of the axial quotient group $\U(\nfl)_A \cong \flatfrac{G_{LR}}{G_V}$.
In mass-independent renormalisation schemes, the ratio
\begin{equation} \label{eq:quark condensate ratio}
\frac{\Sigma_\strange}{\Sigma_\light}
= 1.29 \pm 0.16_{\lat}
\end{equation}
is scale independent \cite{Vermaseren:1997fq,Kaiser:2000gs}.
Second, the \SM Higgs mechanism explicitly breaks the chiral symmetry by inducing the mass term
\begin{align} \label[lag]{eq:Q mass Lagrangian} %
\mathcal L_Q^m &= - \trf{\nonet \m \nonet Q} + \text{h.c.} \ , &
\nonet \m &= \diag(m_\up, m_\down, m_\strange, \dots) \ , &
\nonet Q_a^{\dot a} &= q_a \overline q^{\dot a} \ ,
\end{align}
where $\trf{\param}$ denotes a trace in flavour space and $\nonet Q$ is a scalar quark bilinear.
The formulation of the mass term as a trace of matrices in flavour space is unusual in standard treatments of \QCD, but it is convenient for understanding correspondences between \QCD and \cPT, and serves as preparation for the matching between these two theories, performed in \cref{sec:chiral perturbation theory}.
Third, the axial anomaly explicitly breaks the global axial $\U(1)_A$ flavour symmetry that is part of $\U(\nfl)_A$ \cite{Bell:1969ts,Adler:1969er,Adler:1969gk}.
In general, anomalies appear as a result of the transformation behaviour of the integration measure in the generating functional
\begin{equation} \label{eq:path integral}
\mathcal Z_Q[J] = \mathcal N \int \mD{\varphi} \exp( \i \int (\mathcal L_Q + O_i J_i) \d{x} ) \ ,
\end{equation}
where $\varphi$ collectively denotes the \QCD fields, the $O_i$ are local, gauge-invariant operators composed of \QCD fields, and the $J_i$ are external currents.
The axial anomaly is related to the topologically nontrivial vacuum structure of \QCD, which also causes the existence of a further contribution to the \QCD Lagrangian,
\begin{align} \label[lag]{eq:Q axial anomaly Lagrangian}
\mathcal L_Q^\theta &= - \theta w(x) \ , &
w &= \frac{\trc{\widetilde G_{\mu\nu} G^{\mu\nu}}}{(4\pi)^2}
= \epsilon_{\mu\nu\rho\sigma} \frac{\partial^\mu \omega_0^{\nu\rho\sigma}}{(4\pi)^2} \ , &
\omega_0^{\nu\rho\sigma} &= \trc{G^\nu G^{\rho\sigma} + \frac23\i G^\nu G^\rho G^\sigma}\ ,
\end{align}
where $\widetilde G_{\mu\nu} = \flatfrac{\epsilon_{\mu\nu\rho\sigma} G^{\rho\sigma}}{2}$, and $\theta$ is the \QCD vacuum angle \cite{tHooft:1976rip, tHooft:1976snw}, which is experimentally constrained to be $\abs{\theta}\lesssim 10^{-10}$ \cite{Zyla:2020zbs}.
Although the topological charge density $w(x)$ is a total derivative of the three-dimensional \CS term $\omega_0^{\nu\rho\sigma}(x)$,
its contribution to the \QCD action does not vanish, since the gluon fields remain finite at spatial infinity for field configurations with finite winding number $n_w = \int w(x) \d[4]{x}$ \cite{Belavin:1975fg,Bardeen:1974ry}.
The axial anomaly manifests itself as a shift of the vacuum angle that results from the transformation of the path integral measure $\mD{\varphi}$ under $\U(1)_A$ flavour rotations.
The typical energy scale associated with such a shift is measured by the topological susceptibility \cite{Crewther:1977ce,Bhattacharya:2021lol}
\begin{equation} \label{eq:susceptibility}
\chi = \frac{\vev{n_w^2}}{V} = - \i \int \vev{\T w(x) w(0)} \d[4]{x} = \unit[(66 \pm 13_{\lat})^4]{MeV}^4 \ ,
\end{equation}
where $V$ is a spacetime volume element and $\T$ is the time ordering operator.
The quark contribution to the topological susceptibility is governed by their condensates \eqref{eq:quark condensate} and masses \eqref{eq:Q mass Lagrangian} \cite{DiVecchia:1980yfw, Leutwyler:1992yt},
\begin{align} \label{eq:susceptibility relation}
\frac1\chi &= \frac1{\chi_0} + \frac{\trf{\inv{\nonet m}}}{\Sigma_0} \ , &
\Sigma_0 &= \eval{\Sigma_\light}_{m_\light \to 0} = \eval{\Sigma_\strange}_{m_\strange \to 0} \ ,
\end{align}
where $\chi_0$ is the \enquote{quenched} topological susceptibility obtained in a pure \YM theory without quark fields, and $\Sigma_0$ is the value of the quark condensates in the chiral limit.
Besides the perturbative expansion in the fine-structure constant that breaks down in the vicinity of the \QCD scale \eqref{eq:Landau pole},
one may also expand \QCD in powers of $\inv \Nc$ \cite{tHooft:1973alw}, which corresponds to a semi-classical expansion in an effective theory of weakly interacting mesons and glueballs.
The axial anomaly \eqref{eq:Q axial anomaly Lagrangian} vanishes at zeroth order in the large $\Nc$ limit \cite{Veneziano:1979ec}, which restores the otherwise badly broken $U(\nfl)_A$ flavour symmetry.
Including higher orders, the effect of the axial anomaly is therefore suppressed by factors of $\inv \Nc$.

The large $\Nc$ expansion is defined such that the value of the \QCD scale, which depends on the product $\Nc \inv \omega$, remains finite as $\Nc$ goes to infinity \cite{tHooft:1973alw,Coleman:1985rnk,Manohar:1998xv,tHooft:2002ufq}.
Therefore, the $\Nc$ enhancement of diagrams with additional closed colour loops balances with the suppression due to additional powers of the coupling $\inv \omega \propto \inv \Nc$, and it can be shown that connected diagrams can scale at most as $\Nc^2$,
while disconnected diagrams scale like the product of their connected subdiagrams.
The leading connected diagrams do not contain any closed quark loops or \QCD $\theta$ angle insertions.
Diagrams with $n_q$ quark loops and $n_\theta$ vacuum angle insertions scale at most as \cite{tHooft:1973alw,Coleman:1985rnk,Manohar:1998xv,tHooft:2002ufq}
\begin{equation} \label{eq:large N_c scaling}
\Nc^{2 - n_q - n_\theta} \ .
\end{equation}
Since the leading connected diagrams scale with a positive power of $\Nc$,
correlation functions for operators that can be decomposed into multiple gauge singlets are dominated by contributions from disconnected diagrams.
Hence, renormalised \QCD correlation functions obey the large $\Nc$ factorization rule
\begin{equation} \label{eq:large N factorization}
\vev{O_i O_j} = \vev{O_i} \vev{O_j} \left(1 + \order{\inv \Nc}\right) \ ,
\end{equation}
where the $O_i$ are local colour singlets that cannot be decomposed further into other colour singlets.
This \enquote{vacuum saturation hypothesis} can be used to match certain \QCD observables with their \cPT counterparts.

In addition to the flavour symmetry, the classical theory associated with the kinetic \cref{eq:Q kinetic Lagrangian} is conformally invariant.
The generators of the conformal Poincar\'e group can be expressed via the Hilbert stress-energy tensor
\begin{equation} \label[set]{eq:stress-energy tensor}
\mathcal T^{\mu\nu} = 2 \pdv{\mathcal L}{g_{\mu\nu}} - g^{\mu\nu} \mathcal L \ ,
\end{equation}
which is divergenceless, symmetric, and traceless in the case of conformal theories.
\footnote{The equally conserved canonical stress-energy tensor associated with the Noether current of spacetime translations is generically neither symmetric nor traceless for conformal theories.
This shortcoming can be overcome by adding model dependent improvement terms \cite{Callan:1970ze}, which then must result in the same expression as the Hilbert stress-energy tensor.}
The conformal invariance of \QCD is broken, at the classical level, by the masses of the quarks \eqref{eq:Q mass Lagrangian}, and, at the quantum level, by the conformal anomaly associated with the running of the gauge coupling \eqref{eq:beta function}, as it introduces an additional mass scale.
Consequently, both terms contribute to the trace of the Hilbert stress-energy tensor \cite{Collins:1976yq, Minkowski:1976en, Nielsen:1977sy, Adler:1976zt},
\begin{equation} \label[set]{eq:Q stress-energy tensor}
\mathcal T_Q = - \mathcal L_Q^m + \frac{\beta_\strong}{\omega} \mathcal L_Q^\omega = (\trf{\nonet \m \nonet Q} + \text{h.c.}) - \beta_\strong \gluons(x) \ .
\end{equation}
Notably, the dependence on the inverse fine-structure constant $\omega$ cancels in this expression.
In \cref{sec:chiral perturbation theory}, we use this trace relation to express $\gluons(x)$ as a linear combination of \cPT operators.
Loop corrections associated with the quark masses generate another contribution to the trace of the stress-energy tensor,
\begin{equation}
\gamma_m (\trf{\nonet \m \nonet Q} + \text{h.c.}) \ ,
\end{equation}
where $\gamma_m$ is the anomalous dimension of the \SM quark masses.
However, we do not keep track of this subleading contribution.

\paragraph{Summary}

The complete \QCD Lagrangian without \EW contributions is constructed by adding gauge fixing and ghost Lagrangians to the kinetic \eqref{eq:Q kinetic Lagrangian},
mass \eqref{eq:Q mass Lagrangian}, and axial anomaly \eqref{eq:Q axial anomaly Lagrangian} terms, so that
\begin{equation} \label[lag]{eq:Q Lagrangian}
\mathcal L_Q = \mathcal L_Q^\text{kin} + \mathcal L_Q^m + \mathcal L_Q^\theta + \mathcal L_Q^\xi + \mathcal L_Q^\text{ghost} \ ,
\end{equation}
where, for covariant gauges,
\begin{align}
\mathcal L_Q^\text{\xi} &= \frac1{\text \xi} \trc{\left(\partial_\mu G^\mu \right)^2} \ , &
\mathcal L_Q^\text{ghost} &= 2 \trc{\partial_\mu \overline c D^\mu c} \ ,
\end{align}
with \xi\ being the gauge-fixing parameter while $\overline c$ and $c$ are the \QCD ghost-fields.

\subsection{Electroweak interactions} \label{sec:EW currents}

Besides the quarks and gluons, the \SM at \LEs contains an \EW sector consisting of the photon field, the charged electron and muon fields, and the left-handed \SM neutrino fields.
\QCD couples to the photons $\photon^\mu$ via the left- and right-handed vector current interactions
\begin{align} \label[lag]{eq:Q vector current Lagrangian}
\mathcal L_Q^v &= - \trf{\octet l_\photon^\mu \nonet Q_\mu} - \trf{\octet r_\photon^\mu \overline{\nonet Q}_\mu} \ , &
\octet r_\photon^\mu &= \octet l_\photon^\mu = \octet v_\photon^\mu \ , &
\octet v_\photon^\mu = e \octet \charge \photon^\mu \ ,
\end{align}
where $\octet \charge = \flatfrac{\diag(2, -1, -1)}{3}$ is the quark-charge matrix,
\begin{align} \label[bilinear]{eq:vectorial quark bilinears}
\nonet Q^\mu{}_a^b &= q_a \sigma^\mu q^{b \dagger} \ , &
\overline{\nonet Q}^\mu{}_{\dot a}^{\dot b} &= \overline q_{\dot a}^\dagger \overline \sigma^\mu \overline q^{\dot b} \ ,
\end{align}
are left- and right-handed vectorial quark bilinears, and sans-serif boldface font indicates traceless matrices.
The \EM currents are parity blind ($\octet v_\photon^\mu = \octet l_\photon^\mu = \octet r_\photon^\mu$), traceless, diagonal, and couple identically to the down and strange quarks,
\begin{align}
\octet v_\photon^\mu &= \diag(\octet v_\photon^\mu{}_\up^\up, \octet v_\photon^\mu{}_\down^\down, \octet v_\photon^\mu{}_\strange^\strange) \ , &
\octet v_\photon^\mu{}_\down^\down &= \octet v_\photon^\mu{}_\strange^\strange = - 2 \octet v_\photon^\mu{}_\up^\up \ ,
\end{align}
where individual fermion flavours are indicated by upright font.
The split of the parity blind \EM current into a left- and right-handed current simplifies the generalisation to other spin 1 currents.
However, we will drop this distinction and use $\octet v_\photon^\mu$ when considering the phenomenology of the hidden messengers in \cref{sec:light mesons,sec:models}.

The impact of diagrams at the \EW scale with virtual exchanges of the heavy \SM fields that have been integrated out can be captured at the strong scale by introducing an infinite tower of higher dimensional operators.
As their mass-dimensions are larger than four, these operators are suppressed by powers of
\footnote{
The relevant operators in this paper are generated by contributions with virtual $W$-boson exchanges,
so that they are suppressed by factors $\flatfrac{\partial^2 g_w^2}{m_W^2}$ that involve the mass of the $W$-boson $m_W$ rather than the Higgs \VEV.
We write the ratio of scales in terms of $v^2 = \flatfrac{2 m_W^2}{g_w^2}$ to simplify the shape of the equations that appear throughout this paper.
}
\begin{align} \label{eq:epsilon SM}
\epsilonSM &= \frac{\partial^2}{\Lambda_{\SM}^2} \ , &
\partial^2 &\lesssim m_c^2 \ , &
\Lambda_{\SM} &= 4\pi v \ ,
\end{align}
which measures the ratio between the \EW and \LE momentum scales, where $v = (\sqrt8 G_F)^{-\nicefrac12} = \unit[(174.10358 \pm 0.00004_{\ex})]{GeV}$ is the Higgs \VEV \cite{Zyla:2020zbs}.
\footnote{The subscript $\ex$ indicates an experimental error.}
Since the renormaliseable strong and \EM interactions conserve quark flavour, the higher dimensional operators contribute at \LO to flavour violating processes such as kaon decays.
\LO transitions that violate flavour by one unit, $\Delta \qn f = \pm 1$, are generated by operators with mass dimension five and six.

\begin{figure}
\begin{panels}[t]{3}
\tikzsetnextfilename{feynman-current}\input{feynman-current.pgf}
\caption{Vector current interaction.}
\label{fig:feynman current}
\panel
\tikzsetnextfilename{feynman-dipole-2}\input{feynman-dipole-2.pgf}
\caption{Electromagnetic dipole.}
\label{fig:feynman photon}
\panel
\tikzsetnextfilename{feynman-dipole}\input{feynman-dipole.pgf}
\caption{Chromomagnetic dipole.}
\label{fig:feynman gluon}
\end{panels}
\caption[Feynman diagrams describing charged current and magnetic-dipole interactions]{
Processes that generate higher dimensional operators at the strong scale with two quarks.
Panel \subref{fig:feynman current} shows the tree level diagram that describes the charged current interaction \eqref{eq:Q weak current Lagrangian}.
Panels \subref{fig:feynman photon} and \subref{fig:feynman gluon} show the 1-loop photon and gluon diagrams that describe the dipole interactions \eqref{eq:Q dipole current Lagrangian}.
The cross indicates a mass insertion that can appear at either external fermion leg.
Up type quarks are collectively denoted by $u = \up$, $\charm$, $\topq$.
} \label{fig:feynman diagrams 1}
\end{figure}

At tree level, the contribution depicted in \cref{fig:feynman current} and its Hermitian conjugate
induce the leptonic charged current interactions that couple quarks to charged leptons and neutrinos,
\begin{align} \label[lag]{eq:Q weak current Lagrangian}
\mathcal L_Q^W &= - \trf{\octet l_W^\mu \nonet Q_\mu} \ , &
\octet l_W^\mu &= - \inv[2]{v} \left(V_{\up\down} \flavour \proj_\up^\down + V_{\up\strange} \flavour \proj_\up^\strange\right) \sum_{\ell = \electron, \muon} l_\ell^\dagger \overline \sigma^\mu\nu_\ell + \text{h.c.} \ ,
\end{align}
where the $V_{ij}$ are elements of the \CKM matrix.
We use the matrices
\begin{align}
& \flavour \lambda_a^b \ , &
& (\flavour \lambda_a^b)_i^j = \delta_{ai} \delta^{bj}
\end{align}
to construct an orthonormal basis in flavour space.
The weak leptonic charged current is traceless, Hermitian, and has no neutral contributions, so that
\begin{align}
\octet l_W^\mu{}_\up^\down &= \octet l_W^{\mu\dagger}{}_\down^\up \ , &
\octet l_W^\mu{}_\up^\strange &= \octet l_W^{\mu\dagger}{}_\strange^\up \ ,
\end{align}
while all remaining entries vanish.
In order to prepare for the inclusion of the portal current interactions in \cref{sec:portals}, it is convenient to absorb the charged current interaction into the left-handed external current
\begin{align} \label[cur]{eq:Q vector currents}
\octet l^\mu = \octet l_\photon^\mu + \octet l_W^\mu \ ,
\end{align}
so that the vector current Lagrangian \eqref{eq:Q vector current Lagrangian} accounts for both \EM and weak charged current interactions.

At one-loop, the contributions depicted in \cref{fig:feynman photon,fig:feynman gluon} with a virtual $W$-boson exchange and a light quark mass insertion at one of the external legs further induce the electro- and chromomagnetic-dipole interactions between two quarks and a gauge boson \cite{Buras:1998raa}
\begin{align} \label[lag]{eq:Q dipole current Lagrangian}
\mathcal L_Q^\tau &= - \inv[2]{\Lambda_{\SM}} \trf{\octet \tau^{\mu\nu} \nonet Q_{\mu\nu}} + \text{h.c.}  \ , &
\mathcal L_Q^\chromo &= - \inv[2]{\Lambda_{\SM}} \trf{\octet \chromo_G \widetilde{\nonet Q}} + \text{h.c.} \ ,
\end{align}
where the tensorial and scalar quark bilinears are
\begin{align} \label[bilinear]{eq:EW quark bilinears}
\nonet Q_{\mu\nu}{}_a^{\dot a} &= q_a \sigma_{\mu\nu} \overline q^{\dot a} \ , &
\widetilde{\nonet Q}{}_a^{\dot a} &= q_a \sigma_{\mu\nu} G^{\mu\nu} \overline q^{\dot a} \ .
\end{align}
The tensorial \EM-dipole current and the scalar electro- and chromomagnetic-dipole currents are
\begin{align} \label[cur]{eq:Q dipole current}
\octet \tau^{\mu\nu} &= \frac13 F^{\mu \nu} \octet \chromo_A \ , &
\octet \chromo_V &= \nonet m \left(\flavour \proj_\strange^\down \sum_{\mathclap{u = \up, \charm, \topq}} c_u^V V^\dagger_{\strange u} V_{u\down} + \text{h.c.}\right) \ ,
\end{align}
where the indices $V = G$, $A$ denote either gluon or photon contributions and the $c_u^V$ are known Wilson coefficients \cite{Buras:1998raa}.
In the following, we abbreviate the chromomagnetic-dipole current by $\octet \chromo = \octet \chromo_G$.
The dipole currents are strangeness violating, but not necessarily Hermitian.
The only nonvanishing contributions are
\begin{align}
& \octet \chromo{}_\down^\strange \ , &
& \octet \chromo{}_\strange^\down \ , &
& \octet \tau^{\mu \nu}{}_\down^\strange \ , &
& \octet \tau^{\mu \nu}{}_\strange^\down \ .
\end{align}
The operator $\widetilde{\nonet Q}$ also has nonvanishing condensates
\begin{subequations}\label{eq:quark gluon condensate}
\begin{align}
\Sigma_{G\light} = - \frac12 \vev{\widetilde{\nonet Q}_\up^\up + \widetilde{\nonet Q}_\down^\down}_{\unit[2]{GeV}}^{\MSb} + \text{h.c.} &= \unit[(434 \pm 4_{\lat})^5]{MeV}^5 \ , \\
\Sigma_{G\strange} = - \vev{\widetilde{\nonet Q}_\strange^\strange}_{\unit[2]{GeV}}^{\MSb} + \text{h.c.} &= \unit[(425 \pm 14_{\lat})^5]{MeV}^5 \ ,
\end{align}
\end{subequations}
which are estimated using \QCD sum rules \cite{Belyaev:1982sa,Gubler:2018ctz,Aladashvili:1995zj,Braun:2004vf} or lattice computations \cite{Chiu:2003iw}.
\footnote{For simplicity, we indicate errors for values estimated using \QCD sum rules with the same label as errors for values calculated on the lattice.}
Their ratios with the \VEV of light quark condensate \eqref{eq:chiral quark condensate} are
\begin{align} \label{eq:quark gluon condensate ratio}
\frac{\Sigma_{G\light}}{\Sigma_\light} &= \unit[(875 \pm 31_{\lat})^2]{MeV}^2 \ , &
\frac{\Sigma_{G\strange}}{\Sigma_\strange} &= \unit[(731 \pm 73_{\lat})^2]{MeV}^2 \,.
\end{align}
Note that the ratio between the two quark-gluon condensates
\begin{equation}
\frac{\Sigma_{G\strange}}{\Sigma_{G\light}} = 0.90 \pm 0.15_{\lat} \ .
\end{equation}
is consistent with one.

\paragraph{Four-quark interactions}

\begin{figure}
\begin{panels}{2}
\tikzsetnextfilename{feynman-tree}\input{feynman-tree.pgf}
\caption{Tree-level diagram.}
\label{fig:feynman tree}
\panel
\tikzsetnextfilename{feynman-gluon}\input{feynman-gluon.pgf}
\caption{One-loop penguin diagram.}
\label{fig:feynman penguin}
\end{panels}
\caption[Feynman diagrams describing four-quark interactions]{
Processes that generate higher dimensional operators at the strong scale with four quarks.
The tree level diagram \subref{fig:feynman tree} generates the operators $O_1$ and $O_2$ in \eqref{eq:four-quark operators tree},
while the penguin diagram in \subref{fig:feynman penguin} generates the operators $O_3$ to $O_6$ in \eqref{eq:four-quark operators pinguin,eq:four-quark operators mixed}.
Up-type quarks are collectively denoted by $u = \up$, $\charm$, $\topq$.
} \label{fig:feynman diagrams}
\end{figure}

The diagrams in \cref{fig:feynman diagrams} depict the contributions that generate four-quark interactions of the shape \cite{Altarelli:1974exa, Gaillard:1974nj, Vainshtein:1975sv, Shifman:1975tn}
\begin{equation} \label[lag]{eq:Q four-quark Lagrangian}
\mathcal L_Q^{\fermi} = - \frac{V^\dagger_{\strange\up} V_{\up\down}}{v^2} \sum_{\iota=1}^6 c_\iota O_\iota + \text{h.c.} \ ,
\end{equation}
where $\abs*{V^\dagger_{\strange\up}} \abs{V_{\up\down}} = 0.2186\pm0.00008$ \cite{Zyla:2020zbs} and the $c_\iota$ are known Wilson coefficients \cite{Buras:1998raa}.
After neglecting \EM penguin diagrams, which are suppressed by at least one power of $\alpha_{\EM}$, there are six four-quark operators that violate quark-flavour by one unit \cite{Cirigliano:2011ny},
\begin{subequations} \label{eq:four-quark operators}
\begin{align} \label{eq:four-quark operators tree}
O_1 &= \boson{\strange^\dagger \overline \sigma^\mu \up} \boson{\up^\dagger \overline \sigma_\mu \down} \ , &
O_2 &= \boson{\strange^\dagger \overline \sigma^\mu \down} \boson{\up^\dagger \overline \sigma_\mu \up} \ , \\ \label{eq:four-quark operators pinguin}
O_3 &= \boson{\strange^\dagger \overline \sigma^\mu \down} \boson{q^\dagger \overline \sigma_\mu q} \ , &
O_4 &= \boson{\strange^\dagger \overline \sigma^\mu q} \boson{q^\dagger \overline \sigma_\mu \down} \ , \\ \label{eq:four-quark operators mixed}
O_5 &= \boson{\strange^\dagger \overline \sigma^\mu \down} \boson{\overline q \sigma_\mu \overline q^\dagger} \ , &
O_6 &= \boson{\strange^\dagger \overline q^\dagger} \boson{\overline q \down}\ .
\end{align}
\end{subequations}
Since these operators are necessarily neutral, they can only violate quark-flavour by mediating $\down \leftrightarrow \strange$ transitions and thereby violate strangeness, $\Delta \qn s = \pm 1$.
The operators $O_1$ and $O_2$ \eqref{eq:four-quark operators tree} are generated by the tree-level diagram shown in \cref{fig:feynman tree},
while the operators $O_3$ to $O_6$ \eqref{eq:four-quark operators pinguin,eq:four-quark operators mixed} are generated by one-loop penguin diagrams as shown in \cref{fig:feynman penguin}.
Although the penguin operators are suppressed by loop-factors, the operator $O_6$ is enhanced at low energies due to chirality effects, so that it contributes at \LO to certain transitions.
For a more detailed discussion, see \cref{sec:cpt qcd matching}.
We organise the four-quark operators \eqref{eq:four-quark operators} according to their chirality structure into a scalar-scalar and two vector-vector interaction terms
\begin{equation} \label[lag]{eq:four-quark Lagrangian}
\mathcal L_Q^{\fermi} =
- \inv[2]{v} \left(
 \tensor \fermi_\scalar{}_{a\dot b}^{\dot ab} \nonet Q^\dagger{}_{\dot a}^a \nonet Q_b^{\dot b}
+ \tensor \fermi_r{}_{a\dot a}^{b\dot b} \nonet Q_\mu{}_b^a \overline{\nonet Q}^\mu{}_{\dot b}^{\dot a}
+ \tensor \fermi_l{}_{ac}^{bd} \nonet Q_\mu{}_b^a \nonet Q^\mu{}_d^c
\right) \ ,
\end{equation}
where the parameters $\tensor \fermi_\scalar$, $\tensor \fermi_r$, and $\tensor \fermi$ are four-index tensors in flavour space, which we indicate using symbols in Fraktur font.
Comparing this formulation of the four-quark Lagrangian with the operators listed in \cref{eq:four-quark operators}, the parameters are given as
\begin{subequations} \label{eq:four-quark parameter}
\begin{alignat}{2}
\tensor \fermi_\scalar &= V^\dagger_{\strange\up} V_{\up\down} c_6 \sum_{\mathclap{u = \up, \down, \strange}} \flavour \proj_\strange^u \otimes \flavour \proj_u^\down
+ \text{h.c.} \ , &
\mathllap{\tensor \fermi_r = V^\dagger_{\strange\up} V_{\up\down} c_5 \flavour \proj_\strange^\down \otimes \flavour 1} &
+ \text{h.c.} \ , \\
\tensor \fermi_l &= V^\dagger_{\strange\up} V_{\up\down} \left(
c_1 \flavour \proj_\strange^\up \otimes \flavour \proj_\up^\down
+ c_2 \flavour \proj_\strange^\down \otimes \flavour \proj_\up^\up
+ c_3 \flavour \proj_\strange^\down \otimes \flavour 1
+ c_4 \sum_{\mathclap{u = \up, \down, \strange}} \flavour \proj_\strange^u \otimes \flavour \proj_u^\down\right) & &
+ \text{h.c.} \ ,
\end{alignat}
\end{subequations}
where $\otimes$ denotes a tensor product.

\begin{figure}
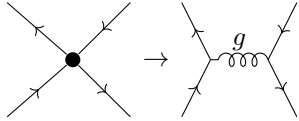

\renewcommand{\width}{10ex}
\newcommand\vcenterpgf[1]{\vcenter{\hbox{\tikzsetnextfilename{#1}\input{#1.pgf}}}}
$\vcenterpgf{feynman-4-vertex} \to \vcenterpgf{feynman-s-channel}$
\caption[Large $\Nc$ scaling of four-quark operators]{
Replacement used to determine the number of closed colour loops in \QCD diagrams with four-quark operators.
Diagrams with a given number of four-quark vertices contain the same number of colour loops as diagrams where each four-quark vertex is replaced by the subdiagram with gluon exchange that is depicted on the right-hand side.
} \label{fig:large Nc scaling}
\end{figure}

Connected diagrams with four-quark vertices in \cref{eq:four-quark Lagrangian} are not included in standard derivations of the large $\Nc$ power counting rule \eqref{eq:large N_c scaling} \cite{tHooft:1973alw,Coleman:1985rnk,tHooft:2002ufq}.
To generalise this counting rule to diagrams with a finite number of four-quark vertices, we use the replacement shown in \cref{fig:large Nc scaling} in order to map a given set of diagrams with four-quark vertices onto an equivalent set of pure \QCD diagrams without four-quark vertices.
This replacement is chosen such that the resulting diagram always contains the same number of closed colour loops as its corresponding original four-quark diagram.
The overall large $\Nc$ scaling of the diagram differs from the scaling of the original diagram in two ways:
First, the two three-point vertices in the pure \QCD diagrams are associated with a total prefactor of $\inv \omega \propto \inv \Nc$, whereas the four quark vertices scale as $\omega^0 \propto 1$,
so that the four-quark diagrams are enhanced by one relative factor of $\Nc$ for each four-quark vertex.
Second, the number of quark loops in the pure \QCD diagrams can be lower than the number of quark loops in the original four-quark diagrams,
even though both diagrams contain the same number of closed colour loops.
Hence, the leading contribution to the infinite series of diagrams with exactly $n_{\fermi}$ four-quark insertions and an arbitrary number of colour loops is given by the subset
for which the equivalent pure \QCD diagram contains exactly one quark loop.
Applying the standard counting formula \eqref{eq:large N_c scaling}, we find that the leading four-quark diagrams scale as
\begin{align}
&\Nc^{1 - n_\theta + n_{\fermi}} \ , & n_{\fermi} &> 0 \ .
\end{align}
Further, the leading diagrams with $n_{\fermi}$ four-quark insertions, as well as $n_q$ simple \QCD quark loops \emph{in addition} to the quark loops associated with the four-quark vertices, scale as
\begin{align} \label{eq:modified large N counting}
&\Nc^{1 - n_q - n_\theta + n_{\fermi}} \ , & n_{\fermi} &> 0 \ ,
\end{align}
which extends the usual scaling behaviour \eqref{eq:large N_c scaling}.

\paragraph{Summary}

The \EW interactions induce the \EW correction to the \QCD \cref{eq:Q Lagrangian}
\begin{equation} \label[lag]{eq:Q EW Lagrangian}
\mathcal L_Q^{\EW} = \mathcal L_Q^{\chromo} + \mathcal L_Q^\tau + \mathcal L_Q^v + \mathcal L_Q^{\fermi} \ ,
\end{equation}
which is given by the \cref{eq:Q four-quark Lagrangian,eq:Q dipole current Lagrangian,eq:Q vector current Lagrangian}, where \cref{eq:Q vector current Lagrangian} includes the full \cref{eq:Q vector currents}.
The \EW interactions in \cref{eq:Q weak current Lagrangian,eq:Q dipole current Lagrangian,eq:Q four-quark Lagrangian} also generate additional contributions to the trace of the Hilbert \cref{eq:Q stress-energy tensor}.
After using the quark field \EOM in the presence of external currents in \cref{eq:quark full eom}, the \EW contribution becomes
\begin{align} \label[set]{eq:Q EW stress-energy tensor}
\mathcal T_Q^{\EW}& = \mathcal L_Q^\chromo + \mathcal L_Q^\tau - \mathcal L_Q^W + 2 \mathcal L_Q^{\fermi} \ .
\end{align}

\subsection{Flavour symmetry} \label{sec:flavour symmetry}

Under the flavour \cref{eq:flavour symmetry} of the kinetic Lagrangian \eqref{eq:Q kinetic Lagrangian}, the quark bilinears \eqref{eq:Q mass Lagrangian}, \eqref{eq:vectorial quark bilinears}, and \eqref{eq:EW quark bilinears} transform as
\begin{subequations} \label{eq:QCD transformation Qs}
\begin{align}
\nonet Q &\to \VL \nonet Q \VR \ , &
\nonet Q_\mu &\to \VL \nonet Q_\mu \VL^\dagger \ , &
\nonet Q_{\mu\nu} &\to \VL \nonet Q_{\mu\nu} \VR \ , \\
\widetilde{\nonet Q} &\to \VL \widetilde{\nonet Q} \VR \ , &
\overline{\nonet Q}_\mu &\to \VR^\dagger \overline{\nonet Q}_\mu \VR \ . &
\end{align}
\end{subequations}
As a consequence, the \QCD path integral \eqref{eq:path integral}
\begin{equation}
\mathcal Z_Q = \mathcal Z_Q[\omega, \theta, \nonet m, \octet l^\mu, \octet r^\mu, \octet h, \octet \tau^{\mu\nu}, \tensor \fermi_\scalar, \tensor \fermi_r, \tensor \fermi_l] \ ,
\end{equation}
is invariant under global $G_{LR}$ flavour rotations that transform the external currents as
\footnote{Being a function of the gauge coupling only, the inverse fine-structure constant $\omega$ is invariant under flavour rotations.}
\begin{subequations} \label{eq:current transformations}
\begin{align} \label{eq:theta transformations}
\theta &\to \theta - \i \trf{\ln \VL \VR} \ , &
\nonet m &\to \VR^\dagger \nonet m \VL{}^\dagger \ , &
\tensor \fermi_\scalar{}_{a\dot c}^{\dot bd} &\to \VL_a^u \VR{}_{\dot y}^{\dot b} \tensor \fermi_\scalar{}_{u\dot x}^{\dot yv} \VR{}^\dagger{}_{\dot c}^{\dot x} \VL^\dagger{}_v^d \ , \\
\octet l^\mu &\to \VL \octet l^\mu \VL^\dagger \ , &
\octet \chromo &\to \VR^\dagger \octet \chromo \VL{}^\dagger \ , &
\tensor \fermi_r{}_{a\dot c}^{b\dot d} &\to \VL_a^u \VL^\dagger{}_v^b \tensor \fermi_r{}_{u\dot x}^{v\dot y} \VR{}^\dagger{}_{\dot c}^{\dot x} \VR{}_{\dot y}^{\dot d} \ , \\
\octet r^\mu &\to \VR^\dagger \octet r^\mu \VR \ , &
\octet \tau^{\mu\nu} &\to \VR^\dagger \octet \tau^{\mu\nu} \VL{}^\dagger \ , &
\tensor \fermi_l{}_{ac}^{bd} &\to \VL_a^u \VL^\dagger{}_v^b \tensor \fermi_l{}_{ux}^{vy} \VL_c^x \VL^\dagger{}_y^d \ .
\end{align}
\end{subequations}
Remarkably, the path-integral is additionally invariant under \emph{local} flavour rotations that transform the left- and right-handed currents in \eqref{eq:Q vector current Lagrangian} as
\begin{align} \label{eq:lr vector currents local rotations}
\octet l^\mu &\to \VL \octet l^\mu \VL^\dagger + \i \VL \partial^\mu \VL^\dagger \ , &
\octet r^\mu &\to \VR^\dagger \octet r^\mu \VR + \i \VR^\dagger \partial^\mu \VR \ ,
\end{align}
while the transformation behaviour of the other external currents is unaltered. This transformation law is analogous to that of gauge fields.
To facilitate the construction of operators that are invariant under the action of $G_{LR}$, it is convenient to define covariant derivatives for the quark fields
\begin{align} \label[cur]{eq:Q left and right handed currents}
\nonet D^\mu q &= \partial^\mu q - \i \octet l^\mu q \ , &
\nonet D^\mu \overline q^\dagger &= \partial^\mu \overline q^\dagger - \i \octet r^\mu \overline q^\dagger \,,
\end{align}
as well as field-strength tensors for the left- and right-handed currents
\begin{align} \label{eq:Q field strength tensors}
\octet l^{\mu\nu} &= \partial^\mu \octet l^\nu - \partial^\nu \octet l^\mu - \i \comm{\octet l^\mu}{\octet l^\nu} \ , &
\octet r^{\mu\nu} &= \partial^\mu \octet r^\nu - \partial^\nu \octet r^\mu - \i \comm{\octet r^\mu}{\octet r^\nu} \ .
\end{align}
While the symmetry of the path integral with respect to \eqref{eq:lr vector currents local rotations} corresponds mathematically to a gauge symmetry, it is important to emphasise that $\octet l^\mu$ and $\octet r^\mu$ are not fields in a physical sense.
In particular, while a gauge symmetry relates different field configurations that correspond to the \emph{same} physical state, the local $G_{LR}$ symmetry relates field configurations that correspond to \emph{different} physical states.

\subsection{Four-quark operators} \label{sec:four-quark operators}

The four-quark operators in \cref{eq:four-quark Lagrangian} transform as singlets under $\U(3)_R$ \cite{Donoghue:1992dd}.
For this reason, we suppress the right-handed indices of the external currents, and define
\begin{align}
\nonet \fermi_\scalar{}_a^d &= \frac1{\nfl} \tensor \fermi_\scalar{}_{a\dot c}^{\dot cd} \ , &
\nonet \fermi_r{}_a^b &= \frac1{\nfl} \tensor \fermi_r{}_{a\dot c}^{b\dot c} \ ,
\end{align}
where the reduced parameters $\nonet \fermi_\scalar$ and $\nonet \fermi_r$ transform under $\U(3)_L$ as
\begin{equation}
\overline{\rep 3} \otimes \rep 3 = \rep 8 \oplus \rep 1 \ .
\end{equation}
The traceless octet contributions are given as
\begin{align} \label{eq:octet parameter}
\octet \fermi_\scalar &= \nonet \fermi_\scalar - \frac{\flavour 1}{\nfl} \fermi_\scalar \ , &
\octet \fermi_r &= \nonet \fermi_r - \frac{\flavour 1}{\nfl} \fermi_r \ ,
\end{align}
where $\fermi_r = \trf{\nonet \fermi_r}$ and $\fermi_\scalar = \trf{\nonet \fermi_\scalar}$.
The corresponding left-handed, traceless octet operators composed of the quark bilinears \eqref{eq:Q mass Lagrangian} and \eqref{eq:vectorial quark bilinears} are
\begin{align} \label{eq:octet operators}
\octet O_\scalar &= \nonet Q^\dagger \nonet Q - \frac{\flavour 1}{\nfl} \trf{\nonet Q^\dagger \nonet Q} \ , &
\octet O_r &= \nonet Q^\mu \overline Q_\mu - \frac{\flavour 1}{\nfl} Q^\mu \overline Q_\mu \ ,
\end{align}
where $Q_\mu = \trf{\nonet Q_\mu}$ and $\overline Q_\mu = \trf{\overline{\nonet Q}_\mu}$.

The purely left-handed vector-vector interaction parameter $\tensor \fermi_{ac}^{bd}$ transforms under $\U(3)_L$ as a member of
\begin{equation}
(\overline{\rep 3} \otimes \rep 3) \otimes (\overline{\rep 3} \otimes \rep 3)
= (\rep 8 \oplus \rep 1) \otimes (\rep 8 \oplus \rep 1)
= \overbrace{\underbrace{\strut \rep 8 \oplus \rep 1}_{\mathllap{\text{totally antisym}}\text{metric}}
\oplus \, (\underbrace{\strut \rep{27} \oplus \rep 8 \oplus \rep 1}_{\mathclap{\text{totally symmetric}}}}^\text{symmetric}
\oplus \underbrace{\overbrace{\strut \overline{\rep{10}} \oplus \rep{10} \oplus \rep 8) \oplus \rep 8}^\text{anti-symmetric}}_\text{mixed symmetric} \ ,
\end{equation}
where the parenthesis on the outermost right-hand side indicate the decomposition of the $\rep 8 \otimes \rep 8$ product.
Furthermore, the symmetry of each representation under exchanges of the quark bilinears and quark spinors is indicated by curly braces above and below the expression, respectively.
Since $\tensor \fermi \nonet Q_\mu \nonet Q^\mu$ is symmetric under exchange of the quark bilinears, only representations that are totally \prefix{anti}{symmetric} under exchanges of the quark spinors can contribute to $\tensor \fermi$.
Therefore, the parameter $\tensor \fermi_{ac}^{bd}$ can be written as
\begin{equation}
\tensor \fermi_l
= \tensor \fermi_l^\sym
+ \frac1{n_8^-} \octet \fermi_l^\asym \wedge \flavour 1
+ \frac1{n_8^+} \octet \fermi_l^\sym \odot \flavour 1
+ \frac1{n_1^-} \fermi_l^\asym \flavour 1 \wedge \flavour 1
+ \frac1{n_1^+} \fermi_l^\sym \flavour 1 \odot \flavour 1 \ ,
\end{equation}
where $\wedge$ and $\odot$ are \prefix{anti}{symmetrised} tensor products and the symmetry prefactors are
\begin{align}
n_8^\pm &= \frac{\nfl \pm 2}{4} \ , &
n_1^\pm &= \frac{n_\fl^2 \pm \nfl}{2} \ .
\end{align}
The totally \prefix{anti}{symmetric} singlet $\fermi_l^\symasym$, octet $\octet \fermi_l^\symasym$, and 27-plet $\tensor \fermi_l^\sym$ contributions are related to the complete tensor via
\footnote{
\prefix{Anti}{symmetrised} tensors are defined as $2 T^{[\mu\nu]} = T^{\mu\nu} - T^{\nu\mu}$ and $2 T^{(\mu\nu)} = T^{\mu\nu} + T^{\nu\mu}$, respectively.
}
\begin{subequations} \label{eq:27-plet parameters}
\begin{align}
\fermi_l^\sym &= \tensor \fermi_{(xy)}^{(xy)} \ , &
\octet \fermi_l^\sym{}_a^b &= \tensor \fermi_{(ax)}^{(bx)} - \frac1{\nfl} \flavour 1_a^b \fermi_l^\sym \ , &
\tensor \fermi_l^\sym{}_{ac}^{bd} &= \tensor \fermi_{(ac)}^{(bd)} - \frac1{n_8^+} \flavour 1_{(a}^{(b} \octet \fermi_l^\sym{}_{c)}^{d)} - \frac1{n_1^+} \flavour 1_{(a}^{(b} \flavour 1_{c)}^{d)} \fermi_l^\sym \ , \\
\fermi_l^\asym &= \tensor \fermi_{[xy]}^{[xy]} \ , &
\octet \fermi_l^\asym{}_a^b &= \tensor \fermi_{[ax]}^{[bx]} - \frac1{\nfl} \flavour 1_a^b \fermi_l^\asym \,.
\end{align}
\end{subequations}
The totally \prefix{anti}{symmetric} octet operators formed by the two traceless pairings of two left-handed quark bilinears \eqref{eq:vectorial quark bilinears} related to the octet parameter $\tensor \fermi$ \eqref{eq:27-plet parameters} are
\begin{equation} \label{eq:27-plet octet operators}
\octet O_l^\symasym = \frac1{2 n_8^\pm}
\left[
\left(\nonet Q^\mu Q_\mu
- \frac{1}{\nfl} Q^\mu Q_\mu \right)
\pm \left(\nonet Q^\mu \nonet Q_\mu
- \frac{1}{\nfl} \trf{\nonet Q^\mu \nonet Q_\mu} \right)
\right] \ ,
\end{equation}
while the (symmetric) 27-plet combination is
\begin{equation} \label{eq:27-plet operator}
\tensor O_l^\sym =
\nonet Q^\mu \odot \nonet Q_\mu
- \flavour 1 \odot \octet O_l^\sym
- \frac1{2 n_1^+} \left(Q^\mu Q_\mu + \trf{\nonet Q^\mu \nonet Q_\mu}\right) \flavour 1 \odot \flavour 1 \ .
\end{equation}
Hence, the complete octet and 27-plet contributions to the four-quark \cref{eq:four-quark Lagrangian} are
\begin{equation} \label[lag]{eq:Q four-quark Lagrangian representations}
\mathcal L_Q^{\fermi} =
- \inv[2]{v}
\trf{
 \octet \fermi_\scalar \octet O_\scalar
+ \octet \fermi_r \octet O_r
+ \octet \fermi_l^\asym \octet O_l^\asym
+ \octet \fermi_l^\sym \octet O_l^\sym}
- \inv[2]{v} \trF{\tensor \fermi_l^\sym \tensor O_l^\sym} \ ,
\end{equation}
where the brackets $\trF{\param}$ denote the complete contraction of the totally symmetric tensors.
Using the symmetry properties of the 27-plet term
\begin{align} \label{eq:27-plet symmetries}
- \tensor \fermi_l^\sym{}_{\strange \up}^{\down \up} &= (\nfl - 1) \tensor \fermi_l^\sym{}_{\strange \down}^{\down \down}
= (\nfl - 1) \tensor \fermi_l^\sym{}_{\strange \strange}^{\down \strange} \ , &
- \tensor O_l^\sym{}_{\strange \up}^{\down \up} &= \tensor O_l^\sym{}_{\strange \down}^{\down \down} + \tensor O_l^\sym{}_{\strange \strange}^{\down \strange} \ ,
\end{align}
the strangeness violating contributions listed in \eqref{eq:four-quark operators} can be extracted via
\begin{equation} \label{eq:Q four-quark Lagrangian projection}
\eval{\mathcal L_Q^{\fermi}}_{\Delta \qn s = \pm 1} =
- \frac1{v^2}
\left(
\octet \fermi_\scalar{}_\strange^\down \octet O_\scalar{}_\down^\strange
+ \octet \fermi_r{}_\strange^\down \octet O_r{}_\down^\strange
+ \octet \fermi_l^\asym{}_\strange^\down \octet O_l^\asym{}_\down^\strange
+ \octet \fermi_l^\sym{}_\strange^\down \octet O_l^\sym{}_\down^\strange
\right)
- 2 \frac{n_{27}}{v^2}
\tensor \fermi_l^\sym{}_{\strange\up}^{\down\up} \tensor O_l^\sym{}_{\down \up}^{\strange \up}
+ \text{h.c.} \ ,
\end{equation}
where the 27-plet symmetry prefactor is
\begin{equation}
n_{27} = \frac{2 \nfl - 1}{\nfl -1} \ .
\end{equation}
In terms of the coefficients in \cref{eq:Q four-quark Lagrangian} the octet and 27-plet coefficients are
\begin{subequations} \label{eq:octet currents from parameters}
\begin{align}
\octet \fermi_l^\sym{}_\strange^\down &= \frac14 V^\dagger_{\strange\up} V_{\up\down} \left(c_{12}^+ + (\nfl + 2) c_{34}^+ \right) \ , &
\octet \fermi_r{}_\strange^\down &= V^\dagger_{\strange\up} V_{\up\down} c_5 \ , &
\tensor \fermi_l^\sym{}_{\strange \up}^{\down \up} &= \frac14 V^\dagger_{\strange\up} V_{\up\down} \frac{\nfl + 1}{\nfl + 2} c_{12}^+ \ , \\
\octet \fermi_l^\asym{}_\strange^\down &= - \frac14 V^\dagger_{\strange\up} V_{\up\down} \left(c_{12}^- + c_{34}^-\right) \ , &
\octet \fermi_\scalar{}_\strange^\down &= V^\dagger_{\strange\up} V_{\up\down} c_6 \ , &
c_{\iota\kappa}^\pm &= c_\iota \pm c_\kappa\,.
\end{align}
\end{subequations}

\paragraph{Summary}

\begin{table}
\begin{tabular}{r*{12}c} \toprule
& $\gluons$ & $w$ & $\nonet Q$ & $\nonet Q_\mu$ & $\overline{\nonet Q}_\mu$ & $\nonet Q_{\mu\nu}$ & $\widetilde{\nonet Q}$ & $\octet O_\scalar$ & $\octet O_r$ & $\octet O_l^\asym$ & $\octet O_l^\sym$ & $\tensor O_l^\sym$ \\
\midrule
$\epsilonSM$ & \multicolumn{5}{c}{0} & \multicolumn{7}{c}{1} \\
\cmidrule(r){1-1} \cmidrule(lr){2-6} \cmidrule(l){7-13}
$d$ & \multicolumn{2}{c}{4} & \multicolumn{3}{c}{3} & 3 & 5 & \multicolumn{4}{c}{6} & 6\\
\cmidrule(r){1-1} \cmidrule(lr){2-3} \cmidrule(lr){4-6} \cmidrule(lr){7-8} \cmidrule(lr){9-12} \cmidrule(l){13-13}
representation & \multicolumn{2}{c}{1} & \multicolumn{3}{c}{$8\oplus1$} & \multicolumn{2}{c}{$8\oplus1$} & \multicolumn{4}{c}{8} & 27 \\
\bottomrule \end{tabular}
\caption[Colour singlets and quark multilinears]{
Colour singlets and quark multilinears at the strong scale.
For each of them, we show respectively their order in $\epsilonSM$, their mass dimension $d$ and their flavour representation.
The composite gluon operators $\gluons$ and $w$ are defined in \eqref{eq:Q kinetic Lagrangian,eq:Q axial anomaly Lagrangian}, the quark bilinears $\nonet Q$ are defined in \eqref{eq:Q mass Lagrangian}, \allowbreak\eqref{eq:vectorial quark bilinears}, and \eqref{eq:EW quark bilinears}, and the quark quadrilinears $\octet O$ and $\tensor O$ are defined in \eqref{eq:octet parameter}, \allowbreak\eqref{eq:27-plet octet operators}, and \eqref{eq:27-plet operator}.
The corresponding external currents including their $\SM$ and $\BSM$ contribution are listed in \cref{tab:strong scale currents}.
} \label{tab:operators}
\end{table}

The \QCD Lagrangian at the strong scale can be written in the compact form
\begin{multline}
\mathcal L_Q
= \theta w
- \omega \gluons
- \trf{(\nonet m \nonet Q + \text{h.c.})
+ \octet l^\mu \nonet Q_\mu
+ \octet r^\mu \overline{\nonet Q}_\mu}
- \inv[2]{\Lambda_{\SM}} \trf{\octet \chromo \widetilde{\nonet Q}
+ \octet \tau^{\mu\nu} \nonet Q_{\mu\nu}
+ \text{h.c.}} \\
- \inv[2]{v} \trf{\octet \fermi_\scalar \octet O_\scalar
+ \octet \fermi_r \octet O_r
+ \octet \fermi_l^\asym \octet O_l^\asym
+ \octet \fermi_l^\sym \octet O_l^\sym}
- \inv[2]{v} \trF{\tensor \fermi_l^\sym \tensor O_l^\sym} \ ,
\end{multline}
where the gluon contributions are defined in \eqref{eq:Q kinetic Lagrangian,eq:Q axial anomaly Lagrangian}, the nonet contributions are defined in \eqref{eq:Q mass Lagrangian},
\allowbreak\eqref{eq:vectorial quark bilinears}, and \eqref{eq:EW quark bilinears}, the octet contributions are defined in \eqref{eq:octet parameter} and \eqref{eq:27-plet octet operators}, and the 27-plet contribution is defined in \eqref{eq:27-plet operator}.
All operators are also listed in \cref{tab:operators}.
Finally, the complete trace of the Hilbert \cref{eq:stress-energy tensor} that includes both strong and \EW contributions is
\begin{equation} \label[set]{eq:Q full stress-energy tensor}
\begin{split}
\mathcal T_Q &
= \frac{\beta_\strong}{\omega} \mathcal L_Q^\omega
- \mathcal L_Q^m
+ \mathcal L_Q^\tau
+ \mathcal L_Q^\chromo
- \mathcal L_Q^W
+ 2 \mathcal L_Q^{\fermi} \\ &
= \begin{multlined}[t]
- \beta_\strong \gluons(x)
+ \trf{(\nonet \m \nonet Q + \text{h.c.})
+ \octet l_W^\mu \nonet Q_\mu}
- \inv[2]{\Lambda_{\SM}} \trf{\octet \chromo \widetilde{\nonet Q}
+ \octet \tau^{\mu\nu} \nonet Q_{\mu\nu} + \text{h.c.}} \\
- 2 \inv[2]{v} \trf{\octet \fermi_\scalar \octet O_\scalar
+ \octet \fermi_r \octet O_r
+ \octet \fermi_l^\asym \octet O_l^\asym
+ \octet \fermi_l^\sym \octet O_l^\sym}
- \inv[2]{v} \trF{\tensor \fermi_l^\sym \tensor O_l^\sym} \ .
\end{multlined}
\end{split}
\end{equation}

\subbib

%% file: feynman-current.pgf
\begin{feyn}[node distance = \height/1 and \width/2]
\vertex (a1) {$l_\ell^\dagger$};
\vertex[right = of a1] (a2);
\vertex[right = of a2] (a3) {$\nu_\ell^\dagger$};
\vertex[below = of a1, align = right] (b1) {$\up$\\$\down$};
\vertex[at = (b1-|a2)] (b2);
\vertex[at = (b2-|a3), align=left] (b3) {$\down$\\$\up$};
\diagram* {
{[edges = fermion]
(a3) -- (a2) -- (a1),
(b1) -- (b2) -- (b3),
},
(a2) -- [boson, edge label' = $W$] (b2)
};
\end{feyn}

%% file: feynman-dipole-2.pgf
\begin{feyn}[node distance = \height/2 and \width/4]
\vertex (a1) {$\strange^\dagger$};
\vertex[right = of a1] (a11);
\vertex[right = of a11] (a2);
\vertex[below = of a2] (c0);
\vertex[right = of c0] (c1);
\vertex[right = of c1] (c2) {$\gamma$};
\vertex[shape = crossed  dot, below left = of c1] (b2);
\vertex[draw, crossed dot, at = (a11|-b2)] (x) {};
\vertex[at = (b2-|a1)] (b1) {$\down$};
\diagram*{
{[edges = fermion]
(b1) -- (x) -- (b2) --[edge label' = $u$] (c1) --[edge label' = $u^\dagger$] (a2) -- (a1),
},
(c2) -- [boson] (c1),
(b2) -- [boson, edge label = $W$] (a2)};
\end{feyn}

%% file: feynman-dipole.pgf
\begin{feyn}[node distance = \height/2 and \width/4]
\vertex (a1) {$\strange^\dagger$};
\vertex[right = of a1] (a11);
\vertex[right = of a11] (a2);
\vertex[below = of a2] (c0);
\vertex[right = of c0] (c1);
\vertex[right = of c1] (c2) {$g$};
\vertex[shape = crossed  dot, below left = of c1] (b2);
\vertex[draw, crossed dot, at = (a11|-b2)] (x) {};
\vertex[at = (b2-|a1)] (b1) {$\down$};
\diagram*{
{[edges = fermion]
(b1) -- (x) -- (b2) --[edge label' = $u$] (c1) --[edge label' = $u^\dagger$] (a2) -- (a1),
},
(c2) -- [gluon] (c1),
(b2) -- [boson, edge label = $W$] (a2)};
\end{feyn}

%% file: feynman-tree.pgf
\begin{feyn}[node distance = \height/1 and \width/2]
\vertex (a1) {$\strange^\dagger$};
\vertex[right = of a1] (a2);
\vertex[right = of a2] (a3) {$\up^\dagger$};
\vertex[below = of a1, align = right] (b1) {$\up$\\$\down$};
\vertex[at = (b1-|a2)] (b2);
\vertex[at = (b2-|a3), align=left] (b3) {$\down$\\$\up$};
\diagram* {
{[edges = fermion]
(a3) -- (a2) -- (a1),
(b1) -- (b2) -- (b3),
},
(a2) -- [boson, edge label' = $W$] (b2)
};
\end{feyn}

%% file: feynman-gluon.pgf
\begin{feyn}[node distance = \height/2 and \width/4]
\vertex (a1) {$\strange^\dagger$};
\vertex[right = of a1] (a2);
\vertex[below = of a2] (c0);
\vertex[right = of c0] (c1);
\vertex[right = of c1] (c2);
\vertex[below left = of c1] (b2);
\vertex[at = (b2-|a1)] (b1) {$\down$};
\vertex[right = of c2] (c3);
\vertex[at = (a2-|c3), align = left] (a3) {$q^\dagger$\\$\overline q^\dagger$};
\vertex[at = (b2-|c3), align = left] (b3) {$q$\\$\overline q$};
\diagram*{
{[edges = fermion]
(b1) -- (b2) --[edge label' = $u$] (c1) --[edge label' = $u^\dagger$] (a2) -- (a1),
(a3) -- (c2) -- (b3),
},
(c2) -- [gluon, edge label' = $g$] (c1),
(b2) -- [boson, edge label = $W$] (a2)};
\end{feyn}

%% file: portals.tex
\glsreset{PET}

\subtoc

\section{Portal interactions between the SM and hidden sectors} \label{sec:portals}

In this section, we present a framework for the construction of general \PETs, and use it to construct \EW and strong scale \PETs that couple \SMEFT and \LEFT to a light messenger of spin 0, \textfrac12, or 1.
The portal \SMEFTs comprise all independent portal operators up to dimension five, and the portal \LEFTs additionally encompass quark-flavour violating portal operators of dimension six and seven.
The latter are necessary to capture quark-flavour violating transitions, which govern for instance hadronic kaon decays.
We use the accidental symmetries of the portal \SMEFTs to further constrain the shape of the corresponding portal \LEFTs, so that these \PETs should be understood as the \LE limit of the portal \SMEFTs,
in which the heavy \SM \DOFs have been integrated out.

For completeness, we provide in \cref{sec:higher spin} a basis of independent portal operators with dimension five or less that couple \SMEFT to hidden particles with spin \textfrac32 and 2.

\subsection{Portal effective theories}

A \PET is an \EFT that couples \SM \DOFs to hidden sectors via messenger fields.
The framework we present is generic and can be used to construct \PETs by starting from any \EFT that either encompasses or is derived from the \SM, such as \SMEFT, \HEFT, \LEFT, \HQET, or \cPT.
The \PET Lagrangian can be cast as
\begin{equation} \label[lag]{eq:SM portal hidden Lagrangian}
\mathcal L = \mathcal L_{\EFT} + \mathcal L_\text{portal} + \mathcal L_\text{hidden} \ ,
\end{equation}
where the original \EFT Lagrangian $\mathcal L_{\EFT}$ and the hidden Lagrangian $\mathcal L_\text{hidden}$ depend only on \SM and hidden fields, respectively.
The portal Lagrangian $\mathcal L_\text{portal}$ contains all available operators that couple the \SM fields to the hidden messenger fields.
Since we aim to capture the physics of the portal Lagrangian while remaining agnostic about the hidden sector, the hidden Lagrangian may be fully general.
In particular, it can contain, in addition to the messenger field, \emph{secluded} fields with arbitrary masses, quantum numbers, and interactions, that do not couple directly to the \SM particles.
This idea is schematically depicted in \cref{fig:PET framework}.
We integrate out all hidden fields with masses well above the characteristic energy scale of the relevant \EFT.
This does not restrict the regime of applicability of the resulting \PET, since the \EFT by itself, even without being coupled to hidden sectors, already becomes invalid at energies well above its characteristic energy scale.
The impact of the heavy particles is captured by an infinite tower of higher dimensional operators in the \EFT, portal, and hidden Lagrangians, which contain only the remaining light \SM and hidden fields.

In the remainder of this section, we construct \PETs that couple the \SM to a single messenger field of spin 0, \textfrac12, and 1.
We begin by constructing \EW scale \PETs that extend \SMEFT, and then use the resulting portal \SMEFTs as a starting point to derive a corresponding set of strong scale \PETs that extend \LEFT.
In the first step, we take the typical energy scale of \SMEFT to be the Higgs \VEV, and in the second step, we take the typical energy scale of \LEFT to be around $\unit[1]{GeV}$, which corresponds roughly the proton mass.
When extending \SMEFT, we assume that the messenger is a singlet under the full \SM gauge group $G_{\SM} = \SU(3)_c \times \SU(2)_L \times \U(1)_Y$ in order to remain consistent with the \SMEFT setup,
but for the \PETs that extend \LEFT we only assume that the messenger field is invariant under the broken \SM gauge group $G_\text{sm} = \SU(3)_c \times \U(1)_V$.
We do not assume that the portal \SMEFTs respect any additional symmetries, such as gauge symmetries or a new parity of the hidden sector.
In particular, we allow for both P and \CP violating portal interactions.
However, we use the accidental symmetries of the portal \SMEFTs to constrain the shape of the corresponding portal \LEFTs.

\subsubsection{Power counting} \label{sec:pet power counting}

The lack of evidence for light sectors at colliders and fixed target experiments \cite{Alekhin:2015byh, Beacham:2019nyx, Agrawal:2021dbo} implies that any portal interaction has to be strongly suppressed.
In order to reflect this suppression, we normalise all portal operators such that they contain at least one explicit degree of smallness $\epsilon_i$, independent of their mass dimension.
Physically, these degrees of smallness can result from a wide variety of mechanisms that do not have to be connected to each other, such as the small breaking of an approximate symmetry of the theory.
At the \EW scale, unitarity implies that higher dimensional portal operators with mass dimension larger than four must be dimensionally suppressed by factors $\epsilon_i^{d-4} = (\flatfrac{v}{f_i})^{d-4}$, where $f_i$ is some \UV scale.
For our purposes, it is not necessary to distinguish between the various degrees of smallness $\epsilon_i$.
Therefore, we define the generic degree of smallness
\begin{align} \label{eq:epsilon UV}
\epsilonUV &= \max\displaylimits_i \epsilon_i = \frac{v}{f_{\UV}} \ , & f_{\UV} &\gg v \ ,
\end{align}
and only count powers of $\epsilonUV$ rather than distinguishing between various sources of smallness for the portal operators.
Using this power counting, portal operators of mass-dimension three, four, and five are suppressed by a single factor of $\epsilonUV$, while higher dimensional portal operators are suppressed by higher powers of $\epsilonUV$, due to the required dimensional suppression.

When constructing the portal \SMEFTs in \cref{sec:EW operators}, we neglect portal operators with mass-dimension six or higher, and in the remainder of this work, we use these \PETs as the starting point for the subsequent construction of the strong scale portal \LEFT and \cPT Lagrangians.
This constraint restricts the types of hidden sectors we are able to describe.
For one, some \SM extensions couple to the \SM only via operators of mass-dimension six or higher.
For example, this is the case of fermionic \DM models that couple to the \SM via four-fermion interactions of dimension six, see \eg \cite{Kavanagh:2018xeh,Arina:2020mxo}.
In addition, higher dimensional portal operators can mediate transitions that are not captured by lower dimensional portal operators.
As we show in \cref{sec:EW operators}, this is the case for baryon-number violating portal interactions, which only appear starting at dimension six.
However, we emphasise that these limitations are not a consequence of the \PET approach as such, but merely a consequence of our choice to only account for portal operators up to dimension five.
We leave the investigation of \PETs with operators of dimension six or higher for future work.

\subsubsection{Mixing between SM and messengers fields} \label{sec:hidden sector mixing}

Generically, the portal sector contains quadratic operators that mix neutral \SM fields with hidden fields.
Even though it is possible to diagonalise the portal Lagrangian such that these quadratic operators are effectively eliminated from the theory, this diagonalisation would induce two new types of portal operators:
First, one would obtain portal operators that mirror \SM interactions, except that one \SM field is replaced by a messenger field.
Second, one would obtain new portal operators that mirror \emph{hidden sector} interactions involving the messenger fields, except that one messenger field is replaced by a neutral \SM field.
This second type of portal operator conflicts with our strategy of being agnostic about the internal structure of the hidden sector, as it introduces direct coupling between the secluded fields and the neutral \SM fields.
Listing all of the corresponding portal operators is impossible without making further assumptions about the hidden sector.
Therefore, we do not diagonalise any of the quadratic portal interactions.

However, in principle, it is necessary to diagonalise the portal mixing in order to construct the proper asymptotic energy eigenstates of the theory.
This can be avoided when performing perturbative calculations at fixed order in $\epsilonUV$, since the undiagonalised fields approximately overlap with the asymptotic energy eigenstates of the theory in the limit of small $\epsilonUV$.
However, it may be necessary to re-sum the quadratic portal interactions in order to describe certain effects that cannot be captured by fixed-order computations in perturbation theory.
For example, consider a type-I seesaw model in which the \SM is augmented by a single \HNL.
In order to capture neutrino oscillations in this model, it is necessary to re-sum the mass-mixing between the \SM neutrinos and the \HNL.
However, this does not affect the computation of $S$-matrix elements for microscopic scattering amplitudes, since these oscillations typically occur over macroscopic distances, \eg over several kilometers in case of neutrinos produced in nuclear reactors \cite{Mariani:2012cb}.

\subsection{Electroweak scale portal effective theories} \label{sec:EW operators}

We explicitly construct the \EW scale \PETs that couple \SMEFT to a single messenger of spin 0, \textfrac12, or 1, and give a complete basis of portal operators with mass dimension five or less for each resulting portal \SMEFT.
We then use these \PETs to define a set of portal currents that parameterise the coupling of \SMEFT to generic hidden sectors, and study the shape of the portal \SMEFTs after \EWSB.

\subsubsection{Minimal bases of portal operators}

\begin{table}
\newcommand{\portal}[2]{$#1$&$#2$}
\newcommand{\qcdportal}[3]{$#1$&$#2#3$}
\newcommand{\field}[2]{\midrule\multirowcell{#1}{#2}}
\newcolumntype{p}{r@{}l}
\begin{tabular}{lc*4p} \toprule
& $d$ & \multicolumn{2}{c}{Higgs} & \multicolumn{2}{c}{$\text{Yukawa}+\text{h.c.}$} & \multicolumn{2}{c}{Fermions} & \multicolumn{2}{c}{Gauge bosons} \\
\field{8}{$s_i$} & 3 & \portal{s_i}{\abs{H}^2} \\
\cmidrule{2-10} & 4 & \portal{s_i s_j}{\abs{H}^2} \\
\cmidrule{2-10} & \multirow{6}{*}{5} & \portal{s_i s_j s_k}{\abs{H}^2} & \qcdportal{s_i}{\qu_a \overline u_b}{\widetilde H^\dagger} & & & \qcdportal{s_i}{G^a_{\mu\nu} G_a^{\mu\nu}}{} \\
& & \portal{s_i}{D^\mu H^\dagger D_\mu H} & \qcdportal{s_i}{\qu_a \overline d_b}{H^\dagger} & & & \portal{s_i}{W^a_{\mu\nu} W_a^{\mu\nu}} \\
& & \portal{s_i}{\abs{H}^4} & \portal{s_i}{\ell_a \overline e_b H^\dagger} & & & \portal{s_i}{B_{\mu\nu} B^{\mu\nu}} \\
& & & & & & & & \qcdportal{s_i}{G^a_{\mu\nu} \widetilde G_a^{\mu\nu}}{} \\
& & & & & & & & \portal{s_i}{W^a_{\mu\nu} \widetilde W_a^{\mu\nu}} \\
& & & & & & & & \portal{s_i}{B_{\mu\nu} \widetilde B^{\mu\nu}} \\
\field{2}{$\lhf_a$\\+\\[-.6ex]h.c.} & \multirow{1}{*}{4} & & & \portal{\lhf_a}{\ell_b \widetilde H^\dagger} \\
\cmidrule{2-10} & \multirow{1}{*}{5} & \portal{\lhf_a \lhf_b}{\abs{H}^2} & \portal{\lhf_a^\dagger}{\overline \sigma^\mu \ell_b D_\mu \widetilde H^\dagger}
& & & \portal{\lhf_a \sigma^{\mu\nu} \lhf_b}{B_{\mu\nu}} \\
\field{5}{$v^\mu$} & \multirow{5}{*}{4} & \portal{v_\mu v^\mu}{\abs{H}^2} & & & \qcdportal{v^\mu}{\qu_a^\dagger \overline \sigma_\mu \qu_b}{} \\
& & \portal{\partial_\mu v^\mu}{\abs{H}^2} & & & \qcdportal{v^\mu}{\overline u_a^\dagger \sigma_\mu \overline u_b}{} \\
& & \portal{v^\mu}{H^\dagger \overleftright D_\mu H}
& & & \qcdportal{v^\mu}{\overline d_a^\dagger \sigma_\mu \overline d_b}{} \\
& & & & & & \portal{v^\mu}{\ell_a^\dagger \overline \sigma_\mu \ell_b} \\
& & & & & & \portal{v^\mu}{\overline e_a^\dagger \sigma_\mu \overline e_b} \\
\bottomrule \end{tabular}
\caption[Portal $\SMEFT$ operators involving messengers with spin 0, \textfrac12, and 1]{
List of all operators up to dimension five with \SM fields and spin 0 ($s_i$ with $i = 1$, $2$), spin \textfrac12 ($\lhf_a$ with $a = 1$, $2$) or spin 1 ($v^\mu$) messengers.
The first column specifies the spin of the messenger field, the second column denotes the dimension $d$ of the operator and the remaining columns label the \SM sectors the messengers interact with.
The left-handed $\SU(2)$ doublets $\ell_a = (\nu_a$, $e_a)^T$ and $\qu_a = (u_a$, $d_a)^T$ and the right-handed singlets $\overline u_a^\dagger$, $\overline d_a^\dagger$, and $\overline e_a^\dagger$ are Weyl fermions.
} \label{tab:pure operators}
\end{table}

In general, a naive listing of all possible portal operators with mass-dimension five or less will contain numerous redundant operators.
In order to obtain a minimal set of independent portal operators for each type of messenger, we use the reduction techniques collected in \cref{sec:redundant}.
The resulting operator basis is presented in \cref{tab:pure operators}.
We consider three types of messengers:
\begin{description}
\item[Spin 0 fields] can be either real \prefix{pseudo}{scalar} or complex scalar fields.
As we do not require portal interactions to conserve parity, pseudoscalar and scalar fields couple to \SMEFT via the same set of portal interactions.
Furthermore, a complex scalar couples to \SMEFT in the same way as two real scalar fields.
Therefore, we can account for all types of spin 0 messengers by considering how \SMEFT couples to two real scalar fields $s_1(x)$ and $s_2(x)$.
These can interact with the \SM fields via a minimal basis of 14 different operators with dimensions ranging from three to five.
There are twelve additional redundant operators.

\item[Spin \textfrac12 fields] can be either Weyl, Majorana, or Dirac fermions.
Without loss of generality, a Dirac fermion can be written as a combination of two left-handed Weyl fermions, while a Majorana fermion can be written as single left-handed Weyl fermion.
Therefore, we can account for all types of fermionic messengers by considering how \SMEFT couples to two left-handed Weyl fermions $\lhf_1(x)$ and $\lhf_2(x)$.
These can interact with the \SM fields via a minimal basis of four portal operators of dimension four and five.
Additionally, there are two redundant operators.
Notice that the operator $\xi_a \sigma^{\mu\nu} \xi_b B_{\mu\nu}$ is antisymmetric under exchange of $a$ and $b$, so that it can only contribute if \SMEFT couples to a Dirac fermion.

\item[Spin 1 fields] can be either vector or axial-vector fields.
As we do not require portal interactions to conserve parity, both of these can couple to \SMEFT via the same portal interactions, and we can account for both possibilities by considering how \SMEFT couples to a vector field $v^\mu(x)$.
These can interact with the \SM fields via a minimal basis of eight independent operators with mass-dimension four.
Notably, there are no operators of dimension five.
There are two additional redundant operators.

\end{description}
For the sake of completeness, we list the redundant operators in \cref{sec:redundant ew portal operators}.
If the internal structure of the hidden sector is known, it is potentially possible to discard further operators by using \eg the \EOMs for the messenger field.
As discussed in \cref{sec:hidden sector mixing}, this may involve other hidden sector fields besides the messenger.
Here and in the following, we refrain from making such model dependent simplifications.

All of the above portal operators conserve baryon number, and portal operators with spin 0 and 1 messengers also conserve lepton number.
Portal operators with spin \textfrac12 messengers can violate lepton number by one unit.
Furthermore, portal operators with spin \textfrac12 messengers do not couple to either the \SM quark fields or any of the right-handed charged lepton fields, and operators with spin 1 messengers only couple to pairs of quarks and leptons with identical chirality,
so that they cannot serve as a separate source of chiral symmetry breaking.
This becomes important when constructing strong scale \PETs, since it implies that some strong scale portal operators are subdominant as a result of chiral suppression due to a light \SM fermion mass insertion.

Further, we note that, although we have focused on the case in which \SMEFT couples only to a single messenger field, the portal sector defined by the operators in \cref{tab:pure operators} already captures interactions between \SMEFT and an arbitrary number of messengers with \emph{identical spin}.
For sets of messengers $s_i$ or $\xi_i$ or $v^\mu_i$, it is sufficient to iterate over all possible values for the index $i$ in the portal operators.
However, we do not account for the possibility of coupling \SMEFT to multiple messengers with \emph{different spin}.

\subsubsection{External current description}

\begin{table}
\begin{tabular}{*2r*8c} \toprule
& & $S_m^H$ & $S_x$ & $\nonet S_x$ & $\lhfc$ & $V_H^\mu$ & $\nonet V_x^\mu$ & $\lhfc_\mu$ & $T_{\mu\nu}$ \\
\midrule
& spin & \multicolumn{3}{c}{0} & $\nicefrac12$ & \multicolumn{2}{c}{1} & $\nicefrac32$ & 2 \\
\cmidrule(r){2-2} \cmidrule(lr){3-5} \cmidrule(lr){6-6} \cmidrule(lr){7-8} \cmidrule(lr){9-9} \cmidrule(l){10-10}
& $d$ & 2 & \multicolumn{2}{c}{0} & $\nicefrac32$ & \multicolumn{2}{c}{1} & $\nicefrac32$ & 2 \\
\cmidrule(r){1-2} \cmidrule(lr){3-3} \cmidrule(lr){4-5} \cmidrule(lr){6-6} \cmidrule(lr){7-8} \cmidrule(lr){9-9} \cmidrule(l){10-10}
\multirow{3}{*}{flavour} & representation & 1 & 1 & $8\oplus1$ & 3 & 1 & $8\oplus1$ & 3 & 1 \\
& symmetry & & & & & & $\nonet V_\mu^\dagger = \nonet V_\mu$ \\
& $\DOFs$ & 1 & 1 & 18 & 3 & 1 & 9 & 3 & 1 \\
\bottomrule \end{tabular}
\caption[Portal $\SMEFT$ currents]{
Properties of the portal $\SMEFT$ currents.
The first two rows list spin and mass dimension $d$, and the remaining rows list the representation and symmetries under flavour transformations as well as the resulting number of $\DOFs$.
} \label{tab:EW currents}
\end{table}

It is convenient to collect all of the operators associated with the three messenger fields into a single portal \cref{eq:SM portal hidden Lagrangian}.
We separate the portal operators into a Higgs $H$, a Yukawa like $Y$, a fermionic $F$, and a gauge $V$ sector
\begin{equation} \label[lag]{eq:EW portal Lagrangian}
\mathcal L_\text{portal} = \mathcal L_{\EW}^H + \mathcal L_{\EW}^Y + \mathcal L_{\EW}^F + \mathcal L_{\EW}^V \ .
\end{equation}
The individual Lagrangians are
\footnote{\label{fn:notation}%
The Higgs doublet is denoted by $H$, and its conjugate is $\widetilde H = - \frac\i2 \sigma_2 H^\dagger$.
We abbreviate $\abs{H}^2 = H^\dagger H$ and the antisymmetrised derivative is $H^\dagger \overleftright \partial^\mu H = (\partial^\mu H)^\dagger H - H^\dagger \partial^\mu H$.
The $G_a^{\mu}$ are the gluon fields, while $W_a^\mu$ and $B^\mu$ denote the \EW gauge bosons.
The field strength tensors are given as $V_a^{\mu\nu} = \partial^\mu V_a^\nu - \partial^\nu V_a^\mu - \i f_{abc} V_b^\mu V_c^\nu $.
}
\begin{subequations}
\begin{align} \label[lag]{eq:higgs Lagrangian}
\mathcal L_{\EW}^H &
= S_m^H \abs{H}^2
+ \frac12 S_\lambda^H \abs{H}^4
+ S_\kappa^H D^\mu H^\dagger D_\mu H
+ \i V_H^\mu \boson{H^\dagger \overleftright D_\mu H}
\ , \\ \label[lag]{eq:yukawa Lagrangian}
\mathcal L_{\EW}^Y &
= \nonet S_m^e \ell \overline e H^\dagger
+ \nonet S_m^d \qu \overline d H^\dagger
+ \nonet S_m^u \qu \overline u \widetilde H^\dagger
+ \lhfc \ell \widetilde H^\dagger
+ \lhfc_\mu \ell D^\mu \widetilde H^\dagger
+ \text{h.c.}
\ , \\ \label[lag]{eq:fermionic Lagrangian}
\mathcal L_{\EW}^F &
= \nonet V_q^\mu \qu^\dagger \overline{\sigma}_\mu \qu
+ \nonet V_\ell^\mu \ell^\dagger \overline{\sigma}_\mu \ell
+ \nonet V_u^\mu \overline u^\dagger \sigma_\mu \overline u
+ \nonet V_d^\mu \overline d^\dagger \sigma_\mu \overline d
+ \nonet V_e^\mu \overline e^\dagger \sigma_\mu \overline e
\ , \\ \label[lag]{eq:gauge Lagrangian}
\mathcal L_{\EW}^V &
\begin{multlined}[t]
= (S_\omega^B B_{\mu\nu} + S_\theta^B \widetilde B_{\mu\nu} + T^B_{\mu\nu}) B^{\mu\nu}
+ (S_\omega^W W_{\mu\nu} + S_\theta^W \widetilde W_{\mu\nu}) W^{\mu\nu} \\
+ (S_\omega G_{\mu\nu} + S_\theta \widetilde G_{\mu\nu}) G^{\mu\nu} \ .
\end{multlined}
\end{align}
\end{subequations}
Lepton and quark doublets are written as left-handed Weyl fermions $\ell_a = (\nu_a, e_a)^T$ and $\qu_a = (u_a, d_a)^T$, and the singlets as conjugated left-handed Weyl fermions $\overline u_a^\dagger$, $\overline d_a^\dagger$, and $\overline e_a^\dagger$.\cref{fn:bar}
\Cref{tab:EW currents} summarises the properties of the scalar $S$, fermionic $\lhfc$, and vectorial $V^\mu$ portal currents.
The \emph{scalar current} of mass-dimension two that appears in the Higgs mass-like term in \cref{eq:higgs Lagrangian} is
\begin{equation}
S_m^H = \epsilonUV \left[
 v c^{S_m^H}_i s_i
+ c^{S_m^H}_{ij} s_i s_j
+ c^{S_m^H}_{v^2} v^\mu v_\mu
+ c^{S_m^H}_{\partial v} \partial^\mu v_\mu
+ \frac{1}{v} \left(c^{S_m^H}_{ijk} s_i s_j s_k + c^{S_m^H}_{ab} \lhf_a^\dagger \lhf_b \right)
\right] \ ,
\end{equation}
where the $c^\text{current}_\text{operator}$ are dimensionless Wilson coefficients.
The other \prefix{\emph{pseudo}}{\emph{scalar}} \emph{currents} of mass dimension zero in \cref{eq:higgs Lagrangian,eq:yukawa Lagrangian,eq:gauge Lagrangian} are
\begin{align}
S_x &= \frac{\epsilonUV}{v} c^{S_x}_i s_i \ , &
\nonet S_x &= \frac{\epsilonUV}{v} \nonet c^{S_x}_i s_i \ ,
\end{align}
where $x$ symbolically labels the different scalar currents.
The \emph{left-handed fermionic currents} in \cref{eq:yukawa Lagrangian} are
\begin{align}
\lhfc &= \epsilonUV c^\lhfc_a \lhf_a \ , &
\lhfc^\mu &= \epsilonUV c^\lhfc_{\partial a} \lhf_a^\dagger \overline \sigma_\mu \ ,
\end{align}
and the \emph{vectorial currents} in \cref{eq:higgs Lagrangian,eq:fermionic Lagrangian} are
\begin{align}
V_x^\mu &= \epsilonUV c^x_v v^\mu \ , &
\nonet V_x^\mu &= \epsilonUV \nonet c^x_v v^\mu \ ,
\end{align}
where the matrix valued vectorial currents and its Wilson coefficient are Hermitian.
The \emph{tensorial current} in \cref{eq:gauge Lagrangian} is
\begin{equation}
T_{\mu\nu} = \frac{\epsilonUV}{v} c^T_{ab} \lhf_a^\dagger \sigma_{\mu\nu} \lhf_b \ .
\end{equation}

\subsubsection{Electroweak symmetry breaking}

After \EWSB, the Higgs field $H$ acquires a finite \VEV $v$, which induces a shift in the currents.
In unitary gauge, the portal \cref{eq:EW portal Lagrangian} becomes
\footnote{In unitary gauge, the Higgs field is given as $H = (0,v + \flatfrac{h}{\sqrt 2})^T$, and $\widetilde H = - (v + \flatfrac{h}{\sqrt 2}, 0)^T$.}
\begin{subequations}
\begin{align} \label[lag]{eq:higgs Lagrangian ewsb}
\mathcal L_{\EWSB}^H &
= \begin{multlined}[t]
\frac12 S_\kappa^H \partial_\mu h \partial^\mu h
+ \frac12 S_\lambda^H \left(v
+ \higgs \right)^4 \\
+ \left(v + \higgs \right)^2 \left(
 \frac12 S_\kappa^H \left(W^+_\mu W^{- \mu} + \frac12 Z^\mu Z_\mu \right)
+ V_H^\mu Z_\mu
+ S_m^H \right) \ ,
\end{multlined} \\ \label[lag]{eq:yukawa Lagrangian ewsb}
\mathcal L_{\EWSB}^Y &
= \frac1{\sqrt2} (v+H) \left(
\widehat{\nonet S}_m^{e\bar e} e \overline e
+ \widehat{\nonet S}_m^{d \bar d} d \overline d
+ \widehat{\nonet S}_m^{u \bar u} u \overline u
- \lhfc \nu \right)
- \frac{\partial^\mu h}{\sqrt2} \lhfc_\mu \nu
+ \text{h.c.} \ , \\ \label[lag]{eq:fermionic Lagrangian ewsb}
\mathcal L_{\EWSB}^F &
= \begin{multlined}[t]
\widehat{\nonet V}_{qu}^\mu u^\dagger \overline \sigma_\mu u
+ \widehat{\nonet V}_{qd}^\mu d^\dagger \overline \sigma_\mu d
+ \widehat{\nonet V}_{u\bar u}^\mu \overline u \sigma_\mu \overline u^\dagger
+ \widehat{\nonet V}^\mu_{d \bar d} \overline d \sigma_\mu \overline d^\dagger \\
+ \widehat{\nonet V}_{\ell e}^\mu e^\dagger \overline \sigma_\mu e
+ \widehat{\nonet V}_{e \bar e}^\mu \overline e \sigma_\mu \overline e^\dagger
+ \nonet V_\ell^\mu \nu^\dagger \overline \sigma_\mu \nu \ ,
\end{multlined} \\ \label[lag]{eq:gauge Lagrangian ewsb}
\mathcal L_{\EWSB}^V &
\begin{multlined}[t]
= \left(S_\omega^Z \overline Z_{\mu \nu} + S_\theta^Z \widetilde{\overline Z}_{\mu \nu} + T^Z_{\mu\nu}\right) \overline Z^{\mu \nu}
+ \left(S_\omega^\photon \photon_{\mu \nu} + S_\theta^\photon \widetilde \photon_{\mu \nu} + T^\photon_{\mu\nu}\right) \photon^{\mu \nu} \\
\hspace{-3em}
+ \left(S_\omega G_{\mu \nu} + S_\theta \widetilde G_{\mu \nu} \right) G^{\mu \nu}
+ \left(S_\omega^{\photon Z} \overline Z_{\mu \nu} + S_\theta^{\photon Z} \widetilde{\overline Z}_{\mu \nu} \right) \photon^{\mu \nu}
+ 2 \left(S_\omega^W \overline W^+_{\mu \nu} + S_\theta^W \widetilde{\overline W}^+_{\mu \nu}\right) \overline W_-^{\mu \nu} \\
\hspace{-2em}
- 2 \i \left(S_\omega^W \partial^{\nu\rho\sigma} - 2 S_\theta^W \epsilon^{\mu \nu \rho \sigma} \partial_\mu \right) W_\nu^3 W_\rho^+ W_\sigma^-
+ 4 S_\omega^W g^{\mu[\nu} g^{\rho]\sigma} \left(2 W^3_\mu W^3_\nu + W^+_\mu W^-_\nu \right) \ ,
\end{multlined}
\end{align}
\end{subequations}
where
\begin{equation}
\partial^{\mu\rho\sigma}
= g^{\rho \sigma} \left(\partial_+ - \partial_- \right)^\mu
+ g^{\sigma \mu} \left(\partial_- - \partial_3 \right)^\rho
+ g^{\mu \rho} \left(\partial_3 - \partial_+ \right)^\sigma
\end{equation}
and we have defined the new \emph{scalar currents}
\begin{align}
S_x^Z &= c_w^2 S_x^W + s_w^2 S_x^B \ , &
S_x^\photon &= s_w^2 S_x^W + c_w^2 S_x^B \ , &
S_x^{\photon Z} &= 2 c_w s_w (S_x^W - S_x^B) \ ,
\end{align}
as well as the new \emph{tensorial currents}
\begin{align}
T^\photon_{\mu\nu} &= c_w T^B_{\mu\nu} \ , &
T^Z_{\mu\nu} &= -s_w T^B_{\mu\nu} \ ,
\end{align}
that couple directly to the photon and $Z$-boson field strength tensors, with $c_w$ and $s_w$ denoting the \prefix{co}{sine} of the \EW mixing angle.
In \cref{eq:higgs Lagrangian ewsb,eq:fermionic Lagrangian ewsb}, we used a singular value decomposition in order to diagonalise the \SM fermion mass matrices $\nonet m_{xy} = \nonet U_x \nonet m_x \nonet U^\dagger_y$ via a unitary rotation of the \SM fermion fields.
The resulting mass-diagonal \SM fermions couple to the rotated portal currents
\begin{align}
\widehat{\nonet S}_m^{xy} &= \nonet U_y^\dagger \nonet S_m^x \nonet U_x \ , &
\widehat{\nonet V}_{xy}^\mu &= \nonet U_y^\dagger \nonet V_x^\mu \nonet U_y \ .
\end{align}
Note that the \CKM matrix $\nonet V_{\CKM} = \nonet U_d^\dagger \nonet U_u$ and the \PMNS matrix that one obtains after diagonalising the neutrino to hidden sector mass mixing are the only combinations of the $\nonet U_x$ constrained by measuring \SM or portal interactions in the broken phase.
This implies that such observations cannot fully constrain the shape of the unrotated portal currents $\nonet S_m^x$ and $\nonet V_x$ that couple to the \SM fermion gauge eigenstates.
This may be of interest when trying to constrain the shape of the portal interactions at high temperatures or in the early universe with collider or fixed-target experiments.

\subsection{Portals at the strong scale} \label{sec:currents}

At the strong scale, which we define to be roughly the scale associated with the \GD contribution $\sim \unit[1]{GeV}$ to the proton mass, the \SM dynamics is captured by \LEFT, which
contains only the massless gauge bosons, electrons, muons, neutrinos, and the light quarks ($\up$, $\down$, and $\strange$).
Starting from the previously constructed portal \SMEFTs, we now derive the strong scale \PETs that couple \LEFT to a single messenger of spin 0, \textfrac12, or 1.
While we have only included portal operators of dimension $d \leq 5$ in the portal \SMEFTs, we now also include quark-flavour violating $d \leq 7$ portal operators.
These operators are generated by diagrams that include virtual $W$-boson exchanges and are necessary to capture quark-flavour violating transitions, such as decays of charged kaons into pions and hidden fields, at \LO in $\epsilonSM$.

\begin{figure}
\tikzsetnextfilename{hidden}\input{hidden.pgf}
\caption[The two possible types of strong scale quark-flavour violating diagrams]{
Schematic representation of the two possible types of quark-flavour violating diagrams at the strong scale, which we distinguish based on the sector in which the flavour violation is located.
We assume that the relevant strong scale \PET is the \LE limit of a corresponding \EW scale portal \SMEFT.
The diagrams show the suppression due to $\NDA$ power counting and the dimension of the operators in the diagram.
\Cref{it:one} diagrams contain a flavour \emph{violating} \SM sub-diagram that scales as $(4\pi)^2$ and contains one $d \leq 6$ Fermi theory operator, as well as one flavour conserving portal operator that scales as $(4\pi)^0$ and has $d \leq 5$.
\Cref{it:two} diagrams contain a flavour \emph{conserving} \SM sub-diagram that scales as $(4\pi)^0$ and contains only renormaliseable $d \leq 4$ operators, as well as one flavour violating portal operators that scales as $(4\pi)^2$ and have $d = 6$, $7$ or $5$, $6$.
\Cref{it:one} diagrams with $d = 5$ portal operators and \cref{it:two} diagrams with $d = 7$ portal operators can appear in strong scale \PETs with spin 1 messengers that are derived from other \EW scale \PETs besides portal \SMEFT.
See also \cref{sec:nda} and \cite{Manohar:1983md, Jenkins:2013sda, Gavela:2016bzc, Manohar:2018aog} for details on the $\NDA$ counting.
} \label{fig:flavour violating processes}
\end{figure}
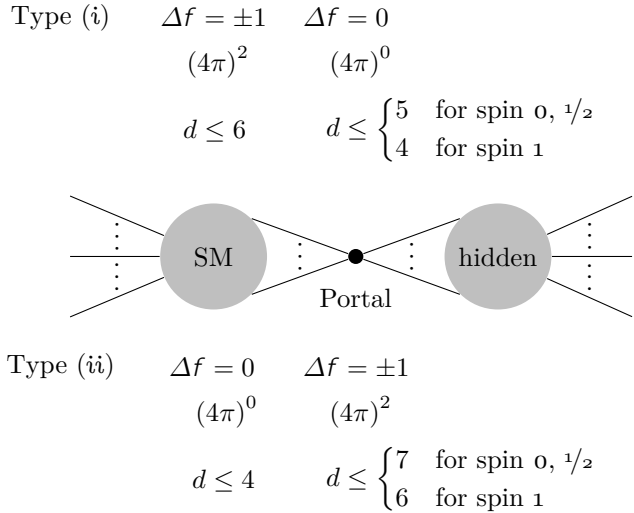

To see why it is necessary to include the higher dimensional operators when constructing a general strong scale \PET, consider a generic quark-flavour violating transition at the strong scale.
Such a transition has to be suppressed by at least one degree of smallness $\epsilonUV$, and another degree of smallness $\epsilonSM \equiv \flatfrac{\partial^2}{\Lambda_{\SM}^2}$, \cf \eqref{eq:epsilon SM}.
At $\order{\epsilonUV \epsilonEW}$, quark-flavour violating processes are described by the two types of diagram depicted in \cref{fig:flavour violating processes}:
\begin{enumerate}[label = (\textui{\roman*})]
\item\label{it:one}%
Diagrams with one quark-flavour \emph{violating} dimension six \SM charged current vertex and one quark-flavour \emph{conserving} strong scale portal vertex.
\item\label{it:two}%
Diagrams with a renormaliseable quark-flavour \emph{conserving} \SM vertex and a quark-flavour \emph{violating} strong scale portal vertex.
\end{enumerate}
To fully capture quark-flavour violating transitions one has to include all portal operators that can appear in either type of diagram.

First, consider the set of portal operators that can appear in \cref{it:one} diagrams:
The \SM charged current interaction that appears in these diagrams is associated with a suppression factor $\epsilonSM$.
Since the overall diagram has to scale as $\epsilonUV \epsilonSM$, portal operators that contribute to the diagrams cannot have a higher mass-dimension than their \EW scale counterparts, as this would imply further suppression by powers of $\sqrt{\epsilonSM}$.
Hence, to capture all \cref{it:one} diagrams, it is sufficient to include quark-flavour conserving portal operators with spin 0 or \textfrac12 messengers
that are at most of dimension five and quark-flavour conserving portal operators with spin 1 messengers that are at most of dimension four.
If the strong scale \PET is the \LE limit of another \EW scale \PET besides \SMEFT, dimension five portal operators with spin 1 messengers can also contribute to \cref{it:one} diagrams.

Next, consider the set of portal operator that can appear in \cref{it:two} diagrams.
Since these diagrams do not contain a \SM four-fermion vertex, they can contain portal operators that are suppressed by a factor $\epsilonUV\epsilonSM$ rather than just a factor $\epsilonUV$.
These portal operators are generated by diagrams in the \EW scale theory that contain a virtual $W$-boson exchange, and they can have a mass-dimension that is at most the mass-dimension of the corresponding \EW scale portal operators plus two.
Therefore, to capture all \cref{it:two} diagrams, one has to include quark-flavour violating portal operators with spin 0 and \textfrac12 messengers that are of dimension seven or less and quark-flavour violating portal operators with spin 1 messengers that are of dimension six or less.
As in the case of \cref{it:one} diagrams, dimension seven portal operators with spin 1 messengers can also contribute to \cref{it:two} diagrams, if the strong scale \PET is the \LE limit of another \EW scale \PET besides \SMEFT.

In order to be phenomenologically viable, any strong scale portal operators have to be invariant under the low energy \SM gauge group $G_\text{sm} = \SU(3)_c \times \U(1)_{\EM}$,
but they do not have to be invariant under the complete \SM gauge group $G_{\text{\SM}}$, which also encompasses weak interactions mediated by the heavy $W$- and $Z$-bosons.
In addition, our operators have to preserve the accidental symmetries obeyed by the relevant portal \SMEFTs.
This implies that all strong scale portal operators have to conserve baryon number and bosonic messenger fields have to conserve lepton number, while operators with spin \textfrac12 messengers can violate lepton number by one unit.
In addition, the portal \SMEFT interactions with spin \textfrac12 and 1 messenger fields do not mix \SM fermions of different chirality,
so that strong scale portal operators with chirality flips are suppressed by an additional factor of $\flatfrac{m_l}{v} \sim \sqrt{\epsilonSM}$, where $m_l$ is the mass of the relevant light \SM fermion.
Portal \SMEFT interactions with scalar messenger fields can induce a single chirality flip, so that only strong scale portal operators with at least two chirality flips are suppressed by such a factor of $\sqrt{\epsilonSM}$.

In addition to the dimensional suppression associated with $\epsilonSM$, the higher dimensional quark-flavour violating portal operators can also be suppressed by loop factors of $\inv[2]{(4\pi)}$.
We keep track of this suppression by using the $4\pi$ power counting scheme of \NDA \cite{Manohar:1983md, Jenkins:2013sda, Gavela:2016bzc, Manohar:2018aog}, see also \cref{sec:nda} for a detailed explanation.
Using \NDA, the most suppressed \cref{it:one} diagrams with spin 0 and \textfrac12 messengers scale as $(4\pi)^2 \epsilonUV \epsilon_{\SM}^3$, while the most suppressed \cref{it:one} diagrams with spin 1 messengers scale as $(4\pi)^2 \epsilonUV \epsilon_{\SM}^2$, see also \cref{fig:flavour violating processes}.
In both cases, the $(4\pi)^2$ enhancement captures the fact that the leading strong-scale Fermi theory interactions are generated by tree-level diagrams at the \EW scale.
When applying \NDA to strong scale \PETs, we discard all quark-flavour violating dimension six and seven \cref{it:two} operators that are even more suppressed than the most suppressed \cref{it:one} operators.
For \PETs with spin 0 and spin \textfrac12 messengers, dimension six operators without chiral suppression are suppressed by a relative factor of $\sqrt{\epsilonSM}$, rather than $\epsilonSM$, compared to the unsuppressed dimension five portal operators in these \PETs.
This means that they are enhanced by a relative factor of $\epsilon_{\SM}^{-\nicefrac12}$ compared to the most suppressed \cref{it:one} diagrams.
Therefore, we only use \NDA to discard operators that are either of dimension seven or of dimension six and chirally suppressed.
For \PETs with spin 1 messengers, we only use \NDA to discard operators that are either of dimension six, or of dimension four or five and sufficiently chirally suppressed.

\subsubsection{Operator list}

\begin{table}
\newcommand{\portal}[2]{$#1$\: &$#2$}
\newcommand{\pair}[2]{\ensuremath{\begin{pmatrix}#1\\#2\end{pmatrix}}}
\newcommand{\field}[2]{\midrule\multirowcell{#1}{#2}}
\newcolumntype{p}{r@{}l}
\begin{panels}[t]{.42}
\begin{tabular}{l@{ }c@{ }*3{@{}p@{}}} \toprule
& $d$ & \multicolumn{2}{c}{Scalar} & \multicolumn{2}{c}{Vector} & \multicolumn{2}{c}{Gauge} \\
\field{5}{$s_i$} & 4 & \portal{s_i}{\overline \psi\psi} \\
\cmidrule{2-8} & \multirow{4}{*}{5} & \portal{s_i s_j}{\overline \psi\psi} & & & \portal{s_i}{F_{\mu\nu} F^{\mu\nu}} \\
& & & & & & \portal{s_i}{F_{\mu\nu} \widetilde F^{\mu\nu}} \\
& & & & & & \portal{s_i}{G_{\mu\nu} G^{\mu\nu}} \\
& & & & & & \portal{s_i}{G_{\mu\nu} \widetilde G^{\mu\nu}} \\
\field{3}{$\lhf_a$\\+\\$\text{h.c.}$} & $3$ & \portal{\lhf_a}{\nu} \\
\cmidrule{2-8} & \multirow{2}{*}{5} & & & & & \portal{\lhf_a \overline \sigma_{\mu\nu} \nu}{F^{\mu\nu}} \\
& & & & & & \portal{\lhf_a \overline \sigma_{\mu\nu} \lhf_b}{F^{\mu\nu}} \\
\field{1}{$v_\mu$} & 4 & & & \portal{v_\mu}{\psi^\dagger \overline \sigma^\mu \psi} \\
\bottomrule \end{tabular}
\caption{\Cref{it:one} quark-flavour conserving portal operators of dimension three, four, and five.}
\label{tab:dominant low scale flavour conserving operators}
\panel{.55}
\begin{tabular}{l@{ }c*3{@{}p@{}}} \toprule
& $d$ & \multicolumn{2}{c}{Two quarks} & \multicolumn{2}{c}{Quark dipole} & \multicolumn{2}{c}{Four fermions} \\
\field{8}{$s_i$}
& \multirow{3}{*}{6} & \portal{s_i s_j s_k}{\overline d d} & \portal{s_i}{F^{\mu \nu} \overline d \sigma_{\mu \nu} d} \\
& & \portal{\partial^2 s_i}{\overline d d} & \portal{s_i}{G^{\mu \nu} \overline d \sigma_{\mu \nu} d} \\
& & \portal{s_i \partial_\mu s_j}{d^\dagger \overline \sigma^\mu d} \\
\cmidrule{2-8} & \multirow{5}{*}{7} & \portal{s_i s_j s_k s_l}{\overline d d} & & & \portal{s_i}{\boson{d^\dagger \overline q^\dagger}\boson{\overline q d}} \\
& & & & & & \portal{s_i}{\boson{q^\dagger \overline \sigma^\mu q}\boson{q^\dagger\overline \sigma_\mu q}} \\
& & & & & & \portal{s_i}{\boson{d^\dagger \overline \sigma^\mu d}\boson{\overline q \sigma_\mu \overline q^\dagger}} \\
& & & & & & \portal{s_i}{\boson{e^\dagger \overline \sigma_\mu \nu} \boson{u^\dagger \overline \sigma^\mu d}} \\
& & & & & & \portal{s_i}{\boson{\nu^\dagger \overline \sigma_\mu \nu} \boson{d^\dagger \overline \sigma^\mu d}} \\
\field{2}{$\lhf_a$\\$\text{h.c.}$} & \multirow{2}{*}{6} & \portal{\lhf_a^\dagger \overline \sigma_\mu}{e \boson{d^\dagger \overline \sigma^\mu u}} \\
& & \portal{\lhf_a^\dagger \overline \sigma_\mu}{\nu \boson{d^\dagger \overline \sigma^\mu d}} \\
\bottomrule \end{tabular}
\caption{\Cref{it:two} dimension six and seven quark-flavour violating portal operators.}
\label{tab:dominant low scale flavour violating operators}
\end{panels}
\caption[Leading portal $\LEFT$ operators]{
List of all \LO strong scale portal operators up to dimension seven that couple \LEFT to messenger fields of spin 0, \textfrac12, or 1.
Panel \subref{tab:dominant low scale flavour conserving operators} shows operators that contribute to \cref{it:one} diagrams and panel \subref{tab:dominant low scale flavour violating operators} shows operators that contribute to \cref{it:two} diagrams.
See also \cref{fig:flavour violating processes} for more details.
The first column specifies the spin of the messenger field, the second column contains the dimension $d$ of the operators and the remaining columns label the \SM sectors they interact with.
A generic \SM fermion is labelled by $\psi = u$, $d$, $e$, $\nu$, the down-type quarks are $d = \down$, $\strange$, the leptons are $e = \electron$, $\muon$ and $\nu = \neutrino_\electron$, $\neutrino_\muon$, $\neutrino_\tau$, and $q$ runs over all three light quarks $\up$, $\down$ and $\strange$.
} \label{tab:dominant low scale operators}
\end{table}

We construct minimal bases of portal operators for each portal \LEFT by combining the restrictions discussed in the previous section with the reduction techniques given in \cref{sec:redundant}.
The complete bases of both quark-flavour conserving and quark-flavour violating operators up to dimension seven are given in \cref{sec:gev_portals}.
\Cref{tab:dominant low scale flavour conserving operators} shows the subset of portal operators with dimension five or less.
This subset mirrors the set of portal operators in the corresponding portal \SMEFTs and contributes at \LO to both quark-flavour conserving and violating transitions.
In the following we focus on the operators appearing only at the strong scale.
\Cref{tab:dominant low scale flavour violating operators} shows the relevant subset of higher dimensional portal operators that contribute to quark-flavour violating transitions at \LO in $\epsilonUV$, $\epsilonSM$, and the $4\pi$ counting of \NDA.
The quark-flavour violating dimension six and seven operators that are sub-leading only due to $4\pi$ loop suppression factors are given in \cref{tab:subleading low scale operators}.
As in the case of the portal \SMEFTs, we consider three types of messenger field:

\begin{table}
\newcommand{\portal}[2]{$#1$\: &$#2$}
\newcolumntype{p}{r@{}l}
\begin{panels}[t]{.45}
\raggedright
\begin{tabular}{@{}c@{ }*2{@{ }p@{ }}} \toprule
$(4\pi)^{-n}\hspace{-1.5em}$ & \multicolumn{2}{r}{Two Quarks} & \multicolumn{2}{c}{Quark Dipole} \\
\midrule
\multirow{3}{*}{1} & \portal{s_i s_j \partial_\mu s_k}{\boson{d^\dagger \overline \sigma^\mu d}} & \portal{\partial_\nu s_i}{\boson{d^\dagger \overline \sigma_\mu V^{\mu \nu} d}} \\
& & & \portal{\partial_\nu s_i}{\boson{d^\dagger \overline \sigma_\mu \widetilde V^{\mu \nu} d}} \\
& & & \portal{s_i s_j}{\boson{\overline d \overline \sigma_{\mu \nu} V^{\mu \nu} d} + \text{h.c.}} \\ \cmidrule{2-5}
\multirow{2}{*}{2} & \portal{\partial^2 s_i}{\overline d d} \\
& \portal{s_i \partial_\mu s_j}{d^\dagger \overline \sigma^\mu d} \\
\bottomrule \end{tabular}
\caption{Scalar.}
\label{tab:subleading low scale operators scalar}
\panel{.55}
\raggedleft
\begin{tabular}{@{}c@{ }*3{@{}p@{}}} \toprule
$(4\pi)^{-n}\hspace{-1.5em}$ & \multicolumn{2}{c}{$\overline dd$} & \multicolumn{2}{c}{$d^\dagger d$} & \multicolumn{2}{c}{$d^\dagger V^{\mu\nu} d$} \\
\midrule
\multirow{2}{*}{1} & & & \portal{v_\mu v^\mu v_\nu}{\boson{d^\dagger \overline \sigma^\nu d}} & \portal{v^\mu}{\boson{d^\dagger \overline \sigma^\nu G_{\mu\nu} d}} \\
& & & & & \portal{v^\mu}{\boson{d^\dagger \overline \sigma^\nu \widetilde G_{\mu\nu} d}} \\
\cmidrule{2-7}
\multirow{8}{*}{2} & \portal{v_\mu v^\mu}{\boson{\overline d d}} & \portal{\boson{\partial_\nu v^\nu} v_\mu}{\boson{d^\dagger \overline \sigma^\mu d}} & \portal{v^\mu}{\boson{d^\dagger \overline \sigma^\nu F_{\mu\nu} d}} \\
& \portal{\boson{\partial_\mu v^\mu}}{\boson{\overline d d}} & \portal{\boson{\partial_\mu v_\nu} v^\nu}{\boson{d^\dagger \overline \sigma^\mu d}} & \portal{v^\mu}{\boson{d^\dagger \overline \sigma^\nu \widetilde F_{\mu\nu} d}} \\
& \portal{\boson{\partial_\mu v_\nu}}{\boson{\overline d \sigma^{\mu \nu} d}} & \portal{\boson{\partial_\nu v_\mu} v^\nu}{\boson{d^\dagger \overline \sigma^\mu d}} \\
& & & \portal{\boson{\partial^2 v_\mu}}{\boson{d^\dagger \overline \sigma^\mu d}} \\
& & & \portal{\epsilon^{\alpha\beta\mu\nu} \boson{\partial_\alpha v_\beta} v_\nu}{\boson{d^\dagger \overline \sigma_\mu d}} \\
& & & \portal{v^\mu v^\nu}{\boson{d^\dagger \overline \sigma_\mu D_\nu d}} \\
& & & \portal{\boson{\partial_{(\nu} v_{\mu)}}}{\boson{d^\dagger \overline \sigma^\mu D^\nu d}} \\
\bottomrule \end{tabular}
\caption{Vector}
\label{tab:subleading low scale operators vector}
\panel{1}
\begin{tabular}{*4{@{ }p@{ }}} \toprule
\multicolumn{2}{@{}c@{}}{$LL\times LL$} & \multicolumn{2}{@{}c@{}}{$LR \times RL$} & \multicolumn{2}{@{}c@{}}{$RL \times RL$} & \multicolumn{2}{@{}c@{}}{$LL \times RL$} \\
\midrule
\portal{\boson{\lhf_a^\dagger \overline \sigma_\mu \lhf_b}}{\boson{d^\dagger \overline \sigma^\mu d}} & \portal{\boson{\nu^\dagger \lhf_a^\dagger}}{\boson{\overline d d}} & \portal{\boson{\nu \lhf_a}}{\boson{\overline d d}} & \portal{\boson{\nu D^\mu \lhf_a}}{\boson{d^\dagger \overline \sigma_\mu d}} \\
\portal{\boson{\overline e \sigma_\mu \lhf_a^\dagger}}{\boson{d^\dagger \overline \sigma^\mu u}} & \portal{\boson{\lhf_a^\dagger \lhf_b^\dagger}}{\boson{\overline d d}} & \portal{\boson{\lhf_a \lhf_b}}{\boson{\overline d d}} & \portal{\boson{\lhf_a D^\mu \lhf_b}}{\boson{d^\dagger \overline \sigma_\mu d}} \\
& & \portal{\boson{e^\dagger \lhf_a^\dagger}}{\boson{\overline u d}} & \portal{\boson{e \lhf_a}}{\boson{\overline d u}} & \portal{\boson{e D^\mu \lhf_a}}{\boson{d^\dagger \overline \sigma^\mu u}} \\
& & & & \portal{\boson{\nu \overline \sigma^{\mu \nu} \lhf_a}}{\boson{\overline d \overline \sigma_{\mu \nu} d}} & \portal{\boson{\nu \overline \sigma^{\mu \nu} D_\nu \lhf_a}}{\boson{d^\dagger \overline \sigma_\mu d}} \\
& & & & \portal{\boson{\lhf_a \overline \sigma^{\mu \nu} \lhf_b}}{\boson{\overline d \overline \sigma_{\mu \nu} d}} & \portal{\boson{\lhf_a \overline \sigma^{\mu \nu} D_\nu \lhf_b}}{\boson{d^\dagger \overline \sigma_\mu d}} \\
& & & & \portal{\boson{e \overline \sigma^{\mu \nu} \lhf_b}}{\boson{\overline d \overline \sigma_{\mu \nu} u}} & \portal{\boson{e \overline \sigma^{\mu \nu} D_\nu \lhf_a}}{\boson{d^\dagger \overline \sigma^\mu u}} \\
\bottomrule \end{tabular}
\caption{Fermion}
\label{tab:subleading low scale operators fermion}
\end{panels}
\caption[Subleading portal $\LEFT$ operators]{
List of all sub-leading quark-flavour changing strong scale operators up to dimension seven that couple \LEFT to messenger fields of spin 0, \textfrac12, or 1.
Panel \subref{tab:subleading low scale operators scalar} shows the operators for spin 0 messenger fields,
panel shows \subref{tab:subleading low scale operators fermion} those for spin \textfrac12 messengers,
and panel \subref{tab:subleading low scale operators vector} shows those for spin 1 messengers.
All fermionic operators are suppressed by factors of $\inv{(4\pi)}$, and the suppression factor for the bosons are given in the tables.
The notation is the same as in \cref{tab:dominant low scale operators}.
} \label{tab:subleading low scale operators}
\end{table}

\paragraph{Spin 0 fields} couple to \LEFT via six operators of dimension five or less.
In addition, there are eleven quark-flavour violating dimension six and seven operators that contribute at \LO in both $\epsilonSM$ and $4\pi$.
At dimension six, there are three leading two-quark operators
\begin{align}
& s_i s_j s_k \boson{\overline d d} \ , &
& \partial^2 s_i \boson{\overline d d} \ , &
& s_i \partial_\mu s_j \boson{d^\dagger \overline \sigma^\mu d} \ ,
\end{align}
and two leading two-quark dipole operators involving the \EM and the gluonic field strength tensor
\begin{align}
& s_i \boson{\overline d \sigma_{\mu \nu} d} F^{\mu \nu} \ ,
& s_i \boson{\overline d \sigma_{\mu \nu} d} G^{\mu \nu} \ .
\end{align}
At dimension seven, there is one leading two-quark operator
\begin{equation}
s_i s_j s_k s_l \boson{\overline d d}
\end{equation}
as well as five leading four-fermion operators
\begin{subequations}
\begin{align}
& s_i \boson{q^\dagger \overline \sigma^\mu q}\boson{q^\dagger\overline \sigma_\mu q} \ , &
& s_i \boson{u^\dagger \overline \sigma^\mu d}\boson{e^\dagger \overline \sigma_\mu \nu} \ , \\ \label{eq:operator scalar penguin box}
& s_i \boson{d^\dagger \overline \sigma^\mu d}\boson{\overline q \sigma_\mu \overline q^\dagger} \ , &
& s_i \boson{d^\dagger \overline \sigma^\mu d} \boson{\nu^\dagger \overline \sigma_\mu \nu} \ , &
& s_i \boson{d^\dagger \overline q^\dagger}\boson{\overline q d} \ .
\end{align}
\end{subequations}
The semi-leptonic neutral current operator $s_i \boson{d^\dagger \overline \sigma^\mu d} \boson{\nu^\dagger \overline \sigma_\mu \nu}$ is generated by the box- and penguin-type diagrams shown in \cref{fig:Feynman scalar box,fig:Feynman scalar penguin}.
These diagrams involve at least two heavy boson exchanges, so that one might expect all of them to be suppressed by an additional factor of $\epsilonSM$ due to the second heavy boson exchange.
However, the analogous \SM four-fermion operators $\boson{d^\dagger \overline \sigma^\mu d} \boson{\nu^\dagger \overline \sigma_\mu \nu}$ scale as $\epsilonSM f(\flatfrac{m_t^2}{v^2})$, with some function $f(x) \sim 1$, so that there is no additional suppression \cite[Section XI.B of][]{Buchalla:1995vs}.
We expect that the same can occur in case of the portal operator $s_i \boson{d^\dagger \overline \sigma^\mu d} \boson{\nu^\dagger \overline \sigma_\mu \nu}$, and we therefore keep this operator as part of the portal Lagrangian.
All of the operators mentioned above are listed in \cref{tab:dominant low scale operators}.

The sub-leading dimension seven operators differ in their suppression.
The four operators
\begin{align}
& s_i s_j \partial_\mu s_k \boson{d^\dagger \overline \sigma^\mu d} \ , &
& \partial_\nu s_i \boson{d^\dagger \overline \sigma_\mu V^{\mu \nu} d} \ , &
& \partial_\nu s_i \boson{d^\dagger \overline \sigma_\mu \widetilde V^{\mu \nu} d} \ , &
& s_i s_j \boson{\overline d \overline \sigma_{\mu \nu} V^{\mu \nu} d} + \text{h.c.} \ ,
\end{align}
with $V^{\mu\nu} \in \left\{F^{\mu\nu}, G^{\mu\nu} \right\}$ are suppressed by factors of $\inv{(4\pi)}$, and the operators
\begin{align}
& s_i \partial^2 s_j \overline d d \ , & \partial_\mu s_i \partial^\mu s_j \overline d d \ ,
\end{align}
and their Hermitian conjugates are suppressed by factors of $\inv[2]{(4\pi)}$.
The above sub-leading dimension seven operators are listed in \cref{tab:subleading low scale operators scalar}.

\paragraph{Spin \textfrac12 fields}

\begin{figure}
\begin{panels}{3}
\tikzsetnextfilename{feynman-scalar-box}\input{feynman-scalar-box.pgf}
\caption{Scalar box diagram.} \label{fig:Feynman scalar box}
\panel
\tikzsetnextfilename{feynman-scalar-penguin}\input{feynman-scalar-penguin.pgf}
\caption{Scalar penguin diagram.} \label{fig:Feynman scalar penguin}
\panel
\tikzsetnextfilename{feynman-fermion-penguin}\input{feynman-fermion-penguin.pgf}
\caption{Fermionic penguin diagram.} \label{fig:Feynman fermion penguin}
\end{panels}
\caption[One-loop portal Feynman diagrams]{
One-loop portal diagrams for some of the portal operators.
Panels \subref{fig:Feynman scalar box} and \subref{fig:Feynman scalar penguin} depict contributions to the scalar portal operator \eqref{eq:operator scalar penguin box},
where the scalar field can couple to any of the heavy \EW bosons.
Panel \subref{fig:Feynman fermion penguin} depicts the contribution to the fermionic portal operator in \eqref{eq:fermionic penguin operator}.
} \label{fig:Feynman scalar}
\end{figure}

couple to \LEFT via three operators of dimension five or less.
In addition, there are two quark-flavour violating dimension six operators
\begin{align} \label{eq:fermionic penguin operator}
& \lhf_a^\dagger{\overline \sigma_\mu e\boson{d^\dagger \overline \sigma^\mu u}} \ , &
& \boson{d^\dagger \overline \sigma^\mu d} \boson{\lhf_a^\dagger \overline \sigma_\mu\nu} \ ,
\end{align}
and their Hermitian conjugates, which contribute at \LO in both $\epsilonSM$ and $4\pi$.
The second operator and its Hermitian conjugate can only be generated by penguin- and box-type diagrams involving at least two heavy \SM bosons.
In analogy to the case of the scalar portal operators in \eqref{eq:operator scalar penguin box}, we expect that the diagrams with a virtual top quark exchange inside the loop can scale as $\epsilonSM f(\flatfrac{m_t^2}{v^2})$, so that there is no additional suppression compared to the first operator.
All of the operators mentioned above are listed in \cref{tab:dominant low scale operators}.

The sub-leading operators can be either of dimension six or seven, and they are suppressed by factors of $\inv{(4\pi)}$ or $\inv[2]{(4\pi)}$.
At dimension six, there are ten operators
\begin{subequations}
\begin{align}
& \boson{\overline d d} \boson{\nu \lhf_a} \ , &
& \boson{\overline d d} \boson{\nu^\dagger \lhf_a^\dagger} \ , &
& \boson{\overline d \overline \sigma_{\mu \nu} d} \boson{\nu \overline \sigma^{\mu \nu} \lhf_a} \ , &
& \boson{\overline d u} \boson{e \lhf_a} \ , &
& \boson{d^\dagger \overline \sigma^\mu u} \boson{\overline e \sigma_\mu \lhf_a^\dagger} \ , \\
& \boson{\overline d d} \boson{\lhf_a \lhf_b} \ , &
& \boson{\overline d d} \boson{\lhf_a^\dagger \lhf_b^\dagger} \ , &
& \boson{\overline d \overline \sigma_{\mu \nu} d} \boson{\lhf_a \overline \sigma^{\mu \nu} \lhf_b} \ , &
& \boson{\overline u d} \boson{e^\dagger \lhf_a^\dagger} \ , &
& \boson{\overline d \overline \sigma_{\mu \nu} u} \boson{e \overline \sigma^{\mu \nu} \lhf_b} \ ,
\end{align}
\end{subequations}
that contain charged right-chiral \SM fermion fields, so that they are suppressed by an additional factor of $\flatfrac{m_\psi}{v} \propto \sqrt{\epsilonSM}$, where $m_\psi$ is the mass of the relevant right-chiral fermion, due to the associated chiral suppression.
As a result, they effectively behave as dimension seven operators.
Applying \NDA, one finds that they are suppressed by factors of $\inv{(4\pi)}$.
In addition, the operator
\begin{align}
& \boson{d^\dagger \overline \sigma^\mu d} \boson{\lhf_a^\dagger \overline \sigma_\mu \lhf_b}
\end{align}
and its Hermitian conjugate, generated by penguin diagrams shown in \cref{fig:Feynman fermion penguin}, contain at least two \SM gauge boson exchanges.
At the \EW scale, the hidden fermion only couples to photons and $Z$-bosons via the dipole-type operator $\lhf_a \overline \sigma^{\mu\nu} \lhf_b B_{\mu\nu}$.
This coupling flips the chirality of the hidden fermion, so that a light mass-insertion is necessary to undo the flip.
Therefore, the operator is suppressed by an additional factor of $\sqrt{\epsilonSM}$, and it effectively counts as a dimension seven operator.
Applying \NDA, one also has to account for the $4\pi$ suppression associated with the \EW gauge couplings, so that the operator is suppressed by at least a factor of $(4\pi)^{-2}$.

Finally, at dimension seven, there are six derivative operators
\begin{subequations}
\begin{align}
& \boson{d^\dagger \overline \sigma_\mu d} \boson{\nu D^\mu \lhf_a} \ , &
& \boson{d^\dagger \overline \sigma^\mu u} \boson{e D^\mu \lhf_a} \ , &
& \boson{d^\dagger \overline \sigma_\mu d} \boson{\lhf_a D^\mu \lhf_b} \ , \\
& \boson{d^\dagger \overline \sigma_\mu d} \boson{\nu \overline \sigma^{\mu \nu} D_\nu \lhf_a} \ , &
& \boson{d^\dagger \overline \sigma^\mu u} \boson{e \overline \sigma^{\mu \nu} D_\nu \lhf_a} \ , &
& \boson{d^\dagger \overline \sigma_\mu d} \boson{\lhf_a \overline \sigma^{\mu \nu} D_\nu \lhf_b} \ ,
\end{align}
\end{subequations}
and their Hermitian conjugates.
We collect all of the above sub-leading operators in \cref{tab:subleading low scale operators}.

\paragraph{Spin 1 fields} couple to \LEFT via one operator of dimension four, see \cref{tab:dominant low scale operators}.
Since there are no dimension five operators that couple spin 1 messengers to \SMEFT, the resulting portal \LEFT contains higher dimensional operators of dimension five and six, but not seven.
None of them contributes at \LO in the $4\pi$ counting.
The dimension six operators
\begin{align}
v_\mu v^\mu v_\nu \boson{d^\dagger \overline \sigma^\nu d} & \ , &
v^\mu \boson{d^\dagger \overline \sigma^\nu G_{\mu\nu} d} & \ , &
v^\mu \boson{d^\dagger \overline \sigma^\nu \widetilde G_{\mu\nu} d} \ ,
\end{align}
are suppressed by factors of $\inv{(4\pi)}$.
The dimension five operators
\begin{align}
v_\mu v^\mu \boson{\overline d d} & \ , &
\boson{\partial_\mu v^\mu} \boson{\overline d d} & \ , &
\boson{\partial_\mu v_\nu} \boson{\overline d \sigma_{\mu \nu} d}
\end{align}
and their Hermitian conjugates are suppressed by a factor of $\sqrt{\epsilonSM}$ associated with each right-chiral light quark insertion, so that they effectively contribute like dimension six operators.
Applying the \NDA rules, one finds that they are suppressed by factors of $\inv[2]{(4\pi)}$.
Finally, the dimension six operators
\begin{equation}
\begin{aligned}
v^\mu F_{\mu\nu} \boson{d^\dagger \overline \sigma^\nu d} & \ , &
\boson{\partial_\nu v^\nu} v_\mu \boson{d^\dagger \overline \sigma^\mu d} & \ , &
v^\mu v^\nu \boson{d^\dagger \overline \sigma_\mu D_\nu d} & \ , &
\boson{\partial^2 v_\mu} \boson{d^\dagger \overline \sigma^\mu d} & \ , \\
v^\mu \widetilde F_{\mu\nu} \boson{d^\dagger \overline \sigma^\nu d} & \ , &
\boson{\partial_\mu v_\nu} v^\nu \boson{d^\dagger \overline \sigma^\mu d} & \ , &
\epsilon^{\alpha\beta\mu\nu} \boson{\partial_\alpha v_\beta} v_\nu \boson{d^\dagger \overline \sigma_\mu d} & \ , &
\boson{\partial_{(\nu} v_{\mu)}} \boson{d^\dagger \overline \sigma^\mu D^\nu d} & \ , \\ & &
\boson{\partial_\nu v_\mu} v^\nu \boson{d^\dagger \overline \sigma^\mu d} & \ .
\end{aligned}
\end{equation}
are also suppressed by factors of $\inv[2]{(4\pi)}$.
We collect all of the above operators in \cref{tab:subleading low scale operators}.

\subsubsection{QCD portal currents} \label{sec:portal currents}

In order to prepare for the derivation of the portal \cPT Lagrangian in the following section, we embed the interactions encompassed by the portal \LEFTs into appropriate portal currents, as we have done for the interactions of the portal \SMEFTs.
These currents contain the leading quark-flavour conserving and violating portal operators collected in \cref{tab:dominant low scale operators}, but we neglect the subleading quark-flavour violating operators collected in \cref{tab:subleading low scale operators}.
Hence, the \QCD sector of the portal Lagrangian is
\begin{multline} \label[lag]{eq:LE portal Lagrangian}
\mathcal L_Q^\text{portal}
= S_\theta w
- S_\omega \gluons
- \trf{\nonet S_m \nonet Q
+ \nonet V_l^\mu \nonet Q_\mu
+ \nonet V_r^\mu \overline{\nonet Q}_\mu}
- \inv[2]{\Lambda_{\SM}} \trf{\octet S_\chromo \widetilde{\nonet Q}
+ \octet T^{\mu\nu} \nonet Q_{\mu\nu}
+ \text{h.c.}} \\
- \inv[2]{v} \trf{\octet S_\scalar \octet O_\scalar
+ \octet S_r \octet O_r
+ \octet S_l^\asym \octet O_l^\asym
+ \octet S_l^\sym \octet O_l^\sym}
- \inv[2]{v} \trF{\tensor S_l^\sym \tensor O_l^\sym} \ ,
\end{multline}
where the composite \QCD gluon operators $w$ and $\gluons$ are defined in \eqref{eq:Q kinetic Lagrangian,eq:Q axial anomaly Lagrangian}, the quark bilinears $\nonet Q$ are defined in \eqref{eq:Q mass Lagrangian}, \allowbreak\eqref{eq:vectorial quark bilinears}, and \eqref{eq:EW quark bilinears}, and the quark quadrilinears $\octet O$ and $\tensor O$ are defined in \eqref{eq:octet parameter}, \allowbreak\eqref{eq:27-plet octet operators}, and \eqref{eq:27-plet operator}.

The \prefix{pseudo}{scalar} portal currents $S_\theta$ and $S_\omega$ couple to \QCD in the same way as the $\theta$ angle and the gluon coupling $\omega$ in \cref{eq:Q kinetic Lagrangian,eq:Q axial anomaly Lagrangian}.
They read
\begin{align}
S_\omega &= \frac{\epsilonUV}{v} c_i^{S_\omega} s_i \ , &
S_\theta &= \frac{\epsilonUV}{v} c^{S_\theta}_i s_i \ .
\end{align}
The \prefix{pseudo}{scalar} portal current $\nonet S_m$ couples to \QCD in the same way as the quark mass matrix in \eqref{eq:Q mass Lagrangian}.
It reads
\begin{equation}
\nonet S_m
= \epsilonUV c^{S_m}_i s_i
+ \frac{\epsilonUV}{v} c^{S_m}_{ij} s_i s_j
+ \frac{\epsilonUV}{v^2}
\left(c^{S_m}_{ijk} s_i s_j s_k
+ c^{S_m}_{\partial^2i} \partial^2 s_i
\right)
+ \frac{\epsilonUV}{v^3} c^{S_m}_{ijkl} s_i s_j s_k s_l \ .
\end{equation}
This current has to be uncharged, so that it obeys
\begin{align}
\nonet S_m &= \octet S_m + \frac{\flavour 1}{\nfl} S_m \ , &
\octet S_\m{}_\down^\up &= \octet S_\m{}_\up^\down = \octet S_\m{}_\strange^\up = \octet S_\m{}_\up^\strange = 0 \ .
\end{align}
The left- and right-handed vector portal currents $\nonet V_l^\mu$ and $\nonet V_r^\mu$ couple to \QCD in the same way as the left- and right-handed \EW currents in \eqref{eq:Q vector current Lagrangian}.
They read
\begin{subequations} \label[cur]{eq:left and right handed portal currents}
\begin{align}
\nonet V_l^\mu &
\begin{multlined}[t]
= \epsilonUV c^L_v v^\mu + \frac{\epsilonUV}{v^2} \left[c^L_{ij} s_i \overleftright \partial^\mu s_j + \left(
\flavour \proj_\up^\strange c^L_{\bar\up\strange \, i} \boson{e^\dagger \overline \sigma^\mu \nu} s_i
+ \flavour \proj_\down^\strange c^L_{\bar\down\strange \, i} \boson{\nu^\dagger \overline \sigma^\mu \nu} s_i
\right.\right.\\\left.\left.
+ \flavour \proj_\up^\strange c^L_{\bar\up\strange \, a} \boson{e^\dagger \overline \sigma^\mu \lhf_a}
+ \flavour \proj_\down^\strange c^L_{\bar\down\strange \, a} \boson{\nu^\dagger \overline \sigma^\mu \lhf_a}
+ \text{h.c.} \right) \right] \ ,
\end{multlined} \\
\nonet V_r^\mu &
= \epsilonUV c^R_v v^\mu \ .
\end{align}
\end{subequations}
The current $\nonet V_l^\mu$ is the only portal current that can carry charge due to the contributions generated by virtual $W$-boson exchanges, which implies
\begin{align}
\octet V_r^\mu{}_\down^\up &= \octet V_r^\mu{}_\up^\down = \octet V_r^\mu{}_\strange^\up = \octet V_r^\mu{}_\up^\strange = 0 \ , &
\nonet V_{l,r}^\mu &= \octet V_{l,r}^\mu + \frac{\flavour 1}{\nfl} V_{l,r}^\mu \ .
\end{align}
$\nonet V_r^\mu$ and $\nonet V_l^\mu$ are also Hermitian, so that
\begin{align}
\octet V_{l,r}^\mu{}_\down^\strange &= (\octet V_{l,r}^\mu{}_\strange^\down)^\dagger \ , &
\octet V_l^\mu{}_\up^\down &= (\octet V_l^\mu{}_\down^\up)^\dagger \ , &
\octet V_l^\mu{}_\up^\strange &= (\octet V_l^\mu{}_\strange^\up)^\dagger \ .
\end{align}
The dipole portal currents $\octet T_\tau^{\mu\nu}$ and $\octet S_{\chromo_V}$ couple to \QCD in the same way as the dipole currents in \cref{eq:Q dipole current Lagrangian}.
They read
\begin{align}
\octet T_\tau^{\mu \nu} &
= - \frac{1}{3} F^{\mu \nu} \octet S_{\chromo_A} \ , &
\octet S_{\chromo_V} &
= \epsilonUV \left(
\flavour \proj_\strange^\down c^{\chromo_V}_{i\overline\strange\down}
+ \flavour \proj_\down^\strange c^{\chromo_V}_{i\overline\down\strange}
\right) s_i \ .
\end{align}
The chromomagnetic and tensor currents $\octet S_{\chromo_G}$ and $\octet T_\tau^{\mu \nu}$ are uncharged and strangeness violating, but not necessarily Hermitian.
Hence, the only non-vanishing contributions are
\begin{align}
&\octet S_{\chromo_G}{}_\down^\strange \ , &
&\octet S_{\chromo_G}{}_\strange^\down \ , &
&\octet T_\tau^{\mu \nu}{}_\down^\strange \ , &
&\octet T_\tau^{\mu \nu}{}_\strange^\down \ .
\end{align}
Finally, the four-quark portal currents mirror the four-quark interactions in \cref{eq:Q four-quark Lagrangian representations}.
They read
\begin{subequations}
\begin{align}
\octet S_\scalar &= \octet \fermi_{\scalar i} \frac{\epsilonUV}{v} s_i \ , &
\octet S_r &= \octet \fermi_{ri} \frac{\epsilonUV}{v} s_i \ , \\
\octet S_l^\asym &= \octet \fermi_{ai} \frac{\epsilonUV}{v} s_i \ , &
\octet S_l^\sym &= \octet \fermi_{si} \frac{\epsilonUV}{v} s_i \ , &
\tensor S_l^\sym &= \tensor \fermi_{si} \frac{\epsilonUV}{v} s_i \ ,
\end{align}
\end{subequations}
where the four-quark portal sector parameters $a_{x 1,2} = a_x(c_{\iota 1,2})$ and Wilson coefficients $c_{\iota 1,2}$ are defined such
that they mirror the \SM four-quark parameters \eqref{eq:octet parameter,eq:27-plet parameters} and Wilson coefficients \eqref{eq:Q four-quark Lagrangian}.
It is convenient to define $a_{x0} = a_x(c_{\iota 0})$ and $c_{\iota0} = c_{\iota}$, so that the generic objects $a_{xi}$ and $c_{\iota i}$ with $i = 0$, $1$, $2$ can be used to collectively refer to the complete set of both \SM and portal sector parameters and Wilson coefficients.

\begin{table}
\begin{tabular}{*3{@{ }r}*{12}{c}} \toprule
& & & $\Omega$ & $\Theta$ & $\nonet M$ & $\octet \Chromo$ & $\octet \Fermi_\scalar$ & $\octet \Fermi_r$ & $\octet \Fermi_l^\asym$ & $\octet \Fermi_l^\sym$ & $\tensor \Fermi_l^\sym$ & $\nonet L^\mu$ & $\nonet R^\mu$ & $\octet T^{\mu\nu}$ \\
\midrule
\multicolumn{2}{l}{\multirow{2}{*}{contribution}} & $\SM$ & $\omega$ & $\theta$& $\nonet m$ & $\octet \chromo$ & $\octet \fermi_\scalar$ & $\octet \fermi_r$ & $\octet \fermi_l^\asym$ & $\octet \fermi_l^\sym$ & $\tensor \fermi_l^\sym$ & $\octet l^\mu$ & $\octet r^\mu$ & $\octet \tau^{\mu\nu}$ \\
& & $\BSM$ & $S_\omega$ & $S_\theta$ & $\nonet S_m$ & $\octet S_\chromo$ & $\octet S_\scalar$ & $\octet S_r$ & $\octet S_l^\asym$ & $\octet S_l^\sym$ & $\tensor S_l^\sym$ & $\nonet V_l^\mu$ & $\nonet V_r^\mu$ & $\octet T_\tau^{\mu\nu}$ \\
\cmidrule(r){1-3} \cmidrule(lr){4-12} \cmidrule(lr){13-14} \cmidrule(l){15-15}
& & spin & \multicolumn{9}{c}{0} & \multicolumn{2}{c}{1} & 2 \\
\cmidrule(lr){3-3} \cmidrule(lr){4-12} \cmidrule(lr){13-14} \cmidrule(l){15-15}
& & $\epsilonSM$ & \multicolumn{3}{c}{0} & \multicolumn{6}{c}{1} & \multicolumn{2}{c}{0} & 1 \\
\cmidrule(lr){3-3} \cmidrule(lr){4-6} \cmidrule(lr){7-12} \cmidrule(lr){13-14} \cmidrule(l){15-15}
& & $d$ & \multicolumn{2}{c}{0} & 1 & 1 & \multicolumn{4}{c}{0} & 0 & \multicolumn{2}{c}{1} & 3 \\
\cmidrule(r){1-3} \cmidrule(lr){4-5} \cmidrule(lr){6-6} \cmidrule(lr){7-7} \cmidrule(lr){8-11} \cmidrule(lr){12-12} \cmidrule(lr){13-14} \cmidrule(l){15-15}
\multirow{5}{*}{flavour} & \multicolumn{2}{r}{representation} & \multicolumn{2}{c}{1} & $8\oplus1$ & 8 & \multicolumn{4}{c}{8} & $27$ & \multicolumn{2}{c}{$8 \oplus 1$} & 8 \\
& \multicolumn{2}{r}{symmetry} & & & & & & & & & & \multicolumn{2}{c}{$\nonet V_\mu^\dagger = \nonet V_\mu$} \\
\cmidrule(lr){2-3} \cmidrule(lr){4-5} \cmidrule(lr){6-6} \cmidrule(lr){7-7} \cmidrule(lr){8-11} \cmidrule(lr){12-12} \cmidrule(lr){13-14} \cmidrule(l){15-15}
& \multirow{2}{*}{$\DOFs$} & & \multicolumn{2}{c}{1} & 18 & 16 & \multicolumn{4}{c}{16} & 54 & \multicolumn{2}{c}{9} & 16 \\
& & $\Delta \qn s = \pm 1$& \multicolumn{2}{c}{0} & 4 & 4 & \multicolumn{4}{c}{4} & 4 & \multicolumn{2}{c}{2} & 4 \\
\bottomrule \end{tabular}
\caption[Portal $\LEFT$ currents]{
List of all external currents interacting with $\QCD$ at the strong scale including both \SM and $\BSM$ contributions.
The first three rows list their spin, the order in $\epsilonSM$ at which they contribute and their mass dimension $d$.
Rows four, five, and six list their representations and symmetries under flavour rotations as well as the resulting $\DOFs$.
The last row counts the number of strangeness violating $\DOFs$, which are the only relevant $\DOFs$ for currents starting contribute at order $\epsilonSM$.
} \label{tab:strong scale currents}
\end{table}

Combining the \SM and \BSM contributions (\cf \cref{tab:strong scale currents}) to the external currents, we define the complete external currents
\begin{subequations} \label[cur]{eq:currents}
\begin{align}
\Theta &= \theta + S_\theta \ , &
\nonet \M &= \nonet \m + \nonet S_m \ , &
\nonet R^\mu &= \octet r^\mu + \nonet V_r^\mu \ , &
\octet T^{\mu\nu} &= \octet \tau^{\mu\nu} + \octet T_\tau^{\mu\nu} \ , \\
\Omega &= \omega + S_\omega \ , &
\octet \Chromo &= \octet \chromo + \octet S_\chromo \ , &
\nonet L^\mu &= \octet l^\mu + \nonet V_l^\mu \ , &
\end{align}
\end{subequations}
and
\footnote{We emphasise that the use of $h$ and $H$ for both the Higgs field and the four quark current can not lead to conflicts as these currents only appear at energy scales at which the Higgs field has been integrated out.}
\begin{subequations} \label[cur]{eq:A currents}
\begin{align}
\octet \Fermi_\scalar &= \octet \fermi_\scalar + \octet S_\scalar \ , &
\octet \Fermi_r &= \octet \fermi_r + \octet S_r \ , \\
\octet \Fermi_l^\asym &= \octet \fermi_l^\asym + \octet S_l^\asym \ , &
\octet \Fermi_l^\sym &= \octet \fermi_l^\sym + \octet S_l^\sym \ , &
\tensor \Fermi_l^\sym &= \tensor \fermi_l^\sym + \tensor S_l^\sym \ .
\end{align}
\end{subequations}
Using these complete external currents in place of the \SM external currents, one obtains the corresponding complete interaction Lagrangians
\begin{subequations}
\begin{align}
\mathcal L_Q^\theta &\to \mathcal L_Q^\Theta \ , &
\mathcal L_Q^\omega &\to \mathcal L_Q^\Omega \ , &
\mathcal L_Q^\m &\to \mathcal L_Q^\M \ , &
\mathcal L_Q^v &\to \mathcal L_Q^V \ , \\
\mathcal L_Q^\chromo &\to \mathcal L_Q^\Chromo \ , &
\mathcal L_Q^\tau &\to \mathcal L_Q^T \ , &
\mathcal L_Q^\fermi &\to \mathcal L_Q^\Fermi \ ,
\end{align}
\end{subequations}
where the original Lagrangians are given in \eqref{eq:Q Lagrangian,eq:Q EW Lagrangian,eq:Q four-quark Lagrangian}.
Hence, the complete external current sector of \QCD including both \SM and hidden contributions is
\begin{multline} \label[lag]{eq:LE portal Lagrangian}
\mathcal L_Q
= \Theta w
- \Omega \gluons
- \trf{\nonet M \nonet Q
+ \nonet L^\mu \nonet Q_\mu
+ \nonet R^\mu \overline{\nonet Q}_\mu}
- \inv[2]{\Lambda_{\SM}} \trf{\octet \Chromo \widetilde{\nonet Q}
+ \octet T^{\mu\nu} \nonet Q_{\mu\nu}
+ \text{h.c.}} \\
- \inv[2]{v} \trf{\octet \Fermi_\scalar \octet O_\scalar
+ \octet \Fermi_r \octet O_r
+ \octet \Fermi_l^\asym \octet O_l^\asym
+ \octet \Fermi_l^\sym \octet O_l^\sym}
- \inv[2]{v} \trF{\tensor \Fermi_l^\sym \tensor O_l^\sym} \ ,
\end{multline}
where the external currents are defined in \eqref{eq:currents} and \eqref{eq:A currents} and summarised in \cref{tab:strong scale currents}.
All of them receive contributions from the \SM.
However, without \NP, the currents $\Theta$, $\Omega$, $\nonet \M$, $\octet H$, $\octet \Fermi_x$, and $\tensor \Fermi_l^\sym$ are constant.
The \SM contributions to the currents $\nonet L^\mu$, $\nonet R^\mu$ and $\octet T^{\mu\nu}$ depend on the photon field, and $\nonet L^\mu$ additionally contains the weak leptonic charged current, \cf \eqref{eq:Q vector currents,eq:Q dipole current}.

\paragraph{Covariant derivatives}

In contrast to the parameters defined in \cref{sec:quantum chromo dynamics}, the external currents defined in this section are spacetime dependent, and the \cPT Lagrangian derived in the next section contains contributions with derivatives acting on the external currents.
To enforce invariance of \cPT under the action of the \emph{local} $G_{LR}$ symmetry \cref{eq:lr vector currents local rotations,eq:current transformations}, these derivatives have to be promoted to covariant derivatives.
The covariant derivative of a generic external current $\tensor J$ is
\begin{equation}
\i D_\mu \tensor J_{A \dot C}^{B \dot D} \equiv
\i \partial_\mu \tensor J_{A \dot C}^{B \dot D}
+ \sum_{i=1}^n \flavour L_\mu{}_{a_i}^x \tensor J_{A_x \dot C}^{B \dot D}
- \sum_{j=1}^m \flavour L_\mu{}_x^{b_j} \tensor J_{A \dot C}^{B_x \dot D}
+ \sum_{k=1}^p \flavour R_\mu{}_{\dot c_k}^{\dot x} \tensor J_{A \dot C_x}^{B \dot D}
- \sum_{l=1}^q \flavour R_\mu{}_{\dot x}^{\dot d_l} \tensor J_{A \dot C}^{B \dot D_x} \ ,
\end{equation}
where the capital indices denote multi-indices
\begin{equation}
\begin{aligned}
A &= a_1 \dots a_n \ , \qquad &
A_x &= a_1 \dots a_{i-1} x a_{i+1} \dots a_n \ , \\
B &= b_1 \dots b_m \ , \qquad &
B_x &= b_1 \dots b_{i-1} x b_{j+1} \dots b_m \ , \\
\dot C &= \dot c_1 \dots \dot c_p \ , \qquad &
\dot C_x &= \dot c_1 \dots \dot c_{k-1} \dot x \dot c_{k+1} \dots \dot c_p \ , \\
\dot D &= \dot d_1 \dots \dot d_q \ , \qquad &
\dot D_x &= \dot d_1 \dots \dot d_{l-1} \dot x \dot d_{l+1} \dots \dot d_q \ .
\end{aligned}
\end{equation}
The current $\Theta$ does not carry any flavour indices, but due to the axial anomaly it transforms like the trace of the logarithm of a unitary matrix $\flavour \vartheta_a^{\dot b} \equiv e^{\i \Theta} \flavour 1_a^{\dot b}$ with two flavour indices.
Hence, its covariant derivative can be defined as
\begin{equation} \label{eq:derivate theta}
D_\mu \Theta \equiv \vartheta_\mu \equiv - \i \trf{\flavour \vartheta^\dagger D_\mu \flavour \vartheta} = \partial_\mu \Theta - L_\mu + R_\mu \ .
\end{equation}
This object is a chiral invariant and therefore not a covariant derivative in the proper sense.
In analogy to gauge fields, the external currents $\nonet L^\mu$ and $\nonet R^\mu$ cannot appear by themselves.
Instead \cPT depends on the left- and right-handed field strength tensors
\begin{align} \label{eq:field strength tensors}
\nonet L^{\mu\nu} &= \partial^\mu \nonet L^\nu - \partial^\nu \nonet L^\mu - \i \comm{\nonet L^\mu}{\nonet L^\nu} \ , &
\nonet R^{\mu\nu} &= \partial^\mu \nonet R^\nu - \partial^\nu \nonet R^\mu - \i \comm{\nonet R^\mu}{\nonet R^\nu} \ .
\end{align}
To prepare for the eventual decomposition of the \cPT Lagrangian into \SM and portal contributions, it is also convenient to define the left- and right-handed portal field strength tensors
\begin{align} \label{eq:hidden field strength tensors}
\nonet V_l^{\mu\nu} &= \partial^\mu \nonet V_l^\nu - \partial^\nu \nonet V_l^\mu - \i \comm{\nonet V_l^\mu}{\nonet V_l^\nu} \ , &
\nonet V_r^{\mu\nu} &= \partial^\mu \nonet V_r^\nu - \partial^\nu \nonet V_r^\mu - \i \comm{\nonet V_r^\mu}{\nonet V_r^\nu} \ .
\end{align}

\subbib

%% file: hidden.pgf
\tikzset{max width/.style args = {#1}{
execute at begin node = {\begin{varwidth}{#1}\narrowragged},
execute at end node = {\end{varwidth}}}}
\begin{tikzpicture}[node distance = 4ex and 3em]
\node (l1) {};
\node (l2) [below = of l1] {};
\node (l3) [below = of l2] {};
\node (left) [fit=(l1) (l2) (l3)] {};
\node (SM) [right= of left, circle, minimum size=4em, fill=gray!50] {$\SM$};
\node (portal) [right= of SM, label={[label distance=-6ex]Portal}, fill, circle, inner sep = 2pt] {};
\node (hidden) [right= of portal, circle, minimum size=4em, fill=gray!50] {hidden};
\node (r1) [right = of hidden] {};
\node (r2) [above = of r1] {};
\node (r3) [below = of r1] {};
\draw (l1) -- (SM) node [midway, below, rotate=90, anchor=east, inner sep = 0pt] {\dots};
\draw (l2) -- (SM);
\draw (l3) -- (SM) node [midway, above, rotate=-90, anchor=east, inner sep = 0pt] {\dots};
\draw (hidden) -- (r1);
\draw (hidden) -- (r2) node [midway, below, rotate=90, anchor=east, inner sep = 0pt] {\dots};
\draw (hidden) -- (r3) node [midway, above, rotate=-90, anchor=east, inner sep = 0pt] {\dots};
\draw (SM.north east) -- (portal) node [midway, below, rotate=90, anchor=east, inner sep = 0pt] {\dots};
\draw (SM.south east) -- (portal);
\draw (portal) -- (hidden.north west) node [midway, below, rotate=90, anchor=east, inner sep = 0pt] {\dots};
\draw (portal) -- (hidden.south west);
\node (d6) [above = 4ex of SM] {$\begin{aligned}\Delta \qn f &= \pm 1\\(&4\pi)^2\\[2.4ex]d&\leq6\end{aligned}$};
\node (d45) [right = of d6.north, anchor=north west] {$\begin{aligned}\Delta \qn f &= 0\\(&4\pi)^0\\d&\leq\begin{cases}5 & \text{for spin 0, \textfrac12}\\4 & \text{for spin 1}\end{cases}\end{aligned}$};
\node (d4) [anchor=north, below = 3ex of SM] {$\begin{aligned}\Delta \qn f &= 0\\(&4\pi)^0\\[2.4ex]d&\leq4\end{aligned}$};
\node (d567) [right = of d4.north, anchor=north west] {$\begin{aligned}\Delta \qn f &= \pm 1\\(&4\pi)^2\\d&\leq\begin{cases}7 & \text{for spin 0, \textfrac12}\\6 & \text{for spin 1}\end{cases}\end{aligned}$};
\node [anchor=north] at (d6.north -| l1) {\Cref{it:one}};
\node [anchor=north] at (d4.north -| l1) {\Cref{it:two}};
\end{tikzpicture}

%% file: feynman-scalar-box.pgf
\begin{feyn}[node distance = \height/1 and \width/4]
\vertex (a1) {$d^\dagger$};
\vertex[right = of a1] (a2);
\vertex[right = \width/2 of a2] (a3);
\vertex[right = of a3] (a4) {$d$};
\vertex[below = of a1] (b1) {$\nu$};
\vertex[below = of a2] (b2);
\vertex[below = of a3] (b3);
\vertex[below = of a4] (b4) {$\nu^\dagger$};
\vertex[below = \height/2 of a4] (h1) {$s$};
\vertex[below = \height/2 of a3] (i1);
\diagram*{
{[edges = fermion]
(a4) -- (a3) --[edge label = $u$] (a2) -- (a1),
(b1) -- (b2) --[edge label = $\ell$] (b3) -- (b4)
},
(b2) -- [boson, edge label' = $W$] (a2),
(b3) -- [boson, edge label = $W$] (a3),
(i1) -- [scalar] (h1)
};
\end{feyn}

%% file: feynman-scalar-penguin.pgf
\begin{feyn}[node distance = \height/2 and \width/4]
\vertex (a1) {$d^\dagger$};
\vertex[right = of a1] (a2);
\vertex[below = of a2] (c0);
\vertex[right = of c0] (c1);
\vertex[right = of c1] (c2);
\vertex[below left = of c1] (b2);
\vertex[at = (b2-|a1)] (b1) {$d$};
\vertex[right = of c2] (c3);
\vertex[at = (a2-|c3), align = left] (a3) {$\nu^\dagger$};
\vertex[at = (b2-|c3), align = left] (b3) {$\nu$};
\vertex[left = of a3] (h1) {$s$};
\vertex[right = \width/8 of c1] (i1);
\diagram*{
{[edges = fermion]
(b1) -- (b2) --[edge label' = $u$] (c1) --[edge label' = $u^\dagger$] (a2) -- (a1),
(a3) -- (c2) -- (b3),
},
(c2) -- [boson, edge label = $Z$] (c1),
(b2) -- [boson, edge label = $W$] (a2),
(i1) -- [scalar] (h1)
};
\end{feyn}

%% file: feynman-fermion-penguin.pgf
\begin{feyn}[node distance = \height/2 and \width/4]
\vertex (a1) {$d^\dagger$};
\vertex[right = of a1] (a2);
\vertex[below = of a2] (c0);
\vertex[right = of c0] (c1);
\vertex[right = of c1] (c2);
\vertex[below left = of c1] (b2);
\vertex[at = (b2-|a1)] (b1) {$d$};
\vertex[right = of c2] (c3);
\vertex[at = (a2-|c3)] (a3) {$\xi^\dagger$};
\vertex[at = (b2-|c3)] (b3) {$\xi$};
\vertex[below left = \height/4 and \width/4 of a3] (i2);
\vertex [draw, crossed dot] at ($(i2)!0.4!(b3)$) (i1) {};
\diagram*{
{[edges = fermion]
(b1) -- (b2) --[edge label' = $u$] (c1) --[edge label' = $u^\dagger$] (a2) -- (a1),
(a3) -- (i2) -- (i1) -- (b3),
},
(i2) -- [boson, edge label = $Z$, edge label' = $\gamma$] (c1),
(b2) -- [boson, edge label = $W$] (a2)
};
\end{feyn}

%% file: chiral-perturbation-theory.tex
\subtoc

\section{Chiral perturbation theory} \label{sec:chiral perturbation theory}

\cPT is an effective theory of the light unflavoured and strange pseudoscalar mesons with masses below roughly \unit[1]{GeV}, which corresponds to the mass scale associated with the \GD contribution to \eg the proton mass.
Experimentally, one observes nine such mesons $\phi$: three pions $\pi^\pm$ and $\pi^0$, four kaons $K^\pm$, $K^0$, and $\overline K^0$, and the two $\eta$- and $\eta^\prime$-mesons.
Neglecting their masses, the typical energy scale of interactions involving these mesons is determined by the meson decay constants, which are defined in terms of the hadronic matrix elements \cite{Zyla:2020zbs}
\begin{align} \label{eq:pion decay matrix element}
f_\phi&= \frac{\i}{2m_\phi^2} \mel**{0}{\partial^\mu \trf{\left(\nonet Q_\mu - \overline{\nonet Q}_\mu\right) \flavour \lambda_\phi} }{\phi(p)} e^{-\i p x} \ ,
\end{align}
where $m_\phi$ is the mass of the meson in question, $(\nonet Q_\mu - \overline{\nonet Q}_\mu)/2$ is axial-vector quark current, and the matrix $\flavour \lambda_\phi$ projects onto the relevant combination of quark flavours.
In particular, the charged pion decay constant $f_\pi = \unit[(65.1\pm0.4_{\ex})]{MeV}$, for which $\flavour \lambda_\phi = \flavour \lambda_\up^\down$, determines the charged pion decay width \cite{Zyla:2020zbs}
\begin{equation}
\Gamma\left(\pi^\pm \to \ell^\pm \nu\right) = \frac{1}{16 \pi m_\pi} \left(1 - \frac{m_\ell^2}{m_\pi^2}\right) \frac{f_\pi^2 m_\ell^2}{v^4} \abs{V_{\up\down}}^2 \left(m_\pi^2 - m_\ell^2\right) \ ,
\end{equation}
where $m_\pi = \unit[(139.57018 \pm 0.00035_{\ex})]{MeV}$ is the mass of the charged pion and $m_\ell$ is the mass of the charged lepton $\ell^\pm = \electron^\pm$, $\muon^\pm$.

The light pseudoscalar mesons can be identified with the \PNGBs of the explicitly broken chiral $G_{LR} = \U(3)_L \times \U(3)_R$ \cref{eq:flavour symmetry} of the kinetic \QCD \cref{eq:Q kinetic Lagrangian}.
\cPT is defined via a perturbative expansion of \QCD around the limit without explicit chiral symmetry breaking, which can be constructed by setting the external currents to zero while keeping only the zeroth order terms in the large $\Nc$ expansion.
In this limit, the quark condensate \eqref{eq:quark condensate} still spontaneously breaks the $G_{LR}$ symmetry to a $G_V = \U(3)_V$ vector symmetry, so that the Goldstone theorem \cite{Nambu:1960tm,Goldstone:1961eq,Goldstone:1962es} implies the existence of nine massless \NGBs, one for each spontaneously broken generator.
Reintroducing the explicit symmetry breaking generated by the light quark masses \eqref{eq:Q mass Lagrangian}, the other external \cref{eq:currents,eq:A currents}, and the axial anomaly \eqref{eq:Q axial anomaly Lagrangian} as small perturbations, one obtains the $\U(3)$ version of \cPT, which contains nine massive \PNGBs.
The \PNGB masses scale as
\begin{equation} \label{eq:pngb mass relation}
m_\phi^2 \propto \mathcal L_\text{broken} \ ,
\end{equation}
where $\mathcal L_\text{broken}$ is the part of the Lagrangian that contains the explicit symmetry breaking terms.
In this version of \cPT, it is necessary to expand \QCD in powers of $\inv \Nc$ in order to control the impact of the axial anomaly.
Without this expansion, the axial anomaly badly breaks the $\U(1)_A$ symmetry of \QCD, and the perturbative expansion in the anomalous contribution to $\mathcal L_\text{broken}$ becomes invalid.
Following this approach, one obtains the $\SU(3)$ version of \cPT, which contains only eight \PNGBs, one for each broken generator of $\SU(3)_L \times \SU(3)_R \subset G_{LR}$.
However, we work in the $\U(3)$ version, since it is better suited for understanding the coupling of the \SM mesons to pseudoscalar hidden mediators such as \ALPs.

\begin{figure}
\begin{panels}{.6}
\tikzsetnextfilename{octet}\input{octet.pgf}
\caption{Meson octet $\octet \Phi$.} \label{fig:octet}
\panel{.4}
\tikzsetnextfilename{singlet}\input{singlet.pgf}
\caption{Meson singlet $\Phi$.} \label{fig:singlet}
\end{panels}
\caption[Quantum numbers of the light scalar meson nonet]{
The light pseudoscalar mesons.
Panel \subref{fig:octet} shows the isospin $\qn I$, strangeness $\qn s$ and electric charge $\qn q$ quantum numbers of the light pseudoscalar meson octet, and panel \subref{fig:singlet} shows the quantum numbers of the singlet.
The three unflavoured mesons ($\pi_8$, $\eta_8$, and $\eta_1$) mix into the neutral mass eigenstates ($\pi^0$, $\eta$, and $\eta^\prime$).
} \label{fig:mesons}
\end{figure}

In $\U(3)$ \cPT, the \PNGBs parameterise the coset $\flatfrac{G_{LR}}{G_V} \cong \U(3)_A$ in terms of a non-linearly realised matrix valued field \cite{DiVecchia:1980yfw,Gasser:1984gg,HerreraSiklody:1996pm,Kaiser:2000gs}
\begin{align} \label{eq:non-linear representation}
\nonet \u(x) &= \exp \frac{\i \nonet \Phi(x)}{\fp} \ ,
\end{align}
where the dimensionful parameter $\fp$ determines the typical energy scale of \cPT.
At \LO in the small momentum expansion of \cPT, the meson decay constants in \eqref{eq:pion decay matrix element} are all identical and equal to $\fp$, that is, $f_\phi = \fp$, but higher order corrections cause the meson decay constants to acquire different values, \cf \cref{sec:SM meson masses}.
Since the impact of higher order corrections is smallest for the pion, it is conventional to fix $\fp$ by matching to the pion decay constant $f_\pi$.
The \PNGB matrix
\begin{align}
\nonet{\Phi}(x) &= \octet \Phi(x) + \frac{\flavour 1}{\nfl} \Phi(x) \ , &
\Phi(x) &= \trf{\nonet{\Phi}(x)} \ ,
\end{align}
transforms as a nonet under $G_V$.
Its trace $\Phi$ transforms as a singlet under $G_V$, while the traceless contribution $\octet \Phi$ transforms as an octet.
Using the \GM matrices
\footnote{
The \GM matrices are
$\begin{aligned}[t]
\flavour \lambda_1 &= \flavour \proj_\up^\down + \flavour \proj_\down^\up \ , &
\flavour \lambda_4 &= \flavour \proj_\up^\strange + \flavour \proj_\strange^\up \ , &
\flavour \lambda_6 &= \flavour \proj_\down^\strange + \flavour \proj_\strange^\down \ , &
\flavour \lambda_3 &= \flavour \proj_\up^\up - \flavour \proj_\down^\down \ , \\
\i \flavour \lambda_2 &= \flavour \proj_\up^\down - \flavour \proj_\down^\up \ , &
\i \flavour \lambda_5 &= \flavour \proj_\up^\strange - \flavour \proj_\strange^\up \ , &
\i \flavour \lambda_7 &= \flavour \proj_\down^\strange - \flavour \proj_\strange^\down \ , &
\sqrt\nfl \flavour \lambda_8 &= \flavour \proj_\up^\up + \flavour \proj_\down^\down - 2 \flavour \proj_\strange^\strange \ .
\end{aligned}$
}
$\flavour \lambda_a$ and the rescaled identity matrix $\flavour \lambda_0 = \sqrt{\flatfrac2{\nfl}} \flavour 1$, which are normalised such that $\trf{\flavour \lambda_a \flavour \lambda_b} = 2 \delta_{ab}$, to parameterise the \PNGB octet and singlet according to
\begin{align} \label{eq:meson matrix}
\octet \Phi & =\sum_{a\neq0} \frac{\phi_a \flavour \lambda_a}{\sqrt2}
= \begin{pmatrix}
\etaoctet + \pion & \pi^+ & K^+ \\
\pi^- & \etaoctet - \pion & K^0 \\
K^- & \overline K^0 & - 2 \etaoctet
\end{pmatrix} \ , &
\Phi &= \trf{\frac{\phi_0 \flavour \lambda_0}{\sqrt2}} = \nfl \etasinglet \ ,
\end{align}
their components can be identified with the light meson flavour eigenstates $\phi_a = \{\pi^\pm$, $K^\pm$, $K^0$, $\overline K^0$, $\pi_8$, $\eta_8$, $\eta_1\}$, whose quantum numbers are depicted in \cref{fig:mesons}.
There is a large mass-mixing between the $\eta_8$- and $\eta_1$-mesons.
After diagonalisation, the two mass eigenstates are denoted as $\eta$ and $\eta^\prime$.
Isospin violating contributions further induce a small mixing between the neutral pion and the two $\eta$-mesons, while \EW corrections induce a feeble kinetic mixing between the charged kaons and pions.

\subsection{Flavour symmetry}

In the absence of explicit symmetry breaking, the \cPT action has to be invariant under the global $G_{LR}$ flavour \cref{eq:flavour symmetry} of the kinetic \QCD \cref{eq:Q kinetic Lagrangian}.
The coset matrix $\nonet g$ and the trace of the pseudoscalar meson matrix $\flatfrac{\Phi}{\fp} = - \i \trf{\ln\nonet g}$ transform under the action of $G_{LR}$ as \cite{DiVecchia:1980yfw,Gasser:1984gg,HerreraSiklody:1996pm,Kaiser:2000gs}
\begin{align} \label{eq:coset transformations}
\nonet \u &\to \VL \nonet \u \VR \ , &
\frac{\Phi}{\fp} &\to \frac{\Phi}{\fp} - \i \trf{\ln \VL \VR} \ .
\end{align}
The transformation behaviour of $\Phi$ mirrors the behaviour of the pseudoscalar external current $\Theta$ \eqref{eq:theta transformations}, which is also a $G_V$ singlet.
When including the external currents $J = \{\Omega$, $\Theta$, $\nonet \M$, $\nonet L_\mu$, $\nonet R_\mu$, $\octet T_{\mu \nu}$, $\octet \Chromo$, $\tensor \Fermi_\scalar$, $\tensor \Fermi_r$, $\tensor \Fermi_l \}$, the \cPT action can be obtained by means of a spurion analysis, which corresponds to enforcing the invariance of the \cPT path integral under the \emph{local} flavour \cref{eq:flavour symmetry} \cite{DiVecchia:1980yfw, Gasser:1984gg, Pich:1995bw, HerreraSiklody:1996pm, Kaiser:2000gs, Scherer:2002tk}.
This entails the promotion of the partial derivative $\partial^\mu \nonet \u$ to a covariant derivative
\begin{align} \label{eq:U covariant derivative}
D^\mu \nonet \u &= \partial^\mu \nonet \u - \i \left(\nonet L^\mu \nonet \u - \nonet \u \nonet R^\mu\right) \ ,
\end{align}
where the left- and right-handed external currents $\nonet L^\mu$ and $\nonet R^\mu$ effectively fulfil the role of gauge fields.
Besides being parts of the covariant derivatives, these two external currents also contribute to the \cPT action via operators involving the left- and right-handed field strength tensors $\nonet L_{\mu\nu}$ and $\nonet R_{\mu\nu}$, \cf definition \eqref{eq:field strength tensors}, while the remaining external currents appear as regular building blocks of the theory.
$G_{LR}$ invariant operators in \cPT are then constructed by taking quark-flavour traces of either purely left- or right-handed products of the coset matrix $\nonet g$, the external currents, and their covariant derivatives.

The spurion analysis is also a standard tool used to embed \cPT into the remainder of the \SM, by parameterising the coupling of \QCD to the \EW sector in terms of the external currents $\Theta$, $\nonet M$, $\nonet L^\mu$, and $\nonet R^\mu$,
which describe \CP-violation, quark masses, and \EM vector current interactions in the \SM, respectively.
For more details, see \eg the general introductions to \cPT in \cite{Pich:1993uq, Pich:1995bw, Ecker:1996yy, Scherer:2002tk}.
In the \SM, the spurion approach neglects contributions to the \cPT Lagrangian that are generated from diagrams with virtual photon exchanges.
Starting at order $\alpha_{\EM} \propto e^2$, one has to include an additional set of \EM operators in order to complete \cPT.
For extensive listings of these operators, see \eg \cite{Bijnens:1983ye,Buras:1987wc,Ecker:1988te,Urech:1994hd,Knecht:1999ag,Ecker:2000zr}.
In particular, they are necessary to obtain the correct \SM estimates for \eg the pion mass splitting and the $\flatfrac{\epsilon^\prime}{\epsilon}$ ratio \cite{Buras:1987qa,Buchalla:1989we,Aebischer:2020jto},
which measures the correlation of \CP-violation in decays of neutral kaons into pairs of charged pions, $K^0 \to \pi^+ \pi^-$, and neutral pions, $K^0 \to \pi^0 \pi^0$.

\subsection{Power counting}

\begin{table}
\begin{tabular}{r*4c} \toprule
& $\epsilonUV$ & $\epsilonSM$ & $\epsilonEW$ & $\delta$ \\ \midrule
Numerator & $v$ & $\partial^2 \lesssim m_\charm^2$ & $\Lambda_{\cPT}^2$ & $\partial^2 \lesssim m_K^2$ \\
Denominator & $f_{\UV}$ & $\Lambda_{\SM}^2$ & $\Lambda_{\SM}^2$ & $\Lambda_{\cPT}^2$ \\
\bottomrule \end{tabular}
\caption[Small parameters appearing in portal $\cPT$]{
Small parameters that are defined as ratios between the relevant $\UV$, \SM, and \cPT scales.
The small parameter $\delta$ also captures the expansion in $\inv \Nc$ of $\U(3)$ \cPT.
$\Lambda_{\SM} = 4\pi v$ and $\Lambda_{\cPT} = 4\pi\fp$ are defined such that they include $\NDA$ loop factors.
For momenta $\partial^2 \lesssim m_K^2$, one has $\epsilonSM = \delta \epsilonEW$.
} \label{tab:small parameters}
\end{table}

When accounting only for the explicit symmetry breaking due to the axial anomaly, $\U(3)$ \cPT is defined via a simultaneous expansion in small momenta $\flatfrac{\partial^2}{\Lambda_{\cPT}^2}$ and $\inv \Nc$,
where $\Lambda_{\cPT} = 4 \pi \fp = \unit[(803\pm15_{\ex}\pm\NNLO)]{MeV}$ is the symmetry breaking scale of \cPT \cite{DiVecchia:1980yfw,Manohar:1983md,Gasser:1984gg,HerreraSiklody:1996pm,Kaiser:2000gs}.
Following \cite{Kaiser:2000gs}, we combine both of these expansions by defining a single degree of smallness $\delta \propto \flatfrac{\partial^2}{\Lambda_{\cPT}^2} \propto \inv \Nc$.
This is appropriate for kaon decays, since $\inv \Nc = \nicefrac13 \simeq \flatfrac{m_K^2}{\Lambda_{\cPT}^2}$.
At lower energies, such as for $\partial^2 \simeq m_\pi^2 \ll m_K^2$, the suppression associated with the small momenta is a much better expansion parameter than $\inv{\Nc}$.
In this case, it is more appropriate to work with $\SU(2)$ or $\SU(3)$ \cPT, so that the large $\Nc$ expansion, which is necessary in $\U(3)$ \cPT, can be avoided.
Besides the expansion in $\delta$, we also track the suppression due to $\epsilonSM$ and $\epsilonUV$, as defined in \eqref{eq:epsilon SM,eq:epsilon UV}, and we eliminate operators that are doubly suppressed in either one of these two parameters.
\Cref{tab:small parameters} summarises the relation between the four expansion parameters.

\paragraph{Momentum expansion}

\cPT can be expanded in powers of $\flatfrac{\partial^2}{\Lambda_{\cPT}^2} = \flatfrac{\partial^2}{(4\pi\fp)^2}$ by adopting the general power counting scheme for \LE \EFTs \cite{Weinberg:1978kz}, which is established by studying the behaviour of individual diagrams under a rescaling $p_i \to x p_i$ of the external momenta $p_i$.
Since $\fp^2 \sim \partial^2$ defines the typical energy scale of \cPT, the resulting power counting in \cPT is equivalent to the $(4\pi)^{-1}$ expansion of \NDA \cite{Manohar:1983md}.
Applying the \NDA power counting rules, derivatives $\partial_\mu$ are suppressed by factors of $\sqrt \delta$, while powers of the \PNGB matrix $\nonet \Phi$ are unsuppressed.
Since the external currents $\nonet L^\mu$ and $\nonet R^\mu$ appear in the covariant derivative \eqref{eq:U covariant derivative}, they also count as $\sqrt \delta \propto \flatfrac{\partial_\mu}{\Lambda_{\cPT}}$.
The external currents $\nonet \M$, $\octet \Chromo$, and $\octet T^{\mu\nu}$ contribute to the \PNGB masses, so relation \eqref{eq:pngb mass relation} implies that all three of them count as $\nonet \M$, $\octet \Chromo$, $\octet T^{\mu\nu} \propto m_\phi^2 \propto \partial^2 \propto \delta$.
In summary, each of these building blocks counts as
\begin{align}
\nonet g &\propto 1 \ , &
\partial, \nonet L, \nonet R &\propto \sqrt \delta \ , &
\nonet \M, \octet \Chromo, \octet T &\propto \delta \ .
\end{align}

\paragraph{Large $\Nc$ expansion}

The standard formula for large $\Nc$ scaling behaviour for diagrams without four-quark operators
\eqref{eq:large N_c scaling} shows that the leading \QCD diagrams with a given number of quark loops are suppressed by one factor of $\inv \Nc$ for each quark-loop.
Since \cPT operators with $n_q$ quark flavour traces have to be generated by contributions in the \QCD path-integral with at least $n_q$ quark loops, each quark-flavour trace in \cPT counts as $\inv \Nc$ \cite[appendix A of][]{Kaiser:2000gs}.
The large $\Nc$ scaling behaviour of the leading \QCD diagrams also directly implies that the external currents $\Theta$ and $\Omega$ count as $\delta \propto \inv \Nc$.

Equation \eqref{eq:modified large N counting} establishes a modified large $\Nc$ scaling for \QCD diagrams with four-quark vertices.
It implies that \cPT operators with one four-quark current insertion, $\tensor \Fermi_x$ or $\octet \Fermi_x$, are enhanced by a relative factor of $\Nc$ associated with the four-quark vertex.
In addition, the leading contributions to the \QCD path-integral with one four-quark insertion contain two quark loops but scale as if they contain only a single quark loop.
Each additional quark loop that is not associated with the four-quark insertion still gives a suppression $\propto \inv{\Nc}$.
In total, this means that \cPT operators with one four-quark current insertion and $n_q =1$, $2$ quark flavour traces scale as $\Nc^2$, while operators with one four-quark current insertion and $n_q > 2$ quark flavour traces scale as $\Nc^{4 - n_q}$.
In summary, each of the above building blocks counts as
\begin{align}
S_\omega, \Theta &\propto \delta \ , &
\tensor \Fermi_x , \octet \Fermi_x &\propto \delta^{-1} \ , &
\trf{\param}^n &\propto \delta^{\max(n-n_\fermi,n_\fermi)} \ , &
n_\fermi &= 0,\,1 \ .
\end{align}
where $n_\fermi =0$, $1$ is the number of four-quark current insertions.

\paragraph{Expansion in powers of $\epsilonSM$ and $\epsilonUV$}

The parameter $\epsilonSM = \flatfrac{\partial^2}{\Lambda_{\SM}^2}$ with $\Lambda_{\SM} = 4\pi v$ measures the degrees of smallness associated with higher-dimensional operators at low energies.
However, it mixes the small momentum expansion of \cPT with the suppression due to virtual $W$-boson exchanges.
In order to separate these two expansions, we define $\epsilonSM = \delta \epsilonEW$, where $\epsilonEW = \flatfrac{\fp^2}{v^2} = \flatfrac{\Lambda_{\cPT}^2}{\Lambda_{\SM}^2}$ is the ratio between the \cPT and \EW scales.
With this definition, the external currents $\octet \Chromo$, $\octet T^{\mu\nu}$, $\tensor \Fermi_x$, and $\octet \Fermi_x$ are all suppressed by one factor of $\epsilonEW$ in \cPT, independent of any additional momentum suppression.
Additionally, the suppression due to factors of $\epsilonUV$ has to be taken into account when considering modifications due to the $\Omega$ current, since the \SM contribution $\omega \propto \inv[2]{g_\strong}$ is integrated out when constructing \cPT, so that only the hidden sector contributions $S_\omega$ remains.
At \LO in both $\epsilonUV$ and $\epsilonEW$, the \cPT action can be at most linear in each of the above currents.

\subsection{Construction of the portal $\cPT$ Lagrangian}

\begin{table}
\begin{tabular}{ccr*4{r@{}l}} \toprule
$n_t$ & $n_t^\prime$ & $\Nc^m$ & \multicolumn{8}{c}{$\delta^n$} \\ \cmidrule{4-11}
& & & \multicolumn{2}{c}{$0$} & \multicolumn{2}{c}{$1$} & \multicolumn{2}{c}{$2$} & \multicolumn{2}{c}{$3$} \\ \midrule
& 1, 2 & 2 & $\partial^0$&$\Nc^2$ & $\partial^2$&$\Nc^2$ & $\partial^4$&$\Nc^2$ & $\partial^6$&$\Nc^2$ \\ \cmidrule(l){6-11}
1 & 3 & 1 & & & $\partial^0$&$\Nc^1$ & $\partial^2$&$\Nc^1$ & $\partial^4$&$\Nc^1$ \\
2 & 4 & 0 & & &&& $\partial^0$&$\Nc^0$ & $\partial^2$&$\Nc^0$ \\
3 & 5 & $-1$ & &&&& & & $\partial^0$&$\Nc^{-1}$ \\ \cmidrule(r){4-5} \cmidrule(l){6-11}
& & & \multicolumn{2}{c}{$\GD$} & \multicolumn{6}{c}{$\cPT$} \\
\bottomrule \end{tabular}
\caption[Impact of flavour traces on the $\delta$-counting of an operator.]{
Impact of flavour traces on the $\delta$-counting of an operator.
$n_t$ counts the number of flavour traces in operators without four-quark current insertions, while $n_t^\prime$ counts the same number in operators with four-quark insertions.
Note that $m=2-n_t= \min(4 - n_t^\prime,2)$ and that \cPT operators proportional to $\Nc^2$ are only possible in the modified four-quark counting scheme
} \label{tab:power counting}
\end{table}

We construct the complete \cPT Lagrangian that couples the light pseudoscalar mesons to generic hidden sectors at \LO.
To this end, we first summarise the shape of the \cPT Lagrangian when neglecting $\epsilonEW$ and $\epsilonUV$ suppressed hidden sector contributions.
In this case, the only non-vanishing external currents are $\nonet L^\mu$, $\nonet R^\mu$, $\nonet \M$, and $\Theta$, and the resulting \cPT Lagrangian is well established, see \eg the discussions in \cite{DiVecchia:1980yfw, Gasser:1984gg, Pich:1993uq, Pich:1995bw, Ecker:1996yy, HerreraSiklody:1996pm, Kaiser:2000gs, Scherer:2002tk, Bijnens:2006zp}.
Afterwards, we consider the $\epsilonEW$ and $\epsilonUV$ suppressed contributions and use the spurion approach to construct the novel contributions
with general spacetime dependent currents $S_\omega$, $\octet \Chromo$, $\tensor \Fermi_x$, and $\octet \Fermi_x$.

The leading contributions to the connected part of the \QCD path integral count as order $\Nc^2$, and determine \GD in the large $\Nc$ limit \cite{tHooft:1973alw, Coleman:1985rnk, Manohar:1998xv, tHooft:2002ufq}.
Since meson dynamics are determined by connected \QCD diagrams with at least one quark loop, which scale at most as order $\Nc$, \cPT operators have to be suppressed by at least a factor of $\delta$ compared to the leading \QCD diagrams.
The only chiral invariant that could contribute at this order is
\begin{equation}
\Left \Theta = \i \left(\Theta - \frac\Phi\fp \right) \ ,
\end{equation}
where the hat indicates a flavour invariant quantity.
However, an operator proportional to $\Left \Theta$ is forbidden by parity conservation \cite{DiVecchia:1980yfw, HerreraSiklody:1996pm, Kaiser:2000gs}.
Hence, the leading contributions to the \cPT action are of order $\delta^2$.
See \cref{tab:power counting} for an overview of the possible orders of an operator.

\paragraph{Order $\delta^2$} \label{sec:U leading order}

Operators that contribute at this order can count either as order $\partial^2 \Nc$ or order $\partial^0 \Nc^0$.
Operators that count as order $\partial^2 \Nc$ contain only a single quark-flavour trace.
In the absence of explicit symmetry breaking due to the mass-like current $\nonet \M$, the only available operator of this type is \cite{DiVecchia:1980yfw, Gasser:1984gg, Pich:1993uq, Pich:1995bw, Ecker:1996yy, HerreraSiklody:1996pm, Kaiser:2000gs, Scherer:2002tk, Bijnens:2006zp}
\begin{equation} \label[lag]{eq:U kinetic Lagrangian}
\mathcal L_U^{D^2} = \frac{\fp^2}{2} \trf{\nonet U_\mu \nonet U^\mu} \ ,
\end{equation}
where the \emph{left-handed} \MC field associated with $\nonet g$ is
\begin{align} \label{eq:Maurer-Cartan field}
\nonet U_\mu &= \nonet \L_\mu - \nonet L_\mu + \Left{\nonet R}_\mu \ , &
\nonet \L_\mu &= \i \nonet \u \partial_\mu \nonet \u^\dagger = - \i (\partial_\mu \nonet \u) \nonet \u^\dagger \ , &
\Left{\nonet R}_\mu &= \nonet \u \nonet R_\mu \nonet \u^\dagger
\end{align}
and $\nonet \L_\mu$ is the \MC field obtained when neglecting the external currents $\nonet L_\mu$ and $\nonet R_\mu$.
Bold hatted operators such as $\Left{\nonet R}_\mu$ are composite operators constructed from an external current and the coset matrix $\nonet g$ such that they transform under $G_{LR}$ in the same way as $\nonet U_\mu$.
\footnote{
$\nonet g$ is defined such that it is adjoined in the mass term \eqref{eq:U mass Lagrangian} whenever the canonical quark mass matrix $\nonet M$ is adjoint.
Furthermore, we define the \MC field to be left-handed (rather than right-handed) in order to simplify the description of $W$-boson induced processes.
However, note that relations that involve only $\nonet U_\mu$, $\vartheta_\mu$, and hatted quantities are invariant under a change of either definition, provided that the hat-operation is first redefined such that it transform external currents into purely right-handed (rather than left-handed) objects and then reapplied appropriately.
}
The \MC field transforms as
\begin{equation}
\nonet U_\mu \to \VL \nonet U_\mu \VL^\dagger
\end{equation}
and corresponds to the \LER of the conserved current associated with \emph{left-handed} chiral quark flavour rotations, \cf \cref{sec:flavour symmetry}.
It obeys the relation
\begin{subequations}
\begin{align}
D_\mu \nonet U_\nu - D_\nu \nonet U_\mu &= \i \comm{\nonet U_\mu}{\nonet U_\nu} - \nonet L_{\mu\nu} + \Left{\nonet R}_{\mu\nu} \ , &
\Left{\nonet R}_{\mu\nu} &= \nonet \u \nonet R_{\mu\nu} \nonet \u^\dagger \ , \\
\partial_\mu \nonet \L_\nu - \partial_\nu \nonet \L_\mu &= \i \comm{\nonet \L_\mu}{\nonet \L_\nu} \ ,
\end{align}
\end{subequations}
and its flavour trace
\begin{equation}
U^\mu = \trf{\nonet U^\mu}
= \frac{D^\mu \Phi}{\fp}
= \frac{\partial^\mu \Phi}{\fp} - L^\mu + R^\mu
\end{equation}
encodes the covariant derivative of the trace of the coset matrix.
Note that the above object $D_\mu \Phi$ is not a covariant derivative in the strict sense, since it remains invariant under chiral rotations rather than following the transformation law for $\Phi$ in \cref{eq:coset transformations}.
When accounting for the quark mass-like current $\nonet \M$, it is possible to construct a second operator that also contributes at order $\partial^2 \Nc$ \cite{Pich:1993uq, Pich:1995bw, Ecker:1996yy, HerreraSiklody:1996pm, Kaiser:2000gs, Scherer:2002tk, Bijnens:2006zp}
\begin{equation} \label[lag]{eq:U mass Lagrangian}
\mathcal L_U^\M = \frac{\fp^2 \B}{2} \Left \M + \text{h.c.} \ ,
\end{equation}
where
\begin{align}
\Left \M &= \trf{\Left{\nonet \M}}, &
\Left{\nonet \M} &= \nonet \u \nonet \M \ , &
\Left{\nonet \M} &\to \VL \Left{\nonet \M} \VL^\dagger \ .
\end{align}
This nonet mass term gives rise to the dominant contribution to the physical masses of the pions, kaons, and the $\eta$-meson.
The mass of the heavy $\eta^\prime$-meson is dominated by the contribution of the third and final term in the \LO \cPT Lagrangian, the \PNGB singlet mass term \cite{Veneziano:1979ec,DiVecchia:1980yfw,HerreraSiklody:1996pm,Kaiser:2000gs}
\begin{equation} \label[lag]{eq:U axial anomaly Lagrangian}
\mathcal L_U^{\Theta^2} = \frac{\fp^2 \mth^2}{2\nfl} \Left \Theta^2 \ .
\end{equation}
This term contains two flavour traces and no derivatives, so that it enters at order $\partial^0 \Nc^0$ rather than $\partial^2 \Nc$.
It is associated with the explicit chiral symmetry breaking due to the axial anomaly \eqref{eq:Q axial anomaly Lagrangian}.
Putting all three contributions together, the complete \LO Lagrangian
\begin{equation} \label{eq:LO cpt lag}
\mathcal L_U^{\delta^2} = \mathcal L_U^{D^2} + \mathcal L_U^\M + \mathcal L_U^{\Theta^2}
\end{equation}
yields the \LO \EOM
\begin{equation} \label{eq:chpt lo eom}
\frac12 \left(\nonet \u D^2 \nonet \u^\dagger - D^2 \nonet \u \nonet \u^\dagger\right)
= \frac\B2 \left(\Left{\nonet \M} - \Left{\nonet \M}^\dagger\right)
+ \frac{\mth^2}{\nfl} \Left \Theta \flavour 1 \ .
\end{equation}
Together with the general identity \eqref{eq:Maurer-Cartan field}, this \EOM implies that, without loss of generality, terms containing $\nonet\u D^2 \nonet \u^\dagger$ and its Hermitian conjugate can always be eliminated from higher order Lagrangians.

\paragraph{Order $\delta^3$} \label{sec:U next-to-leading order}

Starting at this order, the \cPT action can, in principle, contain operators with covariant derivatives acting on $\nonet L^{\mu\nu}$, $\nonet R^{\mu\nu}$, $\nonet \M$, and $\Theta$.
However, up to corrections of order $\delta^4$ or higher, \PI can always be used to eliminate operators with derivatives acting on $\nonet L^{\mu\nu}$, $\nonet R^{\mu\nu}$, $\nonet \M$ in favour of operators with derivatives acting only on $\nonet g$ or $\Theta$.

Operators that contribute at order $\delta^3$ can count either as order $\partial^4 \Nc$, order $\partial^2 \Nc^0$, or order $\partial^0 \inv \Nc$.
Operators that count as order $\partial^4 \Nc$ can contain only a single quark-flavour trace.
In the absence of external currents, the only available operators of this type are \cite{HerreraSiklody:1996pm,Kaiser:2000gs}
\begin{equation} \label[lag]{eq:U NLO kinetic Lagrangians}
\mathcal L_U^{D^4} = (2 L_2 + L_3) \trf{\nonet U^\mu \nonet U_\mu \nonet U^\nu \nonet U_\nu} + L_2 \trf{\nonet U_\mu \nonet U_\nu \nonet U^\mu \nonet U^\nu} \ ,
\end{equation}
where contributions with more than one derivative acting on a single coset matrix $\nonet g$ can be eliminated using \PI, the \EOM \eqref{eq:chpt lo eom}, or identity \eqref{eq:Maurer-Cartan field}.
The quark mass-like current $\nonet \M$ generates the additional contributions \cite{HerreraSiklody:1996pm,Kaiser:2000gs}
\begin{align} \label[lag]{eq:U NLO mass Lagrangians}
\mathcal L_U^{D^2M} &= L_5 \B \trf{\Left{\nonet \M} \nonet U_\mu \nonet U^\mu} + \text{h.c.} \ , &
\mathcal L_U^{M^2} &= L_8 \B^2 \left(\trf{\Left{\nonet \M}^2} + \text{h.c.}\right) + H_2 \B^2 \trf{\Left{\nonet \M}^\dagger \Left{\nonet \M}} \ .
\end{align}
In the operator proportional to $H_2$, the dependence on the coset field $\nonet g$ drops out, so that the term does not contribute to perturbatively computed $S$-matrix elements, but it has to be added to the Lagrangian as a counter term in order to renormalise the theory \cite{Gasser:1984gg,Pich:1995bw,HerreraSiklody:1996pm,Kaiser:2000gs,Scherer:2002tk,Bijnens:2006zp}.
The field strength tensors $\nonet L_{\mu\nu}$ and $\nonet R_{\mu\nu}$ generate the contributions \cite{Gasser:1984gg,HerreraSiklody:1996pm,Kaiser:2000gs}
\begin{subequations} \label[lag]{eq:U field strength Lagrangians}
\begin{align}
\mathcal L_U^{D^2V} &= - \i L_9 \trf{\nonet U^\mu \nonet U^\nu \left(\nonet L_{\mu\nu} + \Left{\nonet R}_{\mu\nu}\right)} \ , \\
\mathcal L_U^{V^2} &= L_{10} \trf{\nonet L^{\mu\nu} \Left{\nonet R}_{\mu\nu}} + H_1 \trf{\nonet L_{\mu\nu} \nonet L^{\mu\nu} + \Left{\nonet R}_{\mu\nu} \Left{\nonet R}^{\mu\nu}} \ ,
\end{align}
\end{subequations}
where the operator proportional to $H_1$ is another counter term.
The operators that count as order $\partial^2 \Nc^0$ contain two flavour traces.
In the absence of external currents, the only available operator of this type is the kinetic term \cite{HerreraSiklody:1996pm,Kaiser:2000gs}
\begin{equation} \label[lag]{eq:U NLO anomaly Lagrangian}
\mathcal L_U^{DD} = \frac{\fp^2 }{2\nfl} \Lambda_1 U_\mu U^\mu \ .
\end{equation}
The external currents $\nonet \M$ and $\Theta$ induce the further mass-like term \cite{HerreraSiklody:1996pm,Kaiser:2000gs}
\begin{equation} \label[lag]{eq:U anomaly mass Lagrangian}
\mathcal L_U^{\M\Theta} = \frac{\fp^2 \B }{2\nfl} \Lambda_2 \Left \M \Left \Theta + \text{h.c.} \ ,
\end{equation}
and a final counter term that depends on the covariant derivative of $\Theta$ defined in \eqref{eq:derivate theta} \cite{HerreraSiklody:1996pm,Kaiser:2000gs}
\begin{equation} \label[lag]{eq:U anomaly kinetic Lagrangian}
\mathcal L_U^{\vartheta^2} = \frac{\fp^2 }{2\nfl} H_0 \vartheta_\mu \vartheta^\mu \ .
\end{equation}
There is no kinetic mixing term proportional to $U_\mu \vartheta^\mu$, since this operator can always be eliminated via a shift of $\Phi$.
There are also no operators of order $\partial^0 \inv \Nc$, since the only candidate operator is proportional to $\Left \Theta^3$, and it is forbidden due to parity conservation in \QCD.

\subparagraph{Wess-Zumino-Witten action}

Since the \NGBs are pseudoscalar fields, a parity transformation corresponds to the combined transformation of spatial inversion $x \leftrightarrow -x$ and meson conjugation $\nonet \u \leftrightarrow \nonet \u^\dagger$.
The contributions derived so far are invariant under both transformations separately, so that the resulting \cPT Lagrangian is more symmetric than \QCD.
In the absence of external currents, there is no four dimensional Lagrangian that breaks this additional symmetry \cite{Wess:1971yu}, but starting at order $\delta^3$ it is possible to construct a so-called \WZW contribution to the \cPT action that takes the form of a five dimensional integral over a sub-manifold of the nine dimensional space of field values that can be assumed by the coset matrix $\nonet \u(x)$ \cite{Witten:1983tw}.
This integral can be connected to an action written in terms of a Lagrangian density by identifying Minkowski space with the four dimensional boundary of this sub-manifold.
Hence, the \WZW term can be written as \cite{Witten:1983tw}
\begin{align} \label[lag]{eq:U WZW Lagrangian}
\Gamma_\L^{\Nc} &= - \frac{\Nc}{(2 \pi)^2} \int \d{x^5} \epsilon_{ijklm} \omega^{ijklm}_0(\nonet u^i) \ , &
\omega^{ijklm}_0(\nonet u^i) &= \frac{2}{5!} \trf{\nonet u^i \nonet u^j \nonet u^k \nonet u^l \nonet u^m} \ ,
\end{align}
where $\omega^{ijklm}_0(\nonet u^i)$ is the pure-gauge \CS term and the $i$, $j$, $k, \dots$ denote coordinate indices of the five-dimensional sub-manifold.
The left- and right-handed external currents $\nonet L^\mu$ and $\nonet R^\mu$ generate additional \WZW contributions that can be written in the form of a conventional four dimensional Lagrangian \cite{Witten:1983tw, Chou:1983qy, Kawai:1984mx, Manes:1984gk}
\begin{equation} \label[lag]{eq:U gauged WZW Lagrangian}
\mathcal L_U^{\Nc} = \frac{\Nc}{(2 \pi)^2} \epsilon_{\mu\nu\rho\sigma} \left(
\rho^{\mu\nu\rho\sigma}(\nonet u^\mu + \Left{\nonet R}^\mu, \nonet L^\mu)
+ \rho^{\mu\nu\rho\sigma}(\Left{\nonet R}^\mu, - \nonet u^\mu)
\right) \ ,
\end{equation}
where the Bardeen counter-term of two vector currents $\nonet V_i^\mu$ is
\begin{equation}
\rho^{\mu\nu\rho\sigma}(\flavour V^\mu_0, \flavour V^\mu_1)
= \frac{2}{4!} \trf{\flavour V^\mu_0 \flavour \sigma_1^{\nu\rho\sigma} + \flavour \sigma_0^{\mu\nu\rho} \flavour V^\sigma_1 + \frac\i2 \flavour V^\mu_0 \flavour V^\nu_1 \flavour V^\rho_0 \flavour V^\sigma_1} \ .
\end{equation}
This term shares a common contribution with the four dimensional gauge transformation of the five dimensional \CS term
\begin{align}
\omega_1^{\mu\nu\rho\sigma} &= \frac1{3!} \trf{\flavour v_\epsilon \partial^\mu \flavour \sigma_i^{\nu\rho\sigma}} \ , &
\flavour \sigma_i^{\mu\nu\rho} &= \frac12 \flavour F_i^{\mu\nu} \flavour V_i^\rho + \frac12 \flavour V_i^\mu \flavour F_i^{\nu\rho} + \i \flavour V_i^\mu \flavour V_i^\nu \flavour V_i^\rho \ ,
\end{align}
where $\flavour v_\epsilon$ is a gauge parameter.
The \WZW action \eqref{eq:U WZW Lagrangian} and the gauged \WZW Lagrangian \eqref{eq:U gauged WZW Lagrangian} constitute the \LO contributions to interactions with an odd number of mesons such as $K^+ K^- \to \pi^+ \pi^- \pi^0$ and $\pi^0 \to \gamma \gamma$.

\paragraph{Low energy coefficients and loops}

\begin{table}
\begin{tabular}{*9cc} \toprule
& \multicolumn{3}{c}{$\delta^2$} & \multicolumn{4}{c}{$\delta^3$} \\ \cmidrule(r){2-4} \cmidrule(l){5-8}
$\Lambda_{\cPT}$ & $\fp$ & $\B$ & $\mth$ & $4L_5$ & $4L_8$ & $2H_2$ & $\Lambda_2$ \\ \midrule
$4 \pi \fp$ & $\sqrt \Nc$ & $\Nc^{-\nicefrac12} \Lambda_{\cPT}$ & $\Nc^{-\nicefrac32} \Lambda_{\cPT}$ & $\Nc (4\pi)^{-2}$ & $\Nc (4\pi)^{-2}$ & $\Nc (4\pi)^{-2}$ & $\inv \Nc$\\
\bottomrule \end{tabular}
\caption[$\NDA$ and large $\Nc$ scaling of selected coefficients]{
$\NDA$ and large $\Nc$ scaling of selected coefficients that appear in the order $\delta^2$ and $\delta^3$ Lagrangians.
Additionally, we have made the omitted symmetry factors explicit.
} \label{tab:coefficients NDA scaling}
\end{table}

The prefactors of operators that contribute at order $\partial^{2n} \Nc^{m}$ scale as
\begin{align}
\text{coefficient} &\propto \Lambda_{\cPT}^{2 - 2n} \fp^2 \Nc^{m+n-2} \ , &
\Lambda_{\cPT}, \fp &\propto \sqrt\Nc \ ,
\end{align}
where $\fp$ scales as $\sqrt \Nc$ in order to reflect the large $\Nc$ counting of the kinetic \cref{eq:U kinetic Lagrangian} in the \LO \cPT Lagrangian.
The standard notation, which we also follow, does not make this scaling explicit.
However, we have summarised the omitted \NDA scaling and symmetry prefactors in \cref{tab:coefficients NDA scaling} and will quote numerical values of the dimensionless \NLO coefficient with symmetry factors and factors of $4\pi$ made explicit.

Diagrams with $n_l$ loops are suppressed by factors $(4\pi \fp)^{-2n_l} \propto (4 \pi)^{-2n_l} \Nc^{-n_l} \propto \delta^{2 n_l}$ compared to tree-level diagrams.
This implies that diagrams with one loop start to contribute at \NNLO.
Since we restrict ourselves to \NLO contributions, we do not consider these loop corrections.
In particular, we fix the values of the \LECs by using tree-level predictions for the light meson observables.
However, it is necessary to emphasise that one-loop contributions are expected to be numerically sizeable due to enhancement from large chiral logarithms that scale as $\propto \ln \flatfrac{\partial^2}{\mu^2}$.
In addition, one has to account for these corrections in order to capture the scale dependence of the $L_i$ and $H_i$ parameters.
As a result, the tree-level estimates for the \LO and \NLO \LECs can only be expected to be order-of-magnitude accurate.
Since the dominant corrections at \NNLO are generated by chiral loops, we expect that our estimates are the most well aligned with the \NNLO estimates that one obtains when working with a relatively small renormalisation scale, such as $\mu^2 = m_{K^\pm}^2$.

\begin{figure}
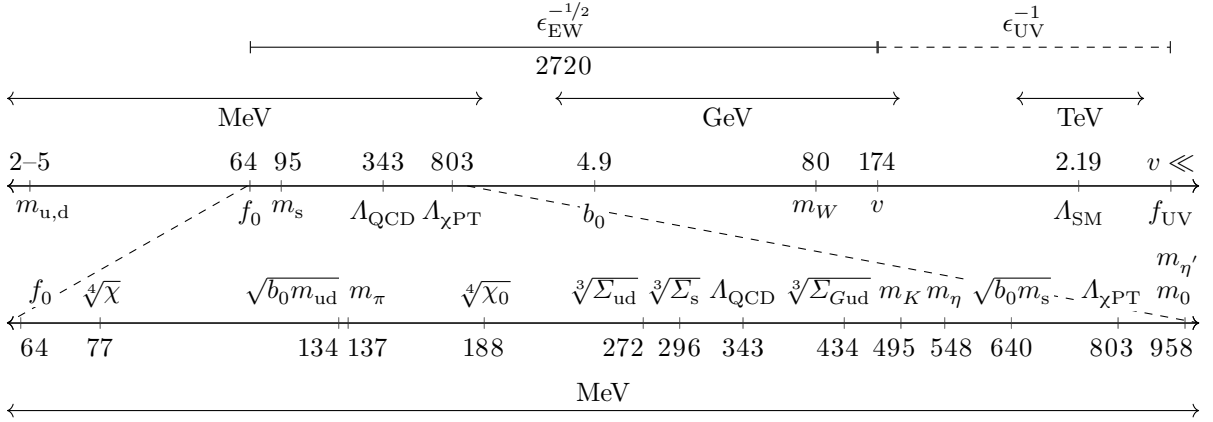

\graphic{energyscale}
\caption[Energy scales appearing in portal $\cPT$]{
Illustration of the energy scales appearing in portal $\cPT$.
The numeric values correspond to the central values, for simplicity we omit all uncertainties.
} \label{fig:scales}
\end{figure}

In total, the \LO and \NLO $\U(3)$ \cPT Lagrangians contain 13 \LECs: Three \LO coefficients $\fp$, $\B$, $\mth$, and ten \NLO coefficients $L_i$, $H_i$, and $\Lambda_i$.
The coefficients $\fp$, $\mth$, and $\Lambda_i$ remain finite even when accounting for loop corrections, but in general the coefficient $\B$, the $L_i$, and the $H_i$ have to be renormalised.
We use the \NLO tree-level estimates derived in \cref{sec:SM meson masses}, which gives
\begin{align}
\fp &= \unit[(63.9 \pm 1.2_{\ex})]{MeV} \pm \NNLO \ , &
\mth &= 4\pi \unit[(76.3 \pm 1.4_{\ex})]{MeV} \pm \NNLO \ ,
\end{align}
and
\begin{subequations}
\begin{align}
\sqrt{\B m_\light} &= 4\pi \unit[(10.68 \pm 0.08_{\ex})]{MeV} \pm \NNLO \ , \\
\sqrt{\B m_\strange} &= 4\pi \unit[(50.95 \pm 0.28_{\ex})]{MeV} \pm \NNLO \ .
\end{align}
\end{subequations}
See \cref{fig:scales} for a comparison of the energy scales involved in this work.
For the subsequent discussion in \cref{sec:cpt qcd matching}, we also require the values of the \NLO parameters $L_5$ and $L_8$.
Using the tree-level results from \cref{sec:SM meson masses}, one obtains the estimates
\begin{subequations} \label{eq:NLO parameter estimates}
\begin{align}
4 (4\pi)^2 L_5 &= 0.66 \pm 0.04_{\ex} \pm \NNLO \ , &
\Lambda_2 &= 0.814 \pm 0.023_{\ex} \pm \NNLO \ , \\
4 (4\pi)^2 L_8 &= 0.215 \pm 0.033_{\ex} \pm \NNLO \ ,
\end{align}
\end{subequations}
which are renormalisation scale independent at this level of accuracy.

\subsubsection{Weak current contributions} \label{sec:weak current contributions}

The weak currents $\octet \Chromo$, $\octet T^{\mu \nu}$, $\tensor \Fermi_l$, $\tensor \Fermi_\scalar$, and $\tensor \Fermi_r$ are suppressed by powers of $\epsilonEW$, so that they are only relevant in quark-flavour violating transitions.
As we have already discarded the quark-flavour conserving contributions to these currents in \cref{sec:quantum chromo dynamics,sec:portals}, the \cPT operators that involve them will automatically violate quark flavour.
We only include the leading contributions for each current.
These contributions can be either of order $\epsilonEW\delta$, $\epsilonEW\delta^2$, or $\epsilonEW\delta^3$.

\paragraph{Dipole contributions}

The dipole current $\octet \Chromo$ transforms under chiral flavour rotations like the mass-like current $\nonet \M$, so it couples to \cPT in the same way.
Hence, there is only one operator with $\octet \Chromo$ at order $\epsilonEW\delta^2$,
\begin{align}
\mathcal L_U^\Chromo &
= \frac{\epsilonEW \fp^2 \B}2 \coupl_\Chromo \Left \Chromo + \text{h.c.} \ , &
\Left \Chromo &= \trf{\Left{\nonet \Chromo}} \ , &
\Left{\nonet \Chromo} &= \nonet \u \octet \Chromo \ .
\end{align}
where $\coupl_\Chromo$ is a free parameter.
For the sake of completeness, we also note that there are three additional contributions with $\octet \Chromo$ that enter at order $\epsilonEW\delta^3$.
These are
\begin{subequations}
\begin{flalign}
\mathcal L_U^{\Chromo D^2} &= \frac{\epsilonEW \B}{4} \coupl_\Chromo^{D^2}\trf{\Left{\nonet \Chromo} \nonet U_\mu \nonet U^\mu} + \text{h.c.} \ , &
\mathcal L_U^{\Chromo \Theta} &= \frac{\epsilonEW \fp^2 \B}{2 \nfl} \coupl_\Chromo^\Theta \Left \Chromo \Left \Theta + \text{h.c.} \ , \\ \label[lag]{eq:U dipole mass Lagrangian}
\mathcal L_U^{\Chromo \M} &= \mathrlap{\frac{\epsilonEW \B^2}{2} \left(\coupl_\Chromo^{M} \trf{\Left{\nonet \Chromo} \Left{\nonet \M}} + \coupl_\Chromo^{M^\prime} \trf{\Left{\nonet \Chromo} \Left{\nonet \M}^\dagger}\right) + \text{h.c.} \ ,}
\end{flalign}
\end{subequations}
The $\coupl_\Chromo^x$ with $x=D^2$, $\Theta$, $M$, and $M^\prime$ are four more free parameters, and the second operator in \eqref{eq:U dipole mass Lagrangian} is another counter term.
The impact of the term $\mathcal L_U^{\Chromo D^2}$ in the \SM, where $\octet \Chromo \to \octet \chromo$ is a constant, has been discussed in \cite{Buras:2018evv}.
The authors also estimate the parameter $\coupl_\Chromo^{D^2}$.
Further operators with covariant derivatives acting on $\Left{\nonet H}$ can be eliminated using \PI.

\paragraph{Tensor contributions}

Without loss of generality, the tensorial current is traceless in Lorentz space, $\octet T^\mu_\mu = 0$, so that its two Lorentz indices have to be contracted by either two covariant derivatives or a field strength tensor.
Hence, the leading contributions with $\octet T^{\mu\nu}$ count as $\order{\epsilonEW\partial^4 \Nc} \sim \order{\epsilonEW\delta^3}$.
The two available operators of this type are
\begin{align} \label[lag]{eq:U tensor Lagrangian}
\mathcal L_U^{T D^2} &
= \frac{\epsilonEW}{4\fp} \coupl_T^{D^2} \trf{\Left{\nonet T}_{\mu\nu} \nonet U^\mu \nonet U^\nu} + \text{h.c.} \ , &
\mathcal L_U^{T V} &
= \frac{\epsilonEW}{2\fp} \coupl_T^{LR} \trf{\Left{\nonet T}_{\mu\nu} (\octet L^{\mu\nu} + \Left{\octet R}^{\mu\nu})} + \text{h.c.} \ ,
\end{align}
where $\Left{\nonet T}_{\mu\nu} = \nonet \u \octet T_{\mu\nu}$ and the $\coupl_T$ are free parameters.
This result is consistent with the list of operators obtained in \cite{Cata:2007ns}, which also includes terms that are quadratic in $\octet T^{\mu\nu}$.

\paragraph{Four-quark contributions}

\begin{table}
\begin{tabular}{r*9r} \toprule
 & \multicolumn{5}{c}{$\QCD$} & \multicolumn{4}{c}{$\cPT$} \\ \cmidrule(r){1-1} \cmidrule(lr){2-6} \cmidrule(l){7-10}
current & $\octet \Fermi_\scalar$ & $\octet \Fermi_r$ & $\octet \Fermi_l^\asym$ & $\octet \Fermi_l^\sym$ & $\tensor \Fermi_l^\sym$ & $\Fermi_1$ & $\Fermi_8$ & $\Fermi_\octetm$ & $\Fermi_{27}$ \\
$\SM$ & $\octet \fermi_\scalar$ & $\octet \fermi_r$ & $\octet \fermi_l^\asym$ & $\octet \fermi_l^\sym$ & $\tensor \fermi_l^\sym$ & $\fermi_1$ & $\fermi_8$ & $\fermi_\octetm$ & $\fermi_{27}$ \\
$\BSM$ & $\octet S_\scalar$ & $\octet S_r$ & $\octet S_l^\asym$ & $\octet S_l^\sym$ & $\tensor S_l^\sym$ & $S_1$ & $S_8$ & $S_\octetm$ & $S_{27}$ \\ \cmidrule(r){1-1} \cmidrule(lr){2-5} \cmidrule(lr){6-6} \cmidrule(l){7-10}
representation & \multicolumn{4}{c}{8} & \multicolumn{1}{c}{27} & \multicolumn{4}{c}{1} \\
\bottomrule \end{tabular}
\caption[Four-quark currents and parameters]{
Currents that couple to the \QCD four-quark operators introduced in \cref{sec:four-quark operators} and the derived parameters we use in \cPT, \cf equations \eqref{eq:U octet current}, \eqref{eq:U 27-plet Lagrangian}, and \eqref{eq:U hidden 4-quark currents}.
The table indicates the names for the \SM and $\BSM$ contributions, as well as their representations under $G_{LR}$.
} \label{tab:four-quark currents}
\end{table}

The leading operators with one four-quark insertion contain either two covariant derivatives or one quark-mass insertion.
According to the modified large $\Nc$ power counting \eqref{eq:modified large N counting}, the contributions to the \QCD path-integral that generate these operators contain two quark loops but scale as $\Nc^2$.
Therefore, the leading operators count as $\order{\epsilonEW\delta} = \order{\epsilonEW\Nc^2 \partial^2}$ and they can contain either one or two quark-flavour traces, where, in stark contrast to operators that are not induced by the four-quark currents, the second flavour trace is \emph{not} associated with a large $\Nc$ suppression factor.
Furthermore, operators with covariant derivatives acting on the four-quark currents can be eliminated using \PI, while operators with two covariant derivatives acting on the same coset-matrix $\nonet g$ can be eliminated using either the identity \eqref{eq:Maurer-Cartan field} or the \EOM \eqref{eq:chpt lo eom}.

We first consider the operators that contain the octet contributions to the four-quark currents $\octet \Fermi_x$ \eqref{eq:A currents} with $x = l$, $r$, $\scalar$ (\cf \cref{tab:four-quark currents}), and then proceed to the operators that contain the 27-plet \cref{eq:A currents} $\tensor \Fermi_l^\sym$.
The only leading octet operators are
\begin{align}
\trf{\octet \Fermi_x \octet U_\mu \octet U^\mu} \ , &&
\trf{\octet \Fermi_x \octet U_\mu} U^\mu \ , &&
\trf{\octet \Fermi_x (\Left{\octet \M} + \text{h.c.})} \ ,
\end{align}
where $\octet U_\mu$ and $\Left{\octet M}$ denote the octet contributions to $\nonet U_\mu$ and $\Left{\nonet M}$.
In order to make contact with the standard form of the four-quark \cPT operators in the \SM, we explicitly extract the quark-flavour violating contributions by replacing $\octet \Fermi_x \to \trsd{\octet \Fermi_x} \flavour \lambda_\strange^\down + \text{h.c.}$ .
The resulting order $\epsilonEW \delta$ octet contributions are
\begin{subequations} \label[lag]{eq:U octet Lagrangian}
\begin{align}
\mathcal L_U^{\Fermi D^2} &= - \frac{\epsilonEW \fp^2}2 \left(\Fermi_8 \trds{\octet U_\mu \octet U^\mu} + \Fermi_1 \trsd{\octet U^\mu} U_\mu\right) + \text{h.c.} \ , \\ \label[lag]{eq:U octet mass Lagrangian}
\mathcal L_U^{\Fermi \M} &= - \frac{\epsilonEW \fp^2 \B}2 \Fermi_\octetm \trds{\Left{\octet \M} + \Left{\octet \M}^\dagger} + \text{h.c.} \ ,
\end{align}
\end{subequations}
where the three parameters $\Fermi_8$, $\Fermi_1$, and $\Fermi_\octetm$ are strangeness violating matrix elements of linear combinations of the \QCD four-quark currents (\cf \cref{tab:four-quark currents})
\begin{equation} \label{eq:U octet current}
\Fermi_y = \trsd{\coupl_y^\sym \octet \Fermi_l^\sym + \coupl_y^\asym \octet \Fermi_l^\asym + \coupl_y^r \octet \Fermi_r + \coupl_y^\scalar \octet \Fermi_\scalar} \ ,
\end{equation}
with $y = \octetm$, $1$, $8$ and twelve free parameters $\coupl_y^x$ with $x = \sym$, $\asym$, $r$, $\scalar$.
The only leading 27-plet operator is
\begin{align}
\trF{\octet U_\mu \tensor \Fermi_l^\sym \octet U^\mu} = \frac12 \left(\octet U_\mu{}_b^a \octet U^\mu{}_d^c + \octet U_\mu{}_d^a \octet U^\mu{}_b^c\right) \tensor \Fermi_l^\sym{}_{ac}^{bd} \ .
\end{align}
Using the first identity of \eqref{eq:27-plet symmetries} to explicitly isolate the quark-flavour violating contributions, one has
\begin{align}
\eval{\trF{\octet U_\mu \tensor \Fermi_l^\sym \octet U^\mu}}_{\Delta \qn s = \pm1}
= \frac2{\nfl -1} \left(\nfl \octet U_\mu{}_\down^\strange \octet U^\mu{}_\up^\up
+ (\nfl - 1) \octet U_\mu{}_\down^\up \octet U^\mu{}_\up^\strange\right) \tensor \Fermi_l^\sym{}_{\down \up}^{\strange \up}
+ \text{h.c.}
\end{align}
The resulting order $\epsilonEW \delta$ 27-plet contribution is
\begin{align} \label[lag]{eq:U 27-plet Lagrangian}
\mathcal L_U^{\tensor \Fermi D^2} &= - \frac{\epsilonEW \fp^2}2 \Fermi_{27} \left(\nfl \octet U_\mu{}_\down^\strange \octet U^\mu{}_\up^\up
+ (\nfl - 1) \octet U_\mu{}_\down^\up \octet U^\mu{}_\up^\strange\right) + \text{h.c.} \ , &
\Fermi_{27} &= \coupl_{27} \tensor \Fermi_l^\sym{}_{\strange \up}^{\down \up} \ .
\end{align}
Hence, the complete \cPT four-quark Lagrangian is
\begin{align}
\mathcal L_U^\Fermi = \mathcal L_U^{\Fermi D^2} + \mathcal L_U^{\Fermi \M} + \mathcal L_U^{\tensor \Fermi D^2} \ ,
\end{align}
where the \SM contribution is consistent with the standard expressions found \eg in \cite{Kambor:1989tz,Pich:1993uq,Donoghue:1992dd}.
Using \eqref{eq:A currents}, we split the scalar currents $\Fermi_y$ \eqref{eq:U octet current} and \eqref{eq:U 27-plet Lagrangian} into \SM and portal contributions
\begin{align} \label{eq:U hidden 4-quark currents}
\Fermi_y &= \fermi_y + S_y \ , &
S_y &= \fermi_{yi} \frac{\epsilonUV}{v} s_i \ , &
i &= 1,2 \ ,
\end{align}
where $y = \octetm$, $1$, $8$, and $27$.
While the \SM parameters $\fermi_8$, $\fermi_1$, and $\fermi_{27}$ are fixed by \SM observations, the \SM parameter $\fermi_\octetm$ and the \BSM parameters $\fermi_{yi}$ with $i=1$, $2$ have to be estimated using non-perturbative methods such as the large $\Nc$ expansion.

\subsubsection{Flavour-singlet current contributions}

\newcommand\test[1]{\left.#1\right|}
\renewcommand\test[1]{#1{}}

The $G_{LR}$ singlet current $\Omega = \omega + S_\omega$ contains a \SM contribution $\omega = \flatfrac{2\pi}{\alpha_\strong}$ and a hidden contribution $S_\omega$, but the \SM contribution is implicitly integrated out when constructing \cPT, so that it cannot appear in the Lagrangian directly.
Accounting for the hidden current $S_\omega$, it is possible to construct additional chiral invariants by multiplying it by each of the chiral invariants that contribute to the previously derived Lagrangians.
Since $S_\omega$ insertions are suppressed by a factor of $\delta \sim \inv \Nc$, the leading strangeness conserving contributions to the resulting sum of invariants count as $\delta^3$, while the leading strangeness violating contributions count as $\epsilonEW \delta^2$ and $\epsilonEW \delta^3$.
The full singlet current Lagrangian is
\begin{align} \label[lag]{eq:U conformal anomaly Lagrangian}
\mathcal L_U^{S_\omega} &
= \test{\mathcal L_U^{S_\omega}}_{\delta^3}
+ \test{\mathcal L_U^{S_\omega}}_{\delta^2}^{\EW}
+ \test{\mathcal L_U^{S_\omega}}_{\delta^3}^{\EW} \ , &
\test{\mathcal L_U^{S_\omega}}_{\delta^n} &
= S_\omega \test{\gluons_U}_{\delta^{n-1}} \ ,
\end{align}
where the strong terms are
\begin{equation}
\test{\gluons_U}_{\delta^2}
= \coupl_\omega^{D^2} \mathcal L_U^{D^2}
+ \coupl_\omega^M \mathcal L_U^M
+ \coupl_\omega^{\Theta^2} \mathcal L_U^{\Theta^2} \ ,
\end{equation}
and the \EW suppressed terms are
\begin{align}
\test{\gluons_U}_\delta^{\EW} &
= \coupl_\omega^{\Fermi D^2} \mathcal L_U^{\Fermi D^2}
+ \coupl_\omega^{\Fermi \M} \mathcal L_U^{\Fermi \M}
+ \coupl_\omega^{\tensor \Fermi D^2} \mathcal L_U^{\tensor \Fermi D^2} \ , &
\test{\gluons_U}_{\delta^2}^{\EW} &
= \coupl_\omega^\Chromo \mathcal L_U^\Chromo \ .
\end{align}
The $\coupl_\omega$ are seven free parameters.
In the following, we abbreviate
\begin{equation}
\gluons_U = \test{\gluons_U}_{\delta^2} + \test{\gluons_U}_{\delta}^{\epsilonEW} + \test{\gluons_U}_{\delta^2}^{\epsilonEW} \ .
\end{equation}
The above result is consistent with the interaction Lagrangian used in \cite{Shifman:1978zn} to capture the coupling of the $\SU(3)$ \cPT to a light Higgs boson.
The treatment in \cite{Shifman:1978zn} neglects the chromo- and electromagnetic interactions captured by $\mathcal L_U^\Chromo$ as well as the 27-plet interactions captured by $\mathcal L_U^{\tensor \Fermi D^2}$.
Furthermore, the $\SU(3)$ \cPT Lagrangian in \cite{Shifman:1978zn} does not contain the contribution $\mathcal L_U^{\Theta^2}$, which only appears in the $\U(3)$ \cPT.

\subsubsection{Stress-energy tensor}

The complete \LO \cPT action contains the Lagrangian contributions
\begin{align}
\test{\mathcal L_U}_{\delta^2} &
= \mathcal L_U^{D^2}
+ \mathcal L_U^\M
+ \mathcal L_U^{\Theta^2} \ , &
\test{\mathcal L_U}_\delta^{\EW} &
= \mathcal L_U^\Fermi \ , &
\test{\mathcal L_U}_{\delta^2}^{\EW} &
= \mathcal L_U^\Chromo + \test{\mathcal L_U^{S_\omega}}_{\delta^2}^{\EW} \ ,
\end{align}
while the \NLO \cPT action contains the Lagrangian contributions
\begin{subequations}
\begin{flalign}
\test{\mathcal L_U}_{\delta^3} &
= \mathcal L_U^{D^4}
+ \mathcal L_U^{D^2M}
+ \mathcal L_U^{M^2}
+ \mathcal L_U^{DD}
+ \mathcal L_U^{M\Theta}
+ \mathcal L_U^{\vartheta^2}
+ \mathcal L_U^{D^2V}
+ \mathcal L_U^{V^2}
+ \test{\mathcal L_U^{S_\omega}}_{\delta^3}
+ \mathcal L_U^{\Nc} \ , \\
\test{\mathcal L_U}_{\delta^3}^{\EW} &
=  \mathcal L_U^{D^2H}
+ \mathcal L_U^{MH}
+ \mathcal L_U^{\Theta H}
+ \mathcal L_U^{D^2T}
+ \mathcal L_U^{VT}
+ \test{\mathcal L_U^{S_\omega}}_{\delta^3}^{\EW}
\end{flalign}
\end{subequations}
as well as the ungauged \WZW term $\Gamma_u^{\Nc}$ \eqref{eq:U WZW Lagrangian}.
In the next section, we use the trace of the \SM Hilbert stress-energy tensor to estimate the novel parameters $\coupl_\omega$ that appear in the singlet current \cref{eq:U conformal anomaly Lagrangian}.
Neglecting the contributions due to the $S_\omega$ current, which does not appear in the \SM, the trace of the Hilbert stress-energy tensor at order $\delta^2$ is
\begin{align} \label[set]{eq:U stress-energy tensor}
\mathcal T_U &
= 2 g^{\mu\nu} \fdv{\mathcal L_U}{g^{\mu\nu}} - 4 \mathcal L_U
= - 2 \mathcal L_U^{D^2}
- 4 \mathcal L_U^M
- 4 \mathcal L_U^{\Theta^2}
+ \mathcal T_U^{\EW} \ ,
\end{align}
where
\begin{align}
\mathcal T_U^{\EW} &
= - \mathcal L_\L^{\partial W}
- 4\mathcal L_U^\Chromo
- 2 \mathcal L_U^{\Fermi D^2}
- 4 \mathcal L_U^{\Fermi \M}
- 2 \mathcal L_U^{\tensor \Fermi D^2}
\end{align}
collects the contributions due to the \EW currents.
The charged-current contribution
\begin{align}
\mathcal L_\L^{\partial W} &
= - \fp^2 \trf{\octet l_W{}_\mu \nonet \L^\mu}
\end{align}
is also a part of the kinetic \cPT \cref{eq:U kinetic Lagrangian}.
This term appears separately because it contains a vierbein $e_a^\mu$ when the theory is embedded into a generic spacetime with background metric tensor $g^{\mu\nu}$.
Due to this vierbein, the derivative contribution for the kinetic Lagrangian
\begin{align}
g^{\mu\nu} \fdv{\mathcal L_U^{D^2}}{g^{\mu\nu}} &
= \mathcal L_U^{D^2} - \frac12 \mathcal L_\L^{\partial W}
\end{align}
picks up a leftover term with a relative prefactor of $-\nicefrac12$.

\subsection{Matching of $\cPT$ to $\QCD$} \label{sec:cpt qcd matching}

So far, we derived the shape of the modified \cPT Lagrangian in the presence of generic external currents $J = \{S_\omega$, $\Theta$, $\nonet \M$, $\nonet L^\mu$, $\nonet R^\mu$, $\octet T^{\mu \nu}$, $\octet \Chromo$, $\tensor \Fermi_\scalar$, $\tensor \Fermi_r$, $\tensor \Fermi_l \}$.
We now aim to provide part of the means necessary to constrain the \QCD portal sector Wilson coefficients at energies above the mass of the charm quark using bounds on hidden sector induced \LE meson transition amplitudes obtained from \cPT.

A key element is that one has to estimate the 27 free parameters $\coupl \in \{ \coupl_\Chromo^x$, $\coupl_T^x$, $\coupl_y^x$, $\coupl_\omega^x\}$, which appear in the $\epsilonUV$ and $\epsilonEW$ suppressed sectors.
This then makes it possible to translate bounds from hidden sector induced meson transitions into constraints on the external currents as they appear in the \cPT Lagrangian.
These currents are defined such that they are identical to the external currents that appear in the \LE \QCD Lagrangian with three light $\up$, $\down$, and $\strange$ quark flavours.
To fully connect \cPT to \QCD in the perturbative regime, it is also necessary to match this version of \QCD to its counterpart that includes dynamical charm and bottom quarks.
We leave the work of matching these two versions of \QCD to future investigations, and instead consider only how to estimate the $\coupl$ parameters in \cPT.
The six parameters $\coupl_\Chromo^{D^2}$, $\coupl_\Chromo^\Theta$, $\coupl_\Chromo^\M$, $\coupl_\Chromo^{\M\prime}$, $\coupl_T^{D^2}$, and $\coupl_T^{LR}$ can be fixed using \SM observations, and we focus on those parameters for which this is not possible:
\begin{enumerate}
\item
We estimate the seven parameters $\coupl_\omega^x$ that couple \cPT to the external current $S_\omega$, which vanishes in the \SM.
These parameters can be quantified using the anomalous trace-relation for the \cref{eq:Q full stress-energy tensor}.
In the past, this technique has already been used to estimate four out of the seven parameters \cite{Leutwyler:1989xj}.
Here, we follow the same strategy to determine the remaining three parameters.
\item
We estimate the free parameter $\coupl_\Chromo$, which couples \cPT to the chromomagnetic dipole current $\octet \Chromo$ at order $\delta^2$, and the combination of parameters $\coupl_\Chromo^\M + \coupl_\Chromo^{\M\prime}$, which couple \cPT to the same current at order $\delta^3$.
In principle, \SM interactions do contribute to both dipole currents $\octet \Chromo$ and $\octet T^{\mu\nu}$, and we expect that \SM observations can be used to constrain the order $\delta^3$ parameters $\coupl_\Chromo^x$ and $\coupl_T^x$ that couple \cPT to the dipole currents.
However, the order $\delta^2$ \SM contribution to the operator associated with the parameter $\coupl_\Chromo$ can be reabsorbed into the quark mass matrix, so that this parameter is not fixed by \SM observations.
Instead, we estimate its value, and the values of $\coupl_\Chromo^\M$ and $\coupl_\Chromo^{\M\prime}$, by matching it to the lattice \QCD prediction for the vacuum condensate of the chromomagnetic operator, which is reasonably well known \cite{Belyaev:1982sa,Chiu:2003iw,Gubler:2018ctz}.
\item
We estimate the thirteen parameters $\coupl_y^x$, which appear in the \cPT four-quark Lagrangian.
These parameters enter into \SM predictions only via the linear combinations that constitute the octet and 27-plet coefficients $\fermi_8$, $\fermi_1$, and $\fermi_{27}$, so that \SM observations do not yield enough information to completely fix their values.
At \LO in the large $\Nc$ expansion, the factorization rule \eqref{eq:large N factorization} can be used to estimate the $\coupl_y^x$ parameters \cite{Bardeen:1986vz,Pich:1990mw,Pich:1995qp,Pallante:2001he,Cirigliano:2003gt,Gerard:2005yk,Buras:2014maa}.
However, this approximation fails to accurately reproduce \eg the $\Delta \qn I = \nicefrac12$ rule in the \SM, which is an approximate selection rule for kaon decays that results from the fact that the octet coefficients $\fermi_{8,1}$, which mediate only $\Delta \qn I = \nicefrac12$ transitions,
are an order of magnitude larger than the 27-plet coefficient $\fermi_{27}$, which mediates both $\Delta \qn I = \nicefrac12$ and $\Delta \qn I = \nicefrac32$ transitions.
For this reason, we expect that one has to include corrections beyond the large $\Nc$ limit to extract order-of-magnitude accurate estimates of the portal sector Wilson coefficients $c_{\iota i}$ from bounds on hidden sector induced meson transitions.
To obtain improved estimates for the $\coupl$ parameters, we adapt the strategies used in \cite{Bardeen:1986vz,Leutwyler:1989xj,Pich:1990mw,Pich:1995qp,Gerard:2005yk}, and neglect the contributions generated by the penguin operators $O_{3 i}$, $O_{4 i}$, and $O_{5 i}$.
Since these operators are generated at 1-loop they are suppressed by factors of $(4\pi)^{-2}$ compared to the tree-level operators $O_{1 i}$ and $O_{2 i}$.
The penguin operator $O_{6 i}$ is also generated at 1-loop, but it is expected to generate the dominant penguin contribution to kaon decay amplitudes \cite{Shifman:1976ge,Bardeen:1986vz,Leutwyler:1989xj,Gerard:2005yk,Buras:2014maa}.
\end{enumerate}

\subsubsection{Scale dependence of the external currents}

Many of the external currents are scale-dependent, and therefore the estimates that one obtains for the \LECs and $\coupl$ parameters in the \cPT Lagrangian depend on the scale at which the external currents are evaluated.
In general, the scale-dependence of the external currents has to cancel with the one of the \LECs and $\coupl$ parameters.
If \cPT is matched to the version of \QCD without the charm quark, so that there are no threshold effects, this implies that the hidden currents can always be evaluated at some arbitrary higher scale, say, $\mu_{\QCD} = \unit[1\text{--}2]{GeV}$, provided that one adjusts the values of the \LECs and $\coupl$ parameters accordingly.

This approach has been used to deal with the scale dependence of the mass-like current $\nonet \M$ and the anomalous axial singlet current $R^\mu - L^\mu = \vartheta^\mu - \partial^\mu \Theta$,
\footnote{\label{fn:invariant currents}%
The currents $J_\text{inv} = \{ \Theta$, $L^\mu + R^\mu$, $\octet L^\mu \pm \octet R^\mu \}$ do not renormalise and are therefore scale-independent in \QCD \cites{Kaiser:2000gs}[section 6.6 in][]{Collins:1984xc}.}
which renormalise according to \cite{Kaiser:2000gs}
\begin{align} \label{eq:current renormalisation}
\nonet \M &= Z_M^{-1} \nonet \M^\text{bare} \ , &
\vartheta_\mu &= \inv Z_\vartheta \vartheta_\mu^\text{bare} \ .
\end{align}
The factors $Z_\vartheta$ and $Z_M$ relate the renormalised quark current corresponding associated with $\nonet \M$ and $\vartheta^\mu$ to their bare counterparts
\begin{align}
\nonet Q &= Z_M \nonet Q^\text{bare} \ , &
Q^\mu - \overline Q^\mu &= Z_\vartheta \left(Q^\mu - \overline Q^\mu\right)^\text{bare} \ .
\end{align}
Extracting the renormalisation of the scalar axial current from \eqref{eq:current renormalisation} gives
\begin{equation}
(L^\mu - R^\mu) = \inv Z_\vartheta (L^\mu - R^\mu)^\text{bare} - \left(1 - \inv Z_\vartheta\right) \partial^\mu \Theta^\text{bare} \ ,
\end{equation}
where $\partial^\mu \Theta = \partial^\mu \Theta^\text{bare}$.\cref{fn:invariant currents}
This equation reflects the fact that the axial anomaly mixes the scalar axial vector current with the derivative of the pseudoscalar current $\partial_\mu \Theta$.
We can see explicitly that the \cPT Lagrangian is invariant under a change of the \QCD renormalisation scale, provided that it is written in terms of the renormalised singlet meson field \cite{Kaiser:2000gs}
\begin{equation}
\Left \Theta = \i \left(\Theta - \frac{\Phi}{\fp}\right) = \i Z_\vartheta^{-1} \left(\Theta - \frac{\Phi^\text{bare}}{\fp}\right)
\end{equation}
as well as the renormalised \LECs
\begin{subequations} \label{eq: qcd renormalizion LECs}
\begin{align}
\B &= Z_M \B^\text{bare} \ , &
\mth &= Z_\vartheta \mth^\text{bare} \ , &
1 + \Lambda_2 &= Z_\vartheta (1 + \Lambda_2^\text{bare}) \ , \\ & &
H_0 &= Z_\vartheta^2 H_0^\text{bare} \ , &
1 + \Lambda_1 &= Z_\vartheta^2 (1 + \Lambda_1^\text{bare}) \ .
\end{align}
\end{subequations}
The scale-dependent values of the renormalised \LECs $\B$, $\mth$, and $\Lambda_{1,2}$ can now be fixed by computing \cPT observables in terms of the renormalised currents $\nonet \M = \nonet \M(\mu_{\QCD})$ and $L^\mu - R^\mu = (L^\mu - R^\mu)(\mu_{\QCD})$.
Of course, this renormalisation procedure only eliminates divergences associated with the strong interaction.
We emphasise that the \cPT action written in terms of the above fields and \LECs still has to be renormalised as usual when accounting for loop corrections starting at \NNLO.
As a result, one has to distinguish between the renormalisation scale of \cPT $\mu_{\cPT}$ and the renormalisation scale of \QCD $\mu_{\QCD}$, which are not necessarily the same.
In general, the renormalised \LECs and parameters depend on \emph{both} scales.
Here and in the remainder of \cref{sec:cpt qcd matching}, we only consider the dependence on $\mu_{\QCD}$, and so we suppress the dependence on $\mu_{\cPT}$ in the notation.

In the following, we apply the above renormalisation procedure to the $\epsilonEW$ or $\epsilonUV$ suppressed currents $J = \{S_\omega$, $\octet \Chromo$, $\octet T^{\mu\nu}$, $\tensor \Fermi_l^+$, $\octet \Fermi_l^\pm$, $\octet \Fermi_{r,s}\}$,
and absorb their scale dependence into the values of the free parameters $\coupl$.
The upshot of this prescription is that, when matching \cPT to \QCD without the charm quark , we can freely choose the renormalisation scale $\mu_{\QCD}$, even choosing a value well above the charm quark mass.
Of course, this would not work if we were to attempt to match \cPT to perturbatively computed low-energy observables in \QCD, since choosing a large renormalisation scale $\mu_{\QCD} \gg m_\charm$ would mean
that we neglect precisely the non-perturbative contributions on the \QCD side that dominate the physics of the strong interaction at low energies.
However, this is not an issue when matching \cPT to the results of \emph{non}-perturbative computations, such as those done in lattice \QCD, where no expansion in $\inv \omega$ is made.
In fact, the scale $\mu_{{\QCD}} = \unit[2]{GeV}$ is a standard choice when computing low-energy observables such as the quark masses and condensates in lattice \QCD with and without the charm quark \cite{Gubler:2018ctz,Aoki:2019cca,Bhattacharya:2021lol}.

\subsubsection{$\cPT$ realizations of $\QCD$ operators}

To establish a point of contact between \cPT and \QCD that does not rely on a perturbative expansion in $\inv \omega$, we use a standard technique employed \eg in \cite{Gasser:1984gg,Leutwyler:1989xj,Pich:1990mw,Kaiser:2000gs,Pallante:2001he,Cirigliano:2003gt,Gerard:2005yk},
and construct a set of well-defined \LERs for \QCD gauge-singlets as functional derivatives of the path integral with respect to the external currents.
For the sake of completeness, we outline the general procedure and then summarise the resulting \cPT \LERs that are relevant to the subsequent discussion.

\paragraph{Constructing low energy realisations}

\begin{table}
\begin{tabular}{l*9c} \toprule
$\delta^n$ & \multicolumn{2}{c}{1} & \multicolumn{3}{c}{2} & \multicolumn{4}{c}{3} \\ \cmidrule(r){1-1} \cmidrule(lr){2-3} \cmidrule(lr){4-6} \cmidrule(l){7-10}
$J$ & $\octet \Fermi_x$ & $\tensor \Fermi_x$ & $\Theta$ & $\nonet L_\mu$ & $\nonet R_\mu$ & $\Omega$ & $\nonet M$ & $\octet \Chromo$ & $\octet T_{\mu\nu}$ \\
$\delta^k$ & \multicolumn{2}{c}{0} & 0 & \multicolumn{2}{c}{$\nicefrac12$} & 0 & \multicolumn{3}{c}{1} \\ \cmidrule(r){1-1} \cmidrule(lr){2-3} \cmidrule(lr){4-4} \cmidrule(lr){5-6} \cmidrule(lr){7-7} \cmidrule(l){8-10}
$O$ & $\octet O_x$ & $\tensor O_x$ & $w$ & $\nonet Q_\mu$ & $\overline{\nonet Q}_\mu$ & $\gluons$ & $\nonet Q$ & $\widetilde{\nonet Q}$ & $\nonet Q_{\mu\nu}$ \\
$\delta^{n-k}$ & \multicolumn{2}{c}{1} & 2 & \multicolumn{2}{c}{$\nicefrac32$} & 3 & \multicolumn{3}{c}{2} \\
\bottomrule \end{tabular}
\caption[Order in $\delta$ at which we evaluate the $\LERs$]{
Order in $\delta$ at which we evaluate the \LERs.
The first row shows the order in $\delta$ at which we evaluate the \cPT generating functional, the second row shows the $\delta$ scaling of the external current, and the final row shows the resulting order in $\delta$ for the \LER.
While a momentum suppression $\propto \partial^2$ counts towards the scaling of the external currents, a large $\Nc$ suppression does not,
because it is associated with the structure of the \QCD diagrams that couple to the external current rather than with the current itself.
The order in $\delta$ at which we evaluate the \cPT generating functional is chosen such that we include the leading nonvanishing contribution for each operator.
For $\nonet Q$, we also include \NLO contributions, since these enter at \LO into the approximate factorised expressions for the four-quark operators.
Note also that the product of operators does not scale as the sum of their individual suppressions.
For instance, $\nonet Q_\mu \nonet Q^\mu \propto \Nc^2 \partial^2 \propto \delta$, rather than $\delta^{\flatfrac32} \times \delta^{\flatfrac32} = \delta^3 $, as one might naively expect.
}
\end{table}

In general, the expectation value of any local, gauge invariant \QCD operator $O_i$ that couples to an external current $J_i$ is
\begin{equation} \label{eq: operator expectation value}
\tr O_i(x) \rho = \fdv{\ln \mathcal Z_Q[J_j]}{J_i(x)} \ ,
\end{equation}
where the von Neumann density matrix $\rho$ encodes the state of the system,
\begin{equation}
\mathcal Z_Q[J_j] = \int \mD{\varphi} \rho \exp(\i S_Q[\varphi] + \i \int \d[4]{x} J_j (x) O_j (x))
\end{equation}
is the generating functional in the presence of external currents $J_i$, and $\varphi$ symbolically denotes the quark and gluon fields.
The \cPT generating functional approximates the \QCD generating functional for small $\delta$,
\begin{equation}
\ln \mathcal Z_Q[J_j] = \ln \mathcal Z_U[J_j] + \order{\delta^n} \ .
\end{equation}
If an external current scales as $J_i(x) \propto \delta^k$, then inserting this relation into the expectation value \eqref{eq: operator expectation value} gives
\begin{equation}
\tr O_i(x) \rho
= \fdv{\ln \mathcal Z_U[J_j]}{J_i(x)} + \order{\delta^{n - k}}
= \tr \fdv{S_U}{J_i} \rho + \order{\delta^{n - k}} \ .
\end{equation}
Since this has to hold for any physical choice of $\rho$, one finds the \LERs
\begin{equation}
O_i(x) = \frac{\delta S_U}{\delta J_i} + \order{\delta^{n - k}} \ .
\end{equation}

\paragraph{Operators}

The \LERs of the colour singlet \eqref{eq:Q axial anomaly Lagrangian} and the quark bilinears \eqref{eq:vectorial quark bilinears} associated with the $\nonet L_\mu$, $\nonet R_\mu$, $\M$, and $\Theta$ currents are well established.
At leading order in $\delta$, they are \cite{Gasser:1984gg,Ecker:1994gg,Pich:1990mw,Pich:1995bw,Kaiser:2000gs}
\begin{align} \label{eq:low energy realizations 1}
\nonet Q_\mu &
= - \fp^2 \nonet U_\mu \ , &
\overline{\nonet Q}_\mu &
= - \fp^2 \nonet \u^\dagger \nonet U_\mu \nonet \u \ , &
\nonet Q &
= - \frac12 \fp^2 \B \nonet \u \ , &
w &
= - \i \fp^2 \frac{\mth^2}{\nfl} \Left \Theta \ .
\end{align}
The \LO contributions to $\nonet Q_\mu$ and $\overline{\nonet Q}_\mu$ count as order $\partial \Nc$, while the \LO contribution to $\nonet Q$ counts as order $\Nc$.
In order to estimate the four-quark coefficients $\coupl_x^s$ at order $\partial^2 \Nc^2$, which is the first non-vanishing order,
it is necessary to also track \NLO corrections to the \LER of $\nonet Q$ that count as $\partial^2 \Nc$.
This gives the expression
\begin{align} \label{eq:low_energy_density}
\nonet Q &= - \frac12 \fp^2 \B \left(1 + \Delta \nonet Q^{\NLO}\right) \nonet \u \ ,
\end{align}
where
\begin{equation}
\Delta \nonet Q^{\NLO}
= \frac{1}{2\fp^2} \left(
4 L_5 \nonet U_\mu \nonet U^\mu
+ 2\B \left(4 L_8 \Left{\nonet \M} + 2 H_2 \Left{\nonet \M}^\dagger\right)
\right)
+ \frac{\Lambda_2}\nfl \Left \Theta
- \epsilonEW \left(\Fermi_\octetm \flavour \lambda_\strange^\down + \text{h.c.}\right) \ .
\end{equation}
Note that this expression differs from it's $\SU(3)$ counterpart by the appearance of the term proportional to $\Lambda_2$, which does not exist in $\SU(3)$ \cPT \cite{Gasser:1984gg,Leutwyler:1989xj,Pich:1990mw}.

We apply the same technique to obtain \LERs for operators associated with the $\epsilonEW$ and $\epsilonUV$ suppressed currents.
We find that the \LERs of the colour singlet \eqref{eq:Q kinetic Lagrangian} and the quark bilinears \eqref{eq:EW quark bilinears} associated with the $\octet T_{\mu\nu}$, $\octet \Chromo$, and $S_\omega$ currents are
\begin{align} \label{eq:low energy realizations 2}
\nonet Q^{\mu\nu} & = - \fp \left(\coupl_T^{D^2} \nonet U^\mu \nonet U^\nu + \coupl_T^{LR} (\nonet L^{\mu\nu} + \Left{\nonet R}^{\mu\nu})\right) \nonet g \ , &
\widetilde{\nonet Q} & = - \frac12 \fp^4 \B \coupl_\Chromo \nonet \u \ , &
\gluons &= - \gluons_U \ .
\end{align}
At \NLO, the \LER of the scalar quark bilinear $\widetilde{\nonet Q}$ is
\begin{equation}
\widetilde{\nonet Q} = - \frac12 \fp^4 \B \left(\coupl_\Chromo + \Delta \widetilde{\nonet Q}^{\NLO}\right) \nonet \u \ ,
\end{equation}
where
\begin{equation}
\Delta \widetilde{\nonet Q}^{\NLO}
= \frac{1}{2\fp^2} \left(\coupl_\Chromo^{D^2} \nonet U_\mu \nonet U^\mu
+ 2 \B \left(\coupl_\Chromo^M \Left{\nonet \M} + \coupl_\Chromo^{M^\prime} \Left{\nonet \M}^\dagger\right)
\right)
+ \frac{\coupl_\Chromo^\Theta}{\nfl} \Left \Theta \ .
\end{equation}
Finally, the \LERs of the octet quark quadrilinear \eqref{eq:octet operators,eq:27-plet octet operators} associated with the $\octet \Fermi_x$ currents are
\begin{equation} \label{eq:low energy realization octet}
\octet O_x{}_\down^\strange
= \frac12 \fp^4 \left(\coupl_8^x \trds{\octet U_\mu \octet U^\mu} + \coupl_1^x \trsd{\octet U^\mu} U_\mu + \coupl_\octetm^x \B \trds{\Left{\nonet \M} + \Left{\nonet \M}^\dagger}\right) \ ,
\end{equation}
where $x = \sym$, $\asym$, $r$, $\scalar$, and the \LER of the 27-plet quark quadrilinear \eqref{eq:27-plet operator} associated with the $\tensor \Fermi_l^\sym$ current is
\begin{equation} \label{eq:low energy realization 27-plet}
2 n_{27} \tensor O_l^\sym{}_{\down \up}^{\strange \up} = \frac12 \fp^4 \coupl_{27} \left(\nfl \octet U^\mu{}_\down^\strange \octet U_\mu{}^\up_\up + (\nfl - 1) \octet U^\mu{}^\up_\down \octet U_\mu{}^\strange_\up\right) \ .
\end{equation}

\paragraph{Lagrangians}

Using the above \LERs, one obtains approximate \cPT expressions for various individual contributions to the \QCD Lagrangian
\begin{align} \label[lag]{eq:low energy realization Lagrangians}
\mathcal L_Q^\M &= \mathcal L_U^\M + \mathcal L_U^{\Fermi \M} \ , &
\mathcal L_Q^\Chromo &= \mathcal L_U^\Chromo \ , &
\mathcal L_Q^W &= \mathcal L_U^{DW} \ , &
\mathcal L_Q^\Fermi &= \mathcal L_U^\Fermi \ .
\end{align}
Note that the mass-like four-quark octet term contributes not only to the \LER of the four-quark Lagrangian, but also to the \LER of the mass Lagrangian.

\subsubsection{Determination of selected parameters}

We now estimate the 22 $\coupl$ parameters $\coupl_\omega^x$, $\coupl_\Chromo$, $\coupl_\Chromo^{\M} + \coupl_\Chromo^{\M\prime}$, and $\coupl_y^x$.
As mentioned in the introduction to this section, we do not estimate the five parameters $\coupl_\Chromo^{D^2}$, $\coupl_\Chromo^{\M} - \coupl_\Chromo^{\M\prime}$, $\coupl_\Chromo^\Theta$ and $\coupl_T^x$, which couple \cPT to the $\octet \Chromo$ and $\octet T^{\mu\nu}$ currents at order $\epsilonEW \delta^3$, and leave this work to future investigations.
To illustrate the use of the \LERs and to prepare for the estimation of the four-quark parameters $\coupl_x^y$, we first discuss two well-known computations that match \cPT to the lattice \QCD predictions for the quark condensates and the topological susceptibility of \QCD.

\paragraph{Quark condensates}

The \LER for the quark bilinear $\nonet Q$ \eqref{eq:low_energy_density} relates the parameters $\B$ and $4 L_8 - 2 H_2$ to the values of the chiral quark condensates \eqref{eq:quark condensate}.
In the isospin conserving limit $m_\up$, $m_\down \to m_\light = \flatfrac{(m_\up + m_\down)}2$, one obtains the \cPT predictions \cite{Gasser:1984gg, Leutwyler:1989xj}
\begin{align} \label{eq:cPT quark condensate}
\Sigma_\light &= \fp^2 \B + \B^2 m_\light \left(4 L_8 + 2 H_2\right) \ , &
\Sigma_\strange &= \fp^2 \B + \B^2 m_\strange \left(4 L_8 + 2 H_2\right) \ .
\end{align}
The quark condensates are proportional to $\B$ and degenerate at \LO.
Their splitting is captured at \NLO by the parameter $4 L_8 + 2 H_2$.
Using the lattice values of condensates in \cref{eq:quark condensate} yields the estimates
\begin{subequations}
\begin{align} \label{eq: b0 condensate estimate}
\B &
= \frac{m_\strange \Sigma_\light - m_\light \Sigma_\strange}{\fp^2(m_\strange - m_\light)}
= 4 \pi \unit[\left(387 \pm 13_{\ex} \pm 23_{\lat}\right)]{MeV} \pm \NNLO \ , \\
4 L_8 + 2 H_2 &
= \frac{\Sigma_\strange - \Sigma_\light}{\B^2 \left(m_\strange - m_\light\right)}
= (4\pi)^{-2} \left(0.48 \pm 0.022_{\ex} \pm 0.26_{\lat}\right) \pm \NNLO \ .
\end{align}
\end{subequations}
While $\B$ depends on the \QCD renormalisation scale in the same way as $\Sigma_\light$ and $\Sigma_\strange$, the dependence cancels in the expression for $4 L_8 + 2 H_2$, which depends only on the renormalisation scale independent ratio of the quark condensates \eqref{eq:quark condensate ratio}.
Since $L_8$ can be estimated from the $\eta$-meson mass splitting and mixing angle, \cf \cref{eq:NLO parameter estimates,eq:eta estimates}, the above expression can be used to estimate the value of the counter-term parameter
\begin{equation}
2 (4\pi)^2 H_2 = 0.27 \pm 0.033_{\ex} \pm 0.26_{\lat} \pm \NNLO \ .
\end{equation}
This parameter does not enter directly into perturbatively computed $S$-matrix elements, but it is needed for the large $\Nc$ estimate of the four-quark parameters $\coupl_x^s$.

\paragraph{Topological susceptibility}

The \LER of the quark condensate can be combined with the \LER of $w$ \eqref{eq:low energy realizations 1} and relation \eqref{eq:susceptibility relation} to express the topological susceptibility \eqref{eq:susceptibility} as a combination of \cPT parameters.
Since diagrams with internal quark loops do not contribute to the \QCD path integral at zeroth order in the large $\Nc$ expansion, \QCD behaves similar to a pure \YM theory with no quark fields in this limit.
Hence, a direct estimate of the topological susceptibility using the \LO \LER \eqref{eq:low energy realizations 1} for $w$ yields an estimate for the \emph{quenched} susceptibility
\cite{Veneziano:1979ec,DiVecchia:1980yfw,Kaiser:2000gs,Bhattacharya:2021lol}
\begin{equation} \label{eq:cPT susceptibility}
\chi_0 = \fp^2 \frac{\mth^2}{\nfl} = \unit[(188.1 \pm 2.4_{\ex})^4]{MeV}^4 \pm \NNLO \ .
\end{equation}
Combining this result with the \LO estimate of the quark condensate \eqref{eq:cPT quark condensate} and relation \eqref{eq:susceptibility relation}, one obtains the estimate
\begin{equation}
\frac{\fp^2}\chi = \frac{\nfl}{\mth^2} + \frac{\trf{\inv{\nonet \m}}}{\B}
= \frac{3}{m_0^2} + \frac{2}{m_\pi^2} + \frac{1}{2 m_K^2-m_\pi^2}
= \frac{\fp^2}{\unit[(76.9 \pm 1.3_{\ex})^4]{MeV}^4 \pm \NNLO} \ ,
\end{equation}
for the topological susceptibility of \QCD, which lies within the error bars of the lattice result \eqref{eq:susceptibility}.
See \cref{sec:SM meson masses} for the definition of the pion and kaon mass parameters $m_\pi$ and $m_K$.

\paragraph{Flavour singlet contribution}

We estimate the three new $\coupl_\omega$ coefficients that appear in the $S_\omega$ contribution to \cPT by following the strategy used in \cite{Voloshin:1980zf, Novikov:1980fa, Leutwyler:1989xj}, where the trace of the \QCD \cref{eq:Q full stress-energy tensor}
\begin{equation}
\beta_\strong \gluons = - \mathcal T_Q - \mathcal L_Q^\M + \mathcal L_Q^\Chromo + \mathcal L_Q^T - \mathcal L_Q^W + 2 \mathcal L_Q^\Fermi
\end{equation}
has been used to express the gluon-kinetic term $\gluons$ as a linear combination of the trace of the stress-energy tensor and the other terms in the \QCD Lagrangian.
Using the \cPT expression for the trace of the \cref{eq:U stress-energy tensor} as well as the \LERs for the other terms in the \cref{eq:low energy realization Lagrangians}, this gives the \LO \LER of the gluon-kinetic term
\begin{equation} \label{eq: trace anomaly low energy realization}
\begin{split}
\beta_\strong \gluons_U &
= \mathcal L_U^\Chromo
- \mathcal L_U^\M
- \mathcal L_U^{D W}
+ 2 \mathcal L_U^\Fermi
- \mathcal T_U \\ &
= 2 \mathcal L_U^{D^2}
+ 3 \mathcal L_U^\M
+ 4 \left(\mathcal L_U^{\Theta^2}
+ \mathcal L_U^{\Fermi D^2}
+ \mathcal L_U^{\tensor \Fermi D^2}\right)
+ 5 \left(\mathcal L_U^\Chromo
+ \mathcal L_U^{\Fermi \M}\right) \ .
\end{split}
\end{equation}
In principle, the contribution to the trace of the stress-energy due to the quark masses receives a further correction associated with their anomalous dimension \cite{Minkowski:1976en,Collins:1976yq,Nielsen:1977sy}, and we expect the same to hold for the contributions due to the other external currents.
However, since the term $\mathcal L_U^{S_\omega}$ has to be independent of $\mu_{\QCD}$, we can choose to evaluate the above relation at a sufficiently large renormalisation scale $\mu_{\QCD} \gg \unit[1]{GeV}$,
where the impact of quantum corrections to the external current contributions to the stress-energy tensor is small due to asymptotic freedom.
With this choice, and provided that we also evaluate $S_\omega (\mu_{\QCD})$ at the same scale, the above relation becomes a valid approximation.
In addition, the $\beta$-function at this scale is well-approximated by its leading term $\beta_\strong = \betacoeff + \order{\inv \omega(\mu_{\QCD})}$.
Hence, choosing to evaluate relation \eqref{eq: trace anomaly low energy realization} at $\mu_{\QCD} \gg 1$, the seven coefficients that appear in \cref{eq:U conformal anomaly Lagrangian} are given as
\begin{align} \label{eq:omega prefactors}
\coupl_\omega^{D^2} &= \frac{2}{\betacoeff} \ , &
\coupl_\omega^M &= \frac{3}{\betacoeff} \ , &
\coupl_\omega^{\Theta^2} &= \coupl_\omega^{\Fermi D^2} = \coupl_\omega^{\tensor \Fermi D^2} = \frac{4}{\betacoeff} \ , &
\coupl_\omega^\Chromo = \coupl_\omega^{\Fermi \M} &= \frac{5}{\betacoeff} \ .
\end{align}
While the coefficients $\coupl_\omega^{D^2}$, $\coupl_\omega^M$, $\coupl_\omega^{\Fermi D^2}$, and $\coupl_\omega^{\Fermi \M}$ are known \cite{Leutwyler:1989xj},
the coefficients $\coupl_\omega^{\tensor \Fermi D^2}$, $\coupl_\omega^{\Theta^2}$, and $\coupl_\omega^\Chromo$ are a new result.

\paragraph{Chromomagnetic contribution and quark gluon condensates}

We estimate the parameter $\coupl_\Chromo$ and the linear combination $\coupl_\Chromo^M + \coupl_\Chromo^{M^\prime}$ by matching the \cPT prediction for the condensate of the chromomagnetic quark bilinear $\widetilde{\nonet Q}$ to the quark-gluon condensates
\eqref{eq:quark gluon condensate}.
Using the \LERs \eqref{eq:low energy realizations 2}, the condensates are given as
\begin{align}
\frac{\Sigma_{G\light}}{(4\pi)^2} &= \fp^4 \B \kappa_\Chromo + \B^2 m_\light \left(\kappa_\Chromo^M + \kappa_\Chromo^{M^\prime}\right) \ , &
\frac{\Sigma_{G\strange}}{(4\pi)^2} &= \fp^4 \B \kappa_\Chromo + \B^2 m_\strange \left(\kappa_\Chromo^M + \kappa_\Chromo^{M^\prime}\right) \ .
\end{align}
The $4\pi$ enhancement of the condensates is a consequence of definition \eqref{eq:Q dipole current Lagrangian}, in which we have not included the loop factor into the operator, but written it as an explicit contribution to the Lagrangian.
Matching this prediction to the lattice and \QCD sum rule values of the condensate from \cref{eq:quark gluon condensate ratio}, one obtains the estimates
\begin{subequations}
\begin{align}
\coupl_\Chromo = \frac{m_\strange \Sigma_{G\light} - m_\light \Sigma_{G\strange}}{\Lambda_{\cPT}^2 \fp^2 \B (m_\strange - m_\light)} &= 1.21 \pm 0.06_{\ex} \pm 0.06_{\lat} \pm \NNLO \ . \\
\coupl_\Chromo^M + \coupl_\Chromo^{M^\prime} = \frac{\Sigma_{G\strange} - \Sigma_{G\light}}{\Lambda_{\cPT}^2 \B^2 (m_\strange - m_\light)} &= - \inv[2]{(4\pi)} \left(0.20 \pm 0.004_{\ex} \pm 0.31_{\lat} \pm \NNLO\right) \ .
\end{align}
\end{subequations}
The negative prefactor of $\coupl_\Chromo^M + \coupl_\Chromo^{M^\prime}$ and the fact that its value is consistent with being zero reflects
that $\Sigma_{G\strange}$ has been estimated to be slightly smaller than $\Sigma_{G\light}$, while its value is consistent with both condensates being equal to each other within their error bars.

\paragraph{Four-quark contributions}

Written in terms of the parameters $\coupl_y^x$ and the Wilson coefficients $c_{\iota i}$, the octet and 27-plet coefficients in the four-quark Lagrangian are given as
\begin{subequations}
\begin{alignat}{4}
\fermi_{yi} &= \frac14 V^\dagger_{\strange\up} V_{\up\down}
\left[\coupl_y^+ \left(c_{12i}^+ + (\nfl + 2)c_{34i}^+\right)
- \coupl_y^- (c_{12i}^- + c_{34i}^+) + 4 \coupl_y^r c_{5i} + 4 \coupl_y^s c_{6i}\right] & \ , \\
\fermi_{27i} &= \frac14 V^\dagger_{\strange\up} V_{\up\down} \coupl_{27} \frac{\nfl + 1}{\nfl + 2} c_{12i}^+ \ , &
\mathllap{c_{\iota\kappa i}^\pm = c_{\iota i} \pm c_{\kappa i}} \ , &
\end{alignat}
\end{subequations}
where $i=0$, $1$, and $2$.
Following the convention introduced in \cref{sec:portal currents}, we denote the \SM Wilson coefficients as $\fermi_x = \fermi_{x 0}$ and $c_\iota = c_{\iota 0}$.
See also \cref{sec:EW currents,sec:weak current contributions}, where we define these coefficients.
Since the coefficients $\fermi_{yi}$ have to be independent of the \QCD renormalisation scale, the scale dependence of the Wilson coefficients cancels with the scale dependence of the thirteen $\coupl_y^x$ parameters.

\newcommand{\largenc}[1]{\overline #1}
The large $\Nc$ factorisation rule \eqref{eq:large N factorization} can be used to estimate the parameters at \LO in $\delta$ \cite{Pallante:2001he,Cirigliano:2003gt}.
The main idea is to combine the vacuum saturation hypothesis \eqref{eq:large N factorization} with the \LERs of the quark bilinears \eqref{eq:low energy realizations 1,eq:low_energy_density} to obtain approximate large $\Nc$ realisations for the octet and 27-plet operators.
These can then be compared with the exact \LERs for the four-quark operators \eqref{eq:low energy realization octet,eq:low energy realization 27-plet} that have been obtained by varying the \cPT with respect to the $\tensor \Fermi_l^\sym$ and $\octet \Fermi_x$ currents.
The resulting approximate large $\Nc$ realisations for the octet operators are
\begin{subequations}
\begin{flalign}
\octet O_\scalar{}_\down^\strange & = \frac14 \fp^2 \B^2 \left(4L_5 \trds{\nonet U^\mu \nonet U_\mu} + \left(4L_8 + 2 H_2\right) \B \trds{\Left{\nonet \M} + \Left{\nonet \M}^\dagger}\right) \ , &
\octet O_r{}_\down^\strange = \fp^4 \octet U^\mu{}_\down^\strange U_\mu & \ , \\
\octet O_l^\asym{}_\down^\strange &
= \fp^4\left(\frac2{\nfl} \octet U^\mu{}_\down^\strange U_\mu - \frac{1}{2n^-_8} \trds{\octet U^\mu \octet U_\mu}\right) \ , &
\mathllap{\octet O_l^\sym{}_\down^\strange = \fp^4 \left(\frac2{\nfl} \octet U^\mu{}_\down^\strange U_\mu + \frac1{2n^+_8} \trds{\octet U^\mu \octet U_\mu}\right) } & \ ,
\end{flalign}
and the approximate large $\Nc$ realizations for the 27-plet operator is
\begin{equation}
\tensor O_l^\sym{}_{\down \up}^{\strange \up}
= \frac12 \fp^4 \left(1 + \frac1{4 n^+_8}\right) \octet U^\mu{}_\down^\strange \octet U_\mu{}^\up_\up
+ \frac12 \fp^4 \left(1 - \frac1{4 n^+_8}\right) \octet U^\mu{}^\up_\down \octet U_\mu{}^\strange_\up \ .
\end{equation}
\end{subequations}
where $\trproj{\param}{i}{j} = \trf{\param \flavour \proj_j^i}$.
Matching these expression to the exact \LERs \eqref{eq:low energy realization octet,eq:low energy realization 27-plet}, one obtains the \LO estimates $\largenc \coupl$ \cite{Pallante:2001he,Cirigliano:2003gt}
\begin{subequations}
\begin{align}
\largenc \coupl{}^r_1 &= 2 \ , &
\largenc \coupl{}^\sym_8 &= \frac1{n_8^+} = \frac45 \ , &
\largenc \coupl{}_{27} &= \frac{n_{27}}{n_8^+} = 2 \ , \\
\largenc \coupl{}^\asym_1 = \largenc \coupl{}^\sym_1 &= \frac4{\nfl} = \frac43 \ , &
- \largenc \coupl{}^\asym_8 &= \frac{1}{n_8^-} = 4 \ ,
\end{align}
\end{subequations}
and
\begin{subequations} \label{eq:four quark parameters large nc estimate}
\begin{align}
\largenc \coupl{}^\scalar_8 = \frac{\nfl}2 \largenc \coupl{}^\scalar_1 = \frac12 \frac{\B^2}{\fp^2} 4 L_5 &= 2.00 \pm 0.16_{\ex} \pm 0.13_{\lat} \ , \\
\largenc \coupl{}^\scalar_\octetm = \frac12 \frac{\B^2}{\fp^2} (4 L_8 + 2 H_2) &= 1.45 \pm 0.10_{\ex} \pm 0.8_{\lat} \ .
\end{align}
\end{subequations}
The remaining parameters $\kappa_y^x$ vanish in the large $\Nc$ limit, $\largenc \kappa = 0$.
In this approximation, the parameters $\coupl_y^{\pm,r}$ and $\coupl_{27}$ are renormalisation scale independent, while the $\coupl_y^s$ run as $\B^2 \sim \flatfrac{\Sigma_\light^2}{\fp^4}$.
This is consistent with the scale dependence of the \SM four-quark Wilson coefficients: In the large $\Nc$ limit, the $c_i$ coefficients with $i\neq6$ are in fact renormalisation scale independent, while $c_6$ remains scale-dependent and runs as $\inv[2]{\B}$ \cite{Bardeen:1986uz,Bardeen:1986vp,Pich:1990mw}.
This running of $c_6$, which we have absorbed into the values of the $\coupl_y^s$, is the physical cause behind the enhancement factors $\flatfrac{\B^2}{\fp^2}$ of the singlet operators and cancels the suppression associated with the factors $L_5$ and $2L_8 + H_2$.

Moving beyond the large $\Nc$ limit, we expect that the resulting corrections to the $\coupl_y^x$ coefficients should depend only on the operator that is being factorised,
\begin{align}
\coupl_y^x &= \kfactor^x \largenc \coupl{}_y^x \ , &
\coupl_{27} = \kfactor_{27} \largenc \coupl{}_{27} \ .
\end{align}
Since the 27-plet contribution proportional to $\coupl_{27}$ is obtained by factorising the same combination of \QCD operators as the symmetric octet contribution proportional to $\coupl_y^+$, we also expect $\kfactor_{27} = \kfactor^+$.
Keeping only contributions from $c_{1 i}$, $c_{2 i}$, and $c_{6 i}$, the resulting predictions for the octet and 27-plet coefficients can be written as
\begin{subequations} \label{eq:p_yi_matching_result}
\begin{align}
\fermi_{8i} &= \frac12 V^\dagger_{\strange\up} V_{\up\down} \left(\frac{\kfactor^+ c_{12i}^+}{2 n^+_8} + \frac{\kfactor^- c_{12i}^-}{2 n^-_8} + 4 L_5 \frac{\B^2}{\fp^2} \kfactor^s c_{6i}\right)
\ , \\
\fermi_{1i} &= \frac1{\nfl} V^\dagger_{\strange\up} V_{\up\down} \left(\kfactor^+ c_{12i}^+ - \kfactor^- c_{12i}^- + 4 L_5 \frac{\B^2}{\fp^2} \kfactor^s c_{6i}\right)
\ , \\
\fermi_{\octetm i} &= \frac12 V^\dagger_{\strange\up} V_{\up\down} \left(4 L_8 + 2 H_2\right) \frac{\B^2}{\fp^2} \kfactor^s c_{6i}
\ , \\
\fermi_{27i} &= \frac14 V^\dagger_{\strange\up} V_{\up\down} \frac{n_{27}}{n_8^+} \frac{\nfl + 1}{\nfl + 2} \kfactor^+ c_{12i}^+
\ .
\end{align}
\end{subequations}
The correction factors $\kfactor^x$ can be fixed by matching them to kaon decay amplitudes.
Neglecting electromagnetic contributions, the experimentally determined amplitudes for $K\to \pi\pi$ decays \cite{Zyla:2020zbs}%
\begin{subequations}\begin{align}
\mathcal A(K^0 \to \pi^+ \pi^-) &= \unit[(277.22\pm0.12_{\exp})]{eV} \ , &
\mathcal A(K^+ \to \pi^+ \pi^0) &= \unit[(18.18\pm0.04_{\exp})]{eV} \ , \\
\mathcal A(K^0 \to \pi^0 \pi^0) &= \unit[(259.18\pm0.22_{\exp})]{eV} \ .
\end{align}\end{subequations}
They can be parameterised as \cite{Cirigliano:2003gt}
\begin{subequations}\begin{align}
\mathcal A(K^0 \to \pi^+ \pi^-) &= \mathcal A_{\nicefrac12} + \mathcal A_{\nicefrac32} \ , &
\mathcal A(K^+ \to \pi^+ \pi^0) &= \frac3{\sqrt2} \mathcal A_{\nicefrac32} \ , \\
\mathcal A(K^0 \to \pi^0 \pi^0) &= \mathcal A_{\nicefrac12} - 2 \mathcal A_{\nicefrac32} \ .
\end{align}\end{subequations}
The amplitudes $\mathcal A_{\nicefrac12}$ and $\mathcal A_{\nicefrac32}$ are associated with $\Delta \qn I = \nicefrac12$ and $\Delta \qn I = \nicefrac32$ transitions, respectively.
In the limit $m_\up$, $m_\down \to m_\light$, they are \cite{Cirigliano:2011ny}
\begin{align}
\mathcal A_{\nicefrac12} &= \epsilonEW \frac{m_K^2 - m_\pi^2}{4\fp} \left(\fermi_8 + \frac13 \fermi_{27}\right) \ , &
\mathcal A_{\nicefrac32} &= \epsilonEW \frac{m_K^2 - m_\pi^2}{4\fp} \frac53 \fermi_{27} \ .
\end{align}
Hence, the absolute values and relative phase of the complex currents are
\begin{subequations}
\begin{align}
\abs{\fermi_8} &= 2.23 \pm 0.09_{\ex} \pm \NLO \ , &
\arg \fermi_8 - \arg \fermi_{27} &= (45.03 \pm 0.77_{\ex})^\circ \pm \NLO \ , \\
\abs{\fermi_{27}} &= 0.0425 \pm 0.0018_{\ex} \pm \NLO \ .
\end{align}
\end{subequations}
The final parameter $\fermi_1$ can be fixed by matching it to $K_L \to \gamma \gamma$ decays \cite{Gerard:2005yk}, which results in
\begin{align}
\fermi_1 &= (0.37 \pm 0.05_{\ex}) \fermi_8 \ , &
\abs{\fermi_1} &= 0.82 \pm 0.12_{\ex} \pm \NLO \ .
\end{align}
Finally, inverting equations \eqref{eq:p_yi_matching_result}, one obtains
\begin{subequations}
\begin{align}
V^\dagger_{\strange\up} V_{\up\down} \kfactor^- c_{12}^- &= \frac23 \fermi_8 + \frac12 \fermi_{27} - \fermi_1 \ , &
V^\dagger_{\strange\up} V_{\up\down} \kfactor^s c_6 = \frac{\fp^2}{4 L_5 \B^2} \left(\frac23 \fermi_8 - 2 \fermi_{27} + 2 \fermi_1 \right) \ , \\
V^\dagger_{\strange\up} V_{\up\down} \kfactor^+ c_{12}^+ &= \frac52 \fermi_{27} \ .
\end{align}
\end{subequations}
Therefore the absolute values are
\begin{subequations}
\begin{align}
\abs{V^\dagger_{\strange\up} V_{\up\down} \kfactor^- c_{12}^-} &= 0.69 \pm 0.13_{\ex} \pm \NLO \ , \\
\abs{V^\dagger_{\strange\up} V_{\up\down} \kfactor^+ c_{12}^+} &= 0.106 \pm 0.005_{\ex} \pm \NLO \ , \\
\abs{V^\dagger_{\strange\up} V_{\up\down} \kfactor^s c_6} &= 0.125 \pm 0.013_{\ex} \pm 0.015_{\lat} \pm \NLO \ .
\end{align}
\end{subequations}
Since the values of the \SM Wilson coefficients are well known even at relatively low scales, such as $\mu_{\QCD} = \unit[1]{GeV}$ \cite{Buchalla:1995vs}, this relation makes it possible to extract estimates for the correction coefficients $\kfactor^x$.
In turn, these can be used to constrain the shape of the portal Wilson coefficients $c_{1i}$, $c_{2i}$, and $c_{6i}$ with $i=1$, $2$ using bounds on the corresponding $\fermi_{yi}$ obtained from searches for hidden sector induced meson transitions.
Keep in mind that we have considered only the leading contributions have for example neglected the impact of the penguin operators associated with $c_{3i}$, $c_{4i}$, and $c_{5i}$.

\subsection{Transition to the physical vacuum}

The two \SM mass like terms \eqref{eq:U octet mass Lagrangian,eq:U dipole mass Lagrangian} contain the tadpole contribution
\begin{align}
\mathcal L_U^{\fermi m} + \mathcal L_U^{hm} &\supset \frac{\i \epsilonEW \fp \B} 2 \vacuum \fermi_\octetm \trds{\comm{\nonet m}{\nonet \Phi}} + \text{h.c.} \ , &
\vacuum \fermi_\octetm &= \fermi_\octetm - \coupl_\Chromo \trsd{\inv{\nonet m_q} \octet \chromo_G} \ ,
\end{align}
which generates a finite \VEV for the \PNGB matrix $\nonet \Phi$.
When computing purely hadronic kaon decay rates in the \SM such as $K\to \pi\pi$ and $K \to \pi\pi\pi$, diagrams that contain tadpole vertices exactly cancel with the other contributions from the mass-like terms \eqref{eq:U octet mass Lagrangian,eq:U dipole mass Lagrangian}, so that the final transition amplitudes do not depend on $\vacuum \fermi_\octetm$ \cite{Cronin:1967jq,Kambor:1989tz,Donoghue:1992dd}.
This reflects the fact that the mass-like terms can be eliminated entirely defining a rotated meson field \cite{Kambor:1989tz}
\begin{align}
\vacuum{\nonet g} &= \nonet W^\dagger \nonet g \overline{\nonet W}^\dagger \ , &
\vev{\vacuum{\nonet g}} = \flavour 1 \ .
\end{align}
Accounting for the impact of the chromomagnetic dipole Lagrangian \eqref{eq:U dipole mass Lagrangian}, which is often neglected \cite{Kambor:1989tz}, the appropriate rotation matrices are
\begin{align}
\nonet W = e^{-\i (\alpha_L \flavour \lambda_7 + \beta_L \flavour \lambda_6)} &= \flavour 1 + \order{\epsilonEW} \ , &
\overline{\nonet W} = e^{\i (\alpha_R \flavour \lambda_7 + \beta_R \flavour \lambda_6)} &= \flavour 1 + \order{\epsilonEW} \ ,
\end{align}
where the angles $\alpha_{L/R}$ and $\beta_{L/R}$ defined by
\begin{subequations}
\begin{align}
\frac{\beta_L}{\alpha_L} = \frac{\beta_R}{\alpha_R} &= - \tan(\arg \vacuum \fermi_\octetm) \ , \\
\abs{\alpha_L + \i \beta_L} \pm \abs{\alpha_R + \i \beta_R} &= \arctan(\epsilonEW \abs{\vacuum \fermi_\octetm} \frac{m_\strange \pm m_\down}{m_\strange \mp m_\down}) \simeq \epsilonEW \abs{\vacuum \fermi_\octetm} \left(1\pm2\frac{m_\down}{m_\strange}\right) \ ,
\end{align}
\end{subequations}
measure the size of \EW contributions to the light quark masses.
After this field redefinition, the entries of the diagonalised quark mass matrix
\begin{align}
\nonet m^\prime &= \overline{\nonet W} \nonet m \, \nonet o(\vacuum \fermi_\octetm) \nonet W \ , &
\nonet o(x) &= \flavour 1 - \epsilonEW \left(x \flavour \lambda_\strange^\down + \text{h.c.}\right)
\end{align}
correspond to the experimentally determined quark masses.
In general, using the redefined external currents
\begin{align}
\nonet \M^\prime &= \nonet m^\prime + \nonet S_m^\prime = \overline{\nonet W} \left(\nonet \M \nonet o\left(\vacuum \fermi_\octetm + S_\octetm\right) + \epsilonEW \coupl_\Chromo \octet S_\chromo\right) \nonet W \ .
\end{align}
and
\begin{align}\label{eq:primed current definition}
\vacuum{\nonet L}_\mu &= \nonet W^\dagger \nonet L_\mu \nonet W \ , &
\vacuum{\nonet R}_\mu &= \overline{\nonet W} \nonet R_\mu \overline{\nonet W}^\dagger \ , &
\vacuum \Theta &= \Theta + \i \trf{\ln \nonet W \overline{\nonet W}} \ ,
\end{align}
in place of the original ones, the net effect of the field redefinition is two-fold:
\begin{enumerate*}[label=\roman*)]%
\item both mass-like terms $\mathcal L_U^{\Fermi \M}$ and $\mathcal L_U^{\Chromo}$ are eliminated from the \cPT Lagrangian, being reabsorbed into $\M^\prime$, and
\item while these mass-like term still contribute to $\gluons_U$, in contrast to \eqref{eq:omega prefactors}, they now contribute with new relative prefactors of
\end{enumerate*}
\begin{equation}
\coupl_\omega^{\Fermi M} - \coupl_\omega^M = \coupl_\omega^H - \coupl_\omega^M = \frac{2}{\betacoeff} \ .
\end{equation}
The rotated mass and octet Lagrangians are
\begin{subequations} \label[lag]{eq:U rotated Lagrangians}
\begin{align} \label[lag]{eq:U rotated kinetic four quark Lagrangian}
\mathcal L_U^{\prime \Fermi} &
=\mathcal L_U^{\Fermi D^2} + \mathcal L_U^{\tensor \Fermi D^2} \ , \\ \label[lag]{eq:U rotated mass Lagrangians}
\mathcal L_U^{\prime \M} &
=\frac{\fp^2 \B}2 \Left \M^\prime + \text{h.c.} \ , &
\mathcal L_U^{\prime \Fermi \M} &
=- \frac{\epsilonEW \fp^2 \B}2 \Fermi_\octetm \trds{\Left{\octet \M}^\prime + \Left{\octet \M}^{\prime\dagger}} + \text{h.c.} \ ,
\end{align}
\end{subequations}
while the rotated $G_{LR}$ singlet contributions to the \cPT Lagrangian are
\begin{align} \label[lag]{eq:U rotated singlet Lagrangians}
\test{\mathcal L_U^{S_\omega}}_{\delta^n} &
= S_\omega \test{\gluons_U}_{\delta^{n-1}} \ , &
\test{\mathcal L_U^{S_\omega}}_{\delta^n}^{\EW} &
= S_\omega \test{\gluons_U}_{\delta^{n-1}}^{\EW} \ ,
\end{align}
where
\begin{align}
\betacoeff \test{\gluons_U^\prime}_{\delta^2} &
= 2 \mathcal L_U^{D^2} + 3 \mathcal L_U^{\prime \M} + 4 \mathcal L_U^{\Theta^2} \ , &
\betacoeff \test{\gluons_U^\prime}_\delta^{\EW} &
= 2 \left(\mathcal L_U^{\prime \Fermi} + \mathcal L_U^{\prime \Fermi \M}\right) \ , &
\betacoeff \test{\gluons_U^\prime}_{\delta^2}^{\EW} &
= 2 \mathcal L_U^\Chromo \ .
\end{align}

\subsection{Expanded Lagrangian} \label{sec:cPT summary}

The final Lagrangian that captures the \LO interactions between the light mesons and each of the external currents is
\begin{align} \label[lag]{eq:U Lagrangian}
\mathcal L_U^{\LO} &
= \test{\mathcal L_U^\prime}_{\delta^2}
+ \test{\mathcal L_U^\prime}_{\delta^3}
+ \test{\mathcal L_U^\prime}_{\delta}^{\EW}
+ \test{\mathcal L_U^\prime}_{\delta^2}^{\EW}
+ \test{\mathcal L_U^\prime}_{\delta^3}^{\EW}
\end{align}
where the strong contributions are
\begin{align}
\test{\mathcal L_U^\prime}_{\delta^2} &
= \mathcal L_U^{D^2}
+ \mathcal L_U^{\prime \M}
+ \mathcal L_U^{\Theta^2} \ , &
\test{\mathcal L_U^\prime}_{\delta^3} &
= \test{\mathcal L_U^{\prime S_\omega}}_{\delta^3}
+ \mathcal L_U^{\Nc} \ ,
\end{align}
and the $\epsilonEW$ suppressed contributions are
\begin{align}
\test{\mathcal L_U^\prime}_\delta^{\EW} &
= \mathcal L_U^{\prime \Fermi} \ , &
\test{\mathcal L_U^\prime}_{\delta^2}^{\EW} &
= \test{\mathcal L_U^{\prime S_\omega}}_{\delta^2}^{\EW} \ , &
\test{\mathcal L_U^\prime}_{\delta^3}^{\EW} &
= \mathcal L_U^{T D^2}
+ \mathcal L_U^{T V}
+ \test{\mathcal L_U^{\prime S_\omega}}_{\delta^3}^{\EW} \ .
\end{align}
The individual terms are given in \cref{eq:U kinetic Lagrangian,eq:U rotated Lagrangians,eq:U rotated singlet Lagrangians,eq:U gauged WZW Lagrangian,eq:U tensor Lagrangian,eq:U axial anomaly Lagrangian}.

To ease the application of this result to phenomenological computations, we decompose the Lagrangian into individual contributions that mediate either purely hadronic meson interactions or the coupling of \cPT to specific combinations of the \SM and portal currents.
Although the final \cPT Lagrangian contains interactions with both one and two photons, we restrict ourselves to explicitly listing interactions with at most a single photon field.
This is sufficient for capturing a large number of interesting hidden sector induces transitions, such as \eg $\pi^0 \to \gamma \gamma_\text{dark}$.

\paragraph{Order $\delta^2$} \label{sec:cPT LO}

The gauged kinetic \cref{eq:U kinetic Lagrangian} $\mathcal L_U^{D^2}$ contains the ungauged kinetic Lagrangian
\begin{align} \label[lag]{eq:u kinetic Lagrangian}
\mathcal L_\L^{\partial^2} &= \frac{\fp^2}2 \trf{\nonet \L_\mu \nonet \L^\mu} \ ,
\end{align}
and couples the mesons to the photon current via the interaction
\begin{align} \label[lag]{eq:u photon Lagrangian}
\mathcal L_\L^{\partial \photon} &= \fp^2 \trf{\nonet \L_\mu (\Left{\octet r}_\photon^\mu - \octet l_\photon^\mu)} \ .
\end{align}
It also couples the mesons to the hidden currents $\nonet V_l^\mu$ and $\Left{\nonet V}_r^\mu$ via the interactions
\begin{subequations} \label[lag]{eq:u Vl Vr Lagrangians}
\begin{align}
\mathcal L_\L^{\partial V_l} &= - \fp^2 \trf{\nonet V_l^{\prime\mu} \nonet \L_\mu} \ , &
\mathcal L_\L^{\photon V_l} &= -\fp^2 \trf{\nonet V_l^{\prime\mu} \Left{\octet r}_{\photon\mu}} \ , \\
\mathcal L_\L^{\partial V_r} &= \fp^2 \trf{\Left{\nonet V}_r^{\prime\mu} \nonet \L_\mu} \ , &
\mathcal L_\L^{\photon V_r} &= - \fp^2 \trf{\Left{\nonet V}_r^{\prime\mu} \octet l_{\photon\mu}} \ .
\end{align}
\end{subequations}
The rotated mass \cref{eq:U rotated mass Lagrangians} and the anomaly \cref{eq:U axial anomaly Lagrangian} contain the purely hadronic mass-terms
\begin{align} \label[lag]{eq:u mass Lagrangians}
\mathcal L_\L^{\prime m} &
= \frac{\fp^2 \B}{2} \Left m^\prime + \text{h.c.} \ , &
\mathcal L_\L^{\theta} &
= - \frac{\fp^2 \mth^2}{2 \nfl} \Left \theta^2
\end{align}
and couple the mesons to the complex scalar $\nonet S_m^\prime$ current and the pseudoscalar $S_\theta$ current via the interactions
\begin{align} \label[lag]{eq:u Sm Stheta Lagrangians}
\mathcal L_\L^{\prime S_m} &
= \frac{\fp^2 \B}{2} \Left S_m^\prime + \text{h.c.} \ , &
\mathcal L_\L^{S_\theta} &
= - \frac{\fp^2 \mth^2}{\nfl} \Left \theta \Left S_\theta \ .
\end{align}

\paragraph{Order $\delta^3$} \label{sec:cPT NLO}

The order $\delta^3$ contribution to the rotated singlet \cref{eq:U rotated singlet Lagrangians} couples mesons to the $S_\omega$ current via the interactions
\begin{align} \label[lag]{eq:u Sw Lagrangian}
\test{\mathcal L_\L^{\prime S_\omega}}_{\delta^3} &
= \frac{S_\omega}{\betacoeff} \left(2 \left(\mathcal L_\L^{\partial^2} + \mathcal L_\L^{\partial \photon}\right) + 3 \mathcal L_\L^{\prime \m} + 4 \mathcal L_\L^{\theta}\right)
\end{align}
and the \WZW \cref{eq:U gauged WZW Lagrangian} couples mesons to the hidden currents $\nonet V_l^\mu$ and $\nonet V_r^\mu$.
The coupling to $\nonet V_l^\mu$ is mediated by the Lagrangians
\begin{subequations}
\begin{align} \label[lag]{eq:u WZW Vl Lagrangian}
\mathcal L_\L^{\Nc V_l} &
= \frac{2 \Nc}{4! (2 \pi)^2} \epsilon_{\mu\nu\rho\sigma} \trf{- \i \nonet V_l^{\prime\mu} \nonet u^\nu \nonet u^\rho \nonet u^\sigma} \ , \\ \label[lag]{eq:u WZW Vl photon Lagrangian}
\mathcal L_\L^{\Nc V_l \photon} &
= \frac{2 \Nc}{4! (2 \pi)^2} \epsilon_{\mu\nu\rho\sigma}
\trf{\nonet V_l^{\prime\mu} \left(
\acomm{\octet l_\photon^{\nu\rho} + \frac12 \Left{\octet r}_\photon^{\nu\rho}}{\nonet u^\sigma}
- \i \nonet u^\nu \octet (\nonet l_\photon^\rho - \Left{\octet r}_\photon^\rho \nonet) \nonet u^\sigma
- \i \acomm{\octet l_\photon^\nu - \Left{\octet r}_\photon^\nu}{\nonet u^\rho \nonet u^\sigma}
\right) } \ ,
\end{align}
\end{subequations}
and the coupling to $\nonet V_r^\mu$ is mediated by the Lagrangians
\begin{subequations}
\begin{align} \label[lag]{eq:u WZW Vr Lagrangian}
\mathcal L_\L^{\Nc V_r} &
= \frac{2 \Nc}{4! (2 \pi)^2} \epsilon_{\mu\nu\rho\sigma} \trf{- \i \Left{\nonet V}_r^{\prime\mu} \nonet u^\nu \nonet u^\rho \nonet u^\sigma} \ , \\ \label[lag]{eq:u WZW Vr photon Lagrangian}
\mathcal L_\L^{\Nc V_r \photon} &
= \frac{2 \Nc}{4! (2 \pi)^2} \epsilon_{\mu\nu\rho\sigma}
\begin{multlined}[t]
\trf{\Left{\nonet V}_r^{\prime\mu} \left(
\acomm{\frac12 \octet l_\photon^{\nu\rho} + \Left{\octet r}_\photon^{\nu\rho}}{\nonet u^\sigma}
- \i \nonet u^\nu \octet (\nonet l_\photon^\rho - \Left{\octet r}_\photon^\rho \nonet) \nonet u^\sigma
- \i \acomm{\octet l_\photon^\nu - \Left{\octet r}_\photon^\nu}{\nonet u^\rho \nonet u^\sigma}
\right)} \ .
\end{multlined}
\end{align}
\end{subequations}

\paragraph{Order $\epsilonEW\delta$} \label{sec:cPT EW LO}

At this order, the kinetic-like \cref{eq:U octet Lagrangian,eq:U 27-plet Lagrangian} that appear in the rotated four-quark \cref{eq:U rotated kinetic four quark Lagrangian} generate additional contributions to the kinetic-like term
\begin{subequations}
\begin{align} \label[lag]{eq:u octet Lagrangian}
\mathcal L_\L^{\fermi\partial^2} &
= - \frac{\epsilonEW \fp^2} 2 \left(\fermi_8 \trds{\octet \L_\mu \octet \L^\mu} + \fermi_1 \octet \L^\mu{}_\down^\strange \L_\mu\right) + \text{h.c.} \ , \\ \label[lag]{eq:u 27-plet Lagrangian}
\mathcal L_\L^{\tensor \fermi\partial^2} &
= - \frac{\epsilonEW \fp^2}2 \fermi_{27} \left(\nfl \octet \L_\mu{}_\down^\strange \octet \L^\mu{}_\up^\up + (\nfl - 1) \octet \L_\mu{}_\down^\up \octet \L^\mu{}_\up^\strange\right) + \text{h.c.} \ ,
\end{align}
\end{subequations}
and couple the mesons to the photon current via the interactions
\begin{subequations}
\begin{align} \label[lag]{eq:u octet photon Lagrangian}
\mathcal L_\L^{\fermi\partial \photon} &
= - \frac{\epsilonEW \fp^2}2 \left(\fermi_8 \trds{\acomm{\octet \L_\mu}{\Left{\octet r}_\photon^\mu - \octet l_\photon^\mu}} + \fermi_1 \Left{\octet r}_\photon^\mu{}_\down^\strange \L_\mu\right) + \text{h.c.} \ , \\ \label[lag]{eq:u 27-plet photon Lagrangian}
\mathcal L_\L^{\tensor \fermi\partial \photon} &
= - \frac{\epsilonEW \fp^2}2 \fermi_{27} \left(\nfl \left(\octet \L_\mu{}_\down^\strange (\Left{\octet r}_\photon^\mu - \octet l_\photon^\mu)_\up^\up
+ \Left{\octet r}_{\photon\mu}{}_\down^\strange \octet \L^\mu{}_\up^\up\right) + (\nfl - 1) \left(\Left{\octet r}_{\photon\mu}{}_\down^\up \octet \L^\mu{}_\up^\strange + \octet \L_\mu{}_\down^\up \Left{\octet r}_\photon^\mu{}_\up^\strange\right)\right) + \text{h.c.} \ .
\end{align}
\end{subequations}
They also couple mesons to the hidden vector currents $\nonet V_l^\mu$ and $\nonet V_r^\mu$ and the hidden scalar currents $S_8$, $S_1$, $S_\octetm$, and $S_{27}$.
Neglecting strangeness conserving contributions generated by interactions involving $\nonet V_l^\mu{}_\down^\strange$, the coupling to $\nonet V_l^\mu$ is mediated by the octet terms
\begin{subequations} \label[lag]{eq:u octet Vl Lagrangians}
\begin{align}
\mathcal L_\L^{\fermi\partial V_l} &
= \frac{\epsilonEW \fp^2}2 \left(\fermi_8 \trds{\acomm{\octet V_l^\mu}{\octet \L_\mu}} + \fermi_1 \octet \L_\mu{}_\down^\strange V_l^\mu\right) + \text{h.c.} \ , \\
\mathcal L_\L^{\fermi\photon V_l} &
= \frac{\epsilonEW \fp^2}2 \left(\fermi_8 \trds{\acomm{\octet V_l^\mu}{\Left{\octet r}_{\photon\mu}}} + \fermi_1 \Left{\octet r}_{\photon\mu}{}_\down^\strange V_l^\mu\right) + \text{h.c.} \ ,
\end{align}
\end{subequations}
and the 27-plet terms
\begin{align} \label[lag]{eq:u 27-plet Vl Lagrangians}
\mathcal L_\L^{\tensor \fermi\partial V_l} &
= \frac{\epsilonEW \fp^2}2 \fermi_{27} \nfl \octet \L_\mu{}_\down^\strange \octet V_l^\mu{}_\up^\up + \text{h.c.} \ , &
\mathcal L_\L^{\tensor \fermi\photon V_l} &
= \frac{\epsilonEW \fp^2}2 \fermi_{27} \nfl \Left{\octet r}_{\photon\mu}{}_\down^\strange \octet V_l^\mu{}_\up^\up + \text{h.c.} \ .
\end{align}
The coupling to $\nonet V_r^\mu$ is mediated by the octet terms
\begin{subequations} \label[lag]{eq:u octet Vr Lagrangians}
\begin{align}
\mathcal L_\L^{\fermi\partial V_r} &
= - \frac{\epsilonEW \fp^2}2
\left(\fermi_8 \trds{\acomm{\Left{\octet V}_r^\mu}{\octet \L_\mu}} + \fermi_1 \left(\Left{\octet V}_r^\mu{}_\down^\strange \L_\mu
+ \octet \L_\mu{}_\down^\strange \Left V_r^\mu\right)\right) + \text{h.c.} \ , \\
\mathcal L_\L^{\fermi\photon V_r} &
= - \frac{\epsilonEW \fp^2}2
\left(\fermi_8 \trds{\acomm{\Left{\octet V}_r^\mu}{ \Left{\octet r}_{\photon\mu} - \octet l_{\photon\mu} }}
+ \fermi_1 \Left{\octet r}_{\photon\mu}{}_\down^\strange \Left V_r^\mu\right) + \text{h.c.} \ ,
\end{align}
\end{subequations}
and the 27-plet terms
\begin{subequations} \label[lag]{eq:u 27-plet Vr Lagrangians}
\begin{align}
\mathcal L_\L^{\tensor \fermi\partial V_r} &
= - \frac{\epsilonEW \fp^2} 2 \fermi_{27}
\begin{multlined}[t]
\left(\nfl \left(\octet \L_\mu{}_\down^\strange \Left{\octet V}_r^\mu{}_\up^\up
+ \Left{\octet V}_r^\mu{}_\down^\strange \octet \L_\mu{}_\up^\up\right)
\right.\\\left.
+ (\nfl - 1) \left(\Left{\octet V}_r^\mu{}_\down^\up \octet \L_\mu{}_\up^\strange
+ \octet \L_\mu{}_\down^\up \Left{\octet V}_r^\mu{}_\up^\strange\right)\right) + \text{h.c.} \ ,
\end{multlined}
\\
\mathcal L_\L^{\tensor \fermi\photon V_r} &
= - \frac{\epsilonEW \fp^2}2 \fermi_{27}
\begin{multlined}[t]
\left(\nfl\left(
\Left{\octet V}_r^\mu{}_\down^\strange (\Left{\octet r}_{\photon\mu} - \octet l_{\photon\mu})_\up^\up
+ \Left{\octet r}_{\photon\mu}{}_\down^\strange \Left{\octet V}_r^\mu{}_\up^\up
\right)
\right.\\\left.
+ (\nfl - 1) \left(\Left{\octet V}_r^\mu{}_\down^\up \Left{\octet r}_{\photon\mu}{}_\up^\strange
+ \Left{\octet r}_{\photon\mu}{}_\down^\up \Left{\octet V}_r^\mu{}_\up^\strange\right)\right) + \text{h.c.} \ .
\end{multlined}
\end{align}
\end{subequations}
Finally, the coupling to the $S_y$ currents with $y = \octetm$, $1$, $8$, $27$ is mediated by the octet terms
\begin{subequations} \label[lag]{eq:u octet Sy Lagrangians}
\begin{align}
\mathcal L_\L^{\partial^2 S} &
= - \frac{\epsilonEW \fp^2}2 \left(S_8 \trds{\octet \L_\mu \octet \L^\mu} + S_1 \octet \L^\mu{}_\down^\strange \L_\mu\right) + \text{h.c.} \ , \\
\mathcal L_\L^{\photon S} &
= - \frac{\epsilonEW \fp^2}2
\left(S_8 \trds{\acomm{\octet \L_\mu}{\Left{\octet r}_\photon^\mu} - \octet l_\photon^\mu} + S_1 \Left{\octet r}_\photon^\mu{}_\down^\strange \L_\mu\right) + \text{h.c.} \ ,
\end{align}
\end{subequations}
and the 27-plet terms
\begin{subequations} \label[lag]{eq:u 27-plet Sy Lagrangians}
\begin{align}
\mathcal L_\L^{\partial^2 \tensor S} &
= - \frac{\epsilonEW \fp^2}2 S_{27} \left(\octet \L_\mu{}_\down^\strange \octet \L^\mu{}_\up^\up + (\nfl - 1) \octet \L_\mu{}_\down^\up \octet \L^\mu{}_\up^\strange\right) + \text{h.c.} \ , \\
\mathcal L_\L^{\photon \tensor S} &
= - \frac{\epsilonEW \fp^2}2 S_{27}
\left(\nfl \left(\octet \L_\mu{}_\down^\strange (\Left{\octet r}_\photon^\mu - \octet l_\photon^\mu)_\up^\up + \Left{\octet r}_{\photon\mu}{}_\down^\strange \octet \L^\mu{}_\up^\up\right)
+ (\nfl - 1) \left(\Left{\octet r}_{\photon\mu}{}_\down^\up \octet \L^\mu{}_\up^\strange + \octet \L_\mu{}_\down^\up \Left{\octet r}_\photon^\mu{}_\up^\strange\right)\right) + \text{h.c.} \ .
\end{align}
\end{subequations}

\paragraph{Order $\epsilonEW\delta^2$} \label{sec:cPT EW NLO}

At this order, the gauged kinetic \cref{eq:U kinetic Lagrangian} couples the mesons to the photon the weak-leptonic charged currents via the interactions
\begin{align} \label[lag]{eq:u weak kinetic Lagrangian}
\mathcal L_\L^{\photon W} &= - \fp^2 \trf{\octet l_W^\mu \Left{\octet r}_{\photon\mu}} \ , &
\mathcal L_\L^{\partial W} &= - \fp^2 \trf{\nonet \L_\mu \octet l_W^\mu} \ .
\end{align}
It also couples the mesons to the hidden current $\Left{\nonet V}_r^\mu$ via the interaction
\begin{align} \label[lag]{eq:u weak Vr Lagrangian}
\mathcal L_\L^{W V_r} &= - \fp^2 \trf{\Left{\nonet V}_r^\mu \octet l_{W\mu}} \ .
\end{align}
The rotated singlet \cref{eq:U rotated singlet Lagrangians} couples mesons to the $S_\omega$ current via the interactions
\begin{align} \label[lag]{eq:u weak NLO Sw Lagrangians}
\test{\mathcal L_\L^{\prime S_\omega}}_{\delta^2}^{\EW} &
= \frac{2 S_\omega}{\betacoeff} \left(\mathcal L_\L^{\fermi\partial^2} + \mathcal L_\L^{\tensor\fermi\partial^2} + \mathcal L_\L^{\fermi\partial\photon} + \mathcal L_\L^{\tensor\fermi\partial\photon} + \mathcal L_\L^{\prime \fermi m}\right) \ ,
\end{align}
where
\begin{align}
\mathcal L_\L^{\prime \fermi m} &
= - \frac{\epsilonEW \fp^2 \B}2 \fermi_\octetm \trds{\Left{\nonet m}^\prime + \Left{\nonet m}^{\prime\dagger}} + \text{h.c.}
\end{align}

\paragraph{Order $\epsilonEW\delta^3$} \label{sec:cPT EW NNLO}

At this order, the rotated singlet \cref{eq:U rotated singlet Lagrangians} couples mesons to the $S_\omega$ current via the interactions
\begin{align} \label[lag]{eq:u weak NNLO Sw Lagrangians}
\test{\mathcal L_\L^{\prime S_\omega}}_{\delta^3}^{\EW} &
= \frac{2 S_\omega}{\betacoeff} \left(\mathcal L_\L^{\partial W} + \mathcal L_\L^{\photon W} + \mathcal L_\L^\chromo\right) \ ,
\end{align}
where
\begin{align}
\mathcal L_\L^\chromo &
= \frac{\epsilonEW \fp^2 \B}2 \coupl_\Chromo \Left \chromo + \text{h.c.}
\end{align}
The dipole \cref{eq:U tensor Lagrangian} couples mesons to the hidden currents $\nonet V_l^\mu$, $\nonet V_r^\mu$, and $\octet T^{\mu \nu}_\tau$ via the interactions
\begin{subequations} \label[lag]{eq:u tensor Lagrangians}
\begin{flalign}
\mathcal L_\L^{T \partial^2} &= \frac{\epsilonEW}{\fp} \coupl_T^{D^2} \trf{\Left{\octet T}_\tau^{\mu\nu} \nonet \L_\mu \nonet \L_\nu} + \text{h.c.} \ , &
\mathcal L_\L^{T V} &= \frac{\epsilonEW}{\fp} \coupl_T^{LR} \trf{\Left{\octet T}_\tau^{\mu\nu} \octet l_{\photon \mu\nu}} + \text{h.c.} \ , \\
\mathcal L_\L^{T \partial V} &= \mathrlap{\frac{\epsilonEW}{\fp} \coupl_T^{D^2} \trf{\Left{\octet T}_\tau^{\mu\nu} \left(\nonet \L_\mu (\octet l_{\photon\nu} - \Left{\octet r}_{\photon\nu}) + (\octet l_{\photon\mu} - \Left{\octet r}_{\photon\mu}) \nonet \L_\nu\right)} + \text{h.c.}}
\end{flalign}
\end{subequations}
Finally, the \WZW \cref{eq:U gauged WZW Lagrangian} couples mesons to the hidden currents $\nonet V_l^\mu$ and $\nonet V_r^\mu$.
The coupling to $\nonet V_l^\mu$ is mediated by the term
\begin{align} \label[lag]{eq:u weak WZW Vl Lagrangian}
\mathcal L_\L^{\Nc V_l W} &
= \frac{2 \Nc}{4! (2 \pi)^2} \epsilon_{\mu\nu\rho\sigma}
\trf{\nonet V_l^\mu \left(\acomm{\octet l_W^{\nu\rho}}{\nonet u^\sigma}
- \i \nonet u^\nu \octet l_W^\rho \nonet u^\sigma
- \i \acomm{\octet l_W^\nu}{\nonet u^\rho \nonet u^\sigma}
\right) } \ ,
\end{align}
and the coupling to $\nonet V_r^\mu$ is mediated by the term
\begin{align} \label[lag]{eq:u weak WZW Vr Lagrangian}
\mathcal L_\L^{\Nc V_r W} &
= \frac{2 \Nc}{4! (2 \pi)^2} \epsilon_{\mu\nu\rho\sigma}
\trf{ \Left{\nonet V}_r^\mu \left(
\frac12 \acomm{\octet l_W^{\nu \rho} }{\nonet u^\sigma}
- \i \nonet u^\nu \octet l_W^\rho \nonet u^\sigma
- \i \acomm{\octet l_W^\nu}{\nonet u^\rho \nonet u^\sigma}
\right)} \ .
\end{align}

\subbib

%% file: octet.pgf
\begin{tikzpicture}[
x = 2.4em, y = 4em,
root/.style = {circle, inner sep = 1.5pt, draw, fill=white},
QN/.style = {pos = 0.5, inner sep = 1ex, circle},
brace/.style = {decoration = {brace}, decorate}
]
\draw (-3,1) -- (1,1);
\draw (-3,0) -- (2,0);
\draw (-3,-1) -- (1,-1);
\draw [dotted] (-1,1) -- (1.5,-1.5);
\draw [dotted] (1,1) -- (3,-1);
\draw [dotted] (-2,0) -- (-0,-2);
\draw [dashed] (0,1.5) -- (0,-1);
\draw [dashed] (-1,1.5) -- (-1,-1);
\draw [dashed] (1,1.5) -- (1,-1);
\draw [dashed] (2,1.5) -- (2,0);
\draw [dashed] (-2,1.5) -- (-2,0);
\node at (0,0) [root, fill=black, label = above right : {$\pi_8$}, label = below left : {$\eta_8$}]{};
\node at (1,1) [root, label = below left : {$K^+$}]{};
\node at (-1,1) [root, label = below left : {$K^0$}]{};
\node at (-2,0) [root, label = above right : {$\pi^-$}]{};
\node at (2,0) [root, label = above right : {$\pi^+$}]{};
\node at (1,-1) [root, label = above right : {$\overline K^0$}]{};
\node at (-1,-1) [root, label = above right : {$K^-$}]{};
\node [label = {[name = q-1] below right : {$-1$}}] at (0,-2){};
\node [label = below right : {$0$}] at (1.5,-1.5){};
\node [label = {[name = q1] below right : {$1$}}] at (3,-1){};
\node [label = {[name = s-1] left : {$-1$}}] at (-3,-1){};
\node [label = left : {$0$}] at (-3,0){};
\node [label = {[name = s1] left : {$\phantom{-}1$}}] at (-3,1){};
\node [label = {[name = i-1] above : {$-1$}}] at (-2,1.58){};
\node [label = above : {$-\frac12$}] at (-1,1.5){};
\node [label = above : {$0$}] at (0,1.58){};
\node [label = above : {$\frac12$}] at (1,1.5){};
\node [label = {[name = i1] above : {$1$}}] at (2,1.58){};
\draw [brace] (q1.south east) -- (q-1.south) node [QN, below right]{$\qn q$};
\draw [brace] (s-1.south west) -- (s1.north west) node [QN, left]{$\qn s$};
\draw [brace] (i-1.north west) -- (i1.north east) node [QN, above]{$\qn I$};
\end{tikzpicture}

%% file: singlet.pgf
\begin{tikzpicture}[
root/.style = {fill=white, draw, circle, inner sep = 1.5pt},
particle/.style = {inner sep = 0pt}
]
\node at (0,0) [root]{};
\node [particle, label = below right : {$\eta_1$}] at (0,0){};
\node at (0,-1) {$\qn I = \qn s = \qn q = 0$};
\node at (0,-3) {};
\end{tikzpicture}

%% file: mesons.tex
\subtoc

\section{Portal interactions of the light pseudoscalar mesons} \label{sec:light mesons}

In this section, we illustrate the information encoded inside the \cPT action derived in the previous section by extracting a set of concrete interactions.
In particular, we expand the \cPT action in terms of the meson matrix $\nonet \Phi$ in order to extract the bilinear and trilinear terms that are induced by the hidden messengers and that contribute to meson decays with at most one \SM meson in the final state.
These decays are among the primary channels for production of hidden particles at fixed target experiments, such as $K^\pm \to \pi^\pm s_i$, $K^\pm \to l^\pm \lhf_a$, and $\pi^0 \to \gamma v_\mu$.
They also include invisible decays of neutral mesons into light hidden fields, which can be constrained with collider or fixed target observations, such as~\cite{Ahn:2018mvc,AmelinoCamelia:2010me}.

In \cref{sec:Phi expanded}, we list the portal interactions that result from expanding the portal \cPT Lagrangian up to quadratic order in the meson matrix $\nonet \Phi$.
Whenever relevant, we additionally show the contributions that originate from the \SM \cPT action.
We refer to \cref{sec:expansions} for a more detailed discussion of the expansion procedure.
In \cref{sec:gauge eigenstates}, we then evaluate the flavour traces extracted in \cref{sec:Phi expanded}, and provide the interactions that couple the individual singlet and octet mesons to flavour blind hidden sectors.

The \SM \cPT Lagrangian mixes the neutral singlet and octet mesons with each other, so that they do not coincide with mass eigenstates of the theory.
The diagonalisation procedure used to construct the mass eigenstates and the corresponding mixing angles is well established and reported in \cref{sec:SM meson masses} for sake of completeness.
In addition, certain one-meson portal interactions mix the \SM mesons with the hidden spin 0 messenger.
At \LO in $\epsilonUV$, it is not necessary to diagonalise these interactions, which can be treated perturbatively when computing microscopic scattering and decay rates.
To facilitate computations in which it is necessary to re-sum the mixing, we present an explicit computation of the mixing angles between SM gauge eigenstates and messengers in \cref{sec:meson to hidden mixing}.

\subsection{One- and two-meson interactions} \label{sec:Phi expanded}

Here we list the one- and two-meson interactions, as described above.
In general, the one-meson interactions mix the \SM mesons with hidden sector particles or mediate non-hadronic decays into some combination of leptons, photons, and hidden particles.
The two-meson interactions mediate semi-hadronic decays with a single meson in the final state.
Due to the mixing between mesons and messenger particles, pure \SM interactions with two or three mesons can also contribute to processes with messenger fields in the final state.
Therefore, whenever relevant, we list the pure \SM terms contributing to such processes.

\paragraph{Order $\delta^2$}

At this order, the photon \cref{eq:u photon Lagrangian} encodes the \emph{\SM two-meson interaction}
\begin{align} \label{eq:photon sm two meson}
\mathcal L_{\Phi^2}^{\partial \photon} &= - \i \trf{\octet v_\photon^\mu \comm{\octet \Phi}{\partial_\mu \octet \Phi}} \ ,
\end{align}
which mediates radiation of virtual photons.
This interaction also contributes to decays with associated photon production, such as $\phi_i \to \phi_j \gamma s_k$ and $\phi_i \to \phi_j \gamma v_\mu$.

The kinetic-like Lagrangians in \eqref{eq:u Vl Vr Lagrangians} couple \cPT to the portal currents $\nonet V_l^\mu$ and $\nonet V_r^\mu$ via the one-meson interactions
\begin{align} \label{eq:kinetic Vl Vr one meson}
\mathcal L_\Phi^{\partial V_l} &= - \fp \trf{\nonet V_l^{\prime\mu} \partial_\mu \nonet \Phi} \ , &
\mathcal L_\Phi^{\partial V_r} &= \fp \trf{\nonet V_r^{\prime\mu} \partial_\mu \nonet \Phi} \ ,
\end{align}
and the two-meson interactions
\begin{subequations} \label{eq:kinetic Vl Vr two meson}
\begin{align} \label{eq:kinetic Vl Vr two meson no photon}
\mathcal L_{\Phi^2}^{\partial V_l} &= -\frac{\i}2 \trf{\octet V_l^{\prime\mu} [\octet \Phi, \partial_\mu \octet \Phi]} \ , &
\mathcal L_{\Phi^2}^{\partial V_r} &= -\frac{\i}2 \trf{\octet V_r^{\prime\mu} [\octet \Phi, \partial_\mu \octet \Phi]} \ , \\ \label{eq:kinetic Vl Vr two meson one photon}
\mathcal L_{\Phi^2}^{\photon V_l} &= \frac12 \trf{\octet V_l^{\prime\mu} [\octet \Phi, [\octet \Phi, \octet v_{\photon\mu}]]} \ , &
\mathcal L_{\Phi^2}^{\photon V_r} &= \frac12 \trf{\octet V_r^{\prime\mu} [\octet \Phi, [\octet \Phi, \octet v_{\photon\mu}]]} \ .
\end{align}
\end{subequations}
The one-meson interactions mediate decays such as $\phi_i \to \ell_a \lhf_b$ and $\phi_i \to \ell_a \nu_b s_j$.
They are also responsible for invisible neutral meson decays into hidden particles.
Even though these channels are not directly measurable experimentally, their relative weights compared to decays with invisible \SM final states constrain the coupling of mesons to \NP, complementing the constraints obtained from decays that feature observable \SM final states and hidden fields.
The two-meson interactions mediate decays such as $\phi_i \to \phi_j s_k s_l$, $\phi_i \to \phi_j \ell_a \lhf_b$, and $\phi_i \to \phi_j \gamma v_\mu$.
The decay $\phi_i \to \phi_j \gamma v_\mu$ producing a photon receives contributions from both \eqref{eq:kinetic Vl Vr two meson no photon,eq:kinetic Vl Vr two meson one photon}.
However, diagrams that contain the interaction \eqref{eq:kinetic Vl Vr two meson no photon}, which does not involve photons directly, also have to contain a \SM interaction \eqref{eq:photon sm two meson}, which radiates the required photon.
If the hidden sector contains secluded neutral particles $X$, which can act \eg as \DM and interact with the \SM only via the hidden field, the two-meson interactions can also give rise to decays mediated by an off-shell messenger exchange, such as $\phi_i \to \phi_j v_\mu^\ast \to \phi_j \overline XX$.

The quark-mass Lagrangian in \eqref{eq:u mass Lagrangians} couples \cPT to the imaginary and real parts of the portal current $S_m^\prime$ via the one- and two-meson interactions
\begin{align} \label{eq:mass Sm one meson two meson}
\mathcal L_\Phi^{\prime S_m} &= - \fp \B \trf{\nonet \Phi \Im \nonet S_m^\prime} \ , &
\mathcal L_{\Phi^2}^{\prime S_m} &= - \frac{\B}{2} \trf{\nonet \Phi^2 \Re \nonet S_m^\prime} \ .
\end{align}
These one-meson interactions are similar to the one in \eqref{eq:kinetic Vl Vr one meson}, which couple \cPT to $\nonet V_l$ and $\nonet V_r$, and mix the \SM mesons with the hidden spin 0 messenger and mediate neutral meson decays into hidden spin 0 particles.
The two-meson interactions mediate decays such as $\phi_i \to \phi_j s_k$ and $\phi_i \to \phi_j s_k s_l$.
Like the interactions \eqref{eq:kinetic Vl Vr two meson}, they can also give rise to decays with photons in the final state, such as $\phi_i \to \phi_j s_k \gamma$, as well as decays into secluded particles $X$ that are mediated by an off-shell messenger exchange, such as $\phi_i \to \phi_j s_k^\ast \to \phi_j \overline XX$.

Finally, the anomaly Lagrangian in \eqref{eq:u mass Lagrangians} couples \cPT to the portal current $S_\theta$ via the one-meson interaction
\begin{align} \label{eq:anomaly Stheta one meson}
\mathcal L_\Phi^{S_\theta} &= \fp \frac{\mth^2}{\nfl} S_\theta \Phi \ ,
\end{align}
which mixes the singlet $\eta_1$-meson with the spin 0 messenger.

\paragraph{Order $\delta^3$}

At this order, the singlet \cref{eq:u Sw Lagrangian} couples \cPT to the portal current $S_\omega$ via the one-meson interactions
\begin{align} \label{eq:singlet LO one meson}
\mathcal L_\Phi^{\prime S_\omega} &
= \frac{S_\omega}{\betacoeff} 4 \mathcal L_\Phi^\theta \ , &
\mathcal L_\Phi^\theta &
= \frac{\fp \mth^2}{\nfl} \theta \Phi \ ,
\end{align}
and the two-meson interactions
\begin{align} \label{eq:singlet LO two meson}
\mathcal L_{\Phi^2}^{\prime S_\omega}
= \frac{S_\omega}{\betacoeff} \left(
2 \left(
 \mathcal L_{\Phi^2}^{\partial^2}
+ \mathcal L_{\Phi^2}^{\partial \photon}
\right)
+ 3 \mathcal L_{\Phi^2}^{\prime m}
+ 4 \mathcal L_{\Phi^2}^\theta
\right) \ ,
\end{align}
where
\begin{align} \label{eq:sm Lagrangians two meson}
\mathcal L_{\Phi^2}^{\partial^2} &
= \frac12 \trf{\partial_\mu \nonet \Phi \partial^\mu \nonet \Phi} \ , &
\mathcal L_{\Phi^2}^\theta &
= - \frac{\mth^2}{2 \nfl} \Phi^2 \ , &
\mathcal L_{\Phi^2}^{\prime m} &
= - \frac{\B}2 \trf{\nonet \Phi^2 \m} \ .
\end{align}
The one-meson interactions \eqref{eq:singlet LO one meson} mix the singlet $\eta_1$ with spin 0 messenger particles, but this mixing is negligible because it is strongly suppressed by the \QCD theta angle.
The two-meson interactions \eqref{eq:singlet LO two meson} are similar to the one in \eqref{eq:mass Sm one meson two meson}.
They mediate decays into spin 0 messengers, such as $\phi_i \to \phi_j s_k$ and $\phi_i \to \phi_j \gamma s_k$, as well as decays into secluded particles $X$, such as $\phi_i \to \phi_j \overline XX$.

The \WZW \cref{eq:u WZW Vl Lagrangian,eq:u WZW Vl photon Lagrangian} couple \cPT to the portal current $\nonet V_l^\mu$ via the one-meson interaction
\begin{align} \label{eq:wzw Vl one meson}
\mathcal L_\Phi^{\Nc V_l \photon} &
= \frac{\epsilon_{\mu\nu\rho\sigma}}{(4\pi)^2 \fp} \frac34
\trf{\nonet\Phi \acomm{\octet v_\photon^{\rho\sigma}}{\nonet V_l^{\prime\mu\nu}}} \ ,
\end{align}
and the two-meson interactions
\begin{align} \label{eq:wzw Vl two meson}
\mathcal L_{\Phi^2}^{\Nc V_l \photon} &
= \frac{\epsilon_{\mu\nu\rho\sigma}}{(4\pi)^2 \fp^2} \frac{\i}4
\trf{\nonet V_l^{\prime\mu} \left(3 \acomm{\octet v_\photon^{\rho\sigma}}{\comm{\octet\Phi}{\partial^\nu\octet\Phi}} + 2 \acomm{\comm{\octet\Phi}{\octet v_\photon^{\rho\sigma}}}{\partial^\nu \nonet\Phi}\right)} \ .
\end{align}
Finally, the \WZW \cref{eq:u WZW Vr Lagrangian,eq:u WZW Vr photon Lagrangian} couple \cPT to the portal current $\nonet V_r^\mu$ via the one-meson interaction
\begin{align} \label{eq:wzw Vr one meson}
\mathcal L_\Phi^{\Nc V_r \photon} &
= \frac{\epsilon_{\mu\nu\rho\sigma}}{(4\pi)^2 \fp} \frac34
\trf{\nonet\Phi \acomm{\octet v_\photon^{\rho\sigma}}{\nonet V_r^{\prime\mu\nu}}} \ ,
\end{align}
and the two-meson interactions
\begin{align} \label{eq:wzw Vr two meson}
\mathcal L_{\Phi^2}^{\Nc V_r \photon} &
= \frac{\epsilon_{\mu\nu\rho\sigma}}{(4\pi)^2 \fp^2} \frac{\i}4
\trf{\nonet V_r^{\prime\mu} \left(
3 \acomm{\octet v_\photon^{\rho\sigma}}{\comm{\octet\Phi}{\partial^\nu\octet\Phi}}
+ 4 \acomm{\comm{\octet\Phi}{\octet v_\photon^{\rho\sigma}}}{\partial^\nu\nonet\Phi}
- 6 \comm{\octet\Phi}{\acomm{\octet v_\photon^{\rho\sigma}}{\partial^\nu\nonet\Phi}}
\right)} \ .
\end{align}
The one-meson interactions \eqref{eq:wzw Vl one meson,eq:wzw Vr one meson} mediate decays such as $\phi_i \to \gamma v_\mu$ and $\phi_i \to \gamma \ell_a \lhf_b$, while the two-meson interactions \eqref{eq:wzw Vl two meson,eq:wzw Vr two meson} mediate decays such as $\phi_i \to \phi_j \gamma v_\mu$.
Notice that the \WZW action is the only contribution that mediates non-hadronic meson decays with a spin 1 messenger particles in the final state.
In particular, the order $\delta^3$ \cref{eq:U NLO kinetic Lagrangians,eq:U NLO mass Lagrangians,eq:U field strength Lagrangians,eq:U NLO anomaly Lagrangian,eq:U anomaly mass Lagrangian,eq:U anomaly kinetic Lagrangian}, which one may expect to do so, do not mediate such transitions.

\paragraph{Order $\epsilonEW \delta$}

At this order, the octet \cref{eq:u octet Lagrangian,eq:u octet photon Lagrangian} encode the strangeness-violating \emph{\SM two-meson interactions}
\begin{subequations} \label{eq:octet sm}
\begin{align}
\mathcal L_{\Phi^2}^{\fermi \partial^2} &
= - \frac{\epsilonEW} 2 \left(\fermi_8 \trds{\partial_\mu \octet \Phi \partial^\mu \octet \Phi} + \fermi_1 \partial^\mu \octet \Phi_\down^\strange \partial_\mu \Phi\right) + \text{h.c.} \ , \\
\mathcal L_{\Phi^2}^{\fermi \partial \photon} &
= - \i \frac{\epsilonEW}2 \fermi_8 \trds{\acomm{\partial_\mu \octet \Phi}{\comm{\octet \Phi}{\octet v_\photon^\mu}}} + \text{h.c.} \ .
\end{align}
\end{subequations}
The 27-plet \cref{eq:u 27-plet Lagrangian,eq:u 27-plet photon Lagrangian} encode the additional strangeness-violating \emph{\SM two-meson interactions}
\begin{subequations} \label{eq:27-plet sm}
\begin{align}
 \mathcal L_{\Phi^2}^{\tensor \fermi \partial^2} &
= - \frac{\epsilonEW}2 \fermi_{27} \left(\nfl \partial^\mu \octet \Phi_\up^\up \partial_\mu \octet \Phi_\down^\strange + (\nfl - 1) \partial_\mu \octet \Phi_\down^\up \partial^\mu \octet \Phi_\up^\strange\right) + \text{h.c.} \ , \\
\mathcal L_{\Phi^2}^{\tensor \fermi \partial \photon} &
= - \i \frac{\epsilonEW}2 \fermi_{27} (\nfl - 1) \left(\langle \comm{\octet \Phi}{\octet v_\photon^\mu}\rangle_\down^\up \partial^\mu \octet \Phi_\up^\strange + \partial_\mu \octet \Phi_\down^\up \langle \comm{\octet \Phi}{\octet v_\photon^\mu}\rangle_\up^\strange\right) + \text{h.c.} \ .
\end{align}
\end{subequations}
These interactions mix kaons with pions and $\eta$-mesons, and also mediate decays such as $\phi_i \to \phi_j \ell_a \ell_a$, where both charged leptons are of the same flavour.
Similarly to \eqref{eq:photon sm two meson}, the latter interactions also contribute to decays with associated photon production, such as $\phi_i \to \phi_j \gamma s_j$ and $\phi_j \to \phi_j \gamma v_\mu$.
The octet \cref{eq:u octet Vl Lagrangians} couples \cPT to the portal current $V_l^\mu$ via the strangeness-violating one-meson interactions
\begin{align} \label{eq:octet Vl one meson}
\mathcal L_\Phi^{\fermi \partial V_l} &
= \frac{\epsilonEW \fp}2 \left(\fermi_8 \left(\octet V_l^\mu{}_\down^\down + \octet V_l^\mu{}_\strange^\strange\right) + \fermi_1 V_l^\mu\right) \partial_\mu \octet \Phi{}_\down^\strange + \text{h.c.}\,,
\end{align}
and the strangeness-violating two-meson interactions
\begin{subequations} \label{eq:octet Vl two meson}
\begin{align}
\mathcal L_{\Phi^2}^{\fermi \partial V_l} &
= \i \frac{\epsilonEW}4 \left(\fermi_8 \left(\octet V_l^\mu{}_\down^\down + \octet V_l^\mu{}_\strange^\strange\right) + \fermi_1 V_l^\mu\right) \trds{\comm{\octet \Phi}{\partial_\mu \octet \Phi}} + \text{h.c.} \ , \\
\mathcal L_{\Phi^2}^{\fermi \photon V_l} &
= - \frac{\epsilonEW}4 \left(\fermi_8 \left(\octet V_l^\mu{}_\down^\down + \octet V_l^\mu{}_\strange^\strange\right) + \fermi_1 V_l^\mu\right) \trds{\comm{\octet \Phi}{\comm{\octet \Phi}{\octet v_{\photon\mu}}}} + \text{h.c.} \,.
\end{align}
\end{subequations}
The 27-plet \cref{eq:u 27-plet Vl Lagrangians} couples \cPT to the portal current $V_l^\mu$ via the strangeness-violating one-meson interaction
\begin{align} \label{eq:27-plet Vl one meson}
\mathcal L_\Phi^{\tensor \fermi \partial V_l} &
= \frac{\epsilonEW \fp}2 \fermi_{27} \nfl \octet V_l^\mu{}_\up^\up \partial_\mu \octet \Phi_\mu{}_\down^\strange + \text{h.c.}\,,
\end{align}
and the strangeness-violating two-meson interactions
\begin{subequations} \label{eq:27-plet Vl two meson}
\begin{align}
\mathcal L_{\Phi^2}^{\tensor \fermi \partial V_l} &
= \i \frac{\epsilonEW}4 \fermi_{27} \nfl \octet V_l^\mu{}_\up^\up \trds{\comm{\octet \Phi}{\partial_\mu \octet \Phi}} + \text{h.c.} \ , \\
\mathcal L_{\Phi^2}^{\tensor \fermi \photon V_l} &
= - \frac{\epsilonEW}4 \fermi_{27} \nfl \octet V_l^\mu{}_\up^\up \trds{\comm{\octet \Phi}{\comm{\octet \Phi}{\octet v_{\photon\mu}}}} + \text{h.c.}\,.
\end{align}
\end{subequations}
The octet \cref{eq:u octet Vr Lagrangians} couples \cPT to the portal current $V_r^\mu$ via the strangeness-violating one-meson interaction
\begin{align} \label{eq:octet Vr one meson}
\mathcal L_\Phi^{\fermi \partial V_r} &
= - \frac{\epsilonEW \fp}2 \left(\fermi_8 \left(\octet V_r^\mu{}_\down^\down + \octet V_r^\mu{}_\strange^\strange\right) + \fermi_1 V_r^\mu\right) \partial_\mu \octet \Phi{}_\down^\strange + \text{h.c.}\,,
\end{align}
and the strangeness-violating two-meson interactions
\begin{subequations} \label{eq:octet Vr two meson}
\begin{align}
\mathcal L_{\Phi^2}^{\fermi \partial V_r} &
= - \i \frac{\epsilonEW}4
\begin{multlined}[t] \left(
2 \fermi_8 \trds{\acomm{\comm{\octet \Phi}{\octet V_r^\mu}}{\partial_\mu \octet \Phi}}
\right.\\\left.
+ \left(\fermi_8 \left(\octet V_r^\mu{}_\down^\down + \octet V_r^\mu{}_\strange^\strange\right) + \fermi_1 V_r^\mu\right) \trds{\comm{\octet \Phi}{\partial_\mu \octet \Phi}} \right) + \text{h.c.} \ ,
\end{multlined} \\
\mathcal L_{\Phi^2}^{\fermi \photon V_r} &
= \frac{\epsilonEW}4
\begin{multlined}[t]
\left(
2 \fermi_8 \trds{\acomm{\comm{\octet \Phi}{\octet V_r^\mu}}{\comm{\octet \Phi}{\octet v_{\photon\mu}}}}
\right.\\\left.
+ \left(\fermi_8 \left(\octet V_r^\mu{}_\down^\down + \octet V_r^\mu{}_\strange^\strange\right) + \fermi_1 V_r^\mu\right) \trds{\comm{\octet \Phi}{\comm{\octet \Phi}{\octet v_{\photon\mu}}}} \right) + \text{h.c.} \,.
\end{multlined}
\end{align}
\end{subequations}
The 27-plet \cref{eq:u 27-plet Vr Lagrangians} couples \cPT to the portal current $V_r^\mu$ via the strangeness-violating one-meson interaction
\begin{align} \label{eq:27-plet Vr one meson}
\mathcal L_\Phi^{\tensor \fermi \partial V_r} &
= - \frac{\epsilonEW \fp} 2 \nfl \fermi_{27} \octet V_r^\mu{}_\up^\up \partial_\mu \octet \Phi_\down^\strange + \text{h.c.}\,,
\end{align}
and the strangeness-violating two-meson interactions
\begin{subequations} \label{eq:27-plet Vr two meson}
\begin{align}
\mathcal L_{\Phi^2}^{\tensor \fermi \partial V_r} &
= - \i \frac{\epsilonEW}4 \fermi_{27}
\begin{multlined}[t] \left(
\nfl \trds{\comm{\octet \Phi}{\partial_\mu \octet \Phi}} \octet V_r^\mu{}_\up^\up
\right.\\\left.
+ 2(\nfl - 1) \left(\langle \comm{\octet \Phi}{\octet V_r^\mu} \rangle_\down^\up \partial_\mu \octet \Phi{}_\up^\strange + \partial_\mu \octet \Phi{}_\down^\up \langle \comm{\octet \Phi}{\octet V_r^\mu} \rangle_\up^\strange\right) \right) + \text{h.c.} \ ,
\end{multlined} \\
\mathcal L_{\Phi^2}^{\tensor \fermi \photon V_r} &
= \begin{multlined}[t]
\frac{\epsilonEW}4 \fermi_{27} \left(
\nfl \trds{\comm{\octet \Phi}{\comm{\octet \Phi}{\octet v_{\photon\mu}}}} \octet V_r^\mu{}_\up^\up
\right.\\\left.
+ 2 (\nfl - 1) \left(\langle \comm{\octet \Phi}{\octet V_{r\mu}} \rangle_\down^\up \langle \comm{\octet \Phi}{\octet v_\photon^\mu} \rangle_\up^\strange
+ \langle \comm{\octet \Phi}{\octet v_{\photon\mu}} \rangle_\down^\up \langle \comm{\octet \Phi}{\octet V_r^\mu} \rangle_\up^\strange\right) \right) + \text{h.c.} \ .
\end{multlined}
\end{align}
\end{subequations}
The one-meson interactions \eqref{eq:octet Vl one meson,eq:27-plet Vl one meson,eq:octet Vr one meson,eq:27-plet Vr one meson} are similar to the one-meson interactions in \eqref{eq:kinetic Vl Vr one meson,eq:mass Sm one meson two meson} and mediate only invisible decays.
The two-meson interactions \eqref{eq:octet Vl two meson,eq:27-plet Vl two meson,eq:octet Vr two meson,eq:27-plet Vr two meson} are similar to the interactions \eqref{eq:kinetic Vl Vr two meson,eq:wzw Vl two meson,eq:wzw Vr two meson}.
They mediate decays with photons in the final state, such as $\phi_i \to \phi_j \gamma v_\mu$, as well as decays into secluded particles $X$, such as $\phi_i \to \phi_j v_\mu^\ast \to \phi_j \overline XX$.

Finally, the octet \cref{eq:u octet Sy Lagrangians} couples \cPT to the portal currents $S_y$ via the strangeness-violating two-meson interactions
\begin{subequations} \label{eq:octet Sy two meson}
\begin{align}
\mathcal L_{\Phi^2}^{\partial^2 S} &
= - \frac{\epsilonEW}2 \left(S_8 \trds{\partial_\mu \octet \Phi \partial^\mu \octet \Phi} + S_1 \partial^\mu \octet \Phi_\down^\strange \partial_\mu \Phi\right) + \text{h.c.} \ , \\
\mathcal L_{\Phi^2}^{\photon S} &
= - \i \frac{\epsilonEW}2 S_8 \trds{\acomm{\partial_\mu \octet \Phi}{\comm{\octet \Phi}{\octet v_\photon^\mu}}} + \text{h.c.} \ ,
\end{align}
\end{subequations}
while the 27-plet \cref{eq:u 27-plet Sy Lagrangians} also couples \cPT to the portal currents $S_y$ via the strangeness-violating two-meson interactions
\begin{subequations} \label{eq:27-plet Sy two meson}
\begin{align}
\mathcal L_{\Phi^2}^{\partial^2 \tensor S} &
= - \frac{\epsilonEW}2 \nfl S_{27} \left(\partial_\mu \octet \Phi{}_\down^\strange \partial^\mu \octet \Phi{}_\up^\up + \frac{\nfl - 1}{\nfl} \partial_\mu \octet \Phi_\mu{}_\down^\up \partial^\mu \octet \Phi{}_\up^\strange\right) + \text{h.c.} \ , \\
\mathcal L_{\Phi^2}^{\photon \tensor S} &
= - \i \frac{\epsilonEW}2 (\nfl-1) S_{27}
\left(\comm{\octet \Phi}{\octet v_\photon^\mu}_\down^\up \partial^\mu \octet \Phi{}_\up^\strange + \partial_\mu \octet \Phi{}_\down^\up \comm{\octet \Phi}{\octet v_\photon^\mu}_\up^\strange\right) + \text{h.c.} \ .
\end{align}
\end{subequations}
These interactions are similar to the two-meson interactions in \eqref{eq:mass Sm one meson two meson}.
They mediate decays such as $\phi_i \to \phi_j s_k$, $\phi_i \to \phi_j s_k \gamma$, and $\phi_i \to \phi_j \overline XX$, with secluded particles $X$ in the final state.

\paragraph{Order $\epsilonEW \delta^2$}

At this order, the kinetic \cref{eq:u weak kinetic Lagrangian} encodes the \emph{\SM one-meson interactions}
\begin{align} \label{eq:weak sm one meson}
\mathcal L_\Phi^{\partial W} &= - \fp \trf{\octet l_W^\mu \partial_\mu \octet \Phi} \ , &
\mathcal L_\Phi^{\photon W} &= - \i \fp \trf{\octet l_W^\mu \comm{\octet \Phi}{\octet v_{\photon\mu}}} \ ,
\end{align}
and the \emph{\SM two-meson interactions}
\begin{align} \label{eq:weak sm two meson}
\mathcal L_{\Phi^2}^{\partial W} &= - \frac{\i}2 \trf{\octet l_W^\mu \comm{\octet \Phi}{\partial_\mu \octet \Phi}} \ , &
\mathcal L_{\Phi^2}^{\photon W} &= \frac12 \trf{\octet l_W^\mu \comm{\octet \Phi}{\comm{\octet \Phi}{\octet v_{\photon\mu}}}} \ .
\end{align}
The one-meson interactions mediate non-hadronic charged meson decays such as $\phi_i \to \ell_a \nu_a$, while the two-meson interactions mediate semi-hadronic three-body decays such as $\phi_i \to \phi_j \ell_a \nu_a$.
The kinetic \cref{eq:u weak Vr Lagrangian} couples \cPT to the portal current $\nonet V_r^\mu$ via the one-meson interaction
\begin{align} \label{eq:weak Vr one meson}
\mathcal L_\Phi^{W V_r} &= \i \fp \trf{\octet V_r^\mu [\octet \Phi, \octet l_{W\mu}]} \ ,
\end{align}
and the two-meson interaction
\begin{align} \label{eq:weak Vr two meson}
\mathcal L_{\Phi^2}^{W V_r} &= \frac12 \trf{\octet V_r^\mu [\octet \Phi, [\octet \Phi, \octet l_{W\mu}]]} \ .
\end{align}
The one-meson interactions mediate decays such as $\phi_i \to \ell_a \nu_b v_\mu$, while the two-meson interactions mediate decays such as $\phi_i \to \phi_j \ell_a \nu_b v_\mu$.
The singlet \cref{eq:u weak NLO Sw Lagrangians} couples \cPT to the portal current $S_\omega$ via the one-meson interaction
\begin{align} \label{eq:singlet weak NLO one meson}
\mathcal L_\Phi^{\prime S_\omega} &
= \frac{S_\omega}{\betacoeff} 2 \mathcal L_\Phi^{\prime \fermi m} \ , &
\mathcal L_\Phi^{\prime \fermi m} &
= - \i \frac{\epsilonEW \fp \B}2 \fermi_\octetm \trds{\comm{\nonet \Phi}{\nonet m^\prime}} + \text{h.c.} \ ,
\end{align}
and the two-meson interactions
\begin{align} \label{eq:singlet weak NLO two meson}
\mathcal L_{\Phi^2}^{\prime S_\omega} &
= \frac{S_\omega}{\betacoeff} 2 \left(
\mathcal L_{\Phi^2}^{\prime \fermi m}
+ 2 \left(
\mathcal L_{\Phi^2}^{\fermi \partial^2}
+ \mathcal L_{\Phi^2}^{\fermi \partial \photon}
+ \mathcal L_{\Phi^2}^{\tensor \fermi \partial^2}
+ \mathcal L_{\Phi^2}^{\tensor \fermi \partial \photon}
\right)
\right) \ , &
\mathcal L_{\Phi^2}^{\prime \fermi m} &
= \frac{\epsilonEW \B}4 \fermi_\octetm \trds{\acomm{\nonet \Phi^2}{\nonet m}} + \text{h.c.} \ .
\end{align}
The one-meson interaction \eqref{eq:singlet weak NLO one meson} is similar to the one meson interaction \eqref{eq:singlet LO one meson} and mixes neutral kaons with the hidden spin 0 messenger.
However, in contrast to interaction \eqref{eq:singlet LO one meson}, the mixing here is not suppressed by the \QCD theta angle, and therefore not in general negligible.
The two-meson interactions \eqref{eq:singlet weak NLO two meson} are similar to the two-meson interactions in \eqref{eq:mass Sm one meson two meson,eq:27-plet Sy two meson} and mediate decays such as $\phi_i \to \phi_j s_k$, $\phi_i \to \phi_j s_k \gamma$, and $\phi_i \to \phi_j \overline XX$.

\paragraph{Order $\epsilonEW \delta^3$}

At this order, the singlet \cref{eq:u weak NNLO Sw Lagrangians} couples \cPT to the portal current $S_\omega$ via the one-meson interactions
\begin{align} \label{eq:singlet weak NNLO one meson}
\mathcal L_\Phi^{\prime S_\omega} &
= \frac{S_\omega}{\betacoeff} 2 \left(
\mathcal L_\Phi^{\partial W}
+ \mathcal L_\Phi^{\photon W}
+ \mathcal L_\Phi^\chromo
\right) \ , &
\mathcal L_\Phi^\chromo &
= - \epsilonEW \fp\B \coupl_\Chromo \trf{\octet \Phi \Im \octet \chromo} \ ,
\end{align}
and the two-meson interactions
\begin{align} \label{eq:singlet weak NNLO two meson}
\mathcal L_{\Phi^2}^{\prime S_\omega} &
= \frac{S_\omega}{\betacoeff} 2 \left(
\mathcal L_{\Phi^2}^{\partial W}
+ \mathcal L_{\Phi^2}^{\photon W}
+ \mathcal L_{\Phi^2}^\chromo
\right) \ , &
\mathcal L_{\Phi^2}^\chromo &
= - \frac{\epsilonEW \B}2 \coupl_\Chromo \trf{\nonet \Phi^2 \Re \octet \chromo} \ .
\end{align}
The one-meson interactions that involve the dipole current $\octet \chromo$ are similar to the interactions \eqref{eq:mass Sm one meson two meson,eq:singlet LO one meson,eq:singlet weak NLO one meson} and mix neutral kaons with the hidden spin 0 messenger.
The one-meson interactions that involve the weak leptonic charged current $l_W^\mu$ mediate decays such as $\phi_i \to \ell_a \nu_b s_j$.
The two-meson interactions \eqref{eq:singlet weak NNLO two meson} mediate decays such as $\phi_i \to \phi_j s_k$, $\phi_i \to \phi_j s_k \gamma$, $\phi_i \to \phi_j \overline XX$, and $\phi_i \to \phi_j \ell_a \nu_b s_k$.
The tensor \cref{eq:u tensor Lagrangians} couples \cPT to the portal current $\octet T^{\mu \nu}_\tau$ via the two-meson interactions
\begin{subequations} \label{eq:dipole two meson}
\begin{align}
\mathcal L_{\Phi^2}^{T \partial^2} &= \frac\epsilonEW{\fp^3} \coupl_T^{D^2} \trf{\octet T_\tau^{\mu\nu} \partial_\mu \nonet \Phi \partial_\nu \nonet \Phi} + \text{h.c.} \ , &
\mathcal L_{\Phi^2}^{T V} &= -\frac{\epsilonEW}{2\fp^3} \coupl_T^{LR} \trf{\octet T_\tau^{\mu\nu} \comm{\octet \Phi}{\comm{\octet \Phi}{\octet v_{\photon\mu\nu}}}} + \text{h.c.} \ , \\
\mathcal L_{\Phi^2}^{T \partial V} &= \mathrlap{\i \frac\epsilonEW{\fp^3} \coupl_T^{D^2} \trf{\octet T_\tau^{\mu\nu} \left(\partial^\mu \octet \Phi \comm{\octet \Phi}{\octet v_{\photon\nu}} + \comm{\octet \Phi}{\octet v_{\photon\mu}} \partial^\nu \octet \Phi\right)} + \text{h.c.} \ .}
\end{align}
\end{subequations}
These interactions mediate decays such as $\phi_i \to \phi_j \gamma s_k$ and $\phi_i \to \phi_j \gamma \gamma s_k$.
The \WZW \cref{eq:u weak WZW Vl Lagrangian} couple \cPT to the portal current $\nonet V_l^\mu$ via the one-meson interactions
\begin{align} \label{eq:wzw weak Vl one meson}
\mathcal L_\Phi^{N V_l W} &
= \frac{\epsilon_{\mu\nu\rho\sigma}}{(4\pi)^2 \fp} \frac12
\trf{\nonet\Phi \acomm{\octet l_W^{\rho\sigma}}{\nonet V_l^{\mu\nu}}} \ ,
\end{align}
and the two-meson interactions
\begin{align} \label{eq:wzw weak Vl two meson}
\mathcal L_{\Phi^2}^{N V_l W} &
= \frac{\epsilon_{\mu\nu\rho\sigma}}{(4\pi)^2 \fp^2} \frac{\i}2
\trf{\nonet V_l^\mu \left(\acomm{\octet l_W^{\rho\sigma}}{\comm{\octet\Phi}{\partial^\nu \octet\Phi}} + 2 \partial^\rho \nonet\Phi \octet l_W^\nu \partial^\sigma \nonet\Phi - 2 \acomm{\partial^\rho \nonet\Phi \partial^\sigma \nonet\Phi}{\nonet l_W^\nu}\right)} \ .
\end{align}
Finally, the \WZW \cref{eq:u weak WZW Vr Lagrangian} couple \cPT to the portal current $\nonet V_r^\mu$ via the one-meson interactions
\begin{align} \label{eq:wzw weak Vr one meson}
\mathcal L_\Phi^{N V_r W} &
= \frac{\epsilon_{\mu\nu\rho\sigma}}{(4\pi)^2 \fp} \frac14
\trf{\nonet\Phi \acomm{\octet l_W^{\rho\sigma}}{\nonet V_r^{\mu\nu}}} \ ,
\end{align}
and the two-meson interactions
\begin{align} \label{eq:wzw weak Vr two meson}
\mathcal L_{\Phi^2}^{N V_r W} &
= \frac{\epsilon_{\mu\nu\rho\sigma}}{(4\pi)^2 \fp^2} \frac{\i}4
\trf{\nonet V_r^\mu \left(\acomm{\octet l_W^{\rho\sigma}}{\comm{\octet\Phi}{\partial^\nu\octet\Phi}} + 4 \partial^\rho \nonet\Phi \octet l_W^\nu \partial^\sigma \nonet\Phi
- 4 \acomm{\partial^\rho \nonet\Phi \partial^\sigma \nonet\Phi}{\nonet l_W^\nu}\right)} \ .
\end{align}
The one-meson interactions \eqref{eq:wzw weak Vl one meson,eq:wzw weak Vr one meson} mediate decays such as and $\phi_i \to v_\mu \ell_a \nu_b$,
while the two-meson interactions \eqref{eq:wzw weak Vl two meson,eq:wzw weak Vr two meson} mediate decays such as $\phi_i \to \phi_j \ell_a \nu_b v_\mu$.

\subsection{Flavour-blind hidden sectors} \label{sec:gauge eigenstates}

In this section, we focus on the coupling of \cPT to flavour-blind hidden sectors and evaluate the \cPT flavour traces to provide the one- and two-meson interactions in terms of the singlet and octet meson eigenstates $\pi_8$, $\eta_8$, $\eta_1$, $\pi^\pm$, $K^\pm$, $K^0$, $\overline K^0$.
Mixing between the $\pi_8$, $\eta_8$, and $\eta_1$ gives rise to the physical mass eigenstates $\pi^0$, $\eta$, and $\eta^\prime$, while $K^0$ and $\overline K^0$ are diagonalised into the two physical mass eigenstates $K_S^0$ and $K_L^0$.

The hidden sector is flavour blind only if the \EW scale \PETs are flavour blind.
After integrating out the heavy \SM particles, the resulting strong scale \PETs can still violate quark-flavour due to virtual $W$-boson exchanges.
Hence, the octet contributions to corresponding strong-scale portal currents are given as
\begin{align}
\octet S_m^\prime &= \flavour \lambda_\down^\strange \octet S_m^\prime{}_\down^\strange + \flavour \lambda_\strange^\down \octet S_m^\prime{}_\strange^\down \ , &
\octet V_l^{\prime\mu} &= \flavour \lambda_\down^\strange \octet V_l^{\prime\mu}{}_\down^\strange + \text{h.c.} \ , &
\octet V_r^{\prime\mu} &= 0 \ , &
\octet T_\tau^{\mu\nu} &= \octet T_\tau^{[\mu\nu]} \ ,
\end{align}
where $\octet S_m^\prime{}_\down^\strange$, $\octet S_m^\prime{}_\strange^\down$, $\octet V_l^\mu{}_\down^\strange$, and $\octet T_\tau^{\mu\nu}$ capture the contributions due to $W$-boson exchanges, so that they are suppressed by a factor of $\epsilonSM$.
This also implies that we can replace the primed currents in \eqref{eq:primed current definition} with their unprimed counterparts.
At order $\epsilonSM$, the right-handed current $\nonet V_r^\mu$ in \eqref{eq:left and right handed portal currents} does not receive any contributions from higher dimensional operators.
Hence, it has to be flavour blind even at the strong scale, and its octet contribution vanishes.

\paragraph{Order $\delta^2$}

After evaluating the flavour traces, the \SM two-meson photon interactions \eqref{eq:photon sm two meson} are
\begin{align} \label{eq:kinmix0}
\mathcal L_{\phi^2}^{\partial \photon} &= - \i e \photon^\mu \left(\pi^+ \overleftright \partial_\mu \pi^- + K^+ \overleftright \partial_\mu K^-\right) \ .
\end{align}
The corresponding kinetic-like interactions \eqref{eq:kinetic Vl Vr one meson,eq:kinetic Vl Vr two meson} that couple \cPT to the portal currents $\nonet V_l^\mu$ and $\nonet V_r^\mu$ become
\begin{subequations}
\begin{align}
\mathcal L_\phi^{\partial V_l} &= - \fp V_l^\mu \partial_\mu \etasinglet - \fp (\octet V_l^\mu{}_\strange^\down \partial_\mu K^0 + \text{h.c.}) \ , &
\mathcal L_\phi^{\partial V_r} &= \fp V_r^\mu \partial_\mu \etasinglet \ ,
\end{align}
\end{subequations}
and
\begin{subequations}
\begin{align}
\mathcal L_{\phi^2}^{\partial V_l} &= - \frac{\i}2 \left(\octet V_l^\mu{}_\strange^\down \left(\pi^- \overleftright \partial_\mu K^+ + K^0 \overleftright \partial_\mu \left(\pion - 3\etaoctet\right)\right) - \text{h.c.}\right) \ , \\
\mathcal L_{\phi^2}^{\photon V_l} &= - e \photon_\mu \left(\octet V_l^\mu{}_\strange^\down K^+ \pi^- + \text{h.c.}\right) \label{eq:rarek1} \ .
\end{align}
\end{subequations}
The mass-like interactions \eqref{eq:mass Sm one meson two meson} that couple \cPT to the portal current $\nonet S_m^\prime$ become
\begin{align}
\mathcal L_\phi^{\prime S_m} &= - \fp \B \Im S_m^\prime \etasinglet - \fp \B \left((\Im \octet S_m^\prime)_\strange^\down K^0 + \text{h.c.}\right) \ ,
\end{align}
and
\begin{multline}
\mathcal L_{\phi^2}^{\prime S_m} = - \frac{\B}{\nfl} \Re S_m^\prime \left(\frac12 \left(\pi_8^2 + \eta_8^2 + \eta_1^2\right) + \pi^+\pi^- + K^+K^- + K^0\overline K^0\right) \\
- \frac{\B}2 \left((\Re \octet S_m^\prime)_\strange^\down \left(K^+\pi^- + K^0 \left(2\etasinglet - \pion - \etaoctet\right)\right) + \text{h.c.}\right) \ .
\end{multline}
Finally, the anomalous interaction \eqref{eq:anomaly Stheta one meson} that couples \cPT to the portal current $S_\theta$ becomes
\begin{align}
\mathcal L_\phi^{S_\theta} &= \fp \mth^2 S_\theta \etasinglet \ .
\end{align}

\paragraph{Order $\delta^3$}

After evaluating the flavour traces, the singlet interactions \eqref{eq:singlet LO one meson,eq:singlet LO two meson} that couple \cPT to $S_\omega$ become
\begin{align}
\mathcal L_\phi^{\prime S_\omega} &
= \frac{S_\omega}{\betacoeff} 4 \mathcal L_\phi^\theta \ , &
\mathcal L_{\phi^2}^{\prime S_\omega} &
= \frac{S_\omega}{\betacoeff} \left(
 2 \left(
 \mathcal L_{\phi^2}^{\partial^2}
+ \mathcal L_{\phi^2}^{\partial \photon}
\right)
+ 3 \mathcal L_{\phi^2}^{\prime m}
+ 4 \mathcal L_{\phi^2}^\theta
\right) \ ,
\end{align}
where the \SM Lagrangians
\begin{align}
\mathcal L_\phi^\theta &
= \fp \mth^2 \theta \etasinglet \ , &
\mathcal L_{\phi^2}^\theta
= - \frac{\mth^2}2 \nfl \left(\etasinglet\right)^2 \ , &
\end{align}
and
\begin{subequations}
\begin{align}
\mathcal L_{\phi^2}^{\partial^2} &
= \begin{multlined}[t]
\frac12\left(\partial_\mu \pi_8 \partial^\mu \pi_8 + \partial_\mu \eta_8 \partial^\mu \eta_8 +
\partial_\mu \eta_1 \partial^\mu \eta_1\right) \\
+ \partial_\mu \pi^+ \partial^\mu \pi^- + \partial_\mu K^+ \partial^\mu K^- + \partial_\mu K^0 \partial^\mu \overline K^0 \ ,
\end{multlined} \\ \label{eq:Phi mass}
\mathcal L_{\phi^2}^{\prime m} &
\begin{multlined}[t] =
- \frac{\B}2 \left(
(m^\prime_\up + m^\prime_\down) \pi^+ \pi^-
+ (m^\prime_\up + m^\prime_\strange) K^+ K^-
+ (m^\prime_\down + m^\prime_\strange) K^0 \overline K^0
\right.\\\left.
+ m^\prime_\up \left(\etasinglet + \etaoctet + \pion\right)^2
+ m^\prime_\down \left(\etasinglet + \etaoctet - \pion\right)^2
+ m^\prime_\strange \left(\etasinglet - 2\etaoctet\right)^2
\right) \ ,
\end{multlined}
\end{align}
\end{subequations}
are identical to the \SM Lagrangians in \eqref{eq:singlet LO one meson,eq:sm Lagrangians two meson}.
The \WZW interactions \eqref{eq:wzw Vl one meson,eq:wzw Vl two meson} that couple \cPT to the portal current $\nonet V_l^\mu$ become
\begin{align}
\mathcal L_\Phi^{N V_l \photon} &
= \frac{2 \Nc e \widetilde F_{\mu\nu}}{3 (4\pi)^2 \fp} \left(
\partial^\mu V_l^\nu \left(\pion + \etaoctet\right) - \left(\partial^\mu \octet V_l^\nu{}_\strange^\down K^0 + \text{h.c.}\right)
\right) \ ,
\end{align}
and
\begin{align}
\mathcal L_{\Phi^2}^{N V_l \photon} &
= \frac{\i \Nc e \widetilde F_{\mu\nu}}{3 (4\pi)^2 \nfl \fp^2}
\left(V_l^\mu \left(\pi^+ \overleftright \partial^\nu \pi^- + K^+ \overleftright \partial^\nu K^-\right)
- \nfl \octet V_l^\mu{}_\strange^\down \left(K^0 \overleftright \partial^\nu \left(\pion - 3 \etaoctet\right)\right)
\right) \ .
\end{align}
Finally, the \WZW interactions \eqref{eq:wzw Vr one meson,eq:wzw Vr two meson} that couple \cPT to the portal current $\nonet V_r^\mu$ become
\begin{align}
\mathcal L_\Phi^{N V_r \photon} &
= \frac{2 \Nc e \widetilde F_{\mu\nu}}{3 (4\pi)^2 \fp} \partial^\mu V_r^\nu \left(\pion + \etaoctet\right) \ ,
\end{align}
and
\begin{align}
\mathcal L_{\Phi^2}^{N V_r \photon} &
= \frac{-\i \Nc e \widetilde F_{\mu\nu}}{3 (4\pi)^2 \nfl \fp^2} V_r^\mu
\left(\pi^+ \overleftright \partial^\nu \pi^- + K^+ \overleftright \partial^\nu K^-\right) \ .
\end{align}

\paragraph{Order $\epsilonEW \delta$}

After evaluating the flavour-traces, the \SM octet interactions \eqref{eq:octet sm} are
\begin{subequations}
\begin{align}
\mathcal L_{\phi^2}^{\fermi \partial^2} &
= - \frac{\epsilonEW} 2
\left(
\fermi_8 \partial^\mu K^+ \partial_\mu \pi^-
+ \partial^\mu K^0 \left(\fermi_8 \partial_\mu \left(\pion + \etaoctet\right) + \nfl \fermi_1 \partial_\mu \etasinglet\right) \right) + \text{h.c.} \ , \\
\mathcal L_{\phi^2}^{\fermi \partial \photon} &
= - \i \frac{e\epsilonEW}2 \fermi_8 \photon^\mu \left(\pi^- \overleftright \partial_\mu K^+\right) + \text{h.c.} \ .
\end{align}
\end{subequations}
The \SM 27-plet interactions \eqref{eq:27-plet sm} are
\begin{subequations}
\begin{align}
\mathcal L_{\phi^2}^{\tensor \fermi \partial^2} &
= - \frac{\epsilonEW}2 \fermi_{27} \left(\nfl \partial^\mu \left(\pion + \etaoctet\right) \partial_\mu K^0 + (\nfl - 1) \partial_\mu \pi^- \partial^\mu K^+\right) + \text{h.c.} \ , \\
\mathcal L_{\phi^2}^{\tensor \fermi \partial \photon} &
= - \i \frac{e\epsilonEW}2 \fermi_{27} (\nfl - 1) \photon_\mu \left(\pi^- \overleftright \partial^\mu K^+\right) + \text{h.c.} \ .
\end{align}
\end{subequations}
Since the octet contribution to the left-handed portal current $\nonet V_l^\mu$ is generated by diagrams that involve virtual $W$-boson exchanges, it counts as $\octet V_l^\mu \propto \epsilonEW$.
At order $\epsilonEW$, the latter and the octet contribution $\octet V_r^\mu$ to the right-handed portal current $\nonet V_r^\mu$ can be both neglected in the 27-plet interactions \eqref{eq:27-plet Vl one meson,eq:27-plet Vl two meson,eq:27-plet Vr one meson,eq:27-plet Vr two meson}, which then vanish.
The octet interactions \eqref{eq:octet Vl one meson,eq:octet Vl two meson} that couple \cPT to the singlet portal current $V_l^\mu$ become
\begin{align}
\mathcal L_\phi^{\fermi \partial V_l} &
= \frac{\epsilonEW \fp}2 \fermi_1 V_l^\mu \partial_\mu K^0 + \text{h.c.} \ ,
\end{align}
and
\begin{subequations}
\begin{align}
\mathcal L_{\phi^2}^{\fermi \partial V_l} &
= \i \frac{\epsilonEW}4 \fermi_1 V_l^\mu \left(\pi^- \overleftright \partial_\mu K^+ + \left(3 \etaoctet - \pion\right) \overleftright \partial_\mu K^0\right) + \text{h.c.} \ , \\ \label{eq:rarek2}
\mathcal L_{\phi^2}^{\fermi \photon V_l} &
= \frac{e \epsilonEW}2 \fermi_1 V_l^\mu \photon_\mu K^+ \pi^- + \text{h.c.} \ .
\end{align}
\end{subequations}
The octet interactions \eqref{eq:octet Vr one meson,eq:octet Vr two meson} that couple \cPT to the portal current $V_r^\mu$ become
\begin{align}
\mathcal L_\phi^{\fermi \partial V_r} &
= - \frac{\epsilonEW \fp}2 \fermi_1 V_r^\mu \partial_\mu K^0 + \text{h.c.} \ ,
\end{align}
and
\begin{subequations}
\begin{align}
\mathcal L_{\phi^2}^{\fermi \partial V_r} &
= - \i \frac{\epsilonEW}4 \fermi_1 V_r^\mu \left(\pi^- \overleftright \partial_\mu K^+ + \left(3 \etaoctet - \pion\right) \overleftright \partial_\mu K^0\right) + \text{h.c.} \ , \\ \label{eq:rarek3}
\mathcal L_{\phi^2}^{\fermi \photon V_r} &
= - \frac{e \epsilonEW}2 \fermi_1 V_r^\mu \photon_\mu K^+ \pi^- + \text{h.c.} \ .
\end{align}
\end{subequations}
The octet interactions \eqref{eq:octet Sy two meson} that couple \cPT to the portal currents $S_y$ become
\begin{subequations}
\begin{align}
\mathcal L_{\phi^2}^{\partial^2 S} &
= - \frac{\epsilonEW} 2
\begin{multlined}[t]
\left(\vphantom{\pion}
S_8 \partial^\mu K^+ \partial_\mu \pi^-
\right.\\\left.
+ \partial^\mu K^0 \left(- S_8 \partial_\mu \left(\pion + \etaoctet\right) + \nfl S_1 \partial_\mu \etasinglet\right) \right) + \text{h.c.} \ ,
\end{multlined} \\ \label{eq:rarek4}
\mathcal L_{\phi^2}^{\photon S} &
= - \i \frac{e\epsilonEW}2 S_8 \photon_\mu \left(\pi^- \overleftright \partial_\mu K^+\right) + \text{h.c.} \ .
\end{align}
\end{subequations}
Finally, the 27-plet interactions \eqref{eq:27-plet Sy two meson} that couple \cPT to the portal currents $S_y$ become
\begin{subequations}
\begin{align}
\mathcal L_{\phi^2}^{\partial^2 \tensor S} &
= - \frac{\epsilonEW}2 \nfl S_{27} \left(\partial_\mu K^0 \partial^\mu \left(\pion + \etaoctet\right) + \frac{\nfl - 1}{\nfl} \partial_\mu \pi^- \partial^\mu K^+\right) + \text{h.c.} \ , \\
\mathcal L_{\phi^2}^{\photon \tensor S} &
= - \i \frac{e \epsilonEW}2 (\nfl-1) S_{27} \photon_\mu \pi^- \overleftright \partial^\mu K^+ + \text{h.c.} \label{eq:rarek5} \ .
\end{align}
\end{subequations}

\paragraph{Order $\epsilonEW \delta^2$}

After evaluating the flavour traces, the \SM one- and two-meson charged-current interactions \eqref{eq:weak sm two meson,eq:weak sm one meson} are
\begin{subequations}
\begin{align}
\mathcal L_\phi^{\partial W} &= - \fp \left(\octet l_W^\mu{}_\down^\up \partial_\mu \pi^+ + \octet l_W^\mu{}_\strange^\up \partial_\mu K^+ + \text{h.c.}\right) \ , \\
\mathcal L_\phi^{\photon W} &= \i \fp e \photon_\mu \left(\octet l_W^\mu{}_\down^\up \pi^+ + \octet l_W^\mu{}_\strange^\up K^+ - \text{h.c.}\right) \ ,
\end{align}
\end{subequations}
and
\begin{subequations}
\begin{align}
\mathcal L_{\phi^2}^{\partial W} &= - \frac{\i}2
\begin{multlined}[t]
\left( \octet l_W^\mu{}_\down^\up \left(2 \pi^+ \overleftright \partial_\mu \pion + K^+ \overleftright \partial_\mu \overline K^0\right)
\right.\\\left.
+ \octet l_W^\mu{}_\strange^\up \left(\pi^+ \overleftright \partial_\mu K^0 - K^+ \overleftright \partial_\mu \left(\pion + 3\etaoctet\right)\right) - \text{h.c.} \right) \ ,
\end{multlined} \\ \label{eq:rarek0}
\mathcal L_{\phi^2}^{\photon W} &
= \frac12 e \photon_\mu
\begin{multlined}[t]
\left( \octet l_W^\mu{}_\down^\up \left(2 \pi^+ \pion + K^+ \overline K^0\right)
\right.\\\left.
+ \octet l_W^\mu{}_\strange^\up \left(K^+ \left(\pion + 3\etaoctet\right) + \pi^+ K^0\right) + \text{h.c.} \right) \ .
\end{multlined}
\end{align}
\end{subequations}
The singlet interactions \eqref{eq:singlet weak NLO one meson,eq:singlet weak NLO two meson} that couple \cPT to the portal current $S_\omega$ become
\begin{align}
\mathcal L_\phi^{\prime S_\omega} &
= \frac{S_\omega}{\betacoeff} 2 \mathcal L_\phi^{\prime \fermi m} \ , &
\mathcal L_{\phi^2}^{\prime S_\omega} &
= \frac{S_\omega}{\betacoeff} 2 \left(
\mathcal L_{\phi^2}^{\prime \fermi m}
+ 2 \left(
+ \mathcal L_{\phi^2}^{\fermi \partial^2}
+ \mathcal L_{\phi^2}^{\fermi \partial \photon}
+ \mathcal L_{\phi^2}^{\tensor \fermi \partial^2}
+ \mathcal L_{\phi^2}^{\tensor \fermi \partial \photon}
\right)
\right) \ ,
\end{align}
where the \SM Lagrangians
\begin{subequations}
\begin{align}
\mathcal L_\phi^{\prime \fermi m} &
= - \i \frac{\epsilonEW \fp \B}2 \fermi_\octetm (m_\strange^\prime - m_\down^\prime) K^0 + \text{h.c.} \ , \\
\mathcal L_{\phi^2}^{\prime \fermi m} &
= \frac{\epsilonEW \B}4 \fermi_\octetm \left(m^\prime_\down + m^\prime_\strange\right) \left(K^+\pi^- - K^0 \left(\pion + \etaoctet\right)\right) + \text{h.c.}
\end{align}
\end{subequations}
are identical to the \SM Lagrangians in \eqref{eq:singlet weak NLO one meson,eq:singlet weak NLO two meson}.

\paragraph{Order $\epsilonEW \delta^3$}

After evaluating the flavour traces, the singlet interactions \eqref{eq:singlet weak NNLO one meson,eq:singlet weak NNLO two meson} that couple \cPT to the portal current $S_\omega$ become
\begin{align}
\mathcal L_\phi^{\prime S_\omega} &
= \frac{S_\omega}{\betacoeff} 2\left(
\mathcal L_\phi^{\partial W}
+ \mathcal L_\phi^{\photon W}
+ \mathcal L_\phi^\chromo
\right) \ , &
\mathcal L_{\phi^2}^{\prime S_\omega} &
= \frac{S_\omega}{\betacoeff} 2 \left(
 \mathcal L_{\phi^2}^{\partial W}
+ \mathcal L_{\phi^2}^{\photon W}
+ \mathcal L_{\phi^2}^\chromo
\right) \ ,
\end{align}
where the \SM Lagrangians
\begin{align}
\mathcal L_\phi^\chromo &
= - \epsilonEW \fp\B \coupl_\Chromo (\Im \octet \chromo)_\strange^\down K^0 + \text{h.c.} \ , \\
\mathcal L_{\phi^2}^\chromo &
= - \frac{\epsilonEW \B}2 \coupl_\Chromo (\Re \octet \chromo)_\strange^\down \left(K^+\pi^- - K^0 \left(\pion + \etaoctet\right)\right) + \text{h.c.} \,,
\end{align}
are identical to the \SM Lagrangians in \eqref{eq:singlet weak NNLO one meson,eq:singlet weak NNLO two meson}.
The tensor interactions \eqref{eq:dipole two meson} that couple \cPT to the portal current $\octet T_\tau^{\mu\nu}$ become
\begin{subequations}
\begin{align}
\mathcal L_{\phi^2}^{T \partial^2} &= 2\frac{\epsilonEW}{\fp^3} \coupl_T^{D^2} (\Re \octet T_\tau^{\mu\nu})_\strange^\down \left(\partial_\mu K^+ \partial_\nu \pi^- - \partial_\mu K^0 \partial_\nu \left(\pion + \etaoctet\right)\right) + \text{h.c.} \ , \\
\mathcal L_{\phi^2}^{T V} &= -\frac{e\epsilonEW}{\fp^3} \coupl_T^{LR} F_{\mu\nu} (\Re \octet T_\tau^{\mu\nu})_\strange^\down 2 K^+ \pi^- + \text{h.c.} \ , \\
\mathcal L_{\phi^2}^{T \partial V} &= \frac{e\epsilonEW}{\fp^3} \coupl_T^{D^2} \photon_\nu (\Im \octet T_\tau^{[\mu\nu]})_\strange^\down \partial_\mu \left(2 K^+ \pi^-\right) + \text{h.c.} \label{eq:rarek6}\ .
\end{align}
\end{subequations}
The \WZW interactions \eqref{eq:wzw weak Vl one meson,eq:wzw weak Vl two meson} that couple \cPT to $\nonet V_l^\mu$ become
\begin{align}
\mathcal L_\Phi^{N V_l W} &
= \frac{2 \Nc \epsilon_{\mu\nu\rho\sigma}}{3(4\pi)^2 \nfl \fp} \partial^\mu V_l^\nu \left(
\octet l_W^{\rho\sigma}{}_\strange^\up K^+ + \octet l_W^{\rho\sigma}{}_\down^\up \pi^+ + \text{h.c.}
\right) \ ,
\end{align}
and
\begin{multline}
\mathcal L_{\Phi^2}^{N V_l W}
= \frac{\i \Nc \epsilon_{\mu\nu\rho\sigma}}{3(4\pi\fp)^2} \frac1{\nfl} V_l^\mu \left(
 \octet l_W^{\rho\sigma}{}_\strange^\up \left(\pi^+\overleftright \partial^\nu K^0 - K^+ \overleftright \partial^\nu \left(\pion + 3 \etaoctet\right)\right)
\right.\\\left.
+ \octet l_W^{\rho\sigma}{}_\down^\up \left(2 \pi^+\overleftright \partial^\nu \pion + K^+ \overleftright \partial^\nu \overline K^0\right)
- 3 \octet l_W^\nu{}_\down^\up \left(2 \partial^\rho \pion \partial^\sigma \pi^+ + \partial^\rho K^+ \partial^\sigma \overline K^0\right)
\right.\\\left.
- 3 \octet l_W^\nu{}_\strange^\up \left(\partial^\rho \left(\pion + 3 \etaoctet\right) \partial^\sigma K^+ + \partial^\rho \pi^+ \partial^\sigma K^0\right)
+ \text{h.c.} \right) \ .
\end{multline}
Finally, the \WZW interactions \eqref{eq:wzw weak Vr one meson,eq:wzw weak Vr two meson} that couple \cPT to $\nonet V_r^\mu$ become
\begin{align}
\mathcal L_\Phi^{N V_r W} &
= \frac{\Nc\epsilon_{\mu\nu\rho\sigma}}{3(4\pi)^2 \nfl \fp} \partial^\mu V_r^\nu \left(
\octet l_W^{\rho\sigma}{}_\strange^\up K^+ + \octet l_W^{\rho\sigma}{}_\down^\up \pi^+ + \text{h.c.}
\right) \ ,
\end{align}
and
\begin{multline}
\mathcal L_{\Phi^2}^{N V_r W}
= \frac{\i\Nc\epsilon_{\mu\nu\rho\sigma}}{3(4\pi\fp)^2} \frac1{\nfl} V_r^\mu \left(
 \frac12 \octet l_W^{\rho\sigma}{}_\strange^\up \left(\pi^+\overleftright \partial^\nu K^0 - K^+ \overleftright \partial^\nu \left(\pion + 3 \etaoctet\right)\right)
\right.\\\left.
+ \frac12 \octet l_W^{\rho\sigma}{}_\down^\up \left(2 \pi^+\overleftright \partial^\nu \pion + K^+ \overleftright \partial^\nu \overline K^0\right)
- 3 \octet l_W^\nu{}_\down^\up \left(2 \partial^\rho \pion \partial^\sigma \pi^+ + \partial^\rho K^+ \partial^\sigma \overline K^0\right)
\right.\\\left.
- 3 \octet l_W^\nu{}_\strange^\up \left(\partial^\rho \left(\pion + 3 \etaoctet\right) \partial^\sigma K^+ + \partial^\rho \pi^+ \partial^\sigma K^0\right) + \text{h.c.}
\right) \ .
\end{multline}

\subbib

%% file: models.tex
\subtoc

\section{Meson interactions of hidden sector models} \label{sec:models}

In this section, we apply the results of \cref{sec:chiral perturbation theory,sec:light mesons} to compute generic transition amplitudes for golden channels used to search for \NP in meson experiments.
This step serves first to validate our results with preexisting computations and second to exemplify their use to compute meson decays involving a hidden particle.
We consider one example for each messenger type that is captured by the \PET framework:
\begin{description}

\item[Spin 0 messengers]
The decay $K^\pm \to \pi^\pm s_i$ is a smoking gun process for \ALP searches at kaon factories, see \eg \cite{CortinaGil:2018fkc, CortinaGil:2020fcx}.
It can be especially relevant within the context of interpreting the recent \KOTO excess \cite{Kitahara:2019lws}.
Scalar, pseudoscalar and complex scalar messengers couple to the \cPT Lagrangian via a large variety of external currents.
As a result, this type of process clearly demonstrates the power of the \PET framework to perform global parameter scans instead of considering only one specific \SM extension at a time.

\item[Spin \textfrac12 messengers]
The decay $K^\pm \to \ell^\pm \lhf_a$ is a key signature for light \HNL searches \cite{CortinaGil:2017mqf, NA62:2020mcv}.
If $\lhf_a$ is a \HNL, the computation of the transition amplitude is straightforward, as the \HNL couples to the \SM only via its mixing with neutrinos \cite{Minkowski:1977sc, GellMann:1980vs, Mohapatra:1979ia, Yanagida:1980xy, Schechter:1980gr, Schechter:1981cv}.
After diagonalising this mixing, the \HNL couples to \QCD via a single operator that mirrors the leptonic charged current interaction in the \SM.
Up to leading order in $\alpha_{\EM}$ and the $4\pi$ counting of \NDA, this operator is also the only one that couples \QCD directly to a completely generic spin \textfrac12 messenger.
Since we do not diagonalise the portal interactions, we keep track of both the mixing and the charged current operator.
As discussed in \cref{sec:hidden sector mixing}, this means that the final decay amplitude also captures hidden sectors that contain a non-trivial secluded sector in addition to the messenger field.
The net-effect is that the mixing angles $\theta_{bi}$ in \cref{eq:neutrino HNL mixing}, which measure the size of the \HNL amplitude, are replaced with effective mixing angles $\theta_{ba}$ in \cref{eq:hnl2}
that measure both the impact of the mixing of $\lhf_a$ with neutrinos and the direct production via the four-fermion operator.

\item[Spin 1 messengers]
The decay $\pi^0 \to \gamma v_\mu$ is a smoking gun process for dark photon searches, see \eg \cite{CortinaGil:2019nuo}.
If $v_\mu$ couples to \cPT like a vector particle in a parity conserving theory, such as in common models of dark photons, the parity-odd \WZW action generates the only contribution to the decay amplitude.
\Apriori, one might expect that the parity-even order $\delta^3$ contributions to the \cPT action in \cref{eq:U NLO kinetic Lagrangians,eq:U NLO mass Lagrangians,eq:U field strength Lagrangians,eq:U NLO anomaly Lagrangian,eq:U anomaly mass Lagrangian,eq:U anomaly kinetic Lagrangian} can mediate neutral pion decays $\pi^0 \to \gamma a_\mu$ into messengers $a_\mu$ that couple to \cPT like axial-vectors in a parity conserving theory.
However, as mentioned below \cref{eq:wzw Vr two meson}, this does not occur.
For this reason, the dark photon decay amplitude actually encompasses the production of generic spin 1 messengers.
\end{description}
To summarise, our decay amplitudes for hidden \prefix{pseudo}{scalar} messengers, \HNLs and dark photons capture the production of generic hidden spin 0, \textfrac12 and 1 messengers to \LO in $\alpha_{\EM}$, $\epsilonEW$, and the \NDA $4\pi$ counting.

\subsection{Charged kaon decay to charged pions and hidden scalars}

We compute the transition amplitude for charged kaon decays $K^\pm \to \pi^\pm s_i$ into spin 0 messengers $s_i$.
These decays can be induced via \emph{seven out of the ten} portal currents that contained in the portal \cPT Lagrangian.
To compute the complete generic decay amplitude, we first consider decays mediated by each of these currents individually, and compute the leading contributions to the corresponding partial decay amplitudes.
We then sum these contributions to obtain a universal expression.

In general, the $\delta$ and $\epsilonEW$ scaling behaviour of each partial amplitude can be different for each of the seven portal currents, and the final result for the decay amplitude will mix contributions of different order in $\delta$ and $\epsilonEW$.
For instance, a quark-flavour violating contribution to the current $\Re \nonet S_m \propto \epsilonUV s_i$ induces an amplitude in \eqref{eq:ktopi re Sm S8+S27 amplitudes} that formally scales as $\epsilonUV \delta^2$, with no suppression due to $\epsilonEW$,
while the currents $\tensor S_x \propto \epsilonUV \flatfrac{s_i}{v}$ induce an amplitude in \eqref{eq:ktopi re Sm S8+S27 amplitudes} that scales as $\epsilonUV \epsilonEW^{\nicefrac32}\delta$,
and the current $S_\omega \propto \epsilonUV \flatfrac{s_i}{v}$ induces an amplitude \eqref{eq:ktopi re Sw amplitude} that scales as $\epsilonUV \epsilonEW^{\nicefrac32}\delta^2$.
In the case of the $S_\omega$ and $\tensor S_x$ currents, the additional $\epsilonEW^{\nicefrac12}$ suppression results from the fact that the underlying \EW scale portal operators are of dimension five rather than dimension four.
When considering a specific \SM extension, it may be possible to neglect the higher order contributions if they appear in conjunction with lower order contributions.
However, to capture the coupling of \cPT to fully generic hidden sectors, it is necessary to keep track of all contributions, since \apriori a hidden sector can couple to \cPT via any one of the portal currents.

\subsubsection{Single scalar portal current contributions}

In \cref{sec:portal currents}, we have given the complete list of portal interactions that contribute to each external current at \LO.
The relevant contributions that mediate $K^\pm \to \pi^\pm s_i$ decays are those with exactly one hidden spin 0 messenger and no other \SM or hidden fields,
\begin{subequations} \label{eq:K decay hidden currents}
\begin{align}
S_\omega &= \frac{\epsilonUV}{v} c_i^{S_\omega} s_i \ , &
\nonet S_\m &\supset \epsilonUV \left(\nonet c^{S_m}_i + \nonet c^{S_m}_{\partial^2i} \frac1{v^2} \partial^2\right) s_i \ , &
\tensor S_\scalar &= \tensor \fermi_{\scalar i} \frac{\epsilonUV}{v} s_i \ , \\
S_\theta &= \frac{\epsilonUV}{v} c^{S_\theta}_i s_i \ , &
\octet S_\chromo &= \epsilonUV \left(\flavour \proj^\strange_\down c^\chromo_{i\overline\strange\down} + \flavour \proj^\down_\strange c^\chromo_{i\overline\down\strange}\right) s_i \ , &
\tensor S_r &= \tensor \fermi_{ri} \frac{\epsilonUV}{v} s_i \ , \\ & & & &
\tensor S_l &= \tensor \fermi_{li} \frac{\epsilonUV}{v} s_i \ .
\end{align}
\end{subequations}
Since $\flatfrac{\partial^2}{v^2} \propto \epsilonEW\delta$, the second term in $\nonet S_m$ induces amplitudes that are suppressed by an additional factor $\epsilonEW\delta$ compared to the contributions generated by the first term.
In the following, we simplify the expressions by approximating $m_\up, m_\down \to m_\light$ and $\epsilon_\strange \equiv \flatfrac{m_\light}{m_\strange} \to 0$.
Matching to \cPT and transitioning to the physical vacuum, this gives the modified currents
\begin{align} \label{eq:K decay modified currents}
\nonet S_m^\prime &= \epsilonUV \left(
\nonet c^{\prime S_m}_i
+ \nonet c^{S_m}_{\partial^2i} \frac1{v^2} \partial^2\right)
s_i + \order{\epsilon_{\EW}^2, \epsilon_\strange^2} \ , &
S_y &= \fermi_{yi} \frac{\epsilonUV}{v} s_i \ ,
\end{align}
where the parameters $\fermi_{yi}$ are given in equation \eqref{eq:p_yi_matching_result}, and
\begin{multline} \label{eq:K decay modified current wilson coeff}
\nonet c^{\prime S_m}_i =
\nonet c^{S_m}_i
+ 2 \epsilonEW
\left[\epsilon_\strange \left(\fermi_\octetm^{\prime\dagger} \flavour\lambda_\down^\strange - \vacuum \fermi_\octetm \flavour\lambda_\strange^\down\right) \nonet c^{S_m}_i
- \fermi_\octetm^{\prime\dagger} \nonet c^{S_m}_i \flavour\lambda_\down^\strange\right] \\
- \frac{\epsilonEW\nonet\m}v\left(\fermi_{\octetm i} \flavour\lambda_\strange^\down+\text{h.c}\right)
+ \epsilonEW\coupl_\Chromo \left(c^\chromo_{i\overline\strange\down}\flavour\proj^\strange_\down+c^\chromo_{i\overline\down\strange}\flavour\proj^\down_\strange\right)
+ \order{\epsilon_{\EW}^{\nicefrac32}, \epsilon_\strange^2} \ .
\end{multline}
The strength of strangeness-violating contributions to $\nonet S_\m^\prime$ is measured by the Wilson coefficients%
\begin{subequations}
\begin{align}
\nonet c^{\prime S_m}_i{}_\strange^\down &= \nonet c^{S_m}_i{}_\strange^\down
- \epsilonEW \left(2\vacuum \fermi_\octetm \epsilon_\strange \nonet c^{S_m}_i{}_\down^\down + \frac{m_\strange}v \fermi_{\octetm i}
- \coupl_\Chromo c^\chromo_{i\overline\down\strange}\right) \ , \\
\nonet c^{\prime S_m}_i{}_\down^\strange &= \nonet c^{S_m}_i{}_\down^\strange
+ \epsilonEW \left(2\fermi_\octetm^{\prime\dagger} \left(\epsilon_\strange \nonet c^{S_m}_i{}_\strange^\strange - \nonet c^{S_m}_i{}_\down^\down\right) - \epsilon_\strange\frac{m_\strange}v \fermi_{\octetm i}^\dagger
+ \coupl_\Chromo c^\chromo_{i\overline\strange\down}\right) \ .
\end{align}
\end{subequations}

\subsubsection{Relevant interactions}

At tree-level, $K^\pm \to \pi^\pm s_i$ decays are mediated by portal interactions with either one or two mesons.
The former give rise to indirect production via mixing of the messenger with the \SM mesons, while the latter give rise to direct production.
Both types of interaction are listed in \cref{sec:Phi expanded}.

\begin{figure}
\begin{panels}{2}
\tikzsetnextfilename{feynman-hidden-scalar-2}\input{feynman-hidden-scalar-2.pgf}
\caption{Production via mass mixing.} \label{fig:feynman hidden scalar mixing}
\panel
\tikzsetnextfilename{feynman-hidden-scalar-1}\input{feynman-hidden-scalar-1.pgf}
\caption{Direct production.} \label{fig:feynman hidden scalar direct}
\end{panels}
\caption{
Feynman diagrams for the $K^\pm \to \pi^\pm s_i$ process.
} \label{fig:feynman hidden scalar}
\end{figure}

We first consider the case of indirect production via the process depicted in the diagram in \cref{fig:feynman hidden scalar mixing}.
The one-meson interactions mix the hidden scalar with the neutral \SM mesons, and contribute to $K^\pm \to \pi^\pm s_i$ decays via off-shell $K^\pm \to \pi^\pm \pi^{0\ast}$, $K^\pm \to \pi^\pm \eta^\ast$, and $K^\pm \to \pi^\pm \eta^{\prime\ast}$ transitions, in which the neutral meson subsequently oscillates into the hidden scalar.
Hence, the diagram in \cref{fig:feynman hidden scalar mixing} contains two vertices:
\begin{inlinelist}
\item\label{it:trilinear SM}%
a trilinear \SM vertex with one $K^\pm$-leg, one $\pi^\mp$-leg, and one neutral meson leg
\item\label{it:bilinear portal}%
a one-meson portal interaction that captures the meson to hidden scalar mixing
\end{inlinelist}
The expressions for the trilinear \SM interactions are known, and can be extracted from the \SM \cPT Lagrangian by following the procedure that we summarise in \cref{sec:trilinear vertices}.
The resulting Lagrangian is
\begin{align}
\mathcal L_{K\pi\Phi} &
= - \frac{\i \epsilonEW}{2\fp} \left(2 V_{K\pi\pi} \neutpion + 3 V_{K\pi\eta} \frac{\eta}{\sqrt 3} + 3 V_{K\pi\eta^\prime} \etaprime \right) K^+ \pi^-
\end{align}
where we have defined the functions
\begin{subequations} \label{eq:type_one_vertices}
\begin{align}
V_{K\pi\pi} &= \frac14 \left[
(\fermi_8 + 7\fermi_{27}) \partial_{\pi^0} \partial_K
- 5\fermi_{27} \partial_{\pi^0} \partial_{\pi^-}
- (\fermi_8 + 2\fermi_{27}) \partial_{\pi^-} \partial_K \right] \ , \\
V_{K\pi\eta} &= \frac1{6\sqrt2} c_\eta
\begin{multlined}[t]
\left[
(3\fermi_8 + 6\fermi_{27}) \partial_\pi \partial_K
\right.\\\left.
- (\fermi_8 + 3\sqrt2 t_\eta \fermi_1 - 3\fermi_{27}) \partial_\eta \partial_K
- (2\fermi_8 - 3\sqrt2 t_\eta \fermi_1 + 9\fermi_{27}) \partial_\eta \partial_\pi
\right] \ ,
\end{multlined} \\
V_{K\pi\eta^\prime} &= \frac1{6\sqrt2} s_\eta
\begin{multlined}[t]
\left[
(3\fermi_8 + 6\fermi_{27}) \partial_\pi \partial_K
\right.\\\left.
- (\fermi_8 - 3\sqrt2 t_\eta^{-1}\fermi_1 - 3\fermi_{27}) \partial_\eta \partial_K
- (2\fermi_8 + 3\sqrt2 t_\eta^{-1} \fermi_1 + 9\fermi_{27}) \partial_\eta \partial_\pi
\right] \ .
\end{multlined}
\end{align}
\end{subequations}
Notice that there is no $K^\pm \pi^\mp K^0$ \SM vertex.
Therefore, we do not have to keep track of the mixing between the neutral kaons and the messenger.
This also means that we can neglect \EW contributions to \cref{it:bilinear portal} interactions.
The hidden currents $\Im \nonet S_m^\prime$ and $S_\theta$ induce the only relevant \cref{it:bilinear portal} vertices, given within the interactions \eqref{eq:mass Sm one meson two meson,eq:anomaly Stheta one meson}.
Extracting the vertices, one obtains
\begin{subequations}
\begin{align} \label{eq:ktopi Stheta}
\mathcal L_\Phi^{S_\theta} &
= \frac{\epsilonUV \mth^2 \fp c^{S_\theta}_i}{v} \left(c_\eta \etaprime - s_\eta \frac{\eta}{\sqrt3}\right) s_i \ , \\ \label{eq: ktopi im Sm}
\mathcal L_\Phi^{\prime S_m} &
\supset - \epsilonUV \fp \B \left(c_{s_i\pi} \neutpion + c_{s_i\eta} \frac\eta{\sqrt3} + c_{s_i\eta^\prime} \etaprime\right) s_i + \order{\epsilonEW} \ ,
\end{align}
\end{subequations}
where the Wilson coefficients are
\begin{subequations}\label{eq:wilson_coeff}
\begin{align}
c_{s_i\pi} &= \Im\nonet c^{S_m}_i{}_\up^\up - \Im\nonet c^{S_m}_i{}_\down^\down \ , \\
c_{s_i\eta} &= - s_\eta \Im c^{S_m}_i + \frac{c_\eta}{\sqrt2} \left(\Im\nonet c^{S_m}_i{}_\up^\up + \Im\nonet c^{S_m}_i{}_\down^\down - 2 \Im\nonet c^{S_m}_i{}_\strange^\strange\right) \ , \\
c_{s_i\eta^\prime} &= c_\eta \Im c^{S_m}_i + \frac{s_\eta}{\sqrt2} \left(\Im\nonet c^{S_m}_i{}_\up^\up + \Im\nonet c^{S_m}_i{}_\down^\down - 2 \Im\nonet c^{S_m}_i{}_\strange^\strange\right) \ .
\end{align}
\end{subequations}

We now move on to the case of direct production via the process depicted in the diagram in \cref{fig:feynman hidden scalar direct}.
This diagram consists of a single trilinear portal vertex with one $K^\pm$-leg, one $\pi^\mp$-leg, and one hidden spin 0 messenger leg.
The hidden current $\Re \nonet S_m^\prime$ induces the vertices
\begin{equation} \label{eq:s_current_LO}
\mathcal L_{\Phi^2}^{\prime S_m}
\supset - \frac{\B}{2} \epsilonUV K^+\pi^-\left(c_{K\pi s_i} + \Re\nonet c^{S_m}_{\partial^2i}{}_\strange^\down \frac{\partial^2}{v^2}\right) s_i + \order{\epsilon_{\EW}^2} \ ,
\end{equation}
which are part of the interactions \eqref{eq:mass Sm one meson two meson}, where
\begin{multline} \label{eq:ktopi re Sm}
c_{K\pi s_i} =
\Re \nonet c^{S_m}_i{}_\strange^\down + \frac{\epsilonEW}2 \left((m_K^2 - m_\pi^2) \Re\nonet c^{S_m}_i{}_\up^\up + m_K^2 \Re\nonet c^{S_m}_i{}_\down^\down - m_\pi^2 \Re\nonet c^{S_m}_i{}_\strange^\strange\right) \theta_{K^\pm \pi^\mp} \\
- \frac{\epsilonEW}2 \left(2\vacuum \fermi_\octetm \left(\epsilon_\strange \nonet c^{S_m}_i{}_\down^\down - \epsilon_\strange {\nonet c^{S_m}_i{}_\strange^\strange}^\dagger + {\nonet c^{S_m}_i{}_\down^\down}^\dagger\right)
+ \frac{m_\light + m_\strange}v \fermi_{\octetm i}
- \coupl_\Chromo \left(c^\chromo_{i\overline\down\strange} + c^{\chromo\dagger}_{i\overline\strange\down}\right)\right) \ .
\end{multline}
The hidden currents $S_y$ induce the vertices
\begin{equation}
\mathcal L_{\Phi^2}^{\partial^2 S} + \mathcal L_{\Phi^2}^{\partial^2 \tensor S}
\supset - \frac{\epsilonUV\epsilonEW}{2v} \left(\fermi_{8i} + (\nfl-1) \fermi_{27i}\right) s_i \partial_\mu \pi^- \partial^\mu K^+ \ ,
\end{equation}
which are encompassed by the interactions \eqref{eq:octet Sy two meson,eq:27-plet Sy two meson}.
Finally, the $S_\omega$ current induces the vertices
\begin{equation}
\mathcal L_{\Phi^2}^{\prime S_\omega}
\supset \frac{\epsilonUV\epsilonEW c_i^{S_\omega}}{v \betacoeff} \left(\fermi_\octetm^\prime m_K^2 K^+ \pi^- - \left(\fermi_8 + (\nfl-1)\fermi_{27}\right) \partial^\mu K^+ \partial_\mu \pi^-\right) s_i \ ,
\end{equation}
which are given within the interactions \eqref{eq:singlet weak NLO two meson}.
These vertices contribute at order $\delta^3$ rather than order $\delta^2$ due to the large $\Nc$ dependence of the $\beta$ function, which scales as $\betacoeff \sim \Nc$.
As mentioned below \cref{eq:singlet LO one meson}, $S_\omega$ induces also a one-meson vertex that mixes the $\eta_1$ singlet with the messenger.
However, this interaction is suppressed by the \QCD $\theta$ angle and is always negligible with respect to the above trilinear portal vertices.

\subsubsection{Partial decay width}

In summary, the hidden currents $\Im S_m^\prime$ and $S_\theta$ couple to \cPT via bilinear one-meson portal interactions,
while the hidden currents $\Re S_m^\prime$, $S_\omega$, and the $S_y$ couple to \cPT via trilinear two-meson portal interactions.
Putting everything together, the complete transition amplitude can be decomposed as
\begin{equation} \label{eq:generic K amplitude}
\mathcal A (K^+ \to \pi^+ s_i) = \mathcal A_\text{direct} + \mathcal A_\text{mixing} \ .
\end{equation}
The amplitude for direct production via the trilinear interactions is
\begin{equation}
\mathcal A_\text{direct} =  \mathcal A_m^{\Re} + \mathcal A_\fermi + \mathcal A_{\omega} \ ,
\end{equation}
where
\begin{subequations}
\label{eq:direct ktopi amplitudes}
\begin{align} \label{eq:ktopi re Sm S8+S27 amplitudes}
\mathcal A_m^{\Re} &
= -\frac{\epsilonUV\B}{2} \left(c_{K\pi s_i} - \Re\nonet c^{S_m}_{\partial^2i}{}_\strange^\down \frac{m_s^2}{v^2}\right)
\ , &%
\mathcal A_\fermi &
= - \frac{\epsilonUV\epsilonEW}{2v} X_i
\ , \\ \label{eq:ktopi re Sw amplitude}
\mathcal A_{\omega} &
= \frac{\epsilonUV\epsilonEW c_i^{S_\omega}}{\betacoeff v} \left(\fermi_\octetm^\prime m_K^2 - X_0
\right)
\ .
\end{align}
\end{subequations}
The quantities
\begin{align}
X_i &= \frac12 \left(\fermi_{8i} + (\nfl-1) \fermi_{27i}\right) \left(m_K^2 + m_\pi^2 - m_s^2\right)
\end{align}
measure the dependence on the octet and 27-plet coefficients $\fermi_{8i}$ and $\fermi_{27 i}$.
Following the discussion in \cref{sec:meson to hidden mixing}, the amplitude for indirect production $\mathcal A_\text{mixing}$ can be written in terms of the generic meson-to-messenger mixing angles
\begin{subequations}\label{eq:meson to hidden mixing angles}
\begin{align}
\theta_{\pi s_i} &= \epsilonUV \fp \frac{\B c_{s_i \pi}}{m_s^2 - m_\pi^2} \ , &
\theta_{\eta s_i} &= \epsilonUV \fp \frac{\B c_{s_i \eta} + c_i^{S_\theta} s_\eta \frac{\mth^2}{v}}{m_s^2 - m_\eta^2} \ , \\ &&
\theta_{\eta^\prime s_i} &= \epsilonUV \fp \frac{\B c_{s_i \eta^\prime} - c_i^{S_\theta} c_\eta \frac{\mth^2}{v}}{m_s^2 - m_{\eta^\prime}^2} \ .
\end{align}\end{subequations}
This results in
\begin{equation}
\mathcal A_\text{mixing} = \mathcal A_m^{\Im} + \mathcal A_{\theta}
= - \i \frac{\epsilonEW}{2\fp} \left(\theta_{\pi s_i} V_{K\pi\pi} + \theta_{\eta s_i} V_{K\pi\eta} +\theta_{\eta^\prime s_i} V_{K\pi\eta^\prime} \right) \ .
\end{equation}
In momentum space, and evaluated on-shell, the functions \eqref{eq:type_one_vertices} become
\begin{subequations}\label{eq:type_one_vertices_on_shell}
\begin{align}
V_{K\pi\pi} &= \frac1{8}
\left[5\fermi_{27} (2 m_K^2 - m_s^2 - m_\pi^2) + (2\fermi_8 + 9\fermi_{27}) (m_s^2 - m_\pi^2) \right] \ , \\
V_{K\pi\eta} &= \frac{c_\eta}{12\sqrt2}
\begin{multlined}[t]
\left[(2\fermi_8 - 3\sqrt2 t_\eta \fermi_1 + 9\fermi_{27}) (2m_K^2 - m_s^2 - m_\pi^2)
\right.\\\left.
- (4 \fermi_8 + 3\sqrt2 t_\eta \fermi_1 + 3\fermi_{27}) (m_s^2 - m_\pi^2) \right] \ ,
\end{multlined} \\
V_{K\pi\eta^\prime} &= \frac{s_\eta}{12\sqrt2}
\begin{multlined}[t]
\left[(2\fermi_8 + 3\sqrt2 t_\eta^{-1} \fermi_1 + 9\fermi_{27}) (2 m_K^2 - m_s^2 - m_\pi^2)
\right.\\\left.
- (4\fermi_8 - 3\sqrt2 t_\eta^{-1}\fermi_1 + 3\fermi_{27}) (m_s^2 - m_\pi^2) \right] \ .
\end{multlined}
\end{align}
\end{subequations}
All of the above amplitudes are determined entirely by $m_K^2$, $m_\pi^2$, and $m_s^2$, with no remaining angular dependence.
The resulting partial decay width is
\begin{equation}
\Gamma(K^+ \to \pi^+ s_i) = \frac1{8 \pi m_K}
\rho(x_\pi, x_s)
\abs{\mathcal A (K^+ \to \pi^+ s_i)}^2
\ ,
\end{equation}
where the phase-space factor is
\begin{align}
\rho(x_\pi, x_s) &= \sqrt{\left(\frac{1 - x_\pi - x_s}2\right)^2 - x_\pi x_s } \ , &
x_i &= \frac{m_i^2}{m_K^2} \ ,
\end{align}
and the squared amplitude is
\begin{align}
\abs{\mathcal A (K^+ \to \pi^+ s_i)}^2 &= \abs{\Re \mathcal A}^2 + \abs{\Im \mathcal A}^2 \ ,
\end{align}
where
\begin{subequations}\begin{align}
\abs{\Re \mathcal A}^2 &
= \frac{\epsilon_{\UV}^2\B^2}4
\left|
\Re\left( c_{K\pi s_i} - \nonet c^{S_m}_{\partial^2i}{}_\strange^\down \frac{m_s^2}{v^2}\right)
+ \frac{\epsilonEW}{\B v}\left(
 X_i
+ 2 \frac{c_i^{S_\omega}}{\betacoeff} \left(X_0
- \fermi_\octetm^\prime m_K^2
\right)
\right)
\right|^2 \ , &\\
\abs{\Im \mathcal A}^2 &
= \frac14 \left|
\epsilonUV \B \Im c_{K\pi s_i} + \frac{\epsilonEW}{\fp} \left(\theta_{\pi s_i} V_{K\pi\pi} + \theta_{\eta s_i} V_{K\pi\eta} +\theta_{\eta^\prime s_i} V_{K\pi\eta^\prime} \right)
\right|^2 \ . &
\end{align}\end{subequations}
Hence, the decay width reads
\begin{multline}
\Gamma(K^+ \to \pi^+ s_i) =
2 \pi m_K \left(\frac{\epsilon_{\UV}}{2} \frac{\B}{4 \pi m_K}\right)^2
\rho(x_\pi, x_s)
\\
\left(
\left|
\Re\left(c_{K\pi s_i}
- \nonet c^{S_m}_{\partial^2i}{}_\strange^\down \frac{m_s^2}{v^2}\right)
+ \frac{\epsilonEW}{\B v}\left(
 X_i
+ 2 \frac{c_i^{S_\omega}}{\betacoeff} \left(X_0
- \fermi_\octetm^\prime m_K^2
\right)
\right)
\right|^2
\right.\\\left.
+ \left|
\Im c_{K\pi s_i} +
\frac{\epsilonEW}{\epsilonUV\fp\B} \left(\theta_{\pi s_i} V_{K\pi\pi} + \theta_{\eta s_i} V_{K\pi\eta} +\theta_{\eta^\prime s_i} V_{K\pi\eta^\prime} \right)
\right|^2
\right)
\ .
\end{multline}

\subsubsection{Flavour-blind hidden sectors}

Starting from the results given in the previous section, we derive the full amplitude squared for $K^\pm \to \pi^\pm s_i$ decays in the case of flavour-blind portal interactions.
For such portal interactions, the Wilson coefficients \eqref{eq:wilson_coeff,eq:ktopi re Sm} simplify to
\begin{align}
c_{s_i\pi} &= 0 \ , &
c_{s_i\eta} &= - s_\eta \Im c^{S_m}_i \ , &
c_{s_i\eta^\prime} &= c_\eta \Im c^{S_m}_i \ ,
\end{align}
and
\begin{subequations}
\begin{align}
\Re c_{K\pi s_i} &= \begin{multlined}[t]
\frac{\epsilonEW}{\nfl} \left(\fermi_8 + (\nfl-1)\fermi_{27} - \vacuum \fermi_\octetm\right) \Re c^{S_m}_i \\
+ \Re \octet c^{S_m}_i{}_\strange^\down - \epsilonEW \left(\frac{m_K^2}{\B v} \Re \fermi_{\octetm i}
- \frac{\coupl_\Chromo}2 \Re \left(c^\chromo_{i\overline\down\strange}
+ c^\chromo_{i\overline\strange\down}\right)\right)
\ , \end{multlined} \\
\Im c_{K\pi s_i} &=
\epsilonEW \left(\frac{\vacuum \fermi_\octetm}{\nfl} \left(1-2\epsilon_\strange\right) \Im c^{S_m}_i
+ \frac{m_K^2}{\B v} \Im \fermi_{\octetm i}
- \frac{\coupl_\Chromo}2 \Im \left(c^\chromo_{i\overline\down\strange} - c^\chromo_{i\overline\strange\down}\right)\right) \ ,
\end{align}
\end{subequations}
while the mixing angles become
\begin{align}
\theta_{\eta s_i} &= - \frac{s_\eta \epsilon_{\eta_1 s_i}}{m_s^2 - m_\eta^2} \ , &
\theta_{\eta^\prime s_i} &= \frac{c_\eta \epsilon_{\eta_1 s_i}}{m_s^2 - m_{\eta^\prime}^2} \ , &
\epsilon_{\eta_1 s_i} &= \epsilonUV\fp \left( \B\Im c^{S_m}_i - c_i^{S_\theta}  \frac{\mth^2}v \right) \ .
\end{align}

\subsubsection{Explicit portal currents for specific hidden sector models}

\PETs including hidden spin 0 fields can be motivated from a broad range of \BSM models and are realised for instance in models of \DM (see \eg \cite{Battaglieri:2017aum, McDonald:2001vt, Hall:2009bx, Bernal:2017kxu, Pospelov:2007mp, Goudelis:2018xqi}), inflation (see \eg \cite{Bezrukov:2013fca, Ballesteros:2016euj}), naturalness (see \eg \cite{Graham:2015cka, Batell:2015fma, Choi:2015fiu, Kaplan:2015fuy, Flacke:2016szy, Giudice:2016yja}) and baryogenesis (see \eg \cite{Curtin:2018mvb} for references).
Spin 0 particles can be grouped into several categories, depending on their portal interactions with the \SM at the \EW scale.
We briefly summarise these categories and describe how the \PET procedure can be applied to each of them.
Additionally, we provide the relevant \PET operators at the GeV scale, and their connection to the hidden currents, for \ALPs and real scalar models, which are among the most studied realisations of light spin 0 messengers.

\paragraph{\ALPs}

\ALPs are \PNGBs associated with the spontaneous breaking of an approximate global symmetry.
Hence, they arise in a multitude of theoretically well motivated models, ranging from string theory (see \eg \cite{Lazarides:1985bj, Derendinger:1985cv, Giddings:1987cg}) to \QCD.
The original axion field is the \PNGB of the Peccei-Quinn symmetry \cite{Peccei:1977ur, Wilczek:1977pj, Weinberg:1977ma, Peccei:1977hh}, which has been introduced in order to solve the strong \CP problem and is broken by the axial anomaly of \QCD.
\footnote{
It has been long thought that axions in the MeV range were excluded, however this might not be the case.
We refer to \cite{Alves:2017avw} for a critical overview of bounds on MeV axions.
}

Depending on the underlying theoretical model, \ALPs can have theoretically unconstrained couplings with the \SM gauge bosons and derivative couplings with the \SM fermions.
The latter couplings can be traded for non-flavour blind Yukawa couplings, as described in \cref{sec:reduction techniques}.
Up to dimension five, the most general Lagrangian before \EWSB is given by \cite{Georgi:1986df, Choi:1986zw, Salvio:2013iaa, Brivio:2017ije, Bauer:2017ris}
\begin{align}
\mathcal L_a &
= \mathcal L_a^\text{hidden} + \mathcal L_a^\text{portal} \ , &
\mathcal L_a^\text{hidden} &= \frac{1}{2} \partial_\mu a \partial^\mu a + \frac{1}{2} m_a^2 a^2 \ .
\end{align}
Here $a$ is the \ALP field and the portal interactions are
\begin{multline} \label[lag]{eq:alps}
\mathcal L_a^\text{portal} = \frac{a}{f_a}
\left(
c_W W_{\mu \nu} \widetilde W^{\mu \nu} + c_B B_{\mu \nu} \widetilde B^{\mu \nu}
+ c_G G_{\mu \nu} \widetilde G^{\mu \nu}
\right.\\\left.
+ \left( \i \nonet c_u q \overline u \widetilde H^\dagger + \nonet c_d q \overline d H^\dagger + \nonet c_e \ell \overline e H^\dagger + \text{h.c.} \right)
\right) \ ,
\end{multline}
where $f_a$ is the energy scale associated with the \ALP and the $c_i$ (with $i = G$, $W$, $B$) and $\nonet c_i$ (with $i = u$, $d$, $e$) are scalar and matrix valued Wilson coefficient in flavour space, respectively.
For models that comply with minimal flavour violation, the coefficient matrices in the Yukawa interactions are aligned with and of comparable strength as the SM Yukawa matrices $\nonet y_i$.
All coefficients have been defined after using the \EOM for the Higgs and fermion fields in order to eliminate the derivative interactions of the \ALP, for details see \cite{Salvio:2013iaa, Brivio:2017ije, Bauer:2017ris}.
For \QCD axions, the mass term is generated by the \QCD quark condensate, so that $f_a m_a \propto m_\pi^2$, while for generic \ALP models, both the scale $f_a$ and the mass term $m_a$ are free parameters of the theory.
The mass term is part of the Lagrangian describing the internal structure of the hidden sector, which we do not need in our procedure, and it is listed here only for completeness.
Considering the portal \cref{eq:alps}, we recognise that all terms can be matched to the spin 0 portal operators defined in \cref{tab:pure operators}.
Hence, the relevant currents that drive the phenomenology of \ALPs at the \EW scale are given by
\begin{align} \label{eq:ALP EW scale currents}
\nonet S_m^X &= \nonet c_X \frac{a}{f_a} \ , &
S_\theta &= c_G \frac{a}{f_a} \ , &
S_\theta^X &= c_X \frac{a}{f_a} \ ,
\end{align}
where we have used $\epsilonUV = \flatfrac{v}{f_a}$, after confronting eq.~\eqref{eq:alps} with the pertinent \PETs in \cref{tab:pure operators}.
Comparing with \cref{eq:K decay hidden currents}, the resulting portal current that couple \QCD to \ALPs at the strong scale are
\begin{subequations}
\begin{align}
\nonet S_\m &\supset \nonet c_{S_m} \frac{v}{f_a} a \ , &
\tensor S_x &= \tensor \fermi_x \frac{a}{f_a} \ , &
S_\theta &= c_{S_\theta} \frac{a}{f_a} \ , &
\octet S_\chromo &= \left(\flavour \proj^\strange_\down c^\chromo_{\overline\strange\down} + \flavour \proj^\down_\strange c^\chromo_{\overline\down\strange}\right) \frac{v}{f_a} a \ ,
\end{align}
\end{subequations}
where we have used the \EOMs for the \ALP to resorb the $\flatfrac{\partial^2}{v^2}$ contribution from the general expression in \eqref{eq:K decay hidden currents} into $\nonet c_{S_m}$.
In addition, the term in \cref{eq:alps} that contains the photon field strength tensor gives rise to the Primakoff effect \cite{Pirmakoff:1951pj}, which our work does not modify.

The axial current $S_\theta$ and the imaginary part of the Yukawa current $\nonet S_\m$ mix the \ALP with pions and $\eta$-mesons, and give rise to \enquote{indirect} production via diagram \eqref{fig:feynman hidden scalar mixing}.
The remaining currents give rise to \enquote{direct} production via diagram \eqref{fig:feynman hidden scalar direct}.
For models that comply with minimal flavour violation, the coefficients $\nonet c_{S_m}$, $c^\chromo_{\overline\strange\down}$, and $c^\chromo_{\overline\down\strange}$ are aligned with and of comparable size as their \SM counterparts,
\begin{align}
v \nonet c_{S_m} &\sim \nonet \m \ , &
v c^\chromo_{\overline\strange\down} &\sim m_\down \ , &
v c^\chromo_{\overline\down\strange} &\sim m_\strange \ .
\end{align}
In \cPT, one finally obtains the currents
\begin{align}
\nonet S_m^\prime &= \nonet c^{\prime S_m} \frac{v}{f_a} a + \order{\epsilon_{\EW}^2, \epsilon_\strange^2} \ , &
S_y &= \fermi_y \frac{a}{f_a} \ ,
\end{align}
where the coefficient $\nonet c^{\prime S_m}$ is defined like its generic counterpart $\nonet c^{\prime S_m}_i$ in \cref{eq:K decay modified current wilson coeff},
except with the generic Wilson coefficients replaced according to
\begin{align}\label{eq: generic to alp coeff replacement}
\left(\nonet c^{S_m}_i + \nonet c^{S_m}_{\partial^2i} \frac1{v^2} \partial^2\right) &\to \nonet c^{S_m} \ , &
c^{S_\theta}_i &\to c^{S_\theta} \ , &
c^\chromo_{i\overline\strange\down} &\to c^\chromo_{\overline\strange\down} \ , &
c^\chromo_{i\overline\down\strange} &\to c^\chromo_{\overline\down\strange} \ , &
\fermi_{yi} &\to \fermi_y \ .
\end{align}
Hence, the complete amplitude for $K^\pm \to \pi^\pm a$ decays is
\begin{equation}
\mathcal A (K^+ \to \pi^+ a) = \mathcal A_\text{direct} + \mathcal A_\text{mixing} \ ,
\end{equation}
where the direct contribution is
\begin{equation}
\mathcal A_\text{direct}
= \mathcal A_m^{\Re} + \mathcal A_\fermi
= -\frac{\B v}{2 f_a} c_{K\pi a} - \frac{\epsilonEW}{2f_a} X_0 \ ,
\end{equation}
while the indirect contribution for production via meson-to-axion mixing is
\begin{equation}
\mathcal A_\text{mixing}
= \mathcal A_m^{\Im} + \mathcal A_{\theta}
= - \i \frac{\epsilonEW}{2\fp} \left(\theta_{\pi a} V_{K\pi\pi} + \theta_{\eta a} V_{K\pi\eta} + \theta_{\eta^\prime a} V_{K\pi\eta^\prime} \right) \ ,
\end{equation}
where the mixing angles are now
\begin{align}
\theta_{\pi a} &= \frac{\fp}{f_a} \frac{\B v c_{a\pi}}{m_a^2 - m_\pi^2} \ , &
\theta_{\eta a} &= \frac{\fp}{f_a} \frac{\B v c_{a\eta} + c_{S_\theta} \mth^2 s_\eta}{m_a^2 - m_\eta^2} \ , &
\theta_{\eta^\prime a} &= \frac{\fp}{f_a} \frac{\B v c_{a\eta^\prime} - c_{S_\theta} \mth^2 c_\eta}{m_a^2 - m_{\eta^\prime}^2} \ .
\end{align}
The coefficients $c_{K\pi a}$ and $c_{aX}$ are defined like their generic counterparts $c_{K\pi i}$ and $c_{s_i X}$ in \cref{eq:wilson_coeff,eq:ktopi re Sm}, except that the Wilson coefficients are replaced according to \eqref{eq: generic to alp coeff replacement}.
If the Wilson coefficients in \cref{eq:alps} are aligned with the SM Yukawa couplings, as it is usually the case, all amplitudes above are of the same order and equally contribute to the decay rate.
However, for flavour-blind \ALPs with $\nonet c_X \sim 1$ in \eqref{eq:ALP EW scale currents}, the amplitudes $\mathcal A_m^{\Re}$ and $\mathcal A_m^{\Im}$ are much bigger than the other two and dominate the decay rate.

We note that the indirect amplitude encompasses \eg the production amplitude of proper \QCD axions given in \cite{Alves:2017avw},
where the authors have neglected the 27-plet contributions $\propto \fermi_{27}$ as well as the finite pion and axion masses $m_\pi^2$, $m_a^2 \to 0$.
In this approximation, the function $V_{K\pi\pi}$ vanishes, and the resulting expression becomes independent of the axion-to-pion mixing angle $\theta_{\pi a}$.

\paragraph{Light real scalar fields}

This type of field can appear in a huge variety of \BSM models, ranging from \DM models, where the scalar is protected by a $\mathbb Z_2$ symmetry (see \eg \cite{Lerner:2009xg, Morrissey:2009tf}),
to models for baryogenesis (see \eg \cite{Curtin:2018mvb}), and \THDMs (see \eg \cite{Barbieri:2006dq, Branco:2011iw}), such as the inert doublet model, see \eg \cite{LopezHonorez:2006gr}.
Additionally, there are interesting candidates in \SUSY with R-parity conservation, such as the saxino, which is the scalar R-odd component of the axion superfield.
The saxino mass is typically of the same order of the gravitino mass, however there are models in which it can be naturally at a low scale, see \ie \cite{Kim:1992eu}.
The most common hidden Lagrangian can be cast as
\begin{align}
\mathcal L_s &
= \mathcal L_s^\text{hidden} + \mathcal L_s^\text{portal} \ , &
\mathcal L_s^\text{hidden} &= \frac{1}{2} \partial_\mu s \partial^\mu s + \lambda s^2 + \lambda^\prime s^3 + \lambda'' s^4 \ ,
\end{align}
where the $\lambda$ denote the self-couplings, however, being part of the hidden Lagrangian they are not relevant for the \PET approach.
The portal interactions are
\begin{multline} \label[lag]{eq:higgs portal}
\mathcal L_s^\text{portal} =
\frac{\alpha_0}{\Lambda} s D^\mu H^\dagger D_\mu H +
\left( \alpha_1 s + \alpha_2 s^2 + \frac{\alpha_3}{\Lambda} s^3 \right) H^\dagger H
+ \frac{\alpha_4}{\Lambda} s \left( H^\dagger H \right)^2 \\
+ \frac{s}{\Lambda} \left( \i \nonet c_u q \overline u \widetilde H^\dagger + \nonet c_d q \overline d H^\dagger + \nonet c_e \ell \overline e H^\dagger + \text{h.c.} \right)
+ \frac{c_W}{\Lambda} s W_{\mu \nu} W^{\mu \nu} + \frac{c_B}{\Lambda} s B_{\mu \nu} B^{\mu \nu}
+ \frac{c_G}{\Lambda} s G_{\mu \nu} G^{\mu \nu} \ ,
\end{multline}
where the $\alpha_i$, the $c_X$ with $X = W, B, G$, and the $\nonet c_x$ with $x=u,d,e$ are dimensionless Wilson coefficients and coefficient matrices, respectively.
The self- and portal-couplings involving an odd number of scalar fields are only present if the scalar field does not obey a $\mathbb Z_2$ symmetry.
The \PET framework is suitable for $n$ equal spin hidden messengers, hence it can describe several cases, such as:
\begin{inlinelist}
\item a single hidden scalar messenger, which is even under the symmetry of the secluded sector and arises for instance in simplified \DM models \cite{Abdallah:2015ter}
\item a \DM candidate which is odd under the $\mathbb Z_2$ symmetry, the typical example being the singlet scalar Higgs portal model \cite{McDonald:1993ex,Burgess:2000yq}
\item models with $\mathbb Z_n$ symmetries (see \eg \cite{Belanger:2014bga} for \DM models)
\end{inlinelist}
Depending on the symmetries of the model, the real scalar $s$ can mix with the \SM Higgs boson or assume a non-zero \VEV, however we will not discuss these possibilities here.
Typically, the scalar portal \cref{eq:higgs portal} only includes terms up to dimension four, while we include here also \EW scale terms of dimension five using the \PET approach.
A term which is especially relevant for light scalar fields is the coupling with the gluon field strength tensor, which is present for instance in theories with a dilaton field, see \eg \cite{Berti:2015itd}.

In order to demonstrate that the generic decay amplitude \eqref{eq:generic K amplitude} encompasses and is consistent with standard computations, we apply this general result to the case of light Higgs production in charged kaon decays $K^\pm \to \pi^\pm h$.
We compare our results with those obtained in \cite{Leutwyler:1989xj}, where $h$ is considered to be the \SM Higgs boson, and \cite{He:2006uu}, where it is taken to be the lightest Higgs particle of a \THDM model.
The computation in \cite{Leutwyler:1989xj} was performed before the discovery of the top-quark and the Higgs boson, so that the Higgs was still allowed to be lighter than the charged kaons.
In general, a light Higgs boson, with a mass $m_h < m_K$, couples to \QCD at the strong scale directly via quark Yukawa interactions, and additionally via effective $h GG$ and $h \overline qq \overline qq$ vertices, which arise after integrating out the heavy \SM \DOFs.
Translating these interactions into the hidden current picture, the only non-vanishing Wilson coefficients in \cref{eq:K decay modified currents,eq:K decay modified current wilson coeff} are \cite{Leutwyler:1989xj,He:2006uu}
\begin{align} \label{eq:light higgs wilson coeff}
\epsilonUV \nonet c_i^{S_m} &= \frac1v \left( \nonet \kappa \nonet \m - m_\strange \kappa_{\down\strange} \flavour \lambda_\strange^\down - m_\light \kappa_{\down\strange}^\dagger \flavour \lambda_\down^\strange\right) \ , &
\epsilonUV c_i^{S_\omega} &= 2 \kappa_G \ , &
\epsilonUV \fermi_{yi} &= - 2 \kappa_W \fermi_y \ ,
\end{align}
where $\nonet \kappa = \diag \left(\kappa_u, \kappa_d, \kappa_d \right)$.
The coefficients $\kappa_\up$ and $\kappa_\down$ measure the coupling of the Higgs-particle to the up-type and down-type quarks in the \SM, respectively.
In the case of a light \THDM Higgs-particle, one has $\kappa_G = \flatfrac{(2\kappa_u + \kappa_d)}3$, while the remaining $\kappa_x$ are free parameters.
In case of the \SM Higgs boson, one has \cite{Leutwyler:1989xj}
\begin{align} \label{eq: SM higgs k param}
\kappa_u = \kappa_d = \kappa_G = \kappa_W = 1 \ .
\end{align}
The constant $\kappa_{\down\strange} \sim \epsilonEW$ is determined by matching the low energy theory to the \EW scale description.
In general, it can be parameterised as \cite{Dawson:1989bm,He:2006uu}
\begin{align}
\kappa_{\down\strange} &= 2 \sum_{\mathclap{u=\up,\charm,\topq}} V_{\down u}^\dagger V_{u\strange} x_u f(x_u) \ , &
x_u &= \frac{m_u^2}{\Lambda_{\SM}^2} \ ,
\end{align}
and $f(x_u)$ is a model dependent function.
For the \SM Higgs-particle, assuming $x_u \ll \inv[2]{(4\pi)}$, and neglecting the running of the Wilson coefficients between the \EW and strong scales, one has $f(x_u) = \nicefrac34$ \cite{Leutwyler:1989xj}.
In the case of the \THDM, the corresponding expression is known, but quite complicated.
It can be found \eg in \cite{Barnett:1984zy,Dawson:1989bm,Lautenbacher:1990fh}.
Using \cref{eq:light higgs wilson coeff}, the coefficient \eqref{eq:ktopi re Sm} becomes
\begin{equation}
\epsilonUV c_{K\pi s_i} = \frac{m_\pi^2}{2v\B} ( \kappa_u - \kappa_d) \epsilonEW  (\fermi_8 + (\nfl - 1) \fermi_{27}) + \frac{ m_K^2}{v\B} \left( 2 \epsilonEW \kappa_W \fermi_\octetm - \kappa_{\down\strange} \right) + \order{\epsilon_\strange^2} \ .
\end{equation}

The overall $K^\pm \to \pi^\pm h$ decay amplitude receives contributions from the partial amplitudes $\mathcal A_m^{\Re}$, $\mathcal A_\omega$, and $\mathcal A_\fermi$, all of which mediate direct production.
There is no meson-to-Higgs mixing because the Higgs is scalar, rather than a pseudoscalar, particle.
One obtains
\begin{subequations}
\begin{align}
\mathcal A_m^{\Re} &
= \frac{m_K^2}{2v} \left(\kappa_{\down\strange} - 2 \kappa_W \epsilonEW \fermi_\octetm\right) - \frac{m_\pi^2}{4v} (\kappa_\up - \kappa_\down) \epsilonEW (\fermi_8 + (\nfl - 1) \fermi_{27})
\ , \\
\mathcal A_\fermi &
= \frac{\epsilonEW m_K^2}{2v} \kappa_W (\fermi_8 + (\nfl-1)\fermi_{27}) \left(1 + \frac{m_\pi^2 - m_s^2}{m_K^2}\right)
\ , \\
\mathcal A_\omega &
= \frac{\epsilonEW m_K^2}{2v} \frac{2 \kappa_G}{\betacoeff} \left(2 \fermi_\octetm^\prime - (\fermi_8 + (\nfl-1)\fermi_{27}) \left(1 + \frac{m_\pi^2 - m_s^2}{m_K^2}\right)\right)
\ .
\end{align}
\end{subequations}
Thus, the full amplitude is
\begin{multline}
\mathcal A(K^+ \to \pi^+ h) = \frac{m_K^2}{v} \left[\left(\frac{\kappa_W}2 - \frac{\kappa_G}{\betacoeff} \right) \epsilonEW (\fermi_8 + (\nfl-1)\fermi_{27}) \left(1 + \frac{m_\pi^2 - m_s^2}{m_K^2}\right)
\right.\\\left.
+ \frac{\kappa_\down - \kappa_\up}{4} \epsilonEW (\fermi_8 + (\nfl-1)\fermi_{27}) \frac{m_\pi^2}{m_K^2} - 2 \epsilonEW \left(\frac{\kappa_W}{2} \fermi_\octetm - \frac{\kappa_G}{\betacoeff} \fermi_\octetm^\prime \right) + \kappa_{\down\strange} \right] \ .
\end{multline}
This result encompasses the one given in \cite{He:2006uu}, where the contributions from the 27-plet and chromomagnetic operators have been neglected, which amounts to replacing $\fermi_{27} \to 0$ and $\fermi_\octetm^\prime \to \fermi_\octetm$.
\footnote{In \cite{He:2006uu} the amplitude is expressed in terms of $2 g_{\mathcal H} = \kappa_{\down\strange}$, $k_G = \flatfrac{2 \kappa_G}{\betacoeff}$, $\gamma_8 = \flatfrac{\epsilonEW \fermi_8}4$, and $\widetilde \gamma_8 = \flatfrac{\epsilonEW \fermi_1}4$.}
Using the values \eqref{eq: SM higgs k param}, one also obtains the result given in \cite{Leutwyler:1989xj}.
\footnote{In \cite{Leutwyler:1989xj} the amplitude is written in terms of the quantities $\xi = \kappa_{\down\strange}^\dagger$, $\kappa \equiv \nicefrac2{\betacoeff}$, $\gamma_1 = \epsilonEW \fermi_8$, and $ \gamma_2 = \epsilonEW \fermi_\octetm$.}

\paragraph{Pseudoscalars}

Pseudoscalar particles are predicted in many extensions to the Higgs sector, see \eg \cite{Andreas:2010ms, Ellwanger:2016qax} and the recently proposed relaxion field (see \eg \cite{Gupta:2015uea, Flacke:2016szy}), and have more general characteristics as compared to \ALPs.
The latter are restricted by being \PNGBs, while generic pseudoscalar particles can couple to the \SM via additional portal operators at the \EW scale, most notably a direct coupling with the Higgs boson.
In this sense, these particle combine features that arise in both \ALPs and light scalar models.

\paragraph{Complex Scalars}

As explained in \cref{sec:portals}, \PETs can describe complex scalars as a combination of two distinct real spin 0 fields that can be either scalar or pseudoscalar.
There are several interesting models with light complex scalar fields, see \eg \cite{Alexander:2016aln}.
Additionally, complex scalars commonly arise in \SUSY models, such as the sgoldstino \cite{Brignole:1998uu, Gorbunov:2000th, Gorbunov:2000cz, Gorbunov:2005nu, Dudas:2012fa, Antoniadis:2012ck, Astapov:2016koq, Astapov:2016zqw},
which can naturally be in the MeV mass range, the sneutrino \cite{Hagelin:1984wv, Ibanez:1983kw, Arina:2007tm, Bobrovskyi:2010ps, Alonso-Alvarez:2019fym}, which appears in the \MSSM, and the additional complex scalar field introduced in the \NMSSM, see \eg \cite{Ellwanger:2009dp} for a review.

\subsection{Charged kaon decay to charged leptons and hidden fermions}

In this section, we compute the transition amplitude for production of a generic fermionic messenger $\lhf_a$ in charged kaon decays $K^\pm \to \ell^\pm \lhf_a$ at \LO in $\delta$.

\subsubsection{Relevant interactions}

\begin{figure}
\begin{panels}{2}
\tikzsetnextfilename{feynman-hidden-fermion-1}\input{feynman-hidden-fermion-1.pgf}
\caption{Direct production.} \label{fig:feynman hidden fermion direct}
\panel
\tikzsetnextfilename{feynman-hidden-fermion-2}\input{feynman-hidden-fermion-2.pgf}
\caption{Production via mass mixing.} \label{fig:feynman hidden fermion mixing}
\end{panels}
\caption{
Feynman diagrams for the $K^\pm \to \ell^\pm \xi_a$ process.
} \label{fig:feynman hidden fermion}
\end{figure}

At tree-level, $K^\pm \to \ell^\pm\lhf_a$ decays are described by the two types of diagrams, depicted in \cref{fig:feynman hidden fermion}:
\begin{inlinelist}
\item\label{it:K lhf diagram one}%
diagrams with a single trilinear one-meson $K^\pm \to \ell^\pm \lhf_a$ portal vertex that directly couples \cPT to hidden sectors
\item\label{it:K lhf diagram two}%
diagrams with one trilinear $K^\pm \to \ell^\pm \nu_{\ell}$ \SM vertex
and a second $\nu_\ell \to \lhf_a$ portal vertex that indirectly couples \cPT to hidden sectors by mixing the \SM neutrinos with the fermionic messenger
\end{inlinelist}
The relevant portal current contributions to \cref{it:K lhf diagram one} diagrams are those with exactly one hidden spin \textfrac12 messenger and one charged lepton.
Using the list of portal currents in \cref{sec:portal currents}, the only such contribution is
\begin{equation} \label{eq:type one direct}
\nonet V_l^\mu \supset
\frac{\epsilonUV}{v^2} c^{L\dagger}_{\bar\up\strange a} \flavour\proj_\strange^\up \boson{\lhf_a^\dagger \overline \sigma_\mu e_b} + \text{h.c.} \ ,
\end{equation}
where $e$ and $c^{L\dagger}_{\bar\up\strange}$ are doublets in flavour space that capture the coupling to both $\electron^\pm$ and $\muon^\pm$.
The corresponding vertex mediating charged kaon decays is encoded inside the kinetic-like one meson portal interactions \eqref{eq:kinetic Vl Vr one meson}, leading to
\begin{equation} \label{eq:type two sm}
\mathcal L_K^{\partial V_l} = - \frac{\epsilonUV \fp}{v^2} c^{L\dagger}_{\bar\up\strange, ba} \boson{\lhf_a^\dagger \overline \sigma_\mu e_b} \partial^\mu K^+ \ .
\end{equation}
To compute diagrams of \cref{it:K lhf diagram two}, we have to specify both the neutrino to hidden fermion mixing vertex and the trilinear \SM vertex.
The mixing vertex is given as
\begin{equation} \label{eq:type two mixing}
\mathcal L_{\nu_b\lhf_a} = - \epsilonUV v \left( c^\nu_{ba} \nu_b \lhf_a + \text{h.c.} \right) \ ,
\end{equation}
where $\nu$ and $c^\nu_a$ are doublets in flavour space that capture the mixing of both $\neutrino_\electron$ and $\neutrino_\muon$.
The trilinear \SM vertex is encoded inside the kinetic-like one meson interactions \eqref{eq:weak sm one meson}, leading to
\begin{equation}
\mathcal L_K^{\partial W} = \frac{\fp V_{\up\strange}}{v^2} \partial_\mu K^+ \sum_{b = \electron, \muon} \nu_b^\dagger \overline \sigma^\mu \ell_b \ .
\end{equation}

\subsubsection{Partial decay width}

The vertices \eqref{eq:type one direct,eq:type two sm,eq:type two mixing} are written in the two-component notation of \cite{Dreiner:2008tw}.
Applying the Feynman rules for the two-component spinor notation \cite{Dreiner:2008tw,Martin:2012us} to compute the two types of diagrams illustrated in \cref{fig:feynman hidden fermion}, one obtains the full decay amplitude
\begin{equation}
\mathcal A(K^+ \to \ell_b^+ \lhf_a) = \mathcal A_\text{direct} + \mathcal A_\text{mixing} \ ,
\end{equation}
where the partial amplitudes are
\begin{subequations}
\begin{align}
\mathcal A_\text{direct} &= - \i \frac{\epsilonUV \fp}{v^2} c^{L\dagger}_{\bar\up\strange, ba} \boson{x^\dagger(p_\lhf,s_\lhf) \overline \sigma_\mu y(p_\ell,s_\ell)} p_K^\mu \ , \\
\mathcal A_\text{mixing} &= \i \frac{\epsilonUV \fp}{v m_\lhf^2} c^\nu_{ba} V_{\up\strange} \boson{ y(p_\lhf,s_\lhf) \sigma_\nu \overline \sigma_\mu y(p_\ell,s_\ell)} p_\lhf^\nu p_K^\mu \ ,
\end{align}
\end{subequations}
and the functions $x(p,s)$ and $y(p,s)$ are the polarisation spinors for two-component fermion fields.
The resulting helicity-summed partial decay width is
\begin{align}
\Gamma(K^+ \to \ell_b^+ \lhf_a) &
= 2 \pi m_K \left(\epsilon_{\UV} \epsilonEW \frac{m_K}{4\pi\fp}\right)^2
\rho(x_\ell, x_\lhf)
\abs{
c^L_{\bar\up\strange, ba} + \frac{c^\nu_{ba} V_{\up\strange} v}{m_\lhf}
}^2
\ , &
x_i &= \frac{m_i^2}{m_K^2} \ ,
\end{align}
where the phase-space factor is
\begin{equation}
\rho(x_\ell, x_\lhf) =
\left(x_\ell + x_\lhf - \left(x_\ell - x_\lhf\right)^2\right)
\sqrt{\left(\frac{1 - x_\ell - x_\lhf}2\right)^2 - x_\ell x_\lhf} \ .
\end{equation}
In terms of the partial decay width for the process $K^+ \to \ell_b^+ \nu_b$, this is
\begin{align}\label{eq:hnl1}
\Gamma(K^+ \to \ell_b^+ \lhf_a) =
\Gamma (K^+ \to \ell_b^+ \nu_b)
\frac{\rho(x_\ell, x_\lhf)}{\rho(x_\ell, 0)}
\abs{\theta_{ba}}^2
\ ,
\end{align}
where the \SM partial decay width and the effective mixing angle are
\begin{align} \label{eq:hnl2}
\Gamma (K^+ \to \ell_b^+ \nu_b) &= 2 \pi m_K \left(\epsilonEW \frac{m_K}{4\pi\fp}\right)^2
\abs{V_{\up\strange} }^2
\rho(x_\ell, 0)
\ , &
\theta_{ba} = \epsilonUV \left( \frac{c^\nu_{ba} v}{m_\lhf} + \frac{c^L_{\bar\up\strange, ba}}{V_{\up\strange}} \right) \ .
\end{align}

\subsubsection{Explicit portal currents for specific hidden sector models}

Gauge singlet fermionic hidden fields are common in \BSM models.
In the \SM, left-handed neutrinos are the only fields without a right-handed partner.
Therefore, it is natural to consider that such fields exist, but have so far not been observed due to their feeble interactions with \SM fields.
One or more right-handed neutrinos can be added to the \SM and can play an important role in several mechanisms of \BSM physics, via their mixing with ordinary neutrinos.
They can be used to generate neutrino masses (via one of the seesaw mechanisms), are required in leptogenesis models, and can act as \DM.
Since the nature of (right-handed) neutrinos is not known, the hidden messengers can be either Majorana or Dirac particles.
The latter case is described in our framework by two hidden Weyl fermions.
For reviews on the plethora of \BSM models with right-handed neutrinos we refer to \eg \cite{Drewes:2013gca, Alekhin:2015byh, Agrawal:2021dbo}.
Many \BSM models with right-handed neutrinos are commonly embedded into \SUSY theories, see various realisation of type-I and inverse seesaw, \eg \cite{Hisano:1995nq, King:2003jb, Arina:2008bb, Boucenna:2014zba, Lindner:2016bgg}.

As an example for a model with \HNLs, we consider the type-I seesaw model.
The minimal type-I seesaw Lagrangian couples the \SM to a pair of two sterile Majorana neutrinos \cite{Minkowski:1977sc, GellMann:1980vs, Mohapatra:1979ia, Yanagida:1980xy, Schechter:1980gr, Schechter:1981cv},
\begin{align}
\mathcal L_\nu &= \mathcal L^\text{portal}_\nu + \mathcal L^\text{hidden}_\nu \ , &
\mathcal L^\text{hidden}_\nu &= \frac12 \left( \overline \nu_i^\dagger \i \slashed \partial \overline \nu_i - M_{ij} \overline \nu_i \overline \nu_j \right) + \text{h.c.} \ ,
\end{align}
where
\begin{equation}
\mathcal L^\text{portal}_\nu = - y_{ia} \overline \nu_i \ell_a \widetilde H^\dagger + \text{h.c.}
\end{equation}
Here, $M_{ij} = M_{ji}$ denotes the sterile neutrino Majorana mass matrix, and $y_{ai}$ is the coupling strength of the sterile neutrino Yukawa interactions.
Without loss of generality, $M_{ij} = \diag(M_1, M_2)$.
The sterile neutrinos do not couple directly to \QCD, and the only contribution to the \EW scale portal currents is
\begin{equation}
\Xi_a = - \overline \nu_i y_{ia} \ .
\end{equation}
At the strong scale, this interaction generates the mass-mixing
\begin{equation}\label{eq:neutrino HNL mixing}
\mathcal L^\text{portal}_\nu \to - y_{ia} \overline \nu_i \nu_a v + \text{h.c.} \ ,
\end{equation}
so that $\epsilonUV v c^\nu_{bi} = v y_{ib}$.
Hence, the effective mixing angle is just the physical mixing angle between the \SM neutrino and the sterile neutrino, $\theta_{bi} = \flatfrac{v y_{ib}}{M_i}$.

Another category of hidden fermionic fields is given by the axinos, which are \SUSY partners of the axions, see \eg \cite{Rajagopal:1990yx, Covi:2001nw}.
They are unrelated to the neutrino sector, unless R-parity violation is allowed.
Axinos can be produced for instance by gluon fusion or in neutralino decays, which are useful mechanisms for searches in beam dump experiments or at colliders, and can be naturally in the MeV mass range, see \eg \cite{Gomez-Vargas:2019vci}.

\subsection{Neutral pion decay to photons and hidden vectors}

In this section, we consider anomalous neutral pion decays into hidden spin 1 messengers, $\pi^0 \to \gamma v_\mu$, at order $\delta^3$.
Unlike in the previous sections, we now include \EM contributions up to order $\alpha_{\EM}$.
However, we neglect all \EW contributions that are suppressed by factors of $\epsilonEW$, as this process is flavour conserving.

\subsubsection{Relevant portal current contributions}

\begin{figure}
\tikzsetnextfilename{feynman-hidden-vector-1}\input{feynman-hidden-vector-1.pgf}
\caption{Feynman diagram for the $\pi^0 \to \gamma v^\mu$ process.} \label{fig:feynman hidden vector}
\end{figure}
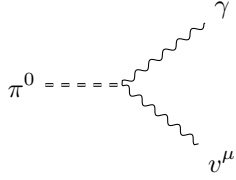

The relevant portal current contributions are those with a single hidden vector field.
Using the list of portal currents in \cref{sec:portal currents}, the only contributions of this type are
\begin{align} \label{eq:one vector hidden currents}
\nonet V_l^\mu &\supset \epsilonUV \nonet c^L_v v^\mu \ , &
\nonet V_r^\mu &= \epsilonUV \nonet c^R_v v^\mu \ .
\end{align}
\Cref{fig:feynman hidden vector} depicts the only relevant Feynman diagram.
In principle the process can be mediated by two types of diagrams:
\begin{inlinelist}
\item\label{it:diagram one}%
diagrams with a trilinear
$\pi^0 \to \gamma \gamma$ \SM vertex and a mixing vertex that makes the \SM photon oscillate into a hidden spin 1 particle
\item\label{it:diagram two}%
diagrams with a direct trilinear
$\pi^0 \to \gamma v_\mu$ portal vertex
\end{inlinelist}
Choosing an appropriate operator basis, there is no \cref{it:diagram one} diagram, since
the kinetic mixing term can always be eliminated from the theory using the \SM \EOM, in favour of a coupling to the \SM fermion fields.
As a result, only the diagram of \cref{it:diagram two} contributes to the decay amplitude $\pi^0 \to \gamma v_\mu$.
This interaction vertex arises from the anomalous \WZW contribution, which enters at order $\delta^3$.

\subsubsection{Partial decay width}

The interaction corresponding to the diagram in \cref{fig:feynman hidden vector} is contained in Lagrangian \eqref{eq:wzw Vl one meson}.
By extracting from it the contribution with a singlet pion, one obtains
\begin{equation}
\mathcal L_{\pi\to\gamma v}
\equiv \frac{\Nc}3\frac1{(4\pi)^2 \fp}
\left( 2 \nonet V_v^{\mu\nu}{}_\up^\up + \nonet V_v^{\mu\nu}{}_\down^\down \right) \pion e \widetilde F_{\mu\nu} \ ,
\end{equation}
where $\nonet V_v^\mu \equiv \nonet V_l^\mu + \nonet V_r^\mu$ and the photon field is canonically normalised.
Using expressions \eqref{eq:one vector hidden currents}, one has
\begin{equation}
\nonet V_v^\mu = \epsilonUV \left(\nonet c^R_v + \nonet c^L_v\right) v^\mu \ .
\end{equation}
The above expression implies that the \WZW does not couple neutral pions to the axial-vector current $\nonet V_a^\mu \equiv \nonet V_l^\mu - \nonet V_r^\mu$.
This is to be expected, since the \WZW mediates parity violating transitions, while neutral pion decays into a photon and a hidden axial-vector would conserve parity.
The partial decay width for $\pi^0 \to \gamma v_\mu$ decays is
\begin{equation}
\Gamma(\pi^0 \to \gamma v) = \frac1{16 \pi m_\pi}
\left(1 - \frac{m_v^2}{m_\pi^2}\right)
\overline{\abs{\mathcal A (\pi^0 \to \gamma v)}^2}
\ ,
\end{equation}
where the square amplitude is
\begin{equation} \label{eq:darkphoton}
\overline{\abs{\mathcal A(\pi^0 \to \gamma v)}^2}
= \left(\frac{\Nc}3 \frac{\epsilonUV}{4\pi\fp}\right)^2 \frac{\alpha_{\EM}}{4\pi} \left[2 (\octet c^R_v + \octet c^L_v)_\up^\up + (\octet c^R_v + \octet c^L_v)_\down^\down\right]^2 \left(m_\pi^2 - m_v^2\right)^2 \ .
\end{equation}
In terms of the partial decay width for the process $\pi^0 \to \gamma \gamma$ it reads
\begin{align}
\Gamma(\pi^0 \to \gamma v) &= 2 \epsilon_\text{eff}^2 \Gamma(\pi^0 \to \gamma \gamma) \left( 1 - \frac{m_v^2}{m_\pi^2} \right)^3 \ ,
\end{align}
where
\begin{align}
\Gamma(\pi^0 \to \gamma \gamma) &= 2\pi m_\pi \left( \frac{\Nc}{3} \frac{\alpha_{\EM}}{4\pi} \frac{m_\pi}{4\pi\fp} \right)^2 \ , &
\epsilon_\text{eff} &= \epsilonUV \frac{2 (\octet c^R_v + \octet c^L_v)_\up^\up + (\octet c^R_v + \octet c^L_v)_\down^\down}{2 e \left(2 \nonet q_\up^\up + \nonet q_\down^\down\right)}  \ ,
\end{align}
are the \SM partial decay width and the effective mixing parameter.

\subsubsection{Explicit portal currents for specific hidden sector models}

Relatively light vectors states (\ie below the GeV scale) that are very weakly coupled to the \SM fields represent attractive physics targets for experimental searches at the cross-over of the intensity and high-energy frontiers.
In the literature there are several proposals, with different motivations, for vector portal models.
The simplest realisations do not charge the \SM fields under the new gauge group related to the hidden vectors, giving rise to kinetic mixing portals.
An attractive alternative is given by gauging certain combinations of \SM fields under the new $\U(1)$, in order to achieve for instance anomaly free or \UV complete models.
Examples of the latter models are the $\qn B - \qn L$ or the $\qn L_\muon - \qn L_\tauon$ anomaly free models, see \eg \cite{He:1990pn, He:1991qd, Baek:2001kca, Ma:2001md, Salvioni:2009jp, Heeck:2011wj}.
For a broad overview of the different models, physics motivations and experimental constraints, we refer to the reviews \cite{Alekhin:2015byh, Alexander:2016aln, Fabbrichesi:2020wbt}.

Here, we consider the simplest dark photon model, which is $\QED$-like, from \cite{Okun:1982xi,Holdom:1986eq}, with a single hidden vector $v_\mu$.
The hidden Lagrangian is given by
\begin{align}
\mathcal L_v &
= \mathcal L_v^\text{hidden} + \mathcal L_v^\text{portal} \ , &
\mathcal L_v^\text{hidden} &
= - \frac{1}{4} F^{\prime\mu\nu}F^\prime_{\mu\nu}+ \frac{1}{2} m^2_v v_\mu v^\mu \ ,
\end{align}
and the portal interaction is
\begin{align} \label[lag]{eq:darkphot}
\mathcal L_v^\text{portal} &
= - \frac{\epsilon}{2} F^{\mu\nu}F^\prime_{\mu\nu} \ .
\end{align}
In this equation, $\epsilon$ is the kinetic mixing parameter between the hidden vector and the photon and $F^\prime_{\mu\nu}$ is the field strength tensor of the hidden vector.
We show part of the hidden Lagrangian, however, this is not needed for our purposes.
First, it is not actually relevant how the dark photon acquires a mass.
This can be achieved by the spontaneous symmetry breaking of the symmetry to which the dark photon is associated, requiring a dark Higgs, or could be achieved via the Stückelberg mechanism \cite{Stueckelberg:1900zz,Kors:2005uz}, if the symmetry is a $\U(1)$.
As long as the dark Higgs is heavier than the \cPT scale and is integrated out, \cref{eq:darkphoton} is not modified by the mass generation mechanism.
Second, we remain agnostic about the remaining particle content of the hidden sector, which might include fermionic states $X$, charged under the new $\U(1)$, that couple only to the dark photon (we already mentioned this possibility in \cref{sec:Phi expanded}).
A model similar to \eqref{eq:darkphot} that couples to the hypercharge instead of the \EM charge is obtained by substituting the \QED $\U(1)$ with the hypercharge $\U(1)$ field in the \SM.

The expected \BR for the process $\pi^0 \to \gamma v_\mu$ is known, see \eg \cite{CortinaGil:2019nuo}, and is equivalent to \cref{eq:darkphoton}, which can be seen by rewriting the kinetic mixing \cref{eq:darkphot} in terms of the portal operators using the \SM \EOM.
Afterwards, the dark photon field couples to \QCD via the neutral current interaction
\begin{align}
\mathcal L_v^\text{portal} &\to - \trf{\octet v^\prime_\mu \left( \nonet Q^\mu + \overline{\nonet Q}^\mu \right) } \ , &
\octet v^\prime_\mu &= \epsilon e \octet q v_\mu \ .
\end{align}
Hence, $\epsilonUV (\nonet c_v^L + \nonet c_v^R) = \epsilon e \octet q$, and therefore $\epsilon_\text{eff} = \epsilon$.

\subbib

%% file: feynman-hidden-scalar-2.pgf
\begin{feyn}[node distance = \height/4 and \width/4]
\vertex (a1) {$K^\pm$};
\vertex[above right = of a1] (c0);
\vertex[above right = of c0] (c2) {$\pi^\pm$};
\vertex[below right = of a1] (c0p);
\vertex[below right = of c0p] (c3) {$s_i$};
\vertex at ($(c0)!0.5!(c3)$) [cross] (c1) {};
\diagram*{
{[edges = {dashed, double}] (a1) -- (c0) -- (c2)},
{[edges = {dashed, double}] (c0) -- [near end, edge label = {$\pi^0$, $\eta$, $\eta^\prime$}] (c1)},
{[edges = scalar] (c1) -- (c3)}
};
\end{feyn}

%% file: feynman-hidden-scalar-1.pgf
\begin{feyn}[node distance = \height/2 and \width/8*3]
\vertex (a1) {$K^\pm$};
\vertex[right = of a1] (c0);
\vertex[right = of c0] (cp);
\vertex[below = of cp] (c2) {$s_i$};
\vertex[above = of cp] (c4) {$\pi^\pm$};
\diagram*{
{[edges = {dashed, double}]
(a1) -- (c0) -- (c4),
},
 (c0) -- [scalar] (c2),
};
\end{feyn}

%% file: feynman-hidden-fermion-1.pgf
\begin{feyn}[node distance = \height/2 and \width/8*3]
\vertex (a1) {$K^\pm$};
\vertex[right = of a1] (c0);
\vertex[right = of c0] (cp);
\vertex[below = of cp] (c2) {$\xi_a$};
\vertex[above = of cp] (c4) {$\ell^\pm$};
\diagram*{
{[edges = {dashed, double}]
(a1) -- (c0),
},{[edges = fermion]
 (c4) -- (c0) -- (c2),},
};
\end{feyn}

%% file: feynman-hidden-fermion-2.pgf
\begin{feyn}[node distance = \height/4 and \width/4]
\vertex (a1) {$K^\pm$};
\vertex[above right = of a1] (c0);
\vertex[above right = of c0] (c2) {$\ell^\pm$};
\vertex[below right = of a1] (c0p);
\vertex[below right = of c0p] (c3) {$\xi_a$};
\vertex at ($(c0)!0.5!(c3)$) [cross] (c1) {};
\diagram*{
{[edges = {dashed, double}] (a1) -- (c0)},
{[edges = fermion] (c2) -- (c0) -- [near end, edge label = $\nu_\ell$] (c1) -- (c3)}
};
\end{feyn}

%% file: feynman-hidden-vector-1.pgf
\begin{feyn}[node distance = \height/2 and \width/8*3]
\vertex (a1) {$\pi^0$};
\vertex[right = of a1] (c0);
\vertex[right = of c0] (cp);
\vertex[below = of cp] (c2) {$v^\mu$};
\vertex[above = of cp] (c4) {$\gamma$};
\diagram*{
{[edges = {dashed, double}]
(a1) -- (c0),
},{[edges = boson]
 (c4) -- (c0) -- (c2),},
};
\end{feyn}

%% file: conclusion.tex
\section{Conclusion} \label{sec:conclusions}

In this paper, we have developed a framework of \PETs, which extend \EFTs associated with the \SM by coupling them to generic hidden messenger fields with masses at or below the characteristic energy scale of the relevant \EFT.
This framework enables the coupling of \SM fields to light hidden sectors while remaining largely agnostic about the internal structure of the hidden sector, which can include secluded particles that do not couple directly to the \SM but interact with each other and the messenger fields.
It also accounts for the coupling to heavier hidden sectors via the inclusion of higher dimensional operators in \PET Lagrangians.
Throughout  the paper, we have focused primarily on hidden fields with masses at or below the strong scale, for which there are extensive searches at intensity frontier experiments.
However, we emphasise that the \PET framework, and in particular the portal \SMEFTs we derived in \cref{sec:portals}, also capture messengers that are much heavier, as long as their mass is within the regime of applicability of the corresponding \EFT.

Using the \PET framework, we have first constructed \EW scale and strong scale \PETs that couple \SMEFT and \LEFT to a messenger of spin 0, \textfrac12, or 1.
The resulting portal \SMEFTs encompass all available portal operators up to dimension five, while the portal \LEFTs additionally encompass all dimension six and seven operators that contribute to quark-flavour violating transitions at \LO in $\epsilonEW$, $\alpha_{\EM}$, and the \NDA $4\pi$ counting scheme.
We have found that all portal \SMEFTs conserve baryon number, and that the spin 0 and 1 messenger portal \SMEFTs conserve lepton number.
In the case of spin \textfrac12 messenger, the portal operators can violate lepton number by one unit, $\abs{\Delta \qn L} \leq 1$.
Additionally, this messenger does not couple to any of the quark fields or the right-chiral charged lepton fields, while the spin 1 messenger only couples to pairs of quarks and leptons with identical chirality, so that it cannot act as a separate source of chiral symmetry breaking.
We used all these properties to constrain the portal \LEFTs, so that the resulting \LEFTs should be understood as a low energy approximation of the corresponding portal \SMEFTs, where the heavy \SM \DOFs have been integrated out.

We have parameterised the coupling of \QCD to hidden sectors at the strong scale in terms of ten external currents $J \in \{ \Omega$, $\Theta$, $\nonet \M$, $\nonet L^\mu$, $\nonet R^\mu$, $\octet T^{\mu\nu}$, $\octet \Chromo$, $\tensor \Fermi_l$, $\tensor \Fermi_r$, $\tensor \Fermi_\scalar \}$, and used a spurion analysis to derive the corresponding \PETs that couple the hidden messengers to the $\U(3)$ version of \cPT, which contains an $\eta_1$ singlet meson in addition to the light pseudoscalar meson octet of $\SU(3)$ \cPT.
The spurion analysis is the standard technique used to embed \cPT in the remainder of the \SM at \LO in $\alpha_{\EM}$. Hence, the coupling of \cPT to the currents $\Theta$, $\nonet \M$, $\nonet L^\mu$, and $\nonet R^\mu$, which capture the impact of photons, the light \SM leptons, and the \QCD theta angle in the \SM, is well understood \cite{DiVecchia:1980yfw, Gasser:1984gg, Pich:1993uq, Pich:1995bw, Ecker:1996yy, HerreraSiklody:1996pm, Kaiser:2000gs, Scherer:2002tk, Bijnens:2006zp}.
Similarly, the coupling of \cPT to $\octet T^{\mu\nu}$ has been studied in \cite{Cata:2007ns}.

Here, we have extended the spurion technique to also account for the space-time dependent external currents $\octet \Chromo$, $\tensor \Fermi_l$, $\tensor \Fermi_r$, $\tensor \Fermi_\scalar$, and $\Omega$.
The \SM contributions to all these currents are constant, and the \SM contribution to the current $\Omega = \omega + S_\omega$ is the inverse fine-structure constant of \QCD $\omega \propto \inv[2]{g_s}$.
Since strong interactions are integrated out when constructing \cPT, only the portal contribution $S_\omega$ can appear in the \cPT action.
$S_\omega$ encompasses \eg the coupling of \cPT to a light Higgs boson $h$ via the interaction $h G_{\mu\nu}G^{\mu\nu}$, previously studied \eg in \cite{Leutwyler:1989xj}.
We generalise that description to account for the coupling of \cPT to a fully generic current $S_\omega$.
The constant \SM contributions to the dipole current $\octet \Chromo$ and the four-quark currents $\tensor \Fermi_x$ are usually included into \cPT by appealing directly to the transformation behaviour of the \QCD dipole and four-quark operators under global quark-flavour rotations \cite{Cronin:1967jq, Kambor:1989tz, Ecker:1992de, Donoghue:1992dd, Buras:2018evv}.
Since it is difficult to generalise this transformation behaviour approach to space-time dependent external currents, we have used the more powerful spurion approach.
In order to include the four-quark currents $\tensor \Fermi_x$ into the power counting for $\U(3)$ \cPT, which is defined via a simultaneous expansion in momenta $\partial^2$ and large $\Nc$, we have generalised the standard \QCD large $\Nc$ counting formula..

The final \cPT Lagrangian contains 27 free coefficients $\coupl \in \{ \coupl_\Chromo^x$, $\coupl_T^x$, $\coupl_y^x$, $\coupl_\omega^x\}$.
In order to make it possible to constrain interactions in the portal \LEFTs using bounds on hidden sector induced meson transitions, we have estimated 22 of these coefficients using a number of well-established techniques for the non-perturbative matching of \cPT to \QCD.
Four of the seven coefficients $\kappa_\omega$, which measure the coupling of \cPT to the $S_\omega$ current, have already been estimated by using the anomalous trace relation of the \QCD \cref{eq:Q full stress-energy tensor} \cite{Leutwyler:1989xj}.
Using this strategy, we have fixed the remaining three coefficients.
The thirteen $\coupl_x^y$ coefficients, which measure the coupling of \cPT to the octet and 27-plet currents $\tensor \Fermi_l$, $\tensor \Fermi_r$, and $\tensor \Fermi_\scalar$,
are well known in the large $\Nc$ limit \cite{Bardeen:1986vz,Pich:1990mw,Pich:1995qp,Pallante:2001he,Cirigliano:2003gt,Gerard:2005yk,Buras:2014maa}.
However, corrections that appear for finite $\inv{\Nc}$ are known to be important when estimating the strength of the four-quark operators in the \SM, and we expect the same to be true for the four-quark operators in the portal sector.
Hence, we have adapted the strategies used in \cite{Bardeen:1986vz, Leutwyler:1989xj, Pich:1990mw, Pich:1995qp, Gerard:2005yk}, and obtained improved estimates for the $\coupl_x^y$ coefficients by matching them to experimental values of the octet and 27-plet coefficients $\fermi_{8,1,27}$.
Finally, we have estimated the coefficients $\coupl_\Chromo$ and $\coupl_\Chromo^{\M} + \coupl_\Chromo^{\M^\prime}$, which measure the coupling of \cPT to the dipole current $\octet \Chromo$, by matching the \cPT prediction for the vacuum condensates of the \QCD dipole \cref{eq:EW quark bilinears} to the corresponding lattice values in \eqref{eq:quark gluon condensate}.

To facilitate the application of our results, we have listed all one- and two-meson interactions that arise from the \LO portal \cPT action.
We have then computed the most general transition amplitudes for three golden channels, which are used to constrain the coupling to hidden sectors in fixed-target experiments:
\begin{inlinelist}
\item $K^\pm \to \pi^\pm s_i$
\item $K^\pm \to \ell^\pm \lhf_a$
\item $\pi^0 \to \gamma v_\mu$
\end{inlinelist}
\footnote{Recall that the fields $s_i$, $\lhf_a$ and $v_\mu$ denote generic spin 0, spin \textfrac12 and spin 1 messengers, respectively.}
For the spin 0 messenger, we have computed a universal decay amplitude and connected it to simple realisations of \ALPs and scalar portal models.
For spin \textfrac12 fields, we have mapped our generic decay amplitude to the case of \HNL by rewriting it in terms of a generalised effective mixing angle.
We have also connected our comprehensive expression for the spin 1 messengers to the case of \QED-like dark photon model by using the photon \EOM to express the kinetic mixing operator in terms of our portal operators.

\subsection*{Outlook}

The work we have presented in this paper opens up several potentially interesting avenues for further investigation, which range from formal improvements of the \PET framework to theoretical work to expand its regime of applicability and further to a number of relevant phenomenological applications.

In this paper, we have focused primarily on completing a minimal version of portal \cPT that can be used to make concrete predictions for meson decays at intensity experiments, and have left open some questions that need to be addressed in order to complete the \PET framework.
For instance, one has to connect the \EW and strong scale \PETs in order to constrain the shape of portal Lagrangians at the \EW scale by means of low-energy experiments.
This connection can be established \eg via an explicit procedure of successive matching and running, where the Wilson coefficients for each portal interaction are run down from the \EW scale ($\mu \sim v$) to the strong scale ($\mu \lesssim m_\charm$), while integrating out each heavy \SM \DOF as it becomes inactive.
Further, it is necessary to complete the matching between the strong scale \PETs and \cPT by determining the remaining $\coupl$ coefficients related to the external currents $\octet \Chromo$ and $\octet T^{\mu\nu}$.
This is an unavoidable procedure to relate meson scattering and decay amplitudes induced by these two currents to the corresponding dipole operators in \QCD.

In addition, there are several avenues that can be pursued to extend the \PET framework by expanding the range of models that it is able to capture.
First, it is possible to include \eg portal operators up to dimension six at the \EW scale, which would allow for describing a larger class of \DM models.
Second, one can construct \PETs for hidden sector models with higher spin messengers or with multiple messengers.
In \cref{sec:higher spin}, we have already constructed portal \SMEFTs for spin \textfrac32 and 2 messengers, but it remains to construct the corresponding portal \LEFTs at the strong scale, as well as the resulting portal \cPT Lagrangian at \LO.
Finally, while the \PETs we have constructed already account for the possibility of multiple messengers with \emph{identical} spin, for a fully general description of models with multiple portals, it might be interesting to add portal operators that encompass hidden fields with \emph{different} spin.

Finally, one can apply the \PET framework to make predictions for various experimental setups besides the low-energy fixed target experiments that have been the focus of this work.
For instance, \EW scale \PETs can be used to constrain hidden sectors at collider experiments, \eg at the \LHC, similarly to how \SMEFT is being used to constrain the coupling to heavy new sectors,
and to make predictions for flavour physics experiments, such as \LHCb \cite{Aaij:2021vac}, or for beam dump experiments, such as \SHiP, which produce an enormous amount of heavy $D$- and $B$-mesons.
In order to apply the \PET approach to heavy meson physics and a wide range of other experimental setups, it will be useful to construct \PETs that extend a large class of \EFTs in the \SM, such as \HEFT, \HQET, \NRQCD, and \SCET.

In the long term, this program of building and linking various \PETs at many different energy scales will make it possible to perform a truly global parameter scan,
which could be used to constrain light hidden sectors in a very general way, as it will combine different observations at the \EW scale, from flavour physics experiments, and from intensity experiments.
This goal will require the ability to compute a large variety of amplitudes for a wide range of distinct hidden-sector induced transitions.
In order to simplify this task, it is thus sensible to implement the various \PETs into tools that automatise Feynman rules, such as \software{FeynRules} \cite{Alloul:2013bka}, and to produce model files for software packages, such as \software{MadGraph} \cite{Alwall:2014hca}, \software{MadDM} \cite{Ambrogi:2018jqj, Arina:2020kko} and \software{MadDump} \cite{Buonocore:2018xjk}, which are able to compute the matrix elements and the necessary theoretical predictions.

\subsection*{Acknowledgement}

We are very grateful to Marco Drewes and Jean-Marc G\'erard for fruitful discussions and the feedback they have provided all along the preparation of this work.
We thank Eduardo Cortina Gil, Anthony Francis, Martin Hoferichter, Ken Mimasu, Urs Wenger and Uwe-Jens Wiese for useful discussions.
Chiara Arina has been supported by the Innoviris ATTRACT 2018 104 BECAP~2 agreement.
Jan Hajer has been supported by the \FSR incoming postdoc fellowship of \UCLouvain and is now supported by the \SNF under the project \no{200020/175502}.
Philipp Klose has been partially supported by the \UCLouvain \FSR funding scheme and the Innoviris ATTRACT 2018 104 BECAP~2 agreement, and is now supported by the \SNF under grant \no{200020B-188712}.

\subbib

\glsunset{SMEFT}
\glsunset{HEFT}
\glsunset{LEFT}
\glsunset{SCET}
\glsunset{HQET}
\glsunset{NRQCD}
\glsunset{NRQED}

%% file: construction.tex
\subtoc

\section{Construction of portal effective theories} \label{sec:redundant}

In this appendix we describe the techniques we use to construct \EW and strong scale \PETs that extend an \EFT of the \SM by coupling the \SM \DOFs to a hidden messenger that is lighter than the characteristic energy scale of the relevant \EFT.
We summarise the \NDA power counting scheme, and give a number of well-known reduction techniques used to obtain a minimal basis of independent portal operators for each \PET.

\subsection{Naive dimensional analysis} \label{sec:nda}

After integrating out the heavy \SM \DOFs, the strong scale \PETs may contain portal operators with dimension larger than five.
The higher dimensional operators are suppressed by powers of $\epsilonSM \equiv \flatfrac{\partial^2}{\Lambda_{\SM}^2}$.
In addition, operators that receive contributions from tree-level diagrams at the \EW scale theory will be suppressed by loop factors of $\inv[1]{(4\pi)}$.
These loop factors can be integrated into the power counting using \NDA \cite{Manohar:1983md, Jenkins:2013sda, Gavela:2016bzc, Manohar:2018aog}.
The \NDA counting scheme assumes that the \EFT Lagrangian can be written as \cite{Jenkins:2013sda}
\begin{equation}
\mathcal L = \mathcal L^{d \leq 4} + \sum_{i} c_i O_i \ ,
\end{equation}
where the $c_i$ are Wilson coefficients and the $O_i$ denote effective operators of dimension $d_i > 4$.
The renormalisable Lagrangian $\mathcal L^{d \leq 4}$ contains gauge interactions with couplings $g_i$, Yukawa interactions with couplings $y_i$, $\phi^3$ interactions with couplings $\kappa_i$, and $\phi^4$ interactions with couplings $\lambda_i$.
Assuming that the kinetic part of the Lagrangian is canonically normalised, \NDA stipulates that the Wilson coefficients $c_i$ are expected to be of order one, or smaller, if the $O_i$ are normalised as
\begin{equation} \label{eq:nda_power_counting}
O_i \propto
\frac{\Lambda^4}{(4\pi)^2}
\left(\frac{g}{4\pi}\right)^{\!n_g}
\left(\frac{y}{4\pi}\right)^{\!n_y}
\left(\frac{\kappa}{4\pi \Lambda}\right)^{\!n_\kappa}
\left(\frac{\lambda}{(4\pi)^2}\right)^{\!n_\lambda}
\left(\frac{p^2}{\Lambda^2}\right)^{\!n_p}
\left(\frac{4\pi \phi}{\Lambda}\right)^{\!n_\phi}
\left(\frac{(4\pi)^2 \psi \psi}{\Lambda^3}\right)^{\!2 n_\psi} \ ,
\end{equation}
where $\Lambda$ is a high-energy scale associated with a small momentum expansion in powers of $\epsilon \propto \flatfrac{\sqrt s}{\Lambda}$,
$\phi$ and $\psi$ denote bosonic and fermionic fields present in the effective theory, and $p^2$ stands for any light mass scale (\ie it includes both derivatives $\partial \propto p$ and light masses $m \propto p$).

The \NDA power counting is self-consistent in the sense that an arbitrary diagram with insertions of
higher dimensional operators normalised according to \eqref{eq:nda_power_counting} is renormalised by operators with the same $4\pi$ normalization.
That is, the Wilson coefficients mix as \cite{Jenkins:2013sda}
\begin{equation}
\delta c_i \propto \prod_j c_j \ ,
\end{equation}
which implies that the Wilson coefficients should satisfy $c_i \lesssim 1$, even if the underlying \UV theory is strongly coupled \cite{Manohar:2018aog}.
If the \UV theory is weakly coupled, the Wilson coefficients may be much smaller than one, $c_i \ll 1$,
so that the $4\pi$ power counting of \NDA can be broken by strongly hierarchical values of the Wilson coefficients,
which could potentially satisfy $4 \pi c_i \ll c_j$ for certain $i \neq j$.

When using the \NDA counting scheme to discriminate between portal operators at the strong scale,
we specifically count $\inv{(4\pi)}$ suppression factors associated with loops in the \EW scale diagrams that generate each strong scale operator.
Since the renormalisable $d = 3$, $4$ operators in the strong scale theory are generated by tree-level diagrams at zeroth order in $\epsilonEW$,
their normalization should not contain any explicit factors of $(4\pi)$.
This requirement implies that the small portal coupling $\epsilonUV$ has to be associated with a factor $\inv{(4\pi)}$,
so that \eg an operator $\epsilonUV \overline qq s_i$ scales as $(4\pi)^0$ rather than $(4\pi)^1$.
This is completely analogous to the $\inv{(4\pi)}$ suppression that has to accompany each \SM Yukawa coupling.
In view of our choice gauge field normalization, which ensures that the covariant derivatives $D_\mu = \partial_\mu - \i \photon_\mu$ are independent of the gauge couplings,
the \SM photon field strength tensor needs also to be associated with a factor of $\inv{(4\pi)}$.
In principle, this reasoning also applies to gluon field strength tensors, but the corresponding factor of $\inv{(4\pi)}$ does not result in any relative suppression,
since $\flatfrac{g_s}{4\pi}$ is not a small parameter in the non-perturbative regime of \QCD.
Therefore, we do not keep track of the factors of $4\pi$ associated with the gluon field strength tensor.
Summarising, we normalise each portal operator at the strong scale as follows,
\begin{multline} \label{eq:strong scale operator normalization}
O_i \propto
\left(\frac{\Lambda_{\SM}^2}{4\pi} \right)^{\!2}
\left(\frac{e}{4\pi}\right)^{\!n_e}
\left(\frac{p^2}{\Lambda_{\SM}^2}\right)^{\!n_p}
\left(\frac{F^{\mu\nu}}{\Lambda_{\SM}^2}\right)^{\!n_F}
\left(\frac{4\pi \psi_{\SM}}{\Lambda_{\SM}^{\flatfrac32}}\right)^{\! n_\psi}
\\ \times
\left( \frac1{4\pi} \right)^{\!\vartheta(n_s + n_v + \flatfrac{n_\lhf}{2})}
\frac{\epsilonUV}{4\pi}
\left(\frac{4\pi s_i}{\Lambda_{\SM}}\right)^{\!n_s}
\left(\frac{4\pi v^\mu_i}{\Lambda_{\SM}}\right)^{\!n_v}
\left(\frac{4\pi \lhf_a}{\Lambda_{\SM}^{\flatfrac32}} \right)^{\!n_\lhf} \ ,
\end{multline}
where $\psi_{\SM}$ stands for \SM fermions, and $p \sim D_\mu, m$ now denotes either a covariant derivative or a light mass scale.
The function
\begin{equation}
\vartheta(x) =
\begin{cases}
1 & x > 1 \\
0 & x \leq 1
\end{cases}
\end{equation}
measures how many hidden fields the operator contains.
If it contains more than one hidden boson or more than two hidden fermions,
the operator has to have been generated by \EW scale diagrams that contain at least one hidden sector interaction,
and according to the general \NDA counting scheme this interactions has to be associated with suppression by at least one factor $\inv{(4\pi)}$.

\subsection{Reduction techniques} \label{sec:reduction techniques}

In general, a naive listing of all available operators at each order in the power counting contains a number of
redundant operators that can be expressed as a linear combination of other operators at the same or higher order in the power counting.
In the following, we list a number of standard reduction techniques that we use to identify minimal bases of portal operators without redundancies:
Further details on these reduction techniques can be found in \cite{Fierz:1937,Nishi:2004st,Grzadkowski:2010es,Klose:2020yhp} and references therein.

\paragraph{Algebraic identities} directly associate operators with each other.
In our analysis, we use
\begin{itemize}[leftmargin=\parindent,labelindent=\parindent]
\item \textbf{Bianchi identities} that relate the covariant derivatives of field strength tensors $V^{\mu\nu}$.
One has
\begin{equation}
D^\mu V^{\nu \rho} + D^\nu V^{\rho \mu} + D^\rho V^{\mu \nu} = 0 \ .
\end{equation}
In particular, these identities imply that
\begin{align}
\sigma_{\mu \nu} (D^\rho V^{\mu \nu} + 2 D^\mu V^{\nu \rho}) &= 0 \ , &
D_\mu \widetilde V^{\mu \nu} &= 0 \ .
\end{align}

\item \textbf{Fierz completeness relations} that relate products of fermion bilinears \cite{Fierz:1937,Nishi:2004st}.
These are often given in terms of four component fermions, see for instance \cite{Nishi:2004st}.
In the two-component notation of \cite{Dreiner:2008tw}, the Fierz identities we use take the form
\begin{subequations} \label{eq:scalar_tensor_fierz}
\begin{align}
\boson{\psi_a \psi_b} \boson{\psi_c \psi_d} &= \frac12 \boson{\psi_a \psi_d} \boson{\psi_c \psi_b} + \frac14 \boson{\psi_a \overline \sigma_{\mu \nu} \psi_d} \boson{\psi_c \overline \sigma^{\mu \nu} \psi_b}
\ , \label{eq:lh_scalar_tensor_fierz} \\
\boson{\psi^\dagger_a \psi^\dagger_b} \boson{\psi^\dagger_c \psi^\dagger_d} &= \frac12 \boson{\psi^\dagger_a \psi^\dagger_d} \boson{\psi^\dagger_c \psi^\dagger_b} + \frac14 \boson{\psi^\dagger_a \sigma_{\mu \nu} \psi^\dagger_d} \boson{\psi^\dagger_c \sigma^{\mu \nu} \psi^\dagger_b}
\ , \label{eq:rh_scalar_tensor_fierz}
\end{align}
\end{subequations}
and
\begin{subequations} \label{eq:vector_vector_fierz}
\begin{align}
\boson{\psi^\dagger_a \overline \sigma_\mu \psi_b} \boson{\psi_c \sigma^\mu \psi^\dagger_d} &= 2 \boson{\psi^\dagger_a \psi^\dagger_d} \boson{\psi_c \psi_b} \ , \\
\boson{\psi^\dagger_a \overline \sigma_\mu \psi_b} \boson{\psi^\dagger_c \overline \sigma^\mu \psi_d} &= - \boson{\psi^\dagger_a \overline \sigma_\mu \psi_d} \boson{\psi^\dagger_c \overline \sigma^\mu \psi_b}
\ , \\
\boson{\psi_a \sigma_\mu \psi^\dagger_b} \boson{\psi_c \sigma^\mu \psi^\dagger_d} &= - \boson{\psi_a \sigma_\mu \psi^\dagger_d} \boson{\psi_c \sigma^\mu \psi^\dagger_b}
\ ,
\end{align}
\end{subequations}
as well as
\begin{equation}
\boson{\psi_a \overline \sigma_{\mu \nu} \psi_d} \boson{\psi^\dagger_c \sigma^{\mu \nu} \psi^\dagger_b} = 0 \ ,
\end{equation}
and finally
\begin{subequations} \label{eq:scalar_vector_fierz}
\begin{align}
\boson{\psi_a \psi_b} \boson{\psi^\dagger_c \overline \sigma^\mu \psi_d} &= \frac12 \boson{\psi_a \psi_d} \boson{\psi^\dagger_c \overline \sigma^\mu \psi_b} - \i \frac38 \boson{\psi_a \overline \sigma^{\mu \nu} \psi_d} \boson{\psi^\dagger_c \overline \sigma_\nu \psi_b}
\ , \\
\boson{\psi_a^\dagger \psi_b^\dagger} \boson{\psi_c^\dagger \overline \sigma^\mu \psi_d} &= \frac12 \boson{\psi_a^\dagger \overline \sigma^\mu \psi_d} \boson{\psi_c^\dagger \psi_b^\dagger} - \i \frac38 \boson{\psi_a^\dagger \overline \sigma_\nu \psi_d} \boson{\psi_c^\dagger \sigma^{\mu \nu} \psi_b^\dagger} \ .
\end{align}
\end{subequations}

\end{itemize}

\paragraph{Partial integration} can be used to rearrange (covariant) derivatives within the operators, assuming that the fields vanish at infinity.

\paragraph{Field redefinitions} of the shape
\begin{equation}
\phi(x) \to \phi(x) - \epsilon^n f[\phi](x) \ ,
\end{equation}
where $\epsilon$ is a small parameter of the theory and $f[\phi](x)$ is a polynomial that depends only on powers of $\phi$ and its derivatives evaluated at $x$,
can be used to eliminate operators proportional to the zeroth order \EOM for the effective \DOFs that appear at order $\epsilon^n$ \cite{Arzt:1993gz, Scherer:2005ri, Grzadkowski:2010es, Manohar:2018aog}.
The repetition of this procedure at each order in $\epsilon$ makes it possible to eliminate operators proportional to the zeroth order \EOM at all orders in $\epsilon$.

\subsection{Standard Model equations of motions}

In this section, we collect the \EOMs for the \SM fields we use throughout this work.

\subsubsection{EOMs at the electroweak scale}

At the \EW scale, the \SM \EOMs for fermions are
\begin{subequations} \label{eq:fermion_eom}
\begin{align}
\i \slashed D \ell &= y_e e H \ , &
\i \slashed D e &= y_e^\dagger H^\dagger \ell \\
\i \slashed D q &= y_u u \widetilde H + y_d d H \ , &
\i \slashed D u &= y_u^\dagger \widetilde H^\dagger q \ , &
\i \slashed D d &= y_d^\dagger H^\dagger q \ ,
\end{align}
\end{subequations}
where the $y_i$ ($i = e$, $u$, $d$) denote the \SM Yukawa coupling matrices.
The \EOMs for \SM bosons are given by
\begin{subequations} \label{eq:phi_eom}
\begin{align}
D^2 H&= \left(\m^2 - \lambda \abs{H}^2 \right) H
- e^\dagger y_e^\dagger \ell - d^\dagger y_d^\dagger q - (\epsilon q)^\dagger y_u u
\ , \\
\partial_\mu B^{\mu \nu} &= \sum_{\text{all } f} y_f \overline \psi_f \gamma_\mu \psi_f
+ \i y_h H^\dagger \overleftright D_\mu H
\ , \\
(D_\mu W^{\mu \nu})^I &= \phi^\dagger \i \overleftright D \vphantom{11pt}^{\nu I} \phi + l_b^\dagger \overline \sigma^\nu T^I l_b + q_b^\dagger \overline \sigma^\nu T^I q_b
\ , \\
2 (D_\mu G^{\mu\nu})^x&= u_a^\dagger \overline \sigma^\nu \lambda^x u_a
+ \overline u_a^\dagger \overline \sigma^\nu \lambda^x \overline u_a
+ \overline d_a^\dagger \overline \sigma^\nu \lambda^x \overline d_a
+ d_a^\dagger \overline \sigma^\nu \lambda^x d_a
\ ,
\end{align}
\end{subequations}
where $\epsilon$ is the $\SU(2)_L$ totally anti-symmetric tensor, the index $x$ denotes objects that transform as members of the adjoint representation of $\SU(3)_C$,
and the $\lambda^x$ are \GM matrices acting on triplets in \emph{colour}-space.

\subsubsection{EOMs at the strong scale}

At the strong scale, the \EOMs for \SM fermions are given by
\begin{subequations}
\begin{align}
\i \slashed D e_i &= m_{e i} \overline e_i^\dagger \ , &
\i \slashed D d_i &= m_{d i} \overline d_i^\dagger \ , &
\i \slashed D u_i &= m_{d i} \overline u_i^\dagger \ , &
\i \slashed D \nu_i &= 0 \ , \\
\i \slashed D \overline e_i &= m_{e i} e_i^\dagger \ , &
\i \slashed D \overline d_i &= m_{d i} d_i^\dagger \ , &
\i \slashed D \overline u_i &= m_{d i} u_i^\dagger \ .
\end{align}
\end{subequations}
At the same scale, the \EOMs for \SM bosons are
\begin{subequations}
\begin{align}
2 \left(D_\mu G^{\mu\nu}\right)^x&= u_a^\dagger \overline \sigma^\nu \lambda^x u_a
+ \overline u_a^\dagger \overline \sigma^\nu \lambda^x \overline u_a
+ \overline d_a^\dagger \overline \sigma^\nu \lambda^x \overline d_a
+ d_a^\dagger \overline \sigma^\nu \lambda^x d_a \ , \\
\partial_\mu F^{\mu\nu}&= \sum_{\text{light } f} q_f \psi_f^\dagger \overline \sigma^\nu \psi_f \ .
\end{align}
\end{subequations}
where $q_f$ is the \EM charge of the fermion in question.

For the \PETs we construct, it is possible to combine the strong scale \EOMs with the other reduction techniques to eliminate all operators with at least one covariant derivative acting on a \SM fermion.
Considering a generic Weyl fermion $\psi_a$ with mass $m_a$ and gauge charges $q_a$, and a field strength tensor $V^{\mu\nu}$, we get
\begin{equation} \label{eq:derived_eliminations_DDpsi}
D_\rho D^\rho \psi_a = \left(\slashed D \slashed D - q_a \frac12 \overline \sigma_{\mu \nu} V^{\mu \nu} \right) \psi_a
\xrightarrow{\EOM} \ - \left(m_a^2 + q_a \frac12 \overline \sigma_{\mu \nu} V^{\mu \nu} \right) \psi_a \ ,
\end{equation}
and further
\begin{equation}\begin{split} \label{eq:derived_eliminations_psiDpsi}
\mathcal O^\mu \left(\psi_a \overleftright D_\mu \psi_b\right) = & \
\frac12 \mathcal O^\mu \left(\psi_a \left[\overleftright{\slashed D} \overline \sigma^\mu + \sigma^\mu \overleftright{\slashed D} \right] \psi_b\right) \\
\xrightarrow{\PI} & \
- \mathcal O^\mu \left(\psi_a \left[\overleft{\slashed D} \overline \sigma^\mu - \sigma^\mu \overright{\slashed D} \right] \psi_b\right)
+ \left(\i D^\nu \mathcal O^\mu\right) \left(\psi_a \overline \sigma_{\nu \mu} \psi_b\right) \\
\xrightarrow{\EOM} & \
- \i \left(m_b + m_a\right) \mathcal O^\mu \left(\overline \psi^\dagger_a \overline \sigma^\mu \psi_b + \psi_a \sigma^\mu \overline \psi^\dagger_b\right)
+ \left(\i D^\nu \mathcal O^\mu\right) \left(\psi_a \overline \sigma_{\nu \mu} \psi_b\right) \ ,
\end{split}\end{equation}
and
\begin{equation}\begin{split} \label{eq:derived_eliminations_psisigmaDpsi}
\mathcal O_\mu \left(\psi_a \overline \sigma^{\mu \nu} \overleftright D_\nu \psi_b\right) = & \
\i \mathcal O_\mu \left(\psi_a \left[\sigma^\mu \slashed{D} + \overleft{\slashed{D}} \overline \sigma^\mu - D^\mu - \overleft D \vphantom{11pt}^\mu \right] \psi_b\right) \\
\xrightarrow{\EOM} & \
\left(m_b + m_a\right) \mathcal O_\mu \left(\overline \psi^\dagger_a \overline \sigma^\mu \psi_b + \psi_a \sigma^\mu \overline \psi^\dagger_b\right) + \left(\i D^\mu \mathcal O_\mu\right) \left(\psi_a \psi_b\right) \ ,
\end{split}\end{equation}
and
\begin{equation}\begin{split} \label{eq:derived_eliminations_DpsiDpsi}
\mathcal O \left(D_\rho \psi_a D^\rho \psi_b\right) = & \
\frac12 \mathcal O \left(D_\rho \psi_a\right) D^\rho \psi_b + \frac12 \mathcal O \left(D_\rho \psi_a\right) D^\rho \psi_b \\
\xrightarrow{\PI} & \
- \frac 12\mathcal O \left(D^2 \psi_a\right) \psi_b - \frac 12\mathcal O \psi_a \left(D^2 \psi_b\right)
- \frac12 \left(D_\rho\mathcal O\right) \partial_\rho \left(\psi_a \psi_b\right) \\
\xrightarrow{\EOM} & \
\frac12 \left(D^2\mathcal O\right) \psi_a \psi_b + \frac 12 \left(m_a^2 + m_b^2\right)\mathcal O \psi_a \psi_b \ ,
\end{split}\end{equation}
and
\begin{equation}\begin{split} \label{eq:derived_eliminations_DpsisigmaDpsi}
\mathcal O \left(D_\mu \psi_a \sigma^{\mu \nu} D_\nu \psi_b\right) = & \
\frac\i2 \mathcal O \left(\psi_a \left[\overleft {\slashed D} \overright {\slashed D} - \overleft D_\mu \gamma^\nu \gamma^\mu \overright D_\nu \right] \psi_b \right) =
\i \mathcal O \left(\psi_a \left[\overleft {\slashed D} \overright {\slashed D} - \overleft D^\rho \overright D_\rho \right] \psi_b \right) \\
\xrightarrow{\EOM} & \
\i m_a m_b \mathcal O \overline \psi_a^\dagger \overline \psi_b^\dagger - \i \mathcal O \left(D^\rho \psi_a D_\rho \psi_b\right) \ ,
\end{split}\end{equation}
as well as
\begin{equation}\begin{split} \label{eq:derived_eliminations_final}
\mathcal O^{[\mu \nu]} \left(\psi_a^\dagger \overline \sigma_\nu \overleftright D_\mu \psi_b\right) = & \
\mathcal O^{[\mu \nu]} \frac\i2 \left(\psi_a^\dagger \left[\sigma_{\mu \nu} \overleftright{\slashed D} - \overleftright{\slashed D} \overline \sigma_{\mu \nu} \right] \psi_b\right) \\
\xrightarrow{\PI} & \
\mathcal O^{[\mu \nu]} \i \left(\psi_a^\dagger \left[\sigma_{\mu \nu} \overright {\slashed D} + \overleft {\slashed D} \overline \sigma_{\mu \nu} \right] \psi_b\right)
+ D^\rho \mathcal O^{\mu \nu} \frac\i2 \left(\psi^\dagger_a \left[\overline \sigma_\rho \overline \sigma_{\mu \nu} + \sigma_{\mu \nu} \overline \sigma_\rho \right] \psi_b\right) \\
\xrightarrow{\EOM} & \
\mathcal O^{[\mu \nu]} \left(m_b \psi_a^\dagger \sigma_{\mu \nu} \overline \psi_b^\dagger - m_a \overline \psi_a \overline \sigma_{\mu \nu} \psi_b\right)
+ \frac\i2 \left(\psi^\dagger_a \overline \sigma_\sigma \psi_b\right) D_\rho \widetilde{\mathcal O}^{\rho \sigma} \ .
\end{split}\end{equation}

\subsubsection{Quark $\EOM$ including external currents} \label{sec:eom_ext_curr}

To compute the trace of the \QCD Hilbert \cref{eq:Q full stress-energy tensor} in the presence of generic external currents, we include these currents into the quark \EOMs.
Therefore, they are
\begin{subequations} \label{eq:quark full eom}
\begin{align}
\i \slashed D q^{\dagger b} &=
\begin{multlined}[t]
\left(
\nonet \M^T \overline q
+ \left(\octet T^T_{\mu\nu} %
+ \octet \Chromo^T G_{\mu\nu} \right) \overline \sigma^{\mu\nu} \overline q
+ \nonet L^T_\mu \sigma^\mu q^\dagger
\right)^b \\
- \frac{\abs{V}^2_{\strange\down}}{v^2} \left(
2 \tensor \Fermi_l{}_{a c}^{bd} \sigma^\mu q^{\dagger a} \boson{q^{\dagger c} \overline \sigma_\mu q_d}
+ \tensor \Fermi_r{}_{a\dot c}^{b\dot d} \sigma^\mu q^{\dagger a} \boson{\overline q^{\dot c} \sigma_\mu \overline q_{\dot d}^\dagger}
- \tensor \Fermi_\scalar{}_{a\dot c}^{\dot db} \overline q^{\dot c} \boson{q^{\dagger a} \overline q^\dagger_{\dot d}}
\right) \ ,
\end{multlined} \\
\i \slashed D \overline q^\dagger_{\dot a} &=
\begin{multlined}[t]
\left(
\nonet \M q - \left(\octet T_{\mu\nu} + \octet \Chromo G_{\mu\nu} \right) \sigma^{\mu\nu} q + \nonet R_\mu \sigma^\mu \overline q^\dagger
\right)_{\dot a} \\
+ \frac{\abs{V}^2_{\strange\down}}{v^2} \left(
\tensor \Fermi_r{}_{c\dot a}^{b\dot d} \sigma^\mu \overline q^\dagger_{\dot d} \boson{q^{\dagger c} \overline \sigma_\mu q_b}
+ \tensor \Fermi_\scalar{}_{c\dot a}^{\dot bd} q_d \boson{q^{\dagger c} \overline q^\dagger_{\dot b}}
\right)
\ .
\end{multlined}
\end{align}
\end{subequations}

\subbib

%% file: electroweak.tex
\subtoc

\section{Electroweak scale portal operators} \label{sec:ewpetsapp}

In this appendix, we first collect the redundant \EW scale portal operators for messengers with spin 0, \textfrac12, and 1, and then present the portal operators for messengers with spin \textfrac32 and 2.

\subsection{Redundant portal operators with messenger field up to spin one} \label{sec:redundant ew portal operators}

The operators listed in %
\cref{tab:pure operators} form a complete basis in the sense that it is impossible to further reduce the number of independent portal operators by using the standard reduction techniques discussed in \cref{sec:reduction techniques}.
In particular, we have used the \SM \EOM to minimise the number of derivatives appearing within each operator,
but for certain applications, it may be more convenient to work with alternative bases of portal operators.
To facilitate this, we list below the redundant \EW portal operators that can be constructed in \PETs based on \SMEFT, and show which techniques were used to trade them.

The redundant \textbf{spin 0} portal operators are
\begin{align}
\partial_\mu \partial_\nu s_i B^{\mu \nu} &
\xleftrightarrow{\PI} \partial_\mu s_i \partial_\nu B^{\mu \nu} \xleftrightarrow{\PI} s_i \partial_\mu \partial_\nu B^{\mu \nu} = 0
\ ,
\end{align}
as well as
\begin{subequations}
\begin{align}
\partial^\mu s_i (H^\dagger \overleftright D_\mu H) &
\xleftrightarrow{\PI} s_i \left(H^\dagger D^2 H - \text{h.c.}\right) \ , \\
\partial^2 s_i \abs{H}^2 \xleftrightarrow{\PI} \partial_\mu s_i \partial^\mu \abs{H}^2 &
\xleftrightarrow{\PI} s_i \left(H^\dagger D^2 H + \text{h.c.}\right) + \text{non-redundant}
\end{align}
and
\begin{align}
\partial_\mu s_i q_a^\dagger \overline \sigma^\mu q_b &\xleftrightarrow{\PI} s_i q_a^\dagger \slashed D q_b \ , &
\partial_\mu s_i u_a^\dagger \sigma^\mu u_b &\xleftrightarrow{\PI} s_i u_a^\dagger \slashed D u_b \ , &
\partial_\mu s_i d_a^\dagger \sigma^\mu d_b &\xleftrightarrow{\PI} s_i d_a^\dagger \slashed D d_b \ , \\
\partial_\mu s_i \ell_a^\dagger \overline \sigma^\mu \ell_b &\xleftrightarrow{\PI} s_i \ell_a^\dagger \slashed D \ell_b \ , &
\partial_\mu s_i e_a^\dagger \sigma^\mu e_b &\xleftrightarrow{\PI} s_i e_a^\dagger \slashed D e_b \ .
\end{align}
\end{subequations}
Notice that the remaining operators $(H^\dagger D^2 H \pm \text{h.c.})$ and $\psi_a^\dagger \slashed D \psi_b$ on the right-hand side of these expressions can be replaced with Yukawa type portal operators using the \SM \EOMs.

The only redundant \textbf{spin \textfrac12} portal operators are
\begin{align}
D_\mu \lhf_a^\dagger \overline \sigma^\mu \ell_a \widetilde H^\dagger + \text{h.c.} &
\xleftrightarrow{\PI}
\lhf_a^\dagger \slashed D \ell_a \widetilde H^\dagger + \text{non-redundant} \ ,
\end{align}
where the remaining operator on the right-hand side can also be replaced with Yukawa type portal operators using the \SM \EOMs.

Finally, the redundant \textbf{spin 1} portal operators are
\begin{align}
\partial_\nu v_\mu \widetilde B^{\mu \nu} &
\xleftrightarrow{\PI} v_\mu \partial_\nu \widetilde B^{\mu \nu} = 0
\ , &
\partial_\nu v_\mu B^{\mu \nu} &
\xleftrightarrow{\PI} v_\mu \partial_\nu B^{\mu \nu}
\ ,
\end{align}
where the only remaining operator $v_\mu \partial_\nu B^{\mu\nu}$ can also be replaced with Yukawa type portal operators using the \SM \EOMs.

Finally, we already argued in \cref{sec:EW operators} that the number of independent portal operators given in \cref{tab:pure operators} can be further reduced by using the \EOMs proper to the hidden sector.
However, these \EOMs depend strongly on the internal structure of the latter, the modelling of which is beyond the scope of this paper.

\subsection{Rarita-Schwinger and Fierz-Pauli fields} \label{sec:higher spin}

\begin{table}
\centering
\newcommand{\portal}[2]{$#1$&$#2$}
\newcommand{\qcdportal}[3]{$#1$&$\mathbf{{#2}}#3$}
\newcommand{\pair}[2]{\ensuremath{\begin{pmatrix}#1\\#2\end{pmatrix}}}
\newcommand{\field}[2]{\midrule\multirowcell{#1}{#2}}
\newcolumntype{p}{@{ }r@{}l@{ }}
\begin{tabular}{lc@{ }*4p} \toprule
& $d$ & \multicolumn{2}{c}{Higgs} & \multicolumn{2}{c}{$\text{Yukawa}+\text{h.c.}$} & \multicolumn{2}{c}{Fermions} & \multicolumn{2}{c}{Gauge bosons} \\
\field{6}{$\lhf_\mu^a$\\+\\h.c.} & \multirow{1}{*}{4} & & & \portal{\lhf_{a\mu}^\dagger\overline \sigma^\mu}{\ell_b \widetilde H^\dagger} \\
\cmidrule{2-10} & \multirow{5}{*}{5} & \portal{\lhf_{a\mu} \lhf_b^\mu}{\abs{H}^2}
& \portal{(\partial^\mu \lhf_{a\mu})}{\ell_b \widetilde H^\dagger} & & & \portal{\lhf_{a\mu} \lhf_{b\nu}}{B^{\mu\nu}} \\
& & \portal{\lhf_{a\mu} \sigma^{\mu\nu} \lhf_{b\nu}}{\abs{H}^2}
& \portal{(\partial_\nu \lhf_{a\mu}) \overline \sigma^{\mu\nu}}{\ell_b \widetilde H^\dagger}
& & & \portal{\lhf_{a\rho} \sigma^{\mu\nu} \lhf_b^\rho}{B_{\mu\nu}} \\
& & & & \portal{\lhf_{a\mu}}{\ell_b D^\mu \widetilde H^\dagger} & & & \portal{\lhf_{a\mu} \sigma^{\mu\rho} \lhf_b^\nu}{B_{\nu\rho}} \\
& & & & \portal{\lhf_{a\mu} \overline \sigma^{\mu\nu}}{\ell_b D_\nu \widetilde H^\dagger}
& & & \portal{\lhf_{a\mu} \sigma^{\mu\rho} \lhf_b^\nu}{\widetilde B_{\nu\rho}} \\
& & & & & & & & \portal {\lhf_{a\alpha} \sigma_\mu^\rho \lhf_{b\beta}} {B_{\nu\rho}\epsilon^{\alpha\beta\mu\nu}} \\
\field{15}{$t_{\mu\nu}$%
} & \multirow{2}{*}{3} & & & & & & & \portal{t_{\mu\nu}}{B^{\mu\nu}} \\
& & & & & & & & \portal{\widetilde t_{\mu\nu}}{B^{\mu\nu}} \\
\cmidrule{2-10} & \multirow{12}{*}{5} & \portal{t^{\mu\nu}}{D_\mu H^\dagger D_\nu H}
& \qcdportal{t^{\mu\nu}}{q_a \sigma_{\mu\nu} u_b}{\widetilde H} & \qcdportal{(\partial^\mu t_{\mu\nu})}{q_a^\dagger \overline \sigma^\nu q_b}{}
& \qcdportal{t_{\mu\nu}}{G^{\mu\rho} G^\nu_\rho}{} \\
& & \portal{\widetilde t^{\mu\nu}}{D_\mu H^\dagger D_\nu H}
& \qcdportal{t^{\mu\nu}}{q_a \sigma_{\mu\nu} d_b}{H} & \qcdportal{(\partial^\mu t_{\mu\nu})}{u_a^\dagger \sigma^\nu u_b}{}
& \qcdportal{\widetilde t_{\mu\nu}}{G^{\mu\rho} G^\nu_\rho}{} \\
& & \portal{(\partial_\mu t^{\mu\nu})}{(H^\dagger \overleftright D_\nu H)}
& \portal{t^{\mu\nu}}{\ell_a \sigma_{\mu\nu} e_b H} & \qcdportal{(\partial^\mu t_{\mu\nu})}{d_a^\dagger \sigma^\nu d_b}{}
& \qcdportal{t_{\mu\nu}}{\widetilde G^{\mu\rho} G^\nu_\rho}{} \\
& & & & & & \portal{(\partial^\mu t_{\mu\nu})}{\ell_a^\dagger \overline \sigma^\nu \ell_b}
& \qcdportal{t_{\mu\nu}}{G^{\mu\rho} \widetilde G^\nu_\rho}{} \\
& & & & & & \portal{(\partial^\mu t_{\mu\nu})}{e_a^\dagger \sigma^\nu e_b}
& \portal{\widetilde t_{\mu\nu}}{W^{\mu\rho} W^\nu_\rho} \\
& & & & & & \qcdportal{(\partial^\mu \widetilde t_{\mu\nu})}{q_a^\dagger \overline \sigma^\nu q_b}{}
& \portal{t_{\mu\nu}}{\widetilde W^{\mu\rho} W^\nu_\rho} \\
& & & & & & \qcdportal{(\partial^\mu \widetilde t_{\mu\nu})}{u_a^\dagger \sigma^\nu u_b}{}
& \portal{t_{\mu\nu}}{W^{\mu\rho} \widetilde W^\nu_\rho} \\
& & & & & & \qcdportal{(\partial^\mu \widetilde t_{\mu\nu})}{d_a^\dagger \sigma^\nu d_b}{}
& \portal{\widetilde t_{\mu\nu}}{B^{\mu\rho} B^\nu_\rho} \\
& & & & & & \portal{(\partial^\mu \widetilde t_{\mu\nu})}{\ell_a^\dagger \overline \sigma^\nu \ell_b} & \portal{t_{\mu\nu}}{\widetilde B^{\mu\rho} B^\nu_\rho} \\
& & & & & & \portal{(\partial^\mu \widetilde t_{\mu\nu})}{e_a^\dagger \sigma^\nu e_b}
& \portal{t_{\mu\nu}}{B^{\mu\rho} \widetilde B^\nu_\rho} \\
& & & & & & & & \portal{t_{\mu\nu}}{B^{\mu\nu} \abs{H}^2} \\
& & & & & & & & \portal{\widetilde t_{\mu\nu}}{B^{\mu\nu} \abs{H}^2} \\
\bottomrule
\end{tabular}
\caption[Portal $\SMEFT$ operators involving messengers with spin \textfrac32 and 2]{
List of all portal operators up to dimension five that couple \SMEFT to hidden spin \textfrac32 fermionic fields $\lhf_a^\mu$ and tensor fields $t^{\mu\nu}$.
The first column specifies the type of portal, the second column denotes the dimension $d$ of the portal operator and the remaining columns label the \SM sectors to which the hidden field couples.
In the case of the vector-fermion \PETs, each operator is supplemented by its Hermitian conjugate.
The bold operators couple to the strong sector of the \SM.
} \label{tab:highspin}
\end{table}

Here we briefly discuss the case in which the \SM couples to hidden \textfrac32 Weyl fields $\lhf_i^\mu(x)$ with $i = 1$, $2$ or to a spin 2 field $t^{\mu\nu}$.
Without loss of generality, we take $t \equiv t^\mu_\mu = 0$, since the scalar \DOF $t$ couples to the \SM via the operators collected in \cref{tab:pure operators}.
\Cref{tab:highspin} collects the complete list of portal operators up to dimension five for both spin \textfrac32 fermion and tensorial messenger fields.

A standard example of hidden spin \textfrac32 fields coupling to the \SM model are the gravitinos appearing in \SUGRA models.
Although their precise mass depends on the details of the model, they can easily be much lighter than the other supersymmetric particles \cite{Ellis:1984kd, Ellis:1984xe}, leading to interesting phenomenology
\cite{Buchmuller:2008vw, Bobrovskyi:2011vx, Bobrovskyi:2012dc}, and placing it into the regime of \PETs.

Broadly speaking, there are two separate energy ranges in which spin 2 messengers constitute viable extensions of the \SM.
On the one hand, in extra-dimension models, see \eg \cite{Han:1998sg, ArkaniHamed:1998rs, ArkaniHamed:1998nn, Randall:1999ee}, besides the massless zero mode, higher order graviton excitations are interesting portals for \NP, and their allowed mass range lies in the TeV scale \cite{Murata:2014nra, Aaboud:2018zhh, Sirunyan:2018exx}.
They can be described by portal \SMEFTs at high-energy colliders \cite{Das:2016pbk, Kraml:2017atm} or for models of TeV scale \DM \cite{Lee:2013bua}.
On the other hand, bimetric theories of gravity \cite{Hassan:2011zd, Schmidt-May:2015vnx}, called bigravity, feature an new massive interacting spin 2 state.
This new boson can be a \DM candidate.
However, either its mass range is beyond the sensitivity of intensity experiments, lying below the eV range, see \cite{Marzola:2017lbt},
or it lies in the MeV range but its interaction strength with ordinary matter is so negligible to make a detection hopeless, see \cite{Murata:2014nra, Chu:2017msm, Albornoz:2017yup}.
Finally, models with hidden spin 2 glueballs have been proposed \cite{Juknevich:2009ji, Juknevich:2009gg}, but in this case the prospects for detection in the mass range of interest would also be low.
However, we do not preclude the possibility of a viable theory involving those fields which can be detected by light meson factories.

\subbib

%% file: strong-scale-operators.tex
\subtoc

\section{Portal operators at the strong scale} \label{sec:gev_portals}

In this appendix, we give a complete basis of both strangeness conserving and violating portal operators at the strong scale that are suppressed by at most a factor of $\flatfrac{\epsilonUV}{v^3}$ while respecting the general restrictions outlined in \cref{sec:currents}.
In particular, we assume that the strong scale \PETs are a \LE limit of a corresponding portal \SMEFT, and also include operators that are sub-leading in the $(4\pi)$ counting of \NDA.
The relevant leading strangeness violating \QCD operators are listed in \cref{tab:dominant low scale operators}, and the sub-leading strangeness violating operators are given in \cref{tab:subleading low scale operators}.
For spin 0 and \textfrac12 mediators, the relevant portal operators may be of dimension $d \leq 7$,
while for spin 1 mediators, the portal operators are of dimension $d \leq 6$.
This basis is constructed using the reduction techniques summarised in \cref{sec:redundant}, see also \cite{Klose:2020yhp} for additional details.

We follow the two component notation in \cite{Dreiner:2008tw} for fermionic fields, and distinguish between portal operators with either zero, two or four fermionic fields.
To list the operators, it is convenient to define stand-ins for various \SM $\SU(3)_c$ colour gauge singlets.
For \SM fermions, we define the following neutral pairs
\begin{subequations} \label{eq:neutral_colour_singlets}
\begin{align}
(q q)_0 & \in \left\{
\overline u_a u_b, \,
\overline d_a d_b
\right\} \ , &
(\psi \psi)_0^\prime & \in \left\{
(q q)_0, \,
\overline e_a e_b
\right\} \ , &
(\psi \psi)_0 & \in \left\{
(\psi \psi)_0^\prime, \,
\nu_a \nu_b
\right\} \ , \\
(q^\dagger q^\dagger)_0 & \in \left\{
\overline u_a^\dagger u_b^\dagger, \,
\overline d_a^\dagger d_b^\dagger
\right\} \ , &
(\psi^\dagger \psi)_0^\prime &
\begin{multlined}[t]
\in \{
u_a^\dagger u_b, \,
d_a^\dagger d_b,
\,
e_a^\dagger e_b,
\!\\
\overline u_a^\dagger \overline u_b, \,
\overline d_a^\dagger \overline d_b, \,
\overline e_a^\dagger \overline e_b
\}\ ,
\end{multlined} &
(\psi^\dagger \psi)_0 & \in \left\{
(\psi^\dagger \psi)_0^\prime, \,
\nu_a^\dagger \nu_b
\right\} \ ,
\end{align} \end{subequations}
and the following charged pairs
\begin{subequations} \label{eq:charged_colour_singlets}
\begin{align}
(q q)_+ & \in \left\{
\overline u_a d_b, \,
\overline d_a u_b
\right\} \ , &
(\psi^\dagger \psi)_+ & \in \left\{
d_a^\dagger u_b, \,
\overline d_a^\dagger \overline u_b, \,
e_a^\dagger \nu_b
\right\} \ , \\
(q^\dagger q^\dagger)_- & \in \left\{
\overline u_a^\dagger d_b^\dagger, \,
\overline d_a^\dagger u_b^\dagger
\right\} \ , &
(\psi^\dagger \psi)_- & \in \left\{
u_a^\dagger d_b, \,
\overline u_a^\dagger \overline d_b, \,
\nu_a^\dagger e_b
\right\} \ ,
\end{align} \end{subequations}
where the indices run over all available flavours at the strong scale
($a$, $b = \up$, $\down$, $\strange$ for quarks, $a$, $b = \electron$, $\muon$ for charged leptons and
$a$, $b = \neutrino_\electron$, $\neutrino_\muon$, $\neutrino_\tauon$ for neutrinos) and the subscript specifies the total electric charge of each fermionic pair.
For the gauge bosons, we indicate their field strength tensors with
\begin{equation}
V^{\mu\nu} \in \left\{ F^{\mu\nu}, G^{\mu\nu} \right\} \ .
\end{equation}
For operators with more that one occurrence of $V^{\mu\nu}$, we adopt the convention that all of these instances  denote the same field strength tensor within each operator.
For instance, the object $V^{\mu\nu}V_{\mu\nu}$ may denote $F^{\mu\nu}F_{\mu\nu}$ or $G^{\mu\nu}G_{\mu\nu}$ but \emph{not} $F^{\mu\nu}G_{\mu\nu}$.
The Fierz completeness relations \eqref{eq:scalar_tensor_fierz,eq:vector_vector_fierz,eq:scalar_vector_fierz} reduce the number of independent four-fermion operators.
We can then restrict ourselves to products of the colour singlets \eqref{eq:neutral_colour_singlets,eq:charged_colour_singlets} without loss of generality.
For operators without quarks, we can further restrict ourselves to products involving only the neutral singlets \eqref{eq:neutral_colour_singlets}.

\subsection{Scalar portal}

At order $\flatfrac{\epsilonUV}{v^3}$, the scalar portal operators can be of dimension seven or less.
We list all portal operators that include at most two hidden real scalar fields ($s_i$ with $i = 1\, 2$).

\paragraph{Zero-fermion operators}

can contain either one, two, or three field strength tensors.
The operators with one field strength tensor are
\begin{align}
& s_i \boson{\partial_\nu s_j} \boson{\partial_\mu s_k} F^{\mu\nu} \ , &
& s_i \boson{\partial_\nu s_j} \boson{\partial_\mu s_k} \widetilde F^{\mu\nu} \ .
\end{align}
The operators with two field strength tensors are
\begin{subequations} \begin{align}
& s_i \trc{V_{\mu\nu} V^{\mu\nu}} \ , &
& s_i s_j \trc{V_{\mu\nu} V^{\mu\nu} } \ , &
& s_i s_j s_k \trc{V_{\mu\nu} V^{\mu\nu} } \ , &
& s_i \trc{D_\rho V_{\mu\nu} D^\rho V^{\mu\nu} } \ , \\
& s_i \trc{V_{\mu\nu} \widetilde V^{\mu\nu} } \ , &
& s_i s_j \trc{V_{\mu\nu} \widetilde V^{\mu\nu} } \ , &
& s_i s_j s_k \trc{V_{\mu\nu} \widetilde V^{\mu\nu} } \ , &
& s_i \trc{D_\rho V_{\mu\nu} D^\rho \widetilde V^{\mu\nu} } \ .
\end{align} \end{subequations}
The operators with three field strength tensors are
\begin{align}
& s_i \trc{G^\nu_\mu G^\mu_\rho G^\rho_\nu } \ , &
& s_i \trc{\widetilde G^\nu_\mu G^\mu_\rho G^\rho_\nu } \ .
\end{align}

\paragraph{Two-fermion operators}

can contain at most a single \SM field strength tensor.
The operators without field strength tensor may contain no more than two derivatives.
The operators with zero derivatives are
\begin{align}
& s_i (\psi \psi)_0 \ , &
& s_i s_j (\psi \psi)_0 \ , &
& s_i s_j s_k (\psi \psi)_0 \ , &
& s_i s_j s_k s_l (\psi \psi)_0 \ .
\end{align}
The operators with one derivative are
\begin{align}\label{eq:oneder}
& \boson{s_i \overleftright \partial_\mu s_j} (\psi^\dagger \overline \sigma^\mu \psi)_0 \ , &
& s_i \boson{s_j \overleftright \partial_\mu s_k} (\psi^\dagger \overline \sigma^\mu \psi)_0 \ .
\end{align}
The operators with two derivatives are
\begin{align}
& \partial^2 s_i (\psi \psi)_0 \ , &
& s_i \partial^2 s_j (\psi \psi)_0 \ , &
& \boson{\partial_\mu s_i} \boson{\partial^\mu s_j} (\psi \psi)_0 \ .
\end{align}
The operators with a single \SM field strength tensor and no derivatives are
\begin{align}
& s_i (\psi \overline \sigma_{\mu\nu} V^{\mu\nu} \psi)_0 \ , &
& s_i s_j (\psi \overline \sigma_{\mu\nu} V^{\mu\nu} \psi)_0 \ .
\end{align}
The operators with a single \SM field strength tensor and one derivative are
\begin{subequations}
\begin{align}
& \boson{\partial_\nu s_i} (\psi^\dagger \overline \sigma_\mu V^{\mu\nu} \psi)_0 \ , &
& \boson{\partial_\nu s_i} (\psi^\dagger \overline \sigma_\mu \widetilde V^{\mu\nu} \psi)_0 \ .
\end{align}
\end{subequations}
All the operators above are accompanied by their Hermitian conjugate.

\paragraph{Four-fermion operators}

cannot contain any derivatives or field strength tensors.
They are
\begin{align}
& s_i (\psi^\dagger \overline \sigma^\mu \psi)_0 (\psi^\dagger \overline \sigma_\mu \psi)_0 \ , &
& s_i (\psi^\dagger \overline \sigma^\mu \psi)_+ (\psi^\dagger \overline \sigma_\mu \psi)_- \ , &
& s_i (q q)_0 (q^\dagger q^\dagger)_0 \ , &
& s_i (q q)_+ (q^\dagger q^\dagger)_- \ , &
\end{align}
plus Hermitian conjugates. The operator $s_i (\psi^\dagger \overline \sigma^\mu \psi)_+ (\psi^\dagger \overline \sigma_\mu \psi)_-$ contains only combinations with either two or four quarks.
Using the Fierz identity \eqref{eq:vector_vector_fierz}, combinations with four leptons can be eliminated in favour of operators contained within $s_i (\psi^\dagger \overline \sigma^\mu \psi)_0 (\psi^\dagger \overline \sigma_\mu \psi)_0$.
At order $\epsilonEW$, there are no operators $s_i (\psi\psi)(\psi\psi)$ or $s_i (\psi \overline \sigma^{\mu\nu} \psi) (\psi \overline \sigma_{\mu\nu} \psi)$,
since these involve at least two chirality flips for the \SM fermions, suppressing them further by an additional  factor of $\sqrt \epsilonEW \propto \flatfrac{m_\psi}v$.

\subsection{Fermionic portal}

At order $\flatfrac{\epsilonUV}{v^3}$, a fermionic portal particle can couple to the \SM via operators up to dimension $d \leq 7$.
These operators can contain either two or four fermions.
As before, it is sufficient to list portal operators with two hidden left-handed Weyl fermions $\lhf_i$ with $i = 1$, $2$ to account for both Majorana and Dirac fermionic fields in general.

\paragraph{Two-fermion operators}

can contain either zero, one, or two \SM field strength tensors.
The sole operator without field strength tensors is
\begin{align}
& \nu_a \lhf_i \ .
\end{align}
The operators with one field strength tensor are
\begin{align}
& \nu_a \overline \sigma_{\mu\nu} \lhf_i F^{\mu\nu} \ , &
& \lhf_i \overline \sigma_{\mu\nu} \lhf_j F^{\mu\nu} \ , &
& \lhf_i \overline \sigma_{\mu\nu} D^2 \lhf_j F^{\mu\nu} \ , &
& \nu_a \overline \sigma_{\mu\nu} D^2 \lhf_i F^{\mu\nu} \ , &
\end{align}
and
\begin{subequations}
\begin{align}
& \nu_a ^\dagger \overline \sigma_\mu D_\nu \lhf_i F^{\mu\nu} \ , &
& \lhf_i^\dagger \overline \sigma_\mu D_\nu \lhf_j F^{\mu\nu} \ , &
& \lhf_i \overline \sigma_{\mu \rho} D_\nu D^\mu \lhf_j F^{\nu \rho} \ , &
& \nu_a \overline \sigma_{\mu \rho} D_\nu D^\mu \lhf_i F^{\nu \rho} \ , \\
& \nu_a^\dagger \overline \sigma_\mu D_\nu \lhf_i \widetilde F^{\mu\nu} \ , &
& \lhf_i^\dagger \overline \sigma_\mu D_\nu \lhf_j \widetilde F^{\mu\nu} \ , &
& \lhf_i \overline \sigma_{\mu \rho} D_\nu D^\mu \lhf_j \widetilde F^{\nu \rho} \ , &
& \nu_a \overline \sigma_{\mu \rho} D_\nu D^\mu \lhf_i \widetilde F^{\nu \rho} \ .
\end{align}
\end{subequations}
The operators with two field strength tensors are
\begin{subequations}
\begin{align}
& \nu_a \lhf_i \trc{V^{\mu\nu} V_{\mu\nu}} \ , &
& \lhf_i \lhf_j \trc{V^{\mu\nu} V_{\mu\nu} } \ , &
& \lhf_i \sigma_{\mu\nu} \lhf_j \trc{V_\rho^\mu V^{\rho\nu} } \ , &
& \nu_a \sigma_{\mu\nu} \lhf_i \trc{V_\rho^\mu V^{\rho\nu} } \ , \\
& \nu_a \lhf_i \trc{V^{\mu\nu} \widetilde V_{\mu\nu} } \ , &
& \lhf_i \lhf_j \trc{V^{\mu\nu} \widetilde V_{\mu\nu} } \ , &
& \lhf_i \sigma_{\mu\nu} \lhf_j \trc{V_\rho^\mu \widetilde V^{\rho\nu} } \ , &
& \nu_a \sigma_{\mu\nu} \lhf_i \trc{V_\rho^\mu \widetilde V^{\rho\nu} } \ ,
\end{align}
\end{subequations}
All operators are accompanied by their Hermitian conjugate.

\paragraph{Four-fermion operators}

can contain at most one derivative.
The operators without derivatives can be either of the scalar-scalar type or of the vector-vector type.
The former are
\begin{subequations}
\begin{align}
& \boson{\lhf_i \lhf_j} \boson{\nu_a \lhf} \ , &
& \boson{\overline d_a u_b} \boson{e_c \lhf_i} \ , &
& (\psi \psi)_0^\prime \boson{\nu_a \lhf_i} \ , &
& (\psi \overline \sigma_{\mu\nu} \psi)_0^\prime \boson{\nu_a \overline \sigma^{\mu\nu} \lhf_i} \ , \\
&&
&&
& (\psi \psi)_0^\prime \boson{\lhf_i \lhf_j} \ , &
& (\psi \overline \sigma_{\mu\nu} \psi)_0^\prime \boson{\lhf_i \overline \sigma^{\mu\nu} \lhf_j} \ ,
\end{align}
and
\begin{align}
& (\psi \psi)_0^\prime \boson{\nu_a^\dagger \lhf_i^\dagger} \ , &
& (\psi \psi)_0^\prime \boson{\lhf_i^\dagger \lhf_j^\dagger} \ , &
& \boson{\overline u_a d_b} \boson{e_a^\dagger \lhf_i^\dagger} \ ,.
\end{align}
\end{subequations}
The vector-vector type operators are
\begin{subequations}
\begin{align}
& \boson{\nu_a^\dagger \overline \sigma^\mu \lhf_i} \boson{\lhf_j^\dagger \overline \sigma_\mu \lhf_k} \ , &
& \boson{d_a^\dagger \overline \sigma^\mu u_b} \boson{\overline e_a^\dagger \overline \sigma_\mu \lhf_i} \ , &
& \boson{u_a^\dagger \overline \sigma^\mu d_b} \boson{e_a^\dagger \overline \sigma_\mu \lhf_i} \ , &
& (\psi^\dagger \overline \sigma_\mu \psi)_0 \boson{\nu_a^\dagger \overline \sigma^\mu \lhf_i} \ , & \\
& & & & & & & (\psi^\dagger \overline \sigma_\mu \psi)_0 \boson{\lhf_i^\dagger \overline \sigma^\mu \lhf_j} \ .
\end{align}
\end{subequations}
The operators with one derivative are
\begin{subequations}
\begin{flalign}
(\psi^\dagger \overline \sigma_\mu \psi)_0 \boson{\nu_a D^\mu \lhf_i} , &&\!
(\psi^\dagger \overline \sigma_\mu \psi)_0 \boson{\nu_a \overline \sigma^{\mu\nu} D_\nu \lhf_i} , &&\!
(\psi^\dagger \overline \sigma_\mu \psi)_0 \boson{\lhf_i D^\mu \lhf_j} , &&\!
(\psi^\dagger \overline \sigma_\mu \psi)_0 \boson{\lhf_i \overline \sigma^{\mu\nu} D_\nu \lhf_j} , \\
\boson{d_a^\dagger \overline \sigma^\mu u_b} \boson{e_c D^\mu \lhf_i} , &&
\boson{d_a^\dagger \overline \sigma^\mu u_b} \boson{e_c \overline \sigma^{\mu\nu} D_\nu \lhf_i} , &&
\boson{\lhf_i^\dagger \overline \sigma^\mu\nu_a} \boson{\lhf_j D_\mu \lhf_k} , &&
\boson{\lhf_i^\dagger \overline \sigma_\mu\nu_a} \boson{\lhf_j \sigma^{\mu\nu} D_\nu \lhf_k} , \\ &&&&
\boson{\nu_a^\dagger \overline \sigma^\mu \lhf_i} \boson{\lhf_j D_\mu \lhf_k} , &&
\boson{\nu_a^\dagger \overline \sigma_\mu \lhf_i} \boson{\lhf_j \overline \sigma^{\mu\nu} D_\nu \lhf_k} \,.
\end{flalign}
\end{subequations}
All operators are accompanied by their Hermitian conjugate.

\subsection{Vector portal}

At the \EW scale, spin 1 messengers do not couple to \SMEFT via operators of dimension five, hence the corresponding \LE portal Lagrangian can only contain interactions that are suppressed at most by a factor of $\flatfrac{\epsilon}{v^2}$ rather than $\flatfrac{\epsilon}{v^3}$.
At order $\flatfrac{\epsilon}{v^2}$, hidden \prefix{axial}{vector} mediators couple to the \SM via operators of dimension $d \leq 6$ only.
Therefore, there are no portal operators with four \SM fermions, since they would be at least of dimension seven.

It is convenient to define
\begin{align}
\partial v & \equiv \partial^\rho v_\rho \ , &
v_{\mu\nu} & \equiv \partial_{[\mu} v_{\nu]} \ , &
\widehat v_{\mu\nu} & \equiv \partial_{\{\mu} v_{\nu\}} \ , &
\widetilde v_{\mu\nu} & \equiv 2 \epsilon_{\mu\nu \rho \sigma} \partial^\rho v^\sigma \ .
\end{align}

\paragraph{Zero-fermion operators}

can contain either one or two \SM field strength tensors.
The operators with one field strength tensor are
\begin{subequations}
\begin{align}
& v^\rho v_\rho v_{\mu\nu} F^{\mu\nu} \ , &
& v_\nu v^\mu \widehat v_{\mu \rho} F^{\nu \rho} \ , &
& v_\nu v^\mu v_{\mu \rho} \widetilde F^{\nu \rho} \ , &
& \widetilde v^{\mu\nu} \widehat v_{\mu \rho} F^\rho_\nu \ , \\
& v^\rho v_\rho \widetilde v_{\mu\nu} F^{\mu\nu} \ , &
& v_\nu v^\mu v_{\mu \rho} F^{\nu \rho} \ , &
& v_\nu v^\mu \widetilde v_{\mu \rho} F^{\nu \rho} \ , &
& \widetilde v^{\mu\nu} v_{\mu \rho} F^\rho_\nu \ ,
\end{align}
\end{subequations}
and
\begin{subequations}
\begin{align}
& \partial v v_{\mu\nu} F^{\mu\nu} \ , &
& v_{\mu\nu} v_\rho \partial^\rho F^{\mu\nu} \ , &
& v_\mu \partial^2 v_\nu F^{\mu\nu} \ , \\
& \partial v v_{\mu\nu} \widetilde F^{\mu\nu} \ , &
& v_{\mu\nu} v_\rho \partial^\rho \widetilde F^{\mu\nu} \ , &
& v_\mu \partial^2 v_\nu \widetilde F^{\mu\nu} \ .
\end{align}
\end{subequations}
The operators with two field strength tensors are
\begin{subequations}
\begin{align}
& v^\rho v_\rho \trc{V^{\mu\nu} V_{\mu\nu} } \ , &
& v_\mu v^\nu \trc{V^{\mu \rho} V_{\nu \rho} } \ , &
& \partial v \trc{V^{\mu\nu} V_{\mu\nu} } \ , \\
& v^\rho v_\rho \trc{V^{\mu\nu} \widetilde V_{\mu\nu} } \ , &
& v_\mu v^\nu \trc{V^{\mu \rho} \widetilde V_{\rho\nu} } \ , &
& \partial v \trc{V^{\mu\nu} \widetilde V_{\mu\nu} } \ , &
& \partial_\mu v^\nu \trc{V^{\mu \rho} \widetilde V_{\rho\nu} } \ .
\end{align}
\end{subequations}

\paragraph{Two-fermion operators}

can contain a scalar-valued, vector-valued, or tensor-valued \SM fermion bilinear.
The operators with scalar- and tensor-valued fermion bilinears are
\begin{align}
& v^\rho v_\rho (\psi \psi)_0 \ , &
& \partial v (\psi \psi)_0 \ , &
& v_{\mu\nu} (\psi \overline \sigma^{\mu\nu} \psi)_0 \ ,
\end{align}
plus Hermitian conjugates.
The operators with vector-valued fermion bilinears are
\begin{subequations}
\begin{align}
v_\mu (\psi^\dagger \overline \sigma^\mu \psi)_0 & \ , &
\widehat v_{\mu\nu} (\psi^\dagger \overline \sigma^\mu D^\nu \psi)_0 & \ , &
v^\mu (\psi^\dagger \overline \sigma^\nu V_{\mu\nu} \psi)_0 & \ , &
v_{\mu\nu} v^\nu (\psi^\dagger \overline \sigma^\mu \psi)_0 & \ , \\
v^\rho v_\rho v_\mu (\psi^\dagger \overline \sigma^\mu \psi)_0 & \ , &
v_\mu v_\nu (\psi^\dagger \overline \sigma^\mu D^\nu \psi)_0 & \ , &
v^\mu (\psi^\dagger \overline \sigma^\nu \widetilde V_{\mu\nu} \psi)_0 & \ , &
\widehat v_{\mu\nu} v^\nu (\psi^\dagger \overline \sigma^\mu \psi)_0 & \ , \\&&&&&&
\widetilde v_{\mu\nu} v^\nu (\psi^\dagger \overline \sigma^\mu \psi)_0 & \ ,
\end{align}
and
\begin{align}
\partial v v_\mu (\psi^\dagger \overline \sigma^\mu \psi)_0 & \ , &
\partial^2 v_\mu (\psi^\dagger \overline \sigma^\mu \psi)_0 & \ .
\end{align}
\end{subequations}

%% file: expansion.tex
\subtoc

\section{Expansion of the $\cPT$ building blocks} \label{sec:expansions}

In this appendix we provide details about the expansion of the chiral Lagrangians in terms of light mesons and hidden particle states.
This material covers the necessary steps to derive the results of \cref{sec:light mesons} and provides the reader with the necessary tools to use the results obtained in \cref{sec:chiral perturbation theory} and \cref{sec:light mesons} for their own calculations.

The matrix $\nonet \L_\mu$ and the hatted external currents $\Left{\nonet X} \in \{\Left{\nonet \M}$, $\Left{\nonet \Chromo}$, $\Left{\nonet T}^{\mu\nu}\}$ and $\Left{\nonet Y} \in \{\Left{\nonet R} _\mu$, $\Left{\nonet R}_{\mu\nu}\}$ can be expanded as
\begin{subequations}
\begin{align}
\Left{\nonet X} = \nonet g \nonet X &
= \left(\flavour 1 + \frac{\i}{\fp} \nonet \Phi - \frac{1}{2 \fp^2} \nonet \Phi^2 - \frac{\i}{6 \fp^3} \nonet \Phi^3 + \dots\right) \nonet X \ , \\
\Left{\nonet Y} = \nonet g \nonet Y \nonet g^\dagger &
= \nonet Y + \frac{\i}{\fp} \comm{\octet \Phi}{\octet Y} - \frac{1}{2 \fp^2} \comm{\octet \Phi}{\comm{\octet \Phi}{\octet Y}} - \frac{\i}{6 \fp^3} \comm{\octet \Phi}{\comm{\octet \Phi}{\comm{\octet \Phi}{\octet Y}}} + \dots \ , \\
\nonet \L_\mu = \i \nonet g \partial_\mu \nonet g^\dagger &
= \frac1{\fp} \partial_\mu \nonet \Phi + \frac{\i}{2 \fp^2} \comm{\octet \Phi}{\partial_\mu \octet \Phi} - \frac1{6 \fp^3} \comm{\octet \Phi}{\comm{\octet \Phi}{\partial_\mu \octet \Phi}} + \dots \ .
\end{align}
\end{subequations}
Using the definition of the meson matrix $\nonet \Phi$
\begin{align}
\nonet \Phi &= \octet \Phi + \frac{\flavour 1}{\nfl} \Phi \ , &
\octet \Phi &
= \begin{pmatrix}
\etaoctet + \pion & \pi^+ & K^+ \\
\pi^- & \etaoctet - \pion & K^0 \\
K^- & \overline K^0 & - 2 \etaoctet
\end{pmatrix} \ , &
\Phi &= \nfl \etasinglet \ ,
\end{align}
one obtains the individual contributions
\begin{subequations}
\begin{flalign}
\octet \Phi^2 &=
\begin{psmallmatrix}
(\etaoctet + \pion)^2 + \pi^+\pi^- + K^+K^- & 2 \pi^+ \etaoctet + K^+ \overline K^0 & \pi^+ K^0 + K^+ (\pion - \etaoctet) \\
2 \pi^-\etaoctet + K^- K^0 & \pi^+\pi^- + (\etaoctet - \pion)^2 + K^0 \overline K^0 & K^+\pi^- - K^0 \left(\pion + \etaoctet\right) \\
\pi^- \overline K^0 + K^- (\pion - \etaoctet) & \pi^+K^- - \overline K^0 \left(\pion + \etaoctet\right) & K^+ K^- + K^0 \overline K^0 + 4 (\etaoctet)^2
\end{psmallmatrix} \ , \\
\comm{\octet \Phi}{\partial_\mu \octet \Phi} &=
\begin{psmallmatrix}
\pi^+ \overleftright \partial_\mu \pi^- + K^+ \overleftright \partial_\mu K^-
 & 2 \pi^+ \overleftright \partial_\mu \pion + K^+ \overleftright \partial_\mu \overline K^0
 & \pi^+ \overleftright \partial_\mu K^0 - K^+ \overleftright \partial_\mu(\pion + 3\etaoctet) \\
2 \pi^- \overleftright \partial_\mu \pion + K^0 \overleftright \partial_\mu K^-
 & - \pi^+ \overleftright \partial_\mu \pi^- + K^0 \overleftright \partial_\mu \overline K^0
 & - K^+ \overleftright \partial_\mu \pi^- + K^0 \overleftright \partial_\mu (\pion - 3\etaoctet) \\
\overline K^0 \overleftright \partial_\mu \pi^- + K^- \overleftright \partial_\mu (\pion + 3\etaoctet)
 & - \pi^+ \overleftright \partial_\mu K^- + \overline K^0 \overleftright \partial_\mu (3\etaoctet-\pion)
 & - K^+ \overleftright \partial_\mu K^- - K^0 \overleftright \partial_\mu \overline K^0
\end{psmallmatrix} \ .
\end{flalign}
\end{subequations}
The interactions involving the \SM photon current are
\begin{subequations}
\begin{align}
\comm{\octet \Phi}{\octet r_\photon^\mu} &
= e \photon^\mu
\begin{pmatrix}
0 & - \pi^+ & - K^+ \\
\pi^- & 0 & 0 \\
K^- & 0 & 0
\end{pmatrix} \ , \\
\comm{\octet \Phi}{\comm{\octet \Phi}{\octet r_\photon^\mu}} &
= - e \photon^\mu
\begin{psmallmatrix}
- 2 (\pi^+ \pi^- + K^+K^-) & 2 \pi^+ \pion + K^+ \overline K^0 & K^+ (3\etaoctet + \pion) + \pi^+ K^0 \\
2\pi^- \pion + K^- K^0 & 2 \pi^+ \pi^- & 2 K^+ \pi^- \\
K^- (3\etaoctet + \pion) + \pi^- \overline K^0 & 2 \pi^+ K^- & 2 K^+ K^-
\end{psmallmatrix}
\ .
\end{align}
\end{subequations}
Finally, for interactions involving the hidden current $\nonet V_r$, one has
\begin{flalign}
\commutator{\octet V_r}{\octet \Phi} &=
\begin{psmallmatrix}
0 &
\pi ^+ \left(V_r{}_{\up\up} - V_r{}_{\down\down}\right)-K^+ V_r{}_{\strange\down} &
K^+ \left(V_r{}_{\up\up}-V_r{}_{\strange\strange}\right)-\pi ^+ V_r{}_{\down\strange} \\
\pi^- \left(V_r{}_{\down\down} - V_r{}_{\up\up}\right) + K^- V_r{}_{\down\strange} &
\overline K^0 V_r{}_{\down\strange} - K^0 V_r{}_{\strange\down} &
K^0 \left(V_r{}_{\down\down} - V_r{}_{\strange\strange}\right) + V_r{}_{\down\strange} \left(\pion - 3 \etaoctet\right) \\
K^- \left(V_r{}_{\strange\strange} - V_r{}_{\up\up}\right) + \pi^- V_r{}_{\strange\down} &
\overline K^0 \left(V_r{}_{\strange\strange} - V_r{}_{\down\down}\right) + \left(3 \etaoctet - \pion\right) V_r{}_{\strange\down} &
K^0 V_r{}_{\strange\down} - \overline K^0 V_r{}_{\down\strange}
\end{psmallmatrix}
\ .
\end{flalign}

\subsection{Standard model meson phenomenology at $\NLO$} \label{sec:SM meson masses}

We summarise the diagonalisation procedure for the $\U(3)$ \cPT mesons  and compute the resulting meson masses and decay constants at \NLO.
We use the approximation $m_\up^\prime$, $m_\down^\prime \to m_\light \equiv \flatfrac{(m_\up^\prime + m_\down^\prime)}2$,
which neglects the mixing between the neutral pion and the two $\eta$-mesons and we also neglect \EM corrections for the charged meson masses, which are of order $\alpha_{\EM} \propto e^2$.
These \EM contributions are given by
\begin{align}
\Delta_\pi^{\EM}  &= m_{\pi^\pm}^2 - m_{\pi^0}^2 \ , &
\Delta_K^{\EM} &= \left(1 + (0.84 \pm 0.25_{\Nc})\right) \Delta_\pi^{\EM}
\end{align}
Where the correction factor captures the impact of \NLO contributions \cite{Bijnens:1996kk,Bijnens:2011tb}.
We use the \EM contributions in combination with the measured values of the meson masses \cite{Zyla:2020zbs}
\begin{subequations} \label{eq:charged meson observables}
\begin{align}
m_{\pi^\pm} &= \unit[(139.57039\pm0.00017_{\ex})]{MeV} \ , &
m_{K^\pm} &= \unit[(493.677\pm0.013_{\ex})]{MeV} \ , \\
m_{\pi^0} &= \unit[(134.9768\pm0.0005_{\ex})]{MeV} \ , &
m_{K^0} &= \unit[(497.611\pm0.013_{\ex})]{MeV} \ .
\end{align}
\end{subequations}

\paragraph{Meson decay constants}

The part of the \NLO Lagrangian that mediates charged meson decays is
\begin{align}
\mathcal L_\phi^{\partial W} + \mathcal L_\phi^{m \partial W} &
= - \fp \trf{\octet l_W^\mu \partial_\mu \octet \Phi} - \frac{2L_5 \B}{\fp} \trf{\octet l_W^\mu \acomm{\nonet m}{\partial_\mu \octet \Phi}} \ .
\end{align}
The resulting predictions for the meson decay constants are
\begin{align}
\frac{f_\pi}{\fp} &= 1 + 4 L_5 \frac{m_\pi^2}{\fp^2} + \order{\delta^3} \ , &
\frac{f_K}{\fp} &= 1 + 4 L_5 \frac{m_K^2}{\fp^2} + \order{\delta^3} \ ,
\end{align}
or equivalently
\begin{align}
\fp &= \frac{m_K^2 f_\pi - m_\pi^2 f_K}{m_K^2 - m_\pi^2} \ , &
4 L_5 &=
\fp \frac{f_\pi - \fp}{m_\pi^2}
= \fp \frac{f_K - \fp}{m_K^2}
\ ,
\end{align}
where
\begin{align}
m_\pi^2 &= m_{\pi^0}^2 \ , &
2 m_K^2 &= m_{K^\pm}^2 + m_{K^0}^2 - \Delta_K^{\EM} \ ,
\end{align}
are
the charged meson masses without electromagnetic contributions.
To fix the values of the parameters $\fp$ and $L_5$, we use the measured values of the meson decay constants \cite{Zyla:2020zbs}
\begin{align} \label{eq:charged meson observables}
f_\pi &= \unit[(65.1\pm0.6_{\ex})]{MeV} \ , &
f_K &= \unit[(77.85\pm0.15_{\ex})]{MeV} \ .
\end{align}
Hence, one obtains the estimates
\begin{align}
\fp &= \unit[(63.9 \pm 1.2_{\ex} \pm \NNLO)]{MeV} \ , &
4 (4\pi)^2 L_5 &= 0.66 \pm 0.04_{\ex} \pm \NNLO \ .
\end{align}

\paragraph{Masses and mixing angles}

After diagonalising the neutral kaon sector via the field redefinition
\begin{align}\label{eq:kskl}
\sqrt 2 K^0_L &= K^0 + \overline K^0 \ , &
- \i \sqrt 2 K^0_S &=  K^0 - \overline K^0 \ ,
\end{align}
one obtains from the Lagrangians \eqref{eq:u mass Lagrangians} the mass term
\begin{equation}
\mathcal L_{\phi^2}^{\prime m} + \mathcal L_{\phi^2}^\theta =
- m_\pi^2 \pi^+ \pi^- - m_K^2 K^+ K^-
- \frac12 \left(m_\pi^2 {\pi^0}^2 + m_K^2 \left({K^0_L}^2 + {K^0_S}^2\right) + \eta_2^T \flavour m_{\eta_2}^2 \eta_2\right)
\ .
\end{equation}
The \NLO predictions for the pion and kaon mass parameters are
\begin{align} \label{eq:charged meson mass prediction}
m_\pi^2 &= \B m_\light \left(1 + 8 L_8 \frac{\B m_\light}{\fp^2}\right) \ , &
m_K^2 &= \frac{\B}{2} (m_\strange^\prime + m_\light) \left(1 + 4 L_8 \frac{\B (m_\strange^\prime + m_\light)}{\fp^2}\right) \ ,
\end{align}
while the prediction for the mass matrix of the the $\eta$-meson doublet $\eta_2 = (\eta_8, \eta_1)^T$ is
\begin{equation}
\flavour m_{\eta_2}^2 =
\begin{pmatrix}
m_{\eta_8}^2 & m_{\eta_8\eta_1}^2 \\
m_{\eta_8\eta_1}^2 & m_{\eta_1}^2
\end{pmatrix} =
\left(M_K^2 - \frac{\Delta_{K\pi}}2\right) \flavour 1_{2\times2}
+ \begin{pmatrix}
\Delta_{K\pi} & - \sqrt 2 \Delta_{K\pi} \\
- \sqrt 2 \Delta_{K\pi} & M_0^2
\end{pmatrix}
+ \order{\delta^3} \ ,
\end{equation}
where the quantities
\begin{subequations}
\begin{align}
M_K^2 &= m_K^2 + \frac23 (m_K^2 - m_\pi^2) \left(\Lambda_2 + 3\frac{4 L_8}{\fp^2} (m_K^2 - m_\pi^2)\right) \ , &
M_0^2 &= m_0^2 - 2 \Lambda_2 m_K^2 \ , \\
\Delta_{K\pi} &= \frac23 \left(m_K^2 - m_\pi^2\right) \left(1 - \Lambda_2 + 4\frac{4 L_8}{\fp^2} m_K^2\right)
\end{align}
\end{subequations}
depend on the kaon and pion masses as well the three parameters $\mth$, $\Lambda_2$, and $L_8$.
The $\eta_2$ mass eigenstates are
\begin{align}
\eta &= c_\eta \eta_8 - s_\eta \eta_1 \ , &
\eta^\prime &= c_\eta \eta_1 + s_\eta \eta_8 \ , &
m_\eta^2 + m_{\eta^\prime}^2 &= \trace \flavour m_{\eta_2}^2 \ , &
m_\eta^2 m_{\eta^\prime}^2 &= \det \flavour m_{\eta_2}^2 \ ,
\end{align}
and their mixing is determined by
\begin{align}
m_{\eta_8}^2 &= \frac{m_\eta^2 + m_{\eta^\prime}^2 t_\eta^4}{1+t_\eta^4} \ , &
m_{\eta_1}^2 &= \frac{m_{\eta^\prime}^2 + m_\eta^2 t_\eta^4}{1+t_\eta^4} \ , &
m_{\eta_8\eta_1}^2 &= \frac{(m_{\eta^\prime}^2 - m_\eta^2 )t_\eta^2}{1+t_\eta^4} \ .
\end{align}
where the sine, cosine, and tangent functions of the $\eta$ mixing angle is indicated by $s_\eta$, $c_\eta$, and $t_\eta$, respectively.
In order to fix the values of the free parameters we fit the above predictions to the experimentally obtained values for the $\eta$-meson masses and mixing angle \cite{Gan:2020aco,Zyla:2020zbs}
\begin{subequations}
\begin{align}
m_\eta &= \unit[(547.862\pm0.018_{\ex})]{MeV} \ , &
t_\eta &= - 0.29\pm0.09_{\ex} \ , \\
m_{\eta^\prime} &= \unit[(957.78\pm0.06_{\ex})]{MeV} \ .
\end{align}
\end{subequations}
Hence, one obtains the estimates
\begin{subequations}\label{eq:eta estimates}\begin{align}
\mth &= 4 \pi \unit[(76.3 \pm 1.4_{\ex} \pm \NNLO)]{MeV} \ , &
\Lambda_2 &= 0.814 \pm 0.023_{\ex} \pm \NNLO\ , \\ &&
4 (4\pi)^2 L_8 &= 0.215 \pm 0.033_{\ex} \pm \NNLO \ .
\end{align} \end{subequations}
Finally, using \cref{eq:charged meson mass prediction} to fix the values of the parameters $\B m_\light$ and $\B m_\strange$, results in
\begin{subequations}
\begin{align}
\sqrt{\B m_\light} &= 4\pi \unit[(10.68 \pm 0.08_{\ex} \pm \NNLO)]{MeV} \ , \\
\sqrt{\B m_\strange} &= 4\pi\unit[(50.95 \pm 0.28_{\ex} \pm \NNLO)]{MeV} \ .
\end{align}
\end{subequations}

\paragraph{Weak interaction induced kinetic mixing}

When computing matrix elements for quark-flavour violating transitions, one also has to account for kinetic mixing due to weak corrections,
which is captured by the quadratic part of the octet and 27-plet Lagrangians
\begin{multline}
\mathcal L_{\phi^2}^{a\partial^2} + \mathcal L_{\phi^2}^{\tensor a\partial^2} =
- \frac{\epsilonEW}2 \left(\left(\fermi_8 + (\nfl - 1) \fermi_{27}\right) \partial_\mu K^+ \partial^\mu \pi^- + \text{h.c.}\right) \\
- \epsilonEW \left(- \Re \fermi_8 + \nfl \Re \fermi_{27}\right) \partial_\mu \Klong \partial^\mu \neutpion \\
- \epsilonEW \left(- \Im \fermi_8 + \nfl \Im \fermi_{27}\right) \partial_\mu \Kshort \partial^\mu \neutpion \\
- \epsilonEW \left[\nfl \Re \fermi_1 2 \varepsilon_{\eta\eta^\prime} + \left(- \Re \fermi_8 + \nfl \Re \fermi_{27}\right)\right] \partial_\mu \Klong \partial^\mu \etaparticle \\
- \epsilonEW \left[\nfl \Im \fermi_1 2 \varepsilon_{\eta\eta^\prime} + \left(- \Im \fermi_8 + \nfl \Im \fermi_{27}\right)\right] \partial_\mu \Kshort \partial^\mu \etaparticle \\
+ \epsilonEW \left[\nfl \Re \fermi_1 + \left(- \Re \fermi_8 + \nfl \Re \fermi_{27}\right) \varepsilon_{\eta\eta^\prime}\right] \partial_\mu \Klong \partial^\mu \etaprime \\
+ \epsilonEW \left[\nfl \Im \fermi_1 + \left(- \Im \fermi_8 + \nfl \Im \fermi_{27}\right) \varepsilon_{\eta\eta^\prime}\right] \partial_\mu \Kshort \partial^\mu \etaprime \ ,
\end{multline}
To \LO in $\epsilonEW$, these interactions can be diagonalised via the field redefinitions
\begin{subequations}
\label{eq:weak induced meson mixing}
\begin{align}
\begin{pmatrix}
\pi^+ \\ K^+
\end{pmatrix} & \to
\left[
\flavour 1_{2\times2} +
\frac{\epsilonEW}2
\begin{pmatrix}
0 & m_{K^\pm}^2 \theta_{K^\pm\pi^\mp} \\
- m_{\pi^\pm}^2 \theta_{K^\pm\pi^\mp}^\dagger & 0
\end{pmatrix}
\right]
\begin{pmatrix}
\pi^+ \\ K^+
\end{pmatrix} \ , \\
\begin{pmatrix}
\pi^0 \\ K^0_L \\ K^0_S \\ \eta \\ \eta^\prime
\end{pmatrix} & \to
\left[
\flavour 1_{5\times5} +
\frac{\epsilonEW}2
\begin{pmatrix}
0 & m_{K^0}^2 \theta_{\pi K}^T & 0 & 0 \\
- m_{\pi^0}^2 \theta_{\pi K} & \flavour 0_{2\times2} & - m_\eta^2 \theta_{\eta K} & - m_{\eta^\prime}^2 \theta_{\eta^\prime K} \\
0 & m_{K^0}^2 \theta_{\eta K}^T & 0 & 0 \\
0 & m_{K^0}^2 \theta_{\eta^\prime K}^T & 0 & 0
\end{pmatrix}
\right]
\begin{pmatrix}
\pi^0 \\ K^0_L \\ K^0_S \\ \eta \\ \eta^\prime
\end{pmatrix} \ ,
\end{align}
\end{subequations}
where the mixing angles are
\begin{subequations}
\begin{align}
\theta_{K^\pm\pi^\mp} &\equiv
\frac{1}{m_{K^\pm}^2 - m_{\pi^\pm}^2} \left(\fermi_8 + (\nfl - 1)\fermi_{27}\right)\ , \\
\theta_{\pi K} &\equiv
\frac1{m_{K^0}^2 - m_{\pi^0}^2}
\begin{pmatrix}
\nfl \Re \fermi_1 2 \varepsilon_{\pi\eta^\prime} + \left(- \Re \fermi_8 + \nfl \Re \fermi_{27}\right) (1+\varepsilon_{\pi\eta}) \\
\nfl \Im \fermi_1 2 \varepsilon_{\pi\eta^\prime} + \left(- \Im \fermi_8 + \nfl \Im \fermi_{27}\right) (1+\varepsilon_{\pi\eta})
\end{pmatrix} \ , \\
\theta_{\eta K} &\equiv
\frac1{m_{K^0}^2 - m_\eta^2}
\begin{pmatrix}
\nfl \Re \fermi_1 2 \varepsilon_{\eta\eta^\prime} + \left(- \Re \fermi_8 + \nfl \Re \fermi_{27}\right) (1-3 \varepsilon_{\pi\eta}) \\
\nfl \Im \fermi_1 2 \varepsilon_{\eta\eta^\prime} + \left(- \Im \fermi_8 + \nfl \Im \fermi_{27}\right) (1-3 \varepsilon_{\pi\eta})
\end{pmatrix} \ , \\
\theta_{\eta^\prime K} &\equiv
\frac1{m_{K^0}^2 - m_{\eta^\prime}^2}
\begin{pmatrix}
\nfl \Re \fermi_1 + \left(- \Re \fermi_8 + \nfl \Re \fermi_{27}\right) (3\varepsilon_{\pi\eta^\prime}+\varepsilon_{\eta\eta^\prime}) \\
\nfl \Im \fermi_1 + \left(- \Im \fermi_8 + \nfl \Im \fermi_{27}\right) (3\varepsilon_{\pi\eta^\prime}+\varepsilon_{\eta\eta^\prime})
\end{pmatrix} \ .
\end{align}
\end{subequations}

\subsection{Mixing between mesons and scalar messengers} \label{sec:meson to hidden mixing}

In \cref{sec:models}, we compute a generic $K^+ \to \pi^+ s_i$ decay amplitude by treating the bilinear portal interactions perturbatively.
In some instances, it may be necessary to resum these bilinear interactions by diagonalising the portal Lagrangian.
Following this strategy, one obtains additional portal interactions generated by both \SM and internal hidden sector interactions,
the size of which is measured by meson to hidden particle mixing angles.

In general, the bilinear interactions that couple \cPT to hidden sectors are
\begin{align}
\mathcal L_\Phi^{\prime S_m} + \mathcal L_\Phi^{S_\theta} &
= - \fp \B \trf{\nonet \Phi \Im \nonet S_m^\prime} + \frac{\fp \mth^2}{\nfl} S_\theta \Phi
= - \frac12 \phi_0^T \epsilon s + \text{h.c.} \ ,
\end{align}
where $\phi_0^T = (\pi$, $K^0_L$, $K^0_S$, $\eta$, $\eta^\prime)$, $s^T= (s_1$, $s_2$, $\dots)$, $\epsilon = (\epsilon_1$, $\epsilon_2$, $\dots)$, and
\begin{equation}
\epsilon_i = \epsilonUV \fp \B
\begin{pmatrix}
\frac1{\sqrt2} c_{s_i\pi} \\
\frac1{\sqrt2} \left(\Im\nonet c^{\prime S_m}_i{}_\strange^\down + \Im\nonet c^{\prime S_m}_i{}_\down^\strange\right) \\
\frac1{\sqrt2\i} \left(\Im\nonet c^{\prime S_m}_i{}_\strange^\down - \Im\nonet c^{\prime S_m}_i{}_\down^\strange\right) \\
\frac1{\sqrt3} \left(c_{s_i\eta} + s_\eta \frac{\mth^2}{v\B} c_i^{S_\theta}\right) \\
\frac1{\sqrt3} \left(c_{s_i\eta^\prime} - c_\eta \frac{\mth^2}{v\B} c_i^{S_\theta}\right)
\end{pmatrix}
\ .
\end{equation}
The coefficients $c_{s_i X}$ and $c^{\prime S_m}_i$  are given in \cref{eq:K decay hidden currents,eq:wilson_coeff}.
For canonical quadratic hidden Lagrangians
\begin{align}
\mathcal L^\text{hidden} &\supset - \frac12 s^T (\partial^2 + m) s \ , &
m &= \diag(m_1^2, m_2^2, \dots)
\end{align}
the mass-mixing matrix is
\begin{align}
\mathcal L &\supset - \frac12
\begin{pmatrix}
\phi_0^T & s^T
\end{pmatrix}
\begin{pmatrix}
M & \epsilon \\ \epsilon^T & m
\end{pmatrix}
\begin{pmatrix}
\phi_0 \\ s
\end{pmatrix} \ , &
M &= \diag \left(m_\pi^2, m_K^2, m_K^2, m_\eta^2, m_{\eta^\prime}^2\right) \ .
\end{align}
This matrix can be diagonalised using a unitary field redefinition
\begin{align}
\begin{pmatrix}
\phi_0 \\ s
\end{pmatrix} &
\to
\begin{pmatrix}
\flavour 1 & \theta \\ - \theta^T & \flavour 1
\end{pmatrix}
\begin{pmatrix}
\phi_0 \\ s
\end{pmatrix}
+ \order{\theta^2} \ , &
\theta &= (\theta_1, \theta_2, \dots) \ ,
\end{align}
where $\theta$ is a solution of the matrix-valued equation
\begin{align}
\epsilon &= \theta m - M \theta \ .
\end{align}
Assuming that all of the $s_i$ share the same mass $m_s = m_i$, one obtains
\begin{align}
\theta_i &= \left(m_s^2 \flavour 1 - M\right)^{-1} \epsilon_i \ .
\end{align}

\subsection{Trilinear Standard Model vertices used in the $K^\pm \to \pi^\pm s_i$ decay} \label{sec:trilinear vertices}

The hidden currents $\Im S_\m$ and $S_\theta$ contribute to the generic $K^\pm \to \pi^\pm s_i$ amplitude via
Feynman diagrams that contain a \SM three-meson vertex with one charged kaon leg, one charged pion leg, and one neutral meson leg, with the neutral meson subsequently oscillating into a hidden scalar.
The \SM three-meson vertices are encoded by the kinetic \cref{eq:u kinetic Lagrangian}, the octet \cref{eq:u octet Lagrangian}, and the 27-plet \cref{eq:u 27-plet Lagrangian},
\begin{equation}
\mathcal L_{\Phi^3}
\equiv \mathcal L^{\partial^2}_{\Phi^3} + \mathcal L^{a\partial^2}_{\Phi^3} + \mathcal L^{\tensor a\partial^2}_{\Phi^3} \ ,
\end{equation}
where
\begin{subequations}
\begin{align}
\mathcal L^{\partial^2}_{\Phi^3} &
= \frac{\i}{2\fp} \trf{\partial_\mu \nonet\Phi \comm{\octet\Phi}{\partial^\mu \octet\Phi}} = 0 \ , \\
\mathcal L^{a\partial^2}_{\Phi^3} &
= - \frac{\i \epsilonEW}{4 \fp} \left(
\fermi_8 \trds{\acomm{\partial_\mu \octet\Phi}{\comm{\octet\Phi}{\partial^\mu \octet\Phi}}}
+ \fermi_1 \trds{\comm{\octet\Phi}{\partial_\mu \octet\Phi}} \partial^\mu \Phi
\right) + \text{h.c.} \ , \\
\mathcal L^{\tensor a\partial^2}_{\Phi^3} &
= - \frac{\i\epsilonEW}{4\fp} \fermi_{27}
\begin{multlined}[t]
\left(
3 \partial_\mu \octet\Phi_\down^\strange \comm{\octet \Phi}{\partial^\mu \octet\Phi}_\up^\up + 2 \partial_\mu \octet\Phi_\down^\up \comm{\octet\Phi}{\partial^\mu \octet\Phi}_\up^\strange
\right.\\\left.
+ 3 \comm{\octet\Phi}{\partial_\mu \octet\Phi}_\down^\strange \partial^\mu \octet\Phi_\up^\up + 2 \comm{\octet\Phi}{\partial_\mu \octet\Phi}_\down^\up \partial^\mu \octet\Phi_\up^\strange
\right)
+ \text{h.c.}
\end{multlined}
\end{align}
\end{subequations}
Evaluating the flavour traces, the relevant terms with one $K^+$, one $\pi^+$, and one neutral meson are
\begin{align}
\mathcal L_{\Phi^3}
\supset \mathcal L_{K\pi\Phi} &
\equiv \mathcal L^{a\partial^2}_{K\pi\Phi} + \mathcal L^{\tensor a\partial^2}_{K\pi\Phi} \ ,
\end{align}
where
\begin{subequations}
\begin{align}
\mathcal L^{a\partial^2}_{K\pi\Phi} &
= - \frac{\i \epsilonEW}{4\fp}
\begin{multlined}[t]
\left[
\fermi_8 \left(3 \partial_{\pi^-} \partial_{K^+} - \partial_{\eta_8} \partial_{K^+} - 2 \partial_{\eta_8} \partial_{\pi^-}\right) \etaoctet K^+ \pi^-
\right.\\\left.
+ \fermi_8 \left(\partial_{K^+} \partial_{\pi_8} - \partial_{K^+} \partial_{\pi^-}\right) \pion \pi^- K^+
\right.\\\left.
+ 3 \fermi_1 \left(\partial_{\eta_1} \partial_{K^+} - \partial_{\eta_1} \partial_{\pi^-}\right) \etasinglet \pi^- K^+
\right] + \text{h.c.} \ ,
\end{multlined}
\\
\mathcal L^{\tensor a\partial^2}_{K\pi\Phi} &
= - \frac{\i\epsilonEW}{4\fp} \fermi_{27}
\begin{multlined}[t]
\left[
\left(7 \partial_{\pi_8} \partial_{K^+} - 5\partial_{\pi_8} \partial_{\pi^-} - 2 \partial_{\pi^-} \partial_{K^+}\right) \pion \pi^- K^+
\right.\\\left.
- 3 \left(3 \partial_{\eta_8} \partial_{\pi^-} - 2 \partial_{\pi^-} \partial_{K^+} - \partial_{\eta_8} \partial_{K^+}\right) \etaoctet \pi^- K^+
\right] + \text{h.c.} \ .
\end{multlined}
\end{align}
\end{subequations}
Diagonalising the Lagrangian, one obtains the final interactions
\begin{align}
\mathcal L_{K\pi\Phi} &
= \mathcal L_{K\pi\pi} + \mathcal L_{K\pi\eta} + \mathcal L_{K\pi\eta^\prime} \ ,
\end{align}
where
\begin{subequations}
\begin{align}
\mathcal L_{K\pi\pi} &
= - \frac{\i\epsilonEW}{4\fp}
\begin{multlined}[t]
\left[ (\fermi_8 + 7\fermi_{27}) \partial_{\pi^0} \partial_{K^+} - 5\fermi_{27} \partial_{\pi^0} \partial_{\pi^-}
\right.\\\left.
- (\fermi_8 + 2\fermi_{27}) \partial_{K^+} \partial_{\pi^-}\right] K^+ \pi^- \frac{\pi^0}{\sqrt2} \ ,
\end{multlined} \\
\mathcal L_{K\pi\eta} &
= - \frac{\i\epsilonEW}{4\fp} c_\eta
\begin{multlined}[t]
\left[ (3\fermi_8 + 6\fermi_{27}) \partial_{K^+} \partial_{\pi^-} - (\fermi_8 + 3\sqrt2 t_\eta \fermi_1 - 3\fermi_{27}) \partial_{\eta} \partial_{K^+}
\right.\\\left.
- (2\fermi_8 - 3\sqrt2 t_\eta \fermi_1 + 9\fermi_{27}) \partial_{\eta} \partial_{\pi^-}\right] K^+ \pi^- \frac{\eta}{\sqrt6} \ ,
\end{multlined} \\
\mathcal L_{K\pi\eta^\prime} &
= - \frac{\i\epsilonEW}{4\fp} s_\eta
\begin{multlined}[t]
\left[ (3\fermi_8 + 6\fermi_{27}) \partial_{K^+} \partial_{\pi^-} - (\fermi_8 - 3\sqrt2 t_\eta^{-1} \fermi_1 - 3\fermi_{27}) \partial_{\eta^\prime} \partial_{K^+}
\right.\\\left.
- (2\fermi_8 + 3\sqrt2 t_\eta^{-1} \fermi_1 + 9\fermi_{27}) \partial_{\eta^\prime} \partial_{\pi^-}\right] K^+ \pi^- \frac{\eta^\prime}{\sqrt6} \ ,
\end{multlined}
\end{align}
\end{subequations}

\subbib